\pdfoutput=1
\documentclass[%
 reprint,
superscriptaddress,
showpacs,preprintnumbers,
 amsmath,amssymb,
 aps,
floatfix,
]{revtex4-1}

\usepackage{graphicx}
\usepackage{dcolumn}
\usepackage{bm}
\usepackage[monochrome]{xcolor}
\usepackage{subfig}
\usepackage{natbib}

\newcommand{\mtxtlbl}[1]{\text{{\scriptsize #1}}}

\begin{document}


\title{The T2K Neutrino Flux Prediction}


\newcommand{\INSTC}{\affiliation{University of Alberta, Centre for Particle Physics, Department of Physics, Edmonton, Alberta, Canada}}
\newcommand{\INSTEE}{\affiliation{University of Bern, Albert Einstein Center for Fundamental Physics, Laboratory for High Energy Physics (LHEP), Bern, Switzerland}}
\newcommand{\INSTFE}{\affiliation{Boston University, Department of Physics, Boston, Massachusetts, U.S.A.}}
\newcommand{\INSTD}{\affiliation{University of British Columbia, Department of Physics and Astronomy, Vancouver, British Columbia, Canada}}
\newcommand{\INSTGA}{\affiliation{University of California, Irvine, Department of Physics and Astronomy, Irvine, California, U.S.A.}}
\newcommand{\INSTI}{\affiliation{IRFU, CEA Saclay, Gif-sur-Yvette, France}}
\newcommand{\INSTCI}{\affiliation{Chonnam National University, Institute for Universe \& Elementary Particles, Gwangju, Korea}}
\newcommand{\INSTGB}{\affiliation{University of Colorado at Boulder, Department of Physics, Boulder, Colorado, U.S.A.}}
\newcommand{\INSTFG}{\affiliation{Colorado State University, Department of Physics, Fort Collins, Colorado, U.S.A.}}
\newcommand{\INSTCJ}{\affiliation{Dongshin University, Department of Physics, Naju, Korea}}
\newcommand{\INSTFH}{\affiliation{Duke University, Department of Physics, Durham, North Carolina, U.S.A.}}
\newcommand{\INSTBA}{\affiliation{Ecole Polytechnique, IN2P3-CNRS, Laboratoire Leprince-Ringuet, Palaiseau, France }}
\newcommand{\INSTEF}{\affiliation{ETH Zurich, Institute for Particle Physics, Zurich, Switzerland}}
\newcommand{\INSTEG}{\affiliation{University of Geneva, Section de Physique, DPNC, Geneva, Switzerland}}
\newcommand{\INSTDG}{\affiliation{H. Niewodniczanski Institute of Nuclear Physics PAN, Cracow, Poland}}
\newcommand{\INSTCB}{\affiliation{High Energy Accelerator Research Organization (KEK), Tsukuba, Ibaraki, Japan}}
\newcommand{\INSTED}{\affiliation{Institut de Fisica d'Altes Energies (IFAE), Bellaterra (Barcelona), Spain}}
\newcommand{\INSTEC}{\affiliation{IFIC (CSIC \& University of Valencia), Valencia, Spain}}
\newcommand{\INSTEI}{\affiliation{Imperial College London, Department of Physics, London, United Kingdom}}
\newcommand{\INSTGF}{\affiliation{INFN Sezione di Bari and Universit\`a e Politecnico di Bari, Dipartimento Interuniversitario di Fisica, Bari, Italy}}
\newcommand{\INSTBE}{\affiliation{INFN Sezione di Napoli and Universit\`a di Napoli, Dipartimento di Fisica, Napoli, Italy}}
\newcommand{\INSTBF}{\affiliation{INFN Sezione di Padova and Universit\`a di Padova, Dipartimento di Fisica, Padova, Italy}}
\newcommand{\INSTBD}{\affiliation{INFN Sezione di Roma and Universit\`a di Roma "La Sapienza", Roma, Italy}}
\newcommand{\INSTEB}{\affiliation{Institute for Nuclear Research of the Russian Academy of Sciences, Moscow, Russia}}
\newcommand{\INSTCC}{\affiliation{Kobe University, Kobe, Japan}}
\newcommand{\INSTCD}{\affiliation{Kyoto University, Department of Physics, Kyoto, Japan}}
\newcommand{\INSTEJ}{\affiliation{Lancaster University, Physics Department, Lancaster, United Kingdom}}
\newcommand{\INSTFC}{\affiliation{University of Liverpool, Department of Physics, Liverpool, United Kingdom}}
\newcommand{\INSTFI}{\affiliation{Louisiana State University, Department of Physics and Astronomy, Baton Rouge, Louisiana, U.S.A.}}
\newcommand{\INSTJ}{\affiliation{Universit\'e de Lyon, Universit\'e Claude Bernard Lyon 1, IPN Lyon (IN2P3), Villeurbanne, France}}
\newcommand{\INSTCE}{\affiliation{Miyagi University of Education, Department of Physics, Sendai, Japan}}
\newcommand{\INSTDF}{\affiliation{National Centre for Nuclear Research, Warsaw, Poland}}
\newcommand{\INSTFJ}{\affiliation{State University of New York at Stony Brook, Department of Physics and Astronomy, Stony Brook, New York, U.S.A.}}
\newcommand{\INSTCF}{\affiliation{Osaka City University, Department of Physics, Osaka,  Japan}}
\newcommand{\INSTGG}{\affiliation{Oxford University, Department of Physics, Oxford, United Kingdom}}
\newcommand{\INSTBB}{\affiliation{UPMC, Universit\'e Paris Diderot, CNRS/IN2P3, Laboratoire de Physique Nucl\'eaire et de Hautes Energies (LPNHE), Paris, France}}
\newcommand{\INSTGC}{\affiliation{University of Pittsburgh, Department of Physics and Astronomy, Pittsburgh, Pennsylvania, U.S.A.}}
\newcommand{\INSTFA}{\affiliation{Queen Mary University of London, School of Physics, London, United Kingdom}}
\newcommand{\INSTE}{\affiliation{University of Regina, Department of Physics, Regina, Saskatchewan, Canada}}
\newcommand{\INSTGD}{\affiliation{University of Rochester, Department of Physics and Astronomy, Rochester, New York, U.S.A.}}
\newcommand{\INSTBC}{\affiliation{RWTH Aachen University, III. Physikalisches Institut, Aachen, Germany}}
\newcommand{\INSTDD}{\affiliation{Seoul National University, Department of Physics and Astronomy, Seoul, Korea}}
\newcommand{\INSTFB}{\affiliation{University of Sheffield, Department of Physics and Astronomy, Sheffield, United Kingdom}}
\newcommand{\INSTDI}{\affiliation{University of Silesia, Institute of Physics, Katowice, Poland}}
\newcommand{\INSTEH}{\affiliation{STFC, Rutherford Appleton Laboratory, Harwell Oxford,  and  Daresbury Laboratory, Warrington, United Kingdom}}
\newcommand{\INSTCH}{\affiliation{University of Tokyo, Department of Physics, Tokyo, Japan}}
\newcommand{\INSTBJ}{\affiliation{University of Tokyo, Institute for Cosmic Ray Research, Kamioka Observatory, Kamioka, Japan}}
\newcommand{\INSTCG}{\affiliation{University of Tokyo, Institute for Cosmic Ray Research, Research Center for Cosmic Neutrinos, Kashiwa, Japan}}
\newcommand{\INSTF}{\affiliation{University of Toronto, Department of Physics, Toronto, Ontario, Canada}}
\newcommand{\INSTB}{\affiliation{TRIUMF, Vancouver, British Columbia, Canada}}
\newcommand{\INSTG}{\affiliation{University of Victoria, Department of Physics and Astronomy, Victoria, British Columbia, Canada}}
\newcommand{\INSTDJ}{\affiliation{University of Warsaw, Faculty of Physics, Warsaw, Poland}}
\newcommand{\INSTDH}{\affiliation{Warsaw University of Technology, Institute of Radioelectronics, Warsaw, Poland}}
\newcommand{\INSTFD}{\affiliation{University of Warwick, Department of Physics, Coventry, United Kingdom}}
\newcommand{\INSTGE}{\affiliation{University of Washington, Department of Physics, Seattle, Washington, U.S.A.}}
\newcommand{\INSTGH}{\affiliation{University of Winnipeg, Department of Physics, Winnipeg, Manitoba, Canada}}
\newcommand{\INSTEA}{\affiliation{Wroclaw University, Faculty of Physics and Astronomy, Wroclaw, Poland}}
\newcommand{\INSTH}{\affiliation{York University, Department of Physics and Astronomy, Toronto, Ontario, Canada}}

\INSTC
\INSTEE
\INSTFE
\INSTD
\INSTGA
\INSTI
\INSTCI
\INSTGB
\INSTFG
\INSTCJ
\INSTFH
\INSTBA
\INSTEF
\INSTEG
\INSTDG
\INSTCB
\INSTED
\INSTEC
\INSTEI
\INSTGF
\INSTBE
\INSTBF
\INSTBD
\INSTEB
\INSTCC
\INSTCD
\INSTEJ
\INSTFC
\INSTFI
\INSTJ
\INSTCE
\INSTDF
\INSTFJ
\INSTCF
\INSTGG
\INSTBB
\INSTGC
\INSTFA
\INSTE
\INSTGD
\INSTBC
\INSTDD
\INSTFB
\INSTDI
\INSTEH
\INSTCH
\INSTBJ
\INSTCG
\INSTF
\INSTB
\INSTG
\INSTDJ
\INSTDH
\INSTFD
\INSTGE
\INSTGH
\INSTEA
\INSTH

\author{K.\,Abe}\INSTBJ
\author{N.\,Abgrall}\INSTEG
\author{H.\,Aihara}\INSTCH
\author{T.\,Akiri}\INSTFH
\author{J.B.\,Albert}\INSTFH
\author{C.\,Andreopoulos}\INSTEH
\author{S.\,Aoki}\INSTCC
\author{A.\,Ariga}\INSTEE
\author{T.\,Ariga}\INSTEE
\author{S.\,Assylbekov}\INSTFG
% ...


\collaboration{The T2K Collaboration}\noaffiliation

\date{\today}

\begin{abstract}
The Tokai-to-Kamioka (T2K) experiment studies neutrino oscillations using
an off-axis muon neutrino beam with a peak energy of about 0.6~GeV that originates at the 
J-PARC accelerator facility.  Interactions of the neutrinos are observed at near detectors placed at 280~m from the production target and at the far detector -- Super-Kamiokande (SK) -- located 295~km away.  
The flux prediction is an essential part of the successful
prediction of neutrino interaction rates at the T2K detectors 
and is an important input to T2K neutrino oscillation and cross section
measurements.
A FLUKA and GEANT3 based simulation models the physical processes involved in the neutrino production,
from the interaction of primary beam protons in the T2K target, to the decay of hadrons and muons that
produce neutrinos.
The simulation uses proton beam monitor measurements
as inputs. The modeling of hadronic interactions is re-weighted using thin target hadron production data,
including recent charged pion and kaon measurements from the NA61/SHINE experiment.  
For the first T2K analyses the uncertainties on the
flux prediction are evaluated to be below 15\% near the flux peak.  
The uncertainty on the ratio of the flux predictions at the far and 
near detectors is less than 2\% near the flux peak. 
\end{abstract}

\maketitle


\section{\label{sec:intro}Introduction}

Predicting the neutrino flux and energy spectrum is an important component of analyses in accelerator 
neutrino experiments~\cite{NOMAD_flux,Ahn:2006zza,PhysRevD.77.072002,MiniBooNE_flux}. However, 
it is difficult to simulate the flux precisely due to uncertainties in the underlying physical processes, 
particularly hadron production in proton-nucleus interactions. To reduce flux-related uncertainties, neutrino 
oscillation experiments are sometimes conducted by comparing measurements between a near detector site and a 
far detector site, allowing for cancellation of correlated uncertainties. Therefore, it is important to correctly 
predict the relationship between the fluxes at the two detector sites, described below as the far-to-near ratio.


T2K (Tokai-to-Kamioka)~\cite{t2kloi}\cite{t2knim} is 
a long-baseline neutrino oscillation experiment
that uses an intense muon neutrino beam
to measure the mixing angle $\theta_{13}$ via 
the $\nu_e$ appearance~\cite{t2knue} and the mixing angle $\theta_{23}$ and 
mass difference $\Delta m^2_{32}$ via the $\nu_\mu$ disappearance~\cite{t2knumu}.
The muon neutrino beam is produced as the decay products of
pions and kaons generated by the interaction of the 30~GeV proton
beam from Japan Proton Accelerator Research Complex (J-PARC)
with a graphite target.
The properties of the generated neutrinos are measured 
at near detectors placed 280~m from the target and at the  
far detector, 
Super-Kamiokande (SK)~\cite{skdet}, which is located 295~km away.  
The effect of oscillation is expected to be negligible at the near detectors 
and significant at SK.

\begin{figure}[h]
  \begin{center}
    \includegraphics[keepaspectratio=true,width=0.44\textwidth]{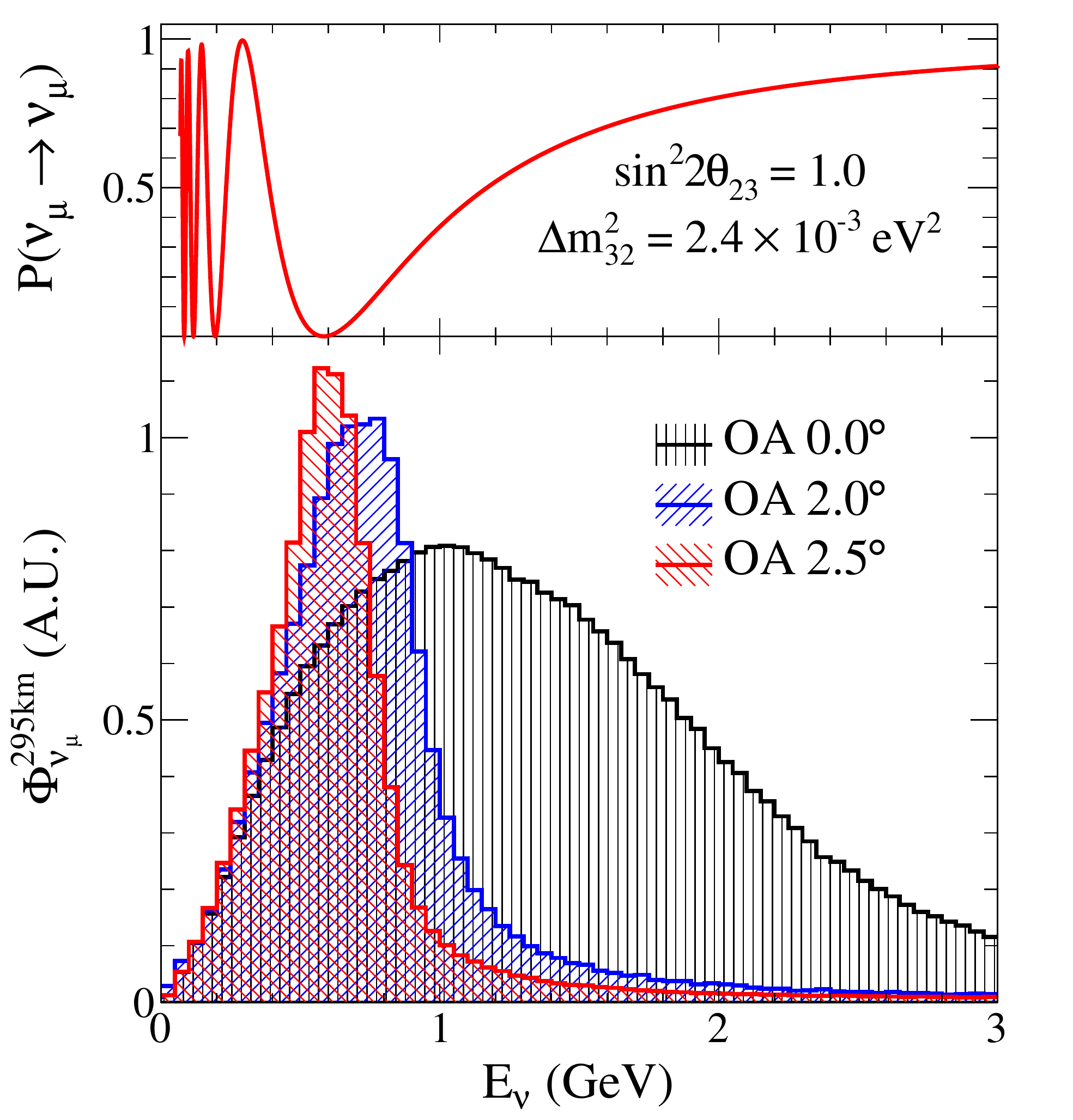}
    \caption{Muon neutrino survival probability at 295 km and neutrino fluxes for different off-axis angles.}
    \label{fig:oaeffect}
  \end{center}
\end{figure}

The T2K experiment employs the off-axis method~\cite{off-axis} 
to generate a narrow-band neutrino beam and this is the first time 
this technique has been used in a search for neutrino oscillations. 
The method utilizes the fact that the energy of a neutrino emitted 
in the two-body pion (kaon) decay, the dominant mode 
for the neutrino production, at an angle relative to 
the parent meson direction is only weakly dependent on the momentum 
of the parent. 
\textcolor{red}{The parent $\pi^{+(-)}$'s are focused parallel to
the proton beam axis to produce the (anti-)neutrino beam.}
By positioning a detector at an angle relative to 
\textcolor{red}{the focusing axis, }
one will, therefore, see neutrinos with a narrow spread in energy. 
The peak energy of the neutrino beam can be varied by changing 
the off-axis angle as illustrated in the lower panel 
of Fig.~\ref{fig:oaeffect}. 
In the case of T2K, the off-axis angle is set at $2.5^{\circ}$ 
so that the neutrino beam at SK has a peak energy at about 0.6~GeV, 
near the expected first oscillation maximum (Fig.~\ref{fig:oaeffect}). 
This maximizes the effect of the neutrino oscillations at 295~km 
as well as reduces background events. 
Since the energy spectrum changes depending on the off-axis angle, 
the neutrino beam direction has to be precisely monitored.

To determine the oscillation parameters,
the expected observables at the far detector are predicted based
on the flux prediction and the neutrino-nucleus interaction model.
To reduce the uncertainty of the prediction, they are
modified based on the near detector measurements. 
For example, the absolute normalization uncertainty is efficiently canceled
by normalizing with the event rate at the near detector.
Then, it is important to reduce the uncertainty on the relation
between the flux at the near detector and that at the far detector.

The physics goals of T2K are to be sensitive to the values
of $\sin^2{2\theta_{13}}$ down to 0.006 and to 
measure the neutrino oscillation parameters with precision of 
$\delta$($\Delta$$m^{2}_{32}$) $\sim$ $10^{-4}$eV$^{2}$ 
and $\delta$(sin$^{2}2\theta_{23})$ $\sim$ 0.01.
To achieve these, the near-to-far extrapolation of the flux, i.e.,
the far-to-near flux ratio as a function of energy has to be known to better than 3\%. 
In addition to this requirement, it is also desirable to reduce 
the absolute flux uncertainty to study the neutrino-nucleus interactions
at the near detector.

For this purpose, the fluxes are calculated and the uncertainties are estimated based on
hadron production measurements including those by the NA61/SHINE experiment~\cite{Abgrall:2011ae}\cite{PhysRevC.85.035210}
and in situ measurements of the primary proton beam properties and 
the neutrino beam direction. 

In this paper, we describe a Monte Carlo based neutrino flux prediction 
as a function of neutrino energy at near and far detectors in T2K 
and the methods to estimate the flux prediction uncertainties.
\textcolor{red}{The neutrino flux treated here is the flux 
for the `neutrino' running mode, in which positive pions are focused.}
Section~\ref{sec:neutrinobeamline} describes the neutrino beamline, 
while Sec.~\ref{sec:beamoperation} summarizes the beam operation history.
Section~\ref{sec:fluxsim} describes a method of neutrino flux prediction 
based on a data-driven simulation. 
Section~\ref{sec:fluxerrors} explains uncertainties on the flux prediction. 
A comparison between the measured and predicted flux is discussed 
in Sec.~\ref{sec:fluxvsdata}.

\section{\label{sec:neutrinobeamline}T2K Neutrino Beamline}

The J-PARC Main Ring (MR) accelerates a 30 GeV proton beam every 2 to 3 seconds.
For each acceleration cycle, the beam is fast-extracted 
to the T2K neutrino beamline as a `spill'. One spill contains eight bunches 
in about 5~$\mu$s.

\begin{figure}[t]
  \begin{center}
    \includegraphics[keepaspectratio=true,width=80mm]{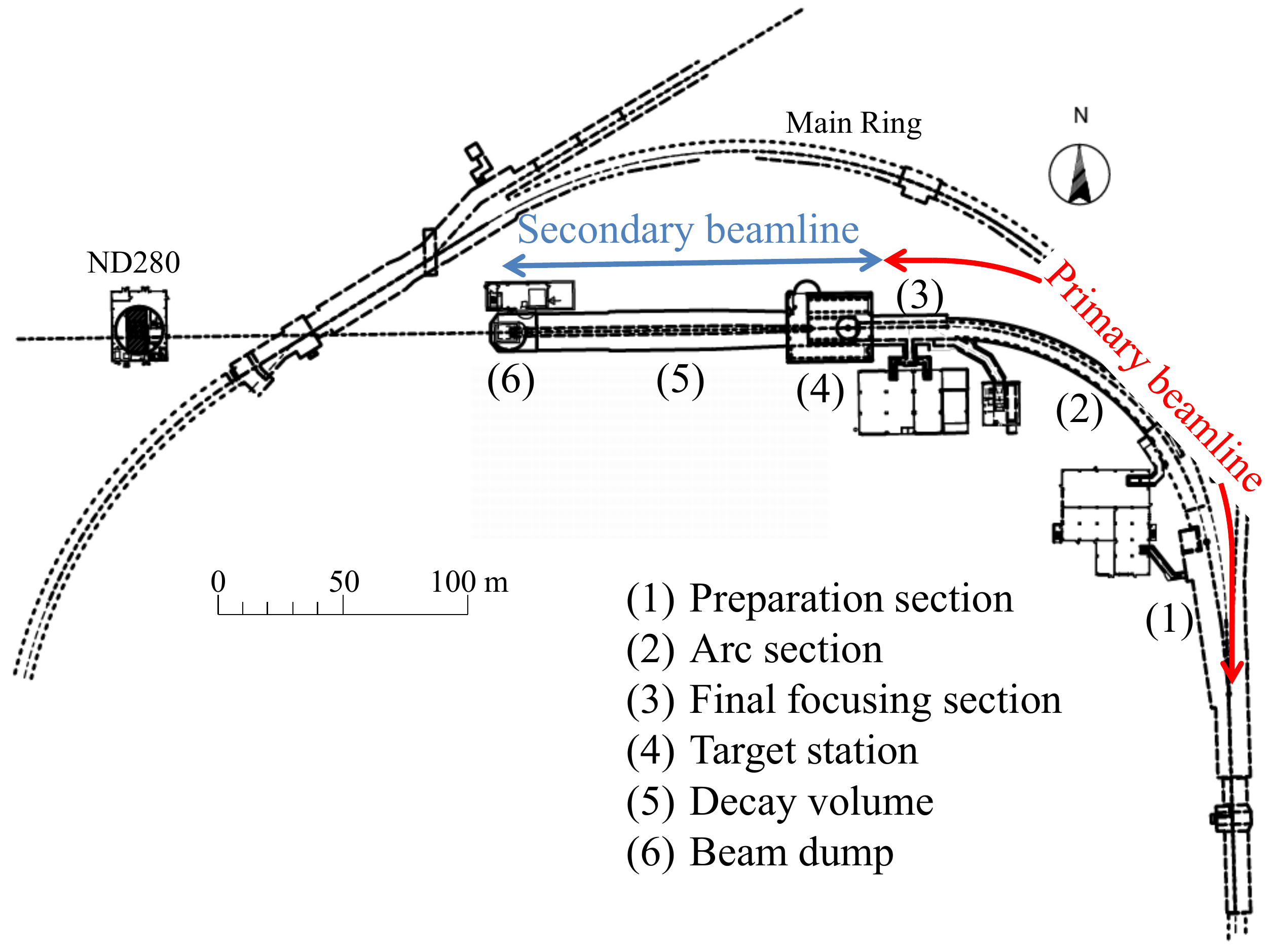}
    \caption[]
    {\label{fig:nu beamline}
    An overview of the T2K neutrino beamline.}
  \end{center}
\end{figure}

The neutrino beamline is composed of two sections: the primary and secondary beamlines.  
In the primary beamline, the extracted proton beam is transported to point 
in the direction of the secondary beamline, and 
focused to have the desired profile at the target.
In the secondary beamline, the proton beam impinges on a target to produce secondary pions and other hadrons, which are focused by magnetic horns and decay into neutrinos.  An overview of the neutrino beamline is shown in Fig.~\ref{fig:nu beamline}. More details of the beamline are described in  \cite{t2knim}.


\subsection{\label{sec:primbeamline}Primary beamline}

The primary beamline consists of the preparation section (54~m long),
arc section (147~m) and final focusing section (37~m). 
In the final focusing (FF) section, ten normal conducting magnets
(four steering, two dipole and four quadrupole magnets) guide and
focus the beam onto the target, while directing the beam downward by
3.64~degrees with respect to the horizontal.  

\begin{figure*}
    \centering
    \includegraphics[width=0.85\textwidth]{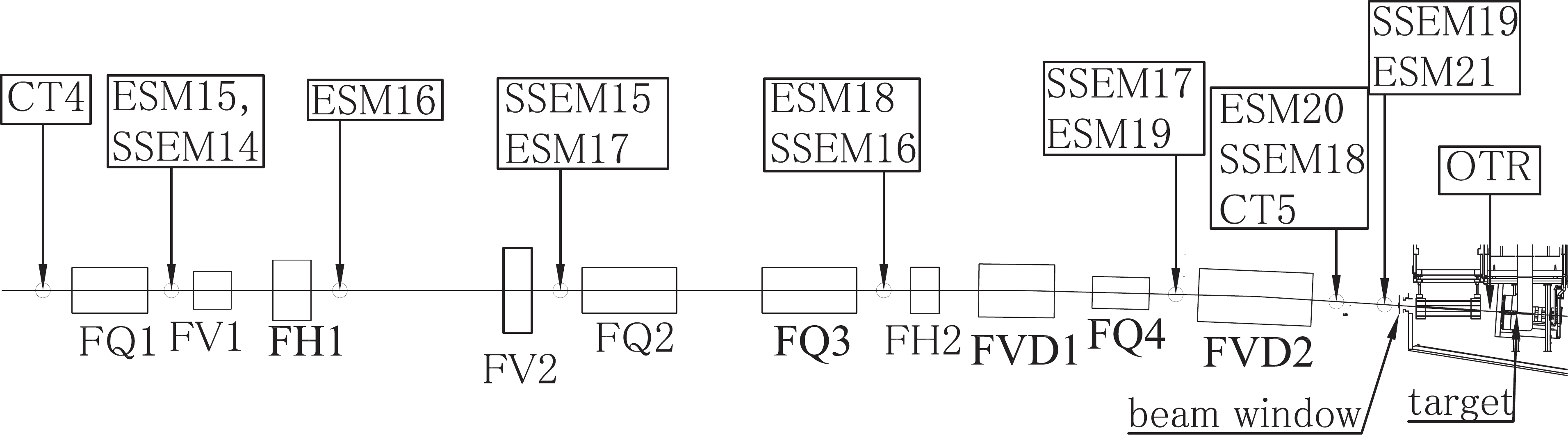}
    \caption[]
    {\label{fig:primary beam line}
    Location of the primary beamline monitors in the final focusing section.}
\end{figure*}

The intensity, position and profile
of the proton beam in the primary sections are precisely monitored by
five current transformers (CTs), 21~electrostatic monitors (ESMs),
19~segmented secondary emission monitors (SSEMs), respectively.  
The monitor locations in FF section are shown in Fig.~\ref{fig:primary beam line}.

%

\subsubsection{\label{sec:pmon}Proton Beam Monitor}

The beam intensity is measured with five CTs. 
Each CT is a 50-turn toroidal coil around a cylindrical ferromagnetic core.
The uncertainty on the beam intensity is 2\%, which
originates from the calibration accuracy (1.7\%),
the effect of secondary electrons produced at the SSEM foils ($<$0.7\%),
the long term stability of the individual CT monitors relative to each
other and the CT monitor measurement from the main ring (0.5\%). 
For the flux prediction, the intensity measured by CT5, 
located most downstream, is used.

The ESMs have four segmented cylindrical electrodes surrounding the
proton beam orbit. 
By measuring the top-bottom and left-right asymmetry of the beam-induced
current on the electrodes, they monitor the proton beam center position
nondestructively (without directly interacting with the beam).
The measurement precision of the \textcolor{red}{projected} beam position 
is better than 450~$\mu$m.

The SSEMs have two 5~$\mu$m thick sets of 
titanium foil strips oriented horizontally and vertically in the plane perpendicular
to the beam axis, and a high voltage anode foil between them. 
They measure the horizontal and vertical projections of the proton beam profile.
The systematic uncertainty of the beam width measurement is 200~$\mu$m.
The uncertainty of the beam center position measurement is dominated
by the monitor alignment uncertainty discussed in Section~\ref{sec:alignment}.
Since each SSEM causes a beam loss (0.005\% loss), they are 
inserted into the beam orbit only during the beam tuning, and
removed from the beam orbit during the continuous beam operation
except for the most downstream SSEM.  

An optical transition radiation (OTR) monitor positioned 30 cm upstream 
of the target measures the two dimensional profiles of the beam 
by imaging transition radiation produced
when the beam crosses a 50~$\mu$m thick titanium alloy foil.  The details of the monitor
have been described elsewhere~\cite{Bhadra:2012st}.

Using the ESMs, SSEMs and OTR measurements, the beam position at the upstream side 
of the baffle (shown in Fig.~\ref{fig:secondarybeamline}) is reconstructed with accuracy better than 0.7~mm
as described in Sec.\ref{sec:pbeamprop}.

\subsection{\label{sec:secbeamline}Secondary beamline}

Pions and kaons are produced by the interaction of protons 
with a graphite target.
They decay in-flight inside a single volume 
of $\sim$1500~$\mathrm{m}^3$ filled with helium gas.
The helium vessel is connected with the primary beamline using a titanium-alloy 
beam window that separates the vacuum in primary beamline and helium 
gas volume in the secondary beamline. 

\begin{figure}[t]
  \begin{center}
    \includegraphics[keepaspectratio=true,width=80mm]{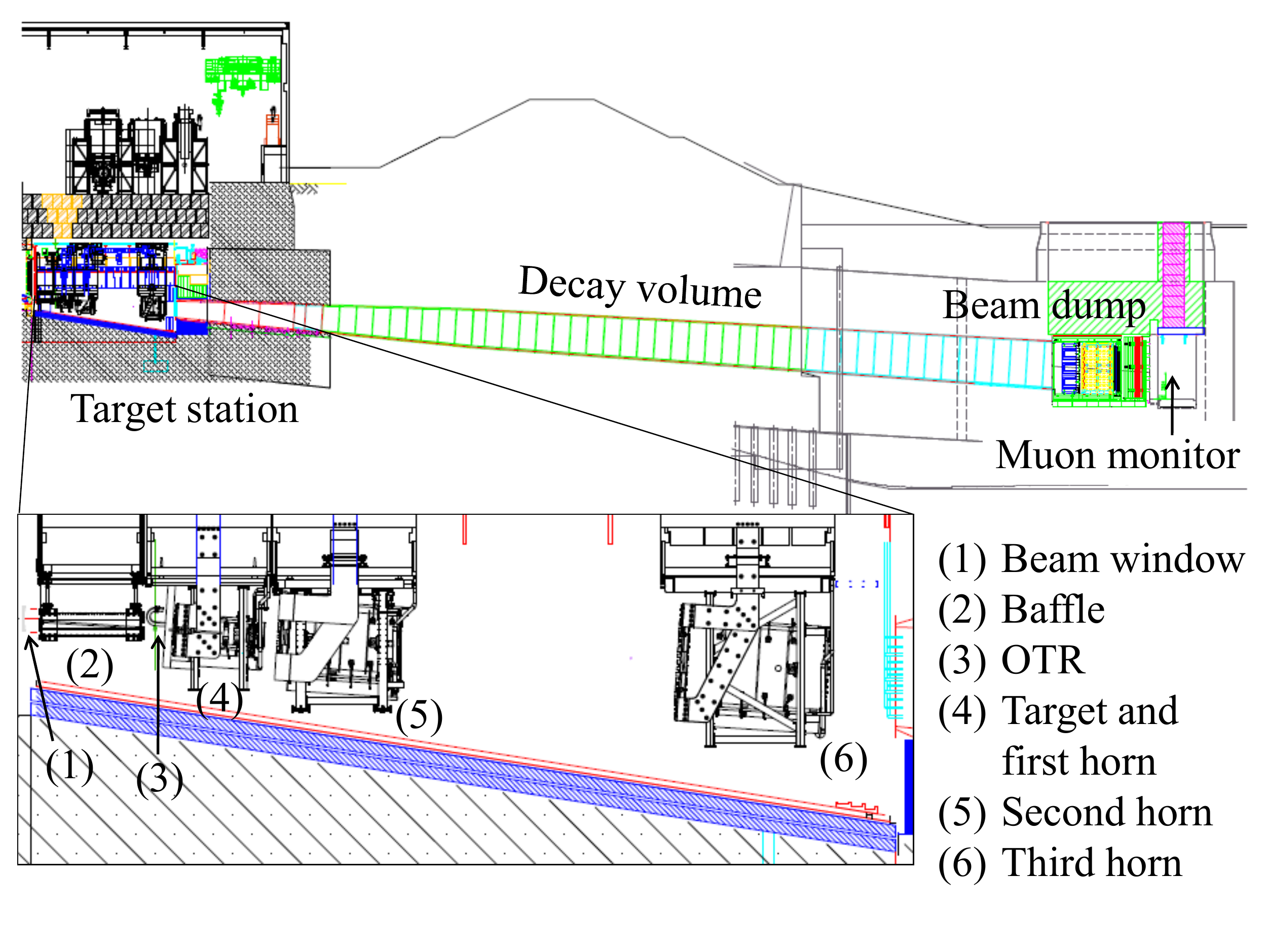}
    \caption[]
    {\label{fig:secondarybeamline}
    Side view of the secondary beamline.}
  \end{center}
\end{figure}

The secondary beamline consists of three sections: the target station,
decay volume and beam dump (Fig.~\ref{fig:secondarybeamline}). 
The helium vessel in the target station is 15 m long, 4 m wide and 11 m high.
The decay volume is a 96 m long steel tunnel. The cross section is 1.4 m 
wide and 1.7 m high at the upstream end, and 3.0 m wide and 5.0 m high at 
the downstream end. The beam dump sits at the end of the decay volume.
The distance between the center of the target and the upstream surface 
of the beam dump is 109 m. 

The target station contains a baffle, the OTR monitor, the target
and three magnetic horns.
The baffle is a collimator to protect the horns.
The 250 kA current pulses magnetize the three horns to focus 
the secondary \textcolor{red}{$\pi^+$'s in `neutrino' running mode.
The $\pi^-$'s are focused in `anti-neutrino' running mode, where
the polarity of the horn current is inverted.}
The produced pions then decay in the decay volume mainly into muons 
and muon neutrinos. 
All the remnants of the decayed pions and other hadrons are stopped 
by the beam dump. The neutrinos pass through the beam dump and are
used for physics experiments. The muons above 5~GeV that also pass
through the beam dump are detected by a muon monitor (MUMON) that monitors the
beam direction and intensity.

\subsubsection{\label{sec:tgthor}Target and Horns}

The target core is a 1.9~interaction length (91.4~cm long), 2.6~cm
diameter graphite rod with a density of 1.8 g/cm$^3$.
The core and a surrounding 2~mm thick graphite tube are sealed inside
a 0.3~mm thick titanium case.  
The target assembly is cantilevered inside the bore of the first horn inner
conductor.

\begin{figure}[t]
  \begin{center}
    \includegraphics[keepaspectratio=true,width=80mm]{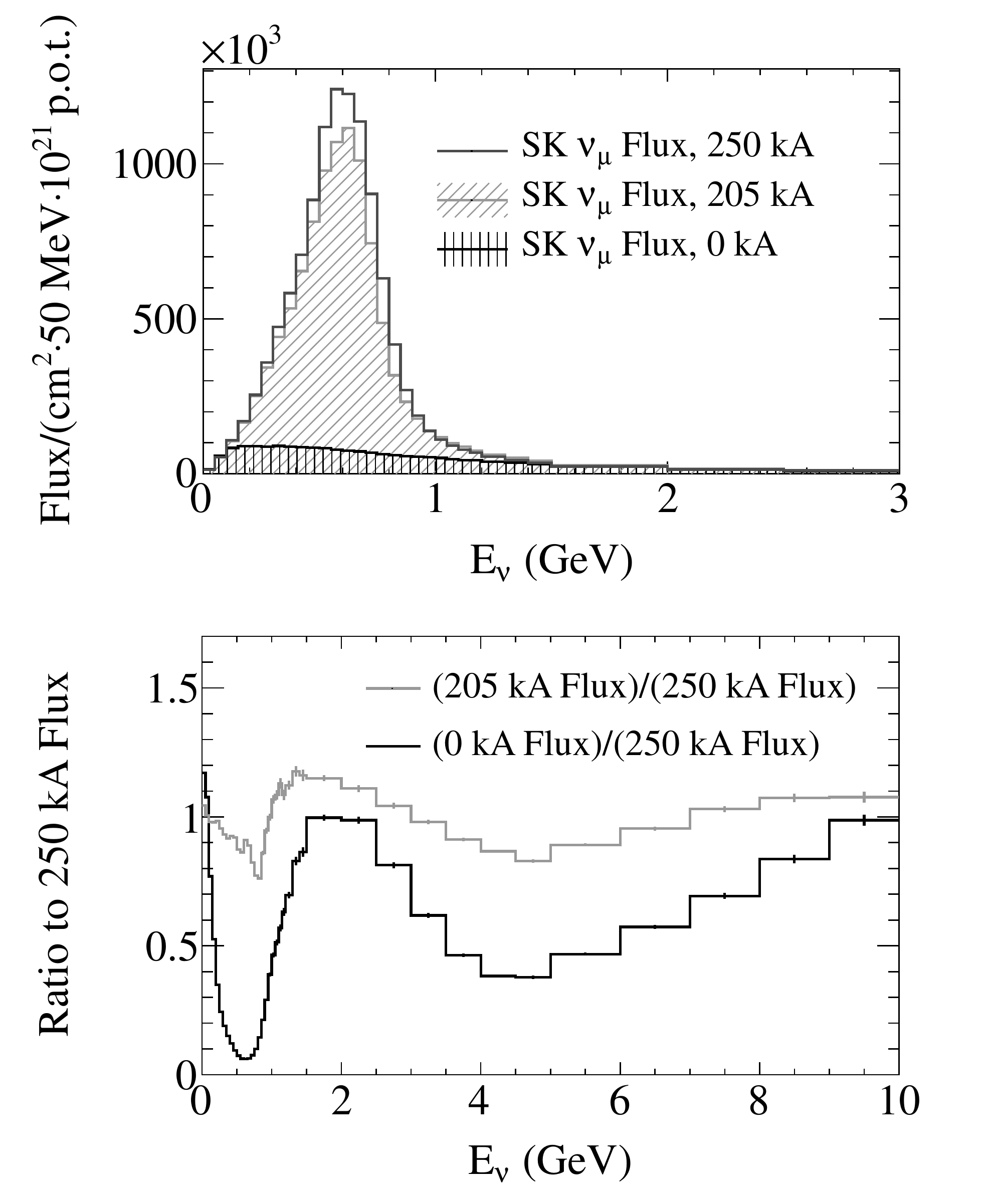}
    \caption[]
    {\label{fig:horn_effect}
    The predicted flux of $\nu_{\mu}$ at the SK far detector for operation at different horn currents. The flux histogram
(top) ranges from 0 to 3 GeV, while the ratios (bottom) range from 0 to 10 GeV.}
  \end{center}
\end{figure}

T2K uses three magnetic horns. Each horn consists of two coaxial
(inner and outer) conductors which encompass a closed
volume~\cite{vanderMeer:1961sk, Palmer:1965zz}.  A toroidal magnetic
field is generated in that volume.  The field varies as $1/r$, where
$r$ is the distance from the horn axis. 
The first horn (Horn~1) collects the
pions that are generated at the target installed in its inner
conductor. The second (Horn~2) and third (Horn~3) horns focus the pions. 
When the horns are operating with a current of 250~kA,
the maximum field is 1.7~T and the neutrino flux at 
SK is increased by a factor of $\sim$17 at the spectrum peak energy ($\sim$0.6~GeV), as 
illustrated in Fig.~\ref{fig:horn_effect}.

\begin{figure}[t]
  \begin{center}
    \includegraphics[keepaspectratio=true,width=80mm]{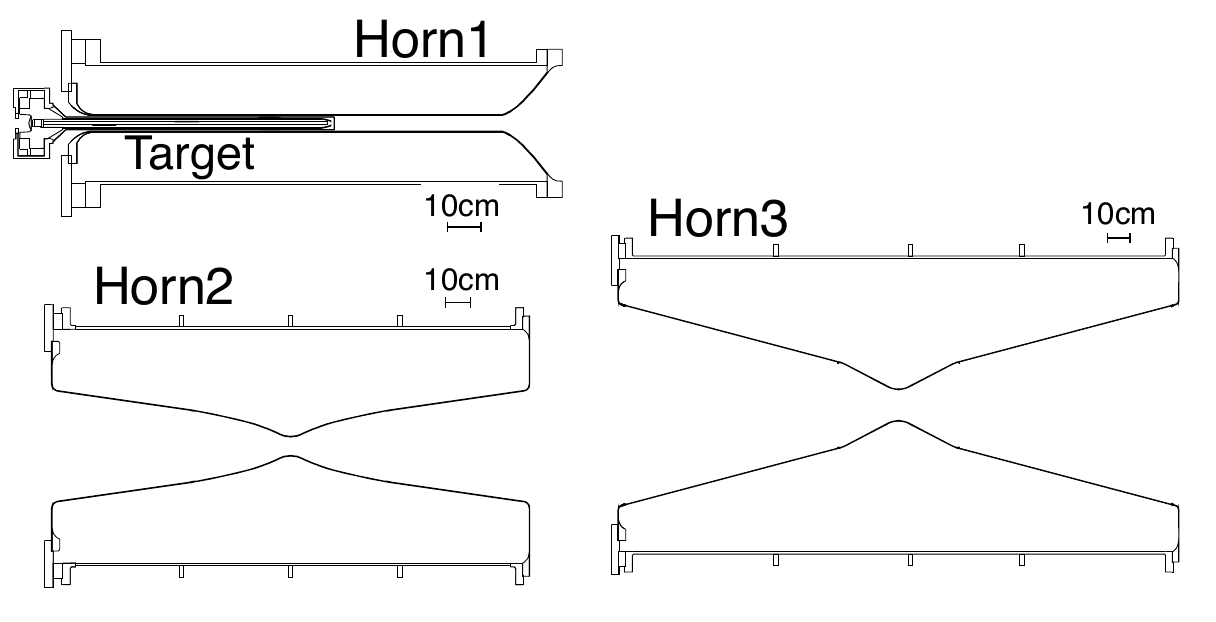}
    \caption[]
    {\label{fig:magnethorn}
    Cross section view of horns. 
}
  \end{center}
\end{figure}

A schematic view of the horns is shown in Fig.~\ref{fig:magnethorn}.
The horn conductor is made of an aluminum alloy.  
Their dimensions are summarized in Table~\ref{tbl:horngeom}.
The thickness of the inner conductors is 3~mm.
They are optimized to maximize the neutrino flux; the inside diameter
is as small as possible to achieve the maximum magnetic field, and the
conductor is as thin as possible to minimize pion absorption while
still being tolerant of the Lorentz force from the 250~kA current and
the thermal shock from the beam~\cite{HORN}.

\begin{table}
\caption{Dimensions of the T2K horns}
\begin{center}
\begin{tabular} {lccc}
\hline
\hline
& Horn1 & Horn2 & Horn3\\
\hline
inner conductor inside diameter (mm) & 54 & 80 & 140 \\
outer diameter (mm) & 400 & 1000 & 1400\\
length (m) & 1.5 & 2 & 2.5\\
\hline
\hline
\end{tabular}
\end{center}
\label{tbl:horngeom}
\end{table}

The electrical currents of the magnetic horns are monitored by Rogowski coils 
whose signal are digitized by 65~MHz FADCs.
Table~\ref{hcur_list} shows the summary of the horn current uncertainties.
The Rogowski coils were calibrated by the production company with $\pm1\%$
precision.
The shape of the ``loop" of the Rogowski coil may cause 
a 1$\%$ change of gain.

FADCs and related electronics are calibrated with better than 1$\%$ precision.


\begin{table}
\caption{Uncertainties on the absolute horn current measurement.
In the total error calculation, 
full width (FW) errors are scaled by $1/\sqrt{12}$ to
estimate 1$\sigma$ uncertainty.}
\begin{center}
\begin{tabular} {lc}
\hline \hline
& uncertainty \\
\hline
coil calibration             & $\pm1\%$ (FW) \\
coil setting                 & $\pm1\%$ (FW) \\
electronics calibration      & $<1\%$ \\
monitor stability            & $2\%$ (FW) \\
\hline
total                         & $1.3\% $\\
\hline \hline
\end{tabular}
\end{center}
\label{hcur_list}
\end{table}


Each horn has several instrumentation ports at various positions along 
the horn axis which permit measurements of the magnetic field
 between the inner and outer conductors.
Multiple magnetic field measurements have been made on the horns to validate the nominal $1/r$ field and to check for the presence of magnetic field asymmetries. 
The magnetic fields generated by Horns 2 and 3 were measured 
using an integrated 3-axis Hall probe inserted between the inner 
and outer conductors via the horns' instrumentation ports.  
The results are summarized in Table~\ref{devtable}. 
The measured field agrees with the expected nominal field  within $2\%$. 

\begin{figure}
\centering
\includegraphics[keepaspectratio=true,width=80mm]{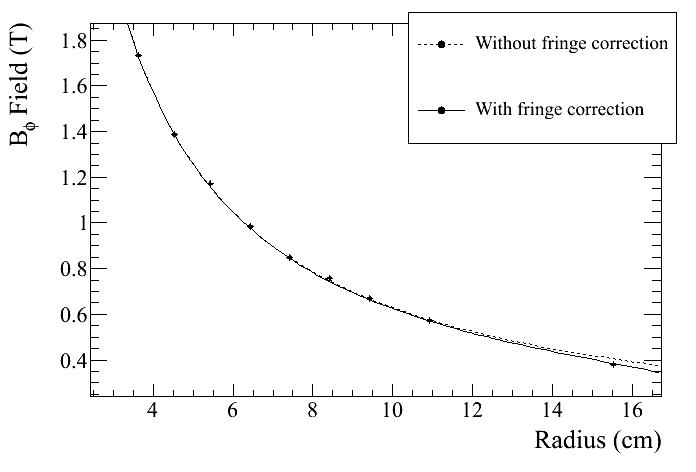}
\caption{Measurements of the magnetic field magnitude taken at the right upstream port of Horn 1. The curve shows the expected field strength, including a small correction to account for fringe effects near the instrumentation port at large radii.}
\label{BvsR}
\end{figure} 

\begin{figure}
\centering
\includegraphics[keepaspectratio=true,width=80mm]{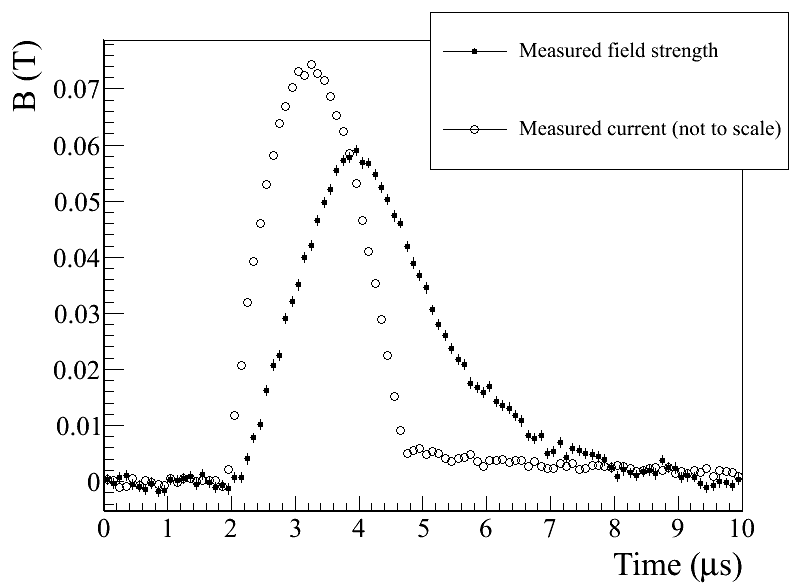}
\caption{A sample of data from the Hall probe showing the field strength in the $x$ direction in the beam coordinates.  This data was taken 100~cm along the axis of the horn. The Rogowski coil output, showing the current time dependence, is drawn with unfilled markers and is not to scale. The peaks are offset by approximately 0.7~ms.}
\label{sampl_pulses}
\end{figure}

Measurements of the magnetic field were also taken on a spare copy of 
the first horn, identical in design to the one currently in use 
in the T2K beamline.  
As with Horns 2 and 3, field measurements were taken via the instrumentation ports using a 3-axis Hall probe. A comparison of the expected field to the data taken at the right upstream port is shown in Fig.~\ref{BvsR}. 
The results agree well with the expected nominal field.  Additional measurements were taken along the horn's axis inside of the inner conductor.  The purpose of these measurements was to detect possible magnetic field asymmetries caused by path length differences between the upper and lower striplines supplying current to the horn.  While no field asymmetry due to path length differences was observed, an on-axis magnetic field with an anomalous time-dependence was detected. 
While the magnitude of the azimuthal fields is always proportional 
to the current, the anomalous on-axis field is characterized 
by a difference of 0.7~ms between maximum field and maximum current, 
as shown in Fig.~\ref{sampl_pulses}.  
This field has a direction perpendicular to the beam axis, 
and reaches a maximum magnitude of 0.065~T near the center of the horn.  
\textcolor{red}{The cause of this anomalous on-axis field is not yet known. 
Therefore, the effect of this field is estimated and added 
to the flux uncertainty.}
The magnitude of this field is 3.7\% of the magnitude of the field 
just outside the inner conductor, and conservative estimates 
from Monte Carlo simulations show the effect on neutrino flux is small, 
as discussed in ~\ref{sec:hcurrerrors}. 


\begin{table}
\begin{center}
\caption{\label{devtable}Magnetic field deviations from expected values at all instrumentation ports.  Blanks in the table are a result of each horn having a different configuration of instrumentation port locations. }
\begin{tabular}{llccccc}
\hline \hline
        &             & Top  & Bottom  & Left & Right  \\
\hline
Horn 1  &Upstream     &   --  &     --   &0.94\%& 0.5\%  \\
        &Downstream   &   --  &     --   & --    & 1.0\% \\
\hline
Horn 2  &Upstream   &0.7\%   & 0.1\%  &1.3\%  &  --     \\
        &Midstream  &0.7\%   &0.6\%   &   --   &0.7\%   \\
\hline
Horn 3  &Upstream  &1.2\%    &    --   &1.2\%  & 1.0\%  \\
        &Downstream&0.7\%    &    --   &0.2\%  & 0.5\%  \\
\hline \hline
\end{tabular}
\end{center}
\end{table}

\subsubsection{\label{sec:secmon}Secondary beam monitoring}
The neutrino beam intensity and direction can be monitored 
on a bunch-by-bunch basis by measuring the profile of muons which 
are produced along with neutrinos from the pion two-body decay.
The neutrino beam direction is measured as the direction from the target
to the center of the muon profile. 
The muon monitor is located just behind the beam dump 
at a distance of 118~m from the target, 
as shown in Fig.~\ref{fig:secondarybeamline}.
It consists of two kinds of detector arrays: ionization chambers and silicon PIN photodiodes.
Each array consists of 49~sensors at 25~cm intervals 
and covers a 150 $\times$ 150~cm$^2$ area.
The precision on the center of the muon profile is 2.95~cm, 
which corresponds to 0.25 mrad precision on the beam direction.
The details of this monitor are described in~\cite{MUMON}.

The neutrino beam intensity and direction are monitored directly 
by measuring the profile of neutrinos 
at the INGRID detector~\cite{Abe:2011xv}, located 280~m away from the target.
It consists of 16 identical neutrino detectors arranged in horizontal and vertical arrays around 
the beam center. Each neutrino detector has a sandwich structure 
of the iron target plates and 
scintillator trackers. 
\textcolor{red}{The intensity and profile of the neutrino beam
are reconstructed from the number of detected neutrino interactions 
in each module. 
At the present beam intensity (about $10^{18}$ protons-on-target/day),
the neutrino event rate is monitored daily with 4\% precision.
The neutrino beam center is measured monthly 
with accuracy better than 0.4~mrad.}

The ND280 detector measures the off-axis neutrino flux at a baseline of 280~m.  
At 280~m, ND280 effectively sees a line source of neutrinos rather than a point source,
therefore it covers a range of off-axis angles. The off-axis angle to ND280
from the target position is 2.04$^{\circ}$.
This angle was chosen 
to make the neutrino spectrum at ND280 as similar as 
possible to the spectrum at SK.
Consisting of electromagnetic
calorimeters, scintillating trackers and time projection chambers in a magnetic field, the ND280 detector
can measure the spectrum of neutrino interactions for use in the extrapolation of the flux prediction to SK.  
Independent neutrino cross section measurements can also be made at ND280, 
for which well-controlled absolute flux uncertainty is a key ingredient.  
The details of the ND280 detector are given in~\cite{t2knim,tpcnim,fgdnim}.

\subsection{\label{sec:alignment}Alignment of the beamline and actual neutrino beam direction}
The neutrino beam direction with respect to SK (``off-axis angle'') and the distance between 
the target and SK are obtained by GPS survey. The distance between the target and the center 
position of SK is 295,335.9$\pm$0.7~m. 
The beam axis is declined by 3.637$^{\circ}$, while SK  direction is 1.260$^{\circ}$ downward
and 0.795$^{\circ}$ to the north from the beam axis.
The off-axis angle is adjusted to 2.50$^\circ$ to maximize 
the neutrino oscillation probability and measured to be 2.504$\pm$0.004$^\circ$.

Based on the surveys, the primary beamline components, target, and horns were aligned in order to send the neutrino beam in the right direction. The muon monitor and the neutrino near detectors were also aligned in order to monitor the neutrino beam direction.

The directional accuracy of a long-baseline GPS survey 
is about 3$\times10^{-6}$ rad. 
The accuracy of a short distance survey (e.g. the direction 
to the near detectors) is about 7$\times10^{-5}$ rad. 

The alignment of the components in the primary beamline was carried out 
with a laser tracker which has a spatial resolution of $50 \mbox{ $\mu$m}$ 
for distances shorter than 20~m. 
The proton monitors were aligned to better than 0.4~mm. 
The OTR monitor, in the secondary beamline, was aligned with a precision of 0.3 mm relative to the 
axis of the first horn. The relative alignment between the OTR and upstream proton monitors is known 
to within 1.0 mm.

The target was surveyed relative to the horn axis after installation. 
A difference of 1.2 mm (0.3 mm) in horizontal (vertical) direction 
between the target and the horn axis was measured at the downstream end, 
while the alignment of the upstream end was found to be centered on the
horn axis to within 0.1 mm. 

The observed displacement at the downstream end of the target translates into 1.3 mrad (0.3 mrad) rotation 
in the horizontal (vertical) plane of the downstream end relative to the target head. The effect of this 
rotation on the predicted neutrino flux is included as a systematic uncertainty (see Section~\ref{sec:alignerrors}).

The position of each horn was surveyed after the installation. 
In the directions horizontally transverse and parallel ($x$ and $z$) 
to the proton beam axis, the horns were aligned relative 
to the beamline survey markers inside the helium vessel. 
The alignment accuracy in these directions are 0.3 mm and 1.0 mm 
for $x$ and $z$, respectively. 
The vertical position, $y$, of the center of each horn was aligned relative 
to the survey markers on one of the magnets in the final section 
of the primary beamline. 
The alignment precision in this direction is dominated 
by an overall uncertainty of 1.0 mm in the vertical alignment 
between the primary and secondary beamlines. 
The precision of the angular alignment of each horn is about 0.2 mrad, 
which is based on the survey of the alignment markers at the upstream 
and downstream end of each horn.
\textcolor{red}{The movement of the horn conductors by the current pulse 
was measured to be less than 0.1~mm.}


After the 2011 Tohoku earthquake, movements of the GPS survey points, 
of the primary beamline tunnel and of the beamline components were observed.
The baseline to SK was increased by 0.9 m, while the beam angle was rotated by
$3\times10^{-5}$ rad.  Both of these shifts have a small effect on the physics performance of the experiment.  
The upstream end of the primary beamline tunnel was observed to have shifted 
by 12~mm horizontally and 5~mm vertically relative to its downstream end, 
which is fixed to the target station and secondary beamline.  
The beamline magnets and monitors were realigned to the same alignment 
accuracy with the fixed point of reference at the most downstream end of the primary beamline.

\begin{figure}[ht]
\centering
\includegraphics[width=0.45\textwidth]{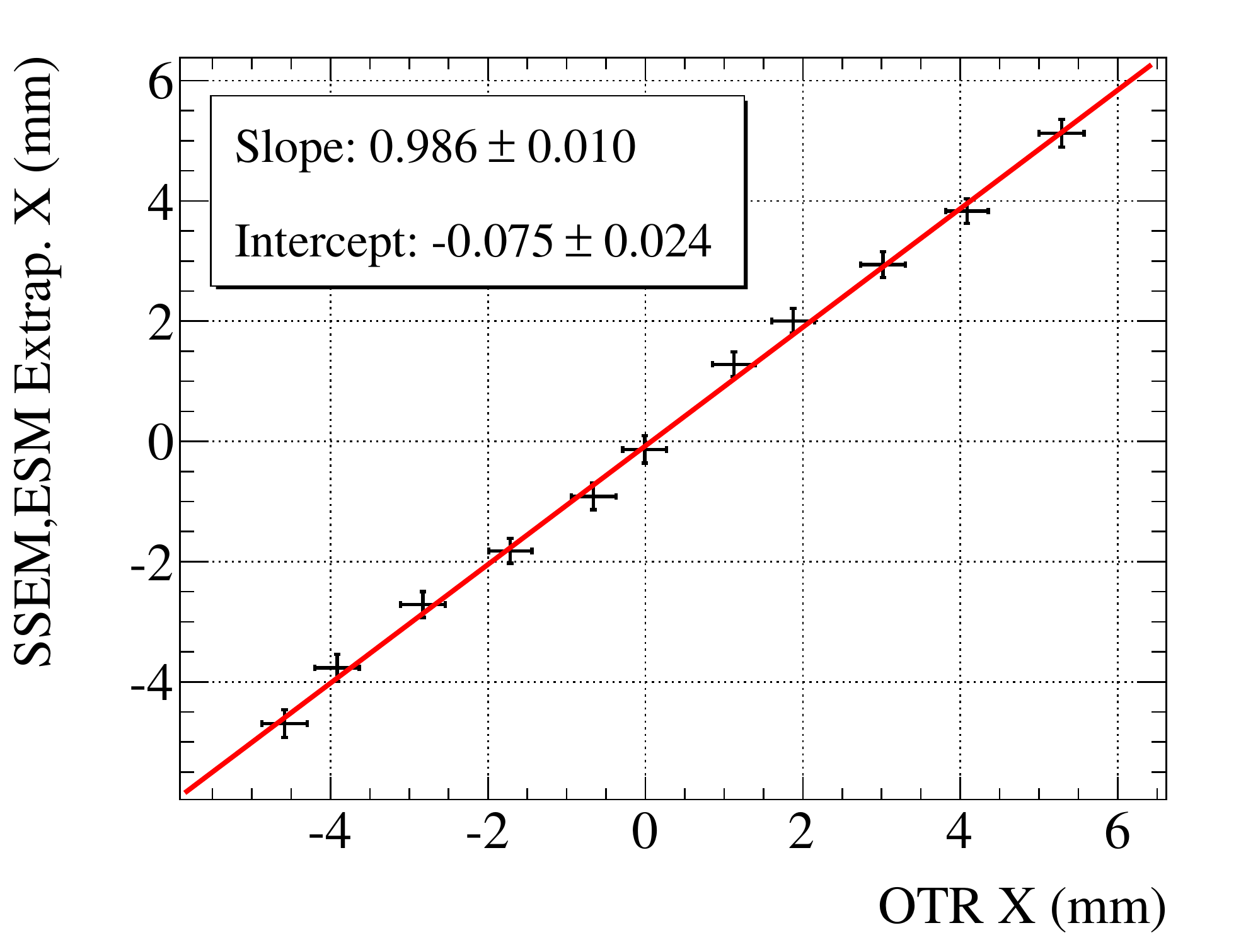}
\includegraphics[width=0.45\textwidth]{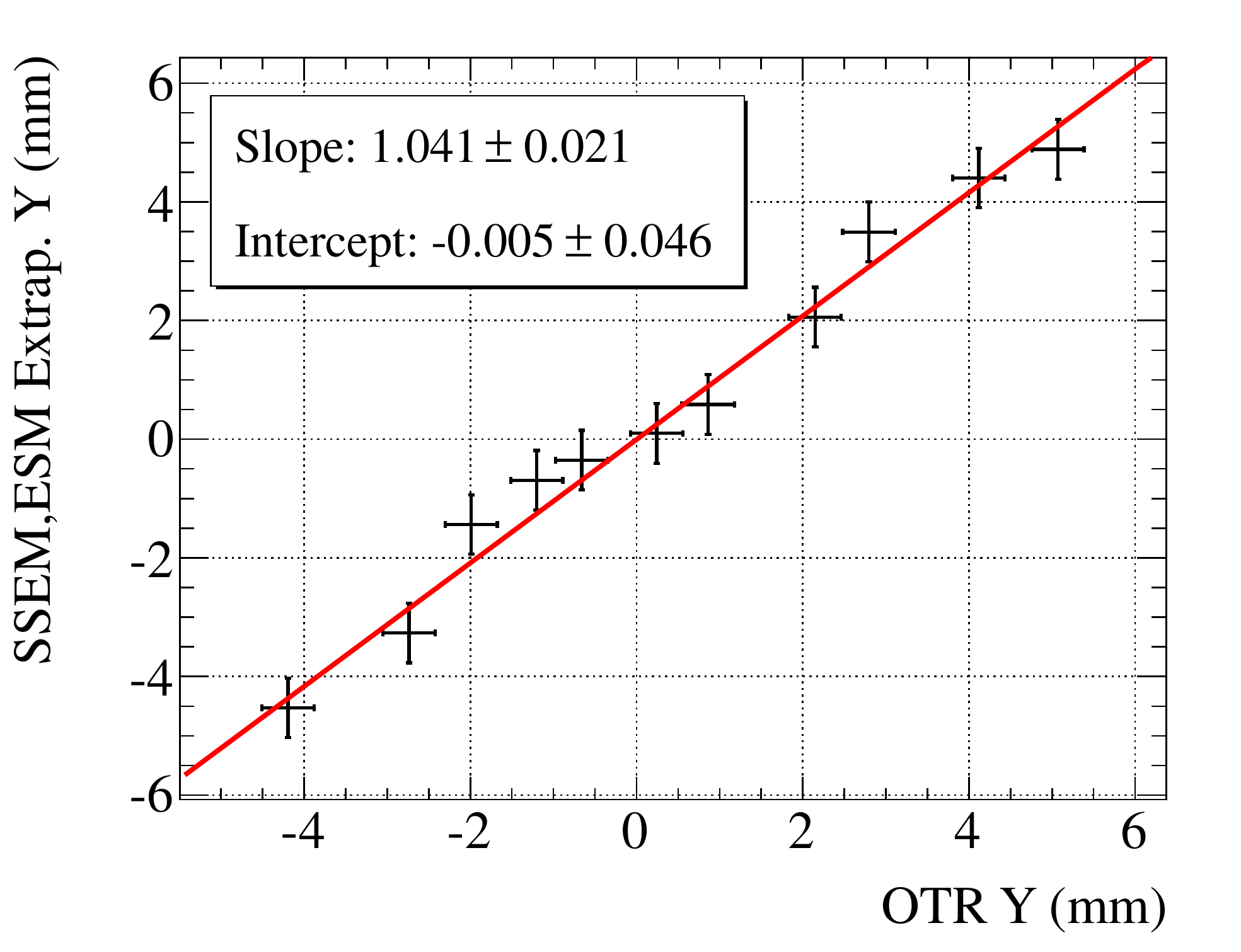}
\caption{The correlations between the beam position measurements in $x$ (top) and $y$ (bottom) by the OTR monitor
and the SSEM monitors extrapolated to the OTR position. The intercept and slope are from a linear fit (red line)
to the measurements.}
\label{fig:otr_ssem_check}
\end{figure}

The horns were observed to have shifted by 3$\sim$9 mm 
according to the survey of alignment markers at the top of the horn 
support modules.  The horns were realigned using the survey markers on the support modules, and
the alignment was confirmed by lowering a rigid frame with a camera and alignment laser into the helium vessel
and checking the position of survey marks on the horns.  The horns were found to be at the expected position, within
the 1 mm accuracy of the survey method.  The alignment of the OTR monitor could not be directly checked since the 
first horn was not removed from the helium vessel.  In situ scans of the 
beam across the target, after realignment of the primary beamline monitors, have shown that the 
measured beam position by the OTR monitor is consistent with the beam position extrapolated from the 
primary beamline SSEM monitors, as shown in Fig.~\ref{fig:otr_ssem_check}.
The MUMON was surveyed after the earthquake and its position relative to the target station shifted by less than
1~mm.

\section{\label{sec:beamoperation}Beam Operation History}

\begin{figure}[tbp]
\centering
\includegraphics[width=0.5\textwidth]{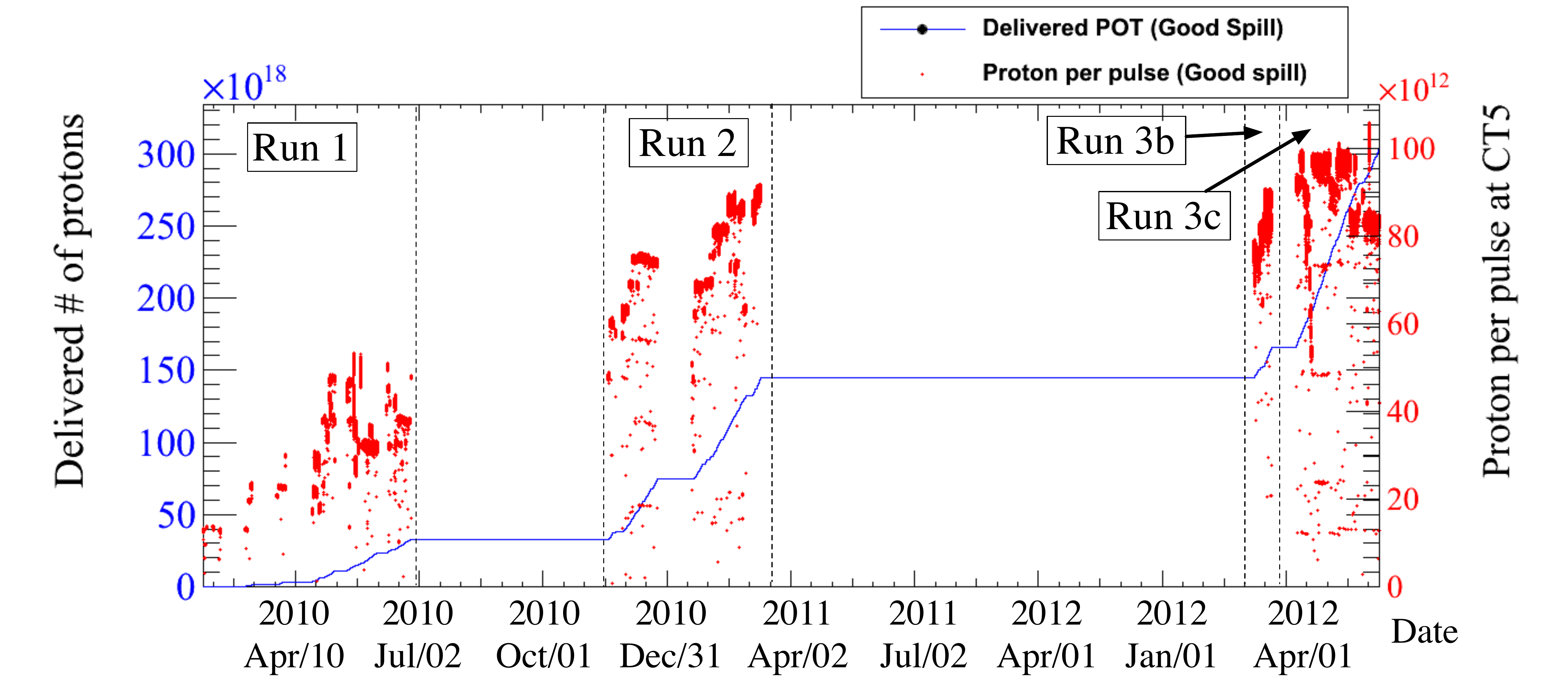}
\caption{History of total accumulated protons and protons per pulse for the good quality beam data. 
The solid line shows the accumulated POT. The dot points show the number of protons per pulse. }
\label{fig:pot_history_run123}
\end{figure}

The proton beam profile, neutrino beam direction and horn current are measured 
during the beam operation and measurement results are reflected 
in the neutrino flux prediction and estimation of uncertainty.
The flux prediction and uncertainty presented in this paper are based 
on three physics runs: Run 1 (January -- June 2010),  
Run 2 (November 2010 -- March 2011), Run 3 (March -- June 2012).
The Run~3 period is divided into three sub periods - Run~3a (March 2012), 
Run~3b (March 2012) and Run~3c (April -- June 2012) -  
according to the horn current settings (with a 0~kA setting in Run~3a 
and a 205~kA setting in Run 3b instead of the nominal 250~kA).
\textcolor{red}{The polarity of the horn current was always set 
to that for the `neutrino' running mode.}
The Run 3a data is not used for the oscillation analysis because the data 
in this period is small (0.3\% of the total) and the horn current was set 
to 0 kA. 
However, it is used for studies of 0~kA operation.
Figure~\ref{fig:pot_history_run123} shows the plot of the accumulated POT and 
protons per pulse for good quality beam data over time. 
The total accumulated number of protons (protons on the target, POT) 
in all run periods is  3.04$\times 10^{20}$ POT, corresponding to 4\% 
of T2K's exposure goal. 
The maximum beam power reached so far is about 200 kW.

\begin{figure}[tbp]
\centering
\includegraphics[width=0.5\textwidth]{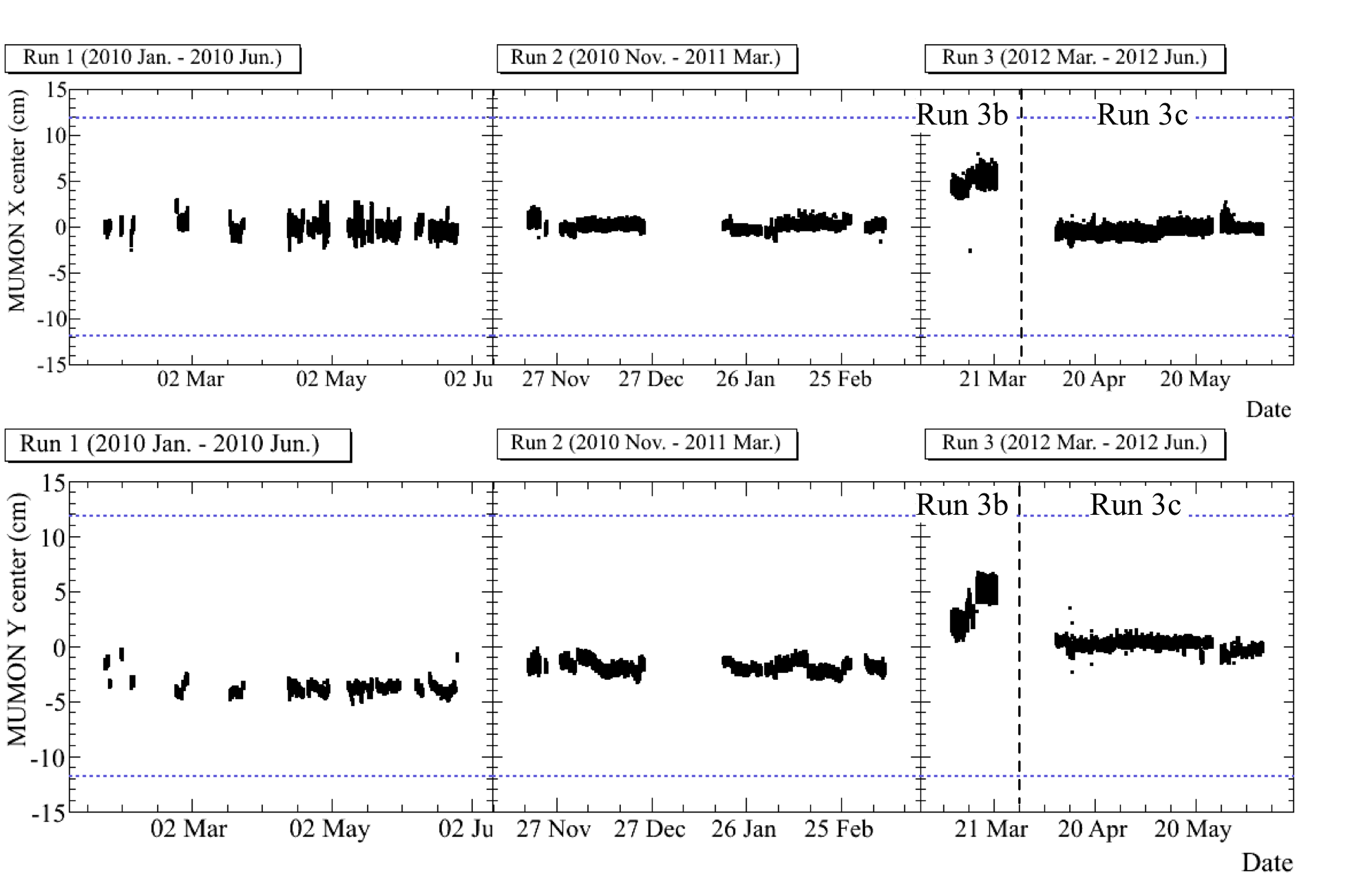}
\caption{Time history of the measured muon profile center at MUMON in all run periods.
A top and bottom figure shows the profile center in the horizontal(X) and the vertical(Y), respectively.
A dashed line corresponds to 1 mrad at MUMON. 
Both directions are controlled within 1 mrad.} 
\label{fig:sicenter_history_run123}
\end{figure}

We select only good quality beam data for physics analysis using the following conditions.
\begin{itemize}
\item Each hardware component works normally.
\item The deviation of all Horns currents from the mean is within $\pm$ 5 kA.
\item The deviation of the beam angle measured by MUMON from the mean is within 1 mrad.
\item The deviation of the total muon yield measured by MUMON from the mean is within $\pm$ 5 \%.
\end{itemize}
The beam data from the beam monitors are checked on a spill-by-spill basis during beam operation 
to ensure data from good quality beam is used.
For example, Fig.~\ref{fig:sicenter_history_run123} shows the history of the muon profile center measured at MUMON. 
In all run periods, this profile center is stable within 1 mrad 
from the beam axis (1 mrad stability is required to achieve 
the physics goal of T2K).
During Run~3b period, \textcolor{red}{the MUMON center deviated from 
the beam-axis in both the X and Y directions.
The possible reason for this deviation is misalignment of the horns.
The beam may be focused in the deviated direction if there is 
a horn misalignment. This deviation can change depending on the horn current.}
As described later, the direction of the neutrino beam in this period had been also measured at INGRID
and it was confirmed to be within 1~mrad (see Table~\ref{tab:bmdir_ingrid}).
After the good quality cut, the fraction of beam data is 99.8\%.
\subsection{\label{sec:pbeamprop}Proton beam properties}

The center position and angle of the proton beam 
at the upstream surface of the baffle are reconstructed by extrapolating 
the center positions measured with ESM20, SSEM19 and OTR for the vertical 
and ESM19, ESM20, SSEM19 and OTR for the horizontal direction for each spill.

Each time the beam conditions change, all of the SSEMs are inserted 
into the beamline and beam profiles are measured for 100 spills.
The Twiss parameters and emittance are obtained by fitting these profiles along with that from the OTR.
The beam width and divergence at the baffle are calculated from the Twiss parameters and emittance. 
After 100 spills, all SSEMs except for SSEM19 are extracted from the beam orbit
and the beam width and divergence are then obtained by scaling the emittance from 
the change of the profile measured at SSEM19 and OTR for each spill.

Proton beam properties for each run period are obtained by reconstructing the
beam profile at the baffle for each spill and summing the spill-by-spill profiles 
weighted by the number of protons in each spill.
Tables~\ref{tab:pbeam_center_x} and \ref{tab:pbeam_center_y} summarize the  
measured center position, angle, width, emittance and Twiss $\alpha$
at the baffle for each run period.

\begin{table}[htb]
  \caption{\label{tab:pbeam_center_x}
Summary of measured proton beam parameters in the horizontal direction at the baffle for each run period : center position ($X$) 
and angle ($\theta_{X}$), Gaussian width ($\sigma$), emittance ($\epsilon$), and Twiss ($\alpha$).}
\begin{center}
\scalebox{1.0}[1.0]{
\begin{tabular} {llllll}
\hline
\hline
Period & $X$ (mm) & $\theta_{X}$ (mrad) &  $\sigma$ (mm) & $\epsilon$ ($\pi$ mm mrad) & $\alpha$ \\
\hline
Run1 & 0.37 & -0.044 & 4.27 & 2.12 & 0.60 \\
Run2 & 0.149 & -0.080 & 4.04 & 5.27 & 0.16 \\
Run3b & 0.087 & 0.020 & 4.13 & 6.50 & 0.16 \\
Run3c & -0.001 & 0.032 & 4.03 & 4.94 & 0.33 \\
\hline
\hline
\end{tabular}
}
\end{center}
\end{table}

\begin{table}[htb]
  \caption{\label{tab:pbeam_center_y}
Summary of measured proton beam parameters in the vertical direction at the baffle for each run period : center position ($Y$) 
and angle ($\theta_{Y}$), Gaussian width ($\sigma$), emittance ($\epsilon$), and Twiss ($\alpha$).}
\begin{center}
\scalebox{1.0}[1.0]{
\begin{tabular} {llllll}
\hline
\hline
Period & $Y$ (mm) & $\theta_{Y}$ (mrad) &  $\sigma$ (mm) & $\epsilon$ ($\pi$ mm mrad) & $\alpha$ \\
\hline
Run1 & 0.84 & 0.004 & 4.17 & 2.29 & -0.09 \\
Run2 & -0.052 & -0.007 & 4.08 & 5.17 & 0.14 \\
Run3b & -0.024 & 0.043 & 3.97 & 5.30 & 0.25 \\
Run3c & -0.366 & 0.068 & 4.22 & 6.02 & 0.34 \\
\hline
\hline
\end{tabular}
}
\end{center}
\end{table}

The following are the sources of uncertainty on the measurements of the average position and angle of the beam:
\begin{itemize}
\item {The alignment uncertainty of the proton beam monitors}
\item {The alignment uncertainty between the primary proton beamline and 
the target station (secondary beamline)}
\item {Systematic errors in the position measurements by the proton beam monitors}
\end{itemize}

These errors are included in the beam orbit fit to the monitor measurements, and the magnitude of the resultant 
errors and their correlations are 
summarized in Table~\ref{tab:pcenter_err}.

\begin{table}[htb]
  \caption{\label{tab:pcenter_err}Systematic errors and correlations
for the position and angle of the beam center at the baffle front surface.
The $X(Y)$ and $\theta_{X}(\theta_{Y})$ stand for horizontal (vertical) position and angle
of the beam center, respectively. }
\begin{center}
\scalebox{0.8}[0.8]{
\begin{tabular} {lllllll}
\hline
\hline
Period & $X$ (mm) & $Y$ (mm) & $\theta_{X}$ (mrad) & $\theta_{Y}$ (mrad) & $\mbox{corr}(X, \theta_{X})$ & $\mbox{corr}(Y, \theta_{Y})$ \\
\hline
Run1 & 0.38 & 0.58 & 0.056 & 0.29 & 0.517  & 0.392 \\
Run2 & 0.27 & 0.62 & 0.064 & 0.32 & 0.752 & 0.398 \\
Run3b & 0.28 & 0.58 & 0.064 & 0.29 & 0.614 & 0.386 \\
Run3c & 0.35 & 0.58 & 0.072 & 0.28 & 0.697 & 0.417 \\
\hline
\hline
\end{tabular}
}
\end{center}
\end{table}

To estimate the systematic uncertainty on the width and divergence of the proton beam,
the following error sources are considered:
\begin{itemize}
\item {The systematic error in the profile measurements by the proton beam monitors.}
\item {Effects of the momentum dispersion $\Delta p/p$, where a conservative estimate of $0.3\%$ is assumed.}
\item {Uncertainties in the quadrupole magnet (FQ2, 3, and 4) field model: 
a conservative estimate of $7\%$ is taken for the uncertainty 
in the magnetic field strength $\Delta B$ and the magnet effective length $\Delta L$.
This is derived from the difference in the field integral of dipole magnets
between the estimate using the actual beam orbit 
and one obtained by multiplying the measured field strength and the pole length.}
\end{itemize}

The resulting uncertainties on the emittance, Twiss $\alpha$  and beam width 
are summarized in Table~\ref{tab:pprof_err}.

\begin{table}[h]
  \caption[]{\label{tab:pprof_err}Uncertainties for 
the emittance $\epsilon$, Twiss $\alpha$ and width $\sigma$ at baffle of the proton beam.}
\begin{center}
\scalebox{0.82}[0.82]{
\begin{tabular}{lllllll}
\hline
\hline
&  $\epsilon_X$  & $\epsilon_Y$   & $\alpha_X$ & $\alpha_Y$ & $\sigma_X$  & $\sigma_Y$  \\
&  ($\pi$ mm mrad) & ($\pi$ mm mrad)  &  &  & (mm) & (mm) \\
\hline
monitor & 0.10 & 0.12 & 0.11 & 0.11 & 0.10 & 0.10 \\
$\Delta p/p = 0.3\%$  & 0.51 & 0.10 & 0.10 & 0.02 & 0.01 & 0.03 \\
$\Delta B = 7\%$, $\Delta L = 7\%$  & 0.56 & 0.52 & 0.28 & 1.68 & 0.06 & 0.97 \\
\hline
\hline
\end{tabular}
}
\end{center}
\end{table}

\subsection{Beam Direction}

\begin{figure}[tbp]
  \begin{center}
    \includegraphics[keepaspectratio=true,width=85mm]{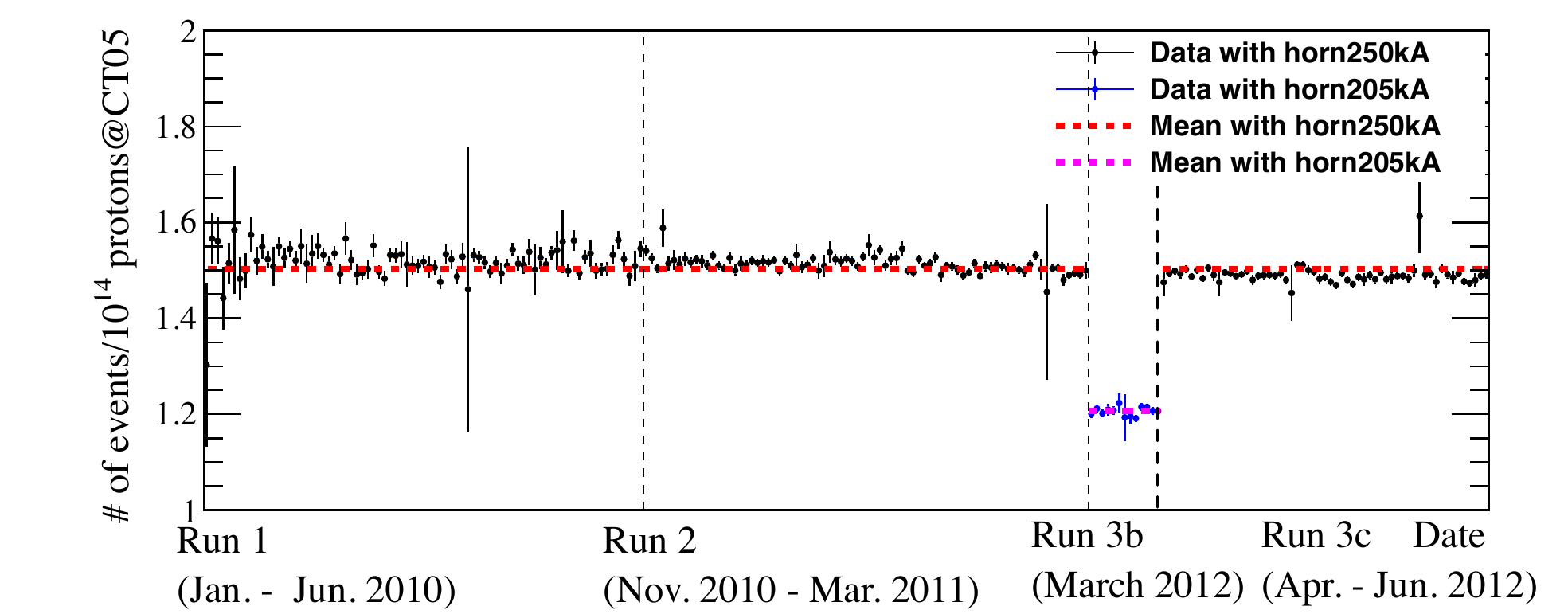}
  \end{center}
  \caption[Neutrino beam direction measured by INGRID.]
  {\label{fig:ingrid_eventrate}
  Neutrino events per 10$^{14}$ POT measured by INGRID (points) overlaid with mean value (dashed lines).
  The error bar represents the statistical error on the INGRID measurement.
  }
\end{figure}

The stability of the beam direction and intensity are also monitored 
by measuring the neutrino beam itself~\cite{Abe:2011xv}. 
Figure~\ref{fig:ingrid_eventrate} shows the stability of the measured neutrino event rate (normalized by protons on target)
measured with INGRID as a function of time. A decrease of $<2\%$ in the rate is observed, but at this time
it has not been determined if the decrease arises from an actual reduction in the neutrino production or beam
intensity dependence in the INGRID event analysis.   
Table~\ref{tab:bmdir_ingrid} summarizes the averaged neutrino beam center measured in each run.
The neutrino beam direction was kept within 1~mrad from the beam axis.

The sources of the systematic error on the neutrino beam direction measurement are: 
systematic error on the INGRID measurement, the observed shift of the beam direction 
from the designed beam axis and the survey error (0.0024 mrad both in horizontal ($x$) and 
vertical ($y$) for SK, and 0.026 mrad in $x$ and 0.038 mrad in $y$ for ND280). 
In total, for Run 1 case, the systematic error on the neutrino beam direction relative to SK and 
ND280 direction is 0.34~mrad in $x$ and 0.45~mrad in $y$.
It corresponds to 0.44~mrad for the uncertainty of the off-axis angle. 

\subsection{Horn Current}
In Run 1, the 2nd and 3rd horns were electrically connected in series and operated with one power supply. 
The 1st horn was operated separately. During Runs 2 and 3, all three horns were connected in series and operated with one power supply.
All horns were usually operated at 250~kA except for Run 3b, during which the horns were operated at 205~kA.
During the data taking periods, the monitored values of 
the horn current drifted within 2\%.
This drift is most likely due to temperature dependence in the
operation of the monitoring hardware, but variations of the actual
horn current have not been definitively ruled out.

\begin{table}[]
\caption[Neutrino beam direction measured by INGRID]{
Neutrino beam direction measured by INGRID in each period. 
Each X and Y position includes the statistical error (first error term) 
and systematic error (second error term).}
\label{tab:bmdir_ingrid}
\begin{center}
\begin{tabular}{lll}
\hline 
\hline 
Period & X center [mrad] & Y center [mrad] \\
\hline 
RUN1  &  0.009$\pm$0.052$\pm$0.336 & -0.314$\pm$0.055$\pm$0.373 \\
RUN2  & -0.028$\pm$0.027$\pm$0.333 &  0.050$\pm$0.030$\pm$0.374 \\
RUN3b & -0.110$\pm$0.085$\pm$0.385 &  0.152$\pm$0.100$\pm$0.472 \\
RUN3c & -0.001$\pm$0.026$\pm$0.331 &  0.232$\pm$0.029$\pm$0.378 \\
\hline
\hline
\end{tabular}
\end{center}
\end{table}


%

\section{\label{sec:fluxsim}The neutrino flux simulation}


The prediction of the flux and spectrum of neutrinos at the T2K detectors (INGRID, ND280 and SK) is based on 
a simulation that begins with the primary proton beam upstream of the baffle 
and ends with the decay of hadrons or muons that produce neutrinos. 
The simulation and its associated uncertainties are driven by 
primary proton beam profile measurements, measurements of the horns' magnetic fields, and hadron 
production data, including NA61/SHINE measurements~\cite{Abgrall:2011ae,PhysRevC.85.035210}. 

FLUKA2008~\cite{Battistoni:2007zzb,Ferrari:2005zk} is used to simulate the hadronic interactions in the target and baffle where the primary
proton beam first interacts and produces the majority of the secondary pions, since FLUKA is found to have the
best agreement with external hadron production data.
Kinematic information for particles emitted from the target is saved and transferred to the JNUBEAM simulation.  
JNUBEAM is a GEANT3~\cite{GEANT3} Monte Carlo simulation of the baffle, target, horn magnets, helium vessel, 
decay volume, beam dump, and muon monitor.  The geometry of these components is based on the final mechanical 
drawings of the constructed beamline. JNUBEAM also includes the INGRID, ND280, and SK detectors, 
which are positioned in the simulation according to the latest survey results. 
Hadronic interactions are modeled by GCALOR model~\cite{GCALOR,Fasso:1993kr} 
in JNUBEAM.

In JNUBEAM, particles are propagated through the horn magnetic field, 
and may interact with the horn material in the target station.
Particles are propagated through the decay volume until they interact or decay.  As described in Sec.~\ref{subsubsec:nuproduct}, neutrinos from particle decays
are forced to point to SK or a randomly chosen point in the near detector plane.
The neutrino kinematic variables and the probability based on the decay phase-space density and branching fraction are saved. 
The flux and energy spectrum are obtained from these simulated events by weighting according to the saved probabilities.
In addition, the kinematic information of the initial proton 
and full interaction chain producing the neutrino are saved 
to allow for re-weighting of the proton beam profile and hadron interactions.

\begin{figure}[]
  \begin{center}
    \includegraphics[keepaspectratio=true,width=70mm]{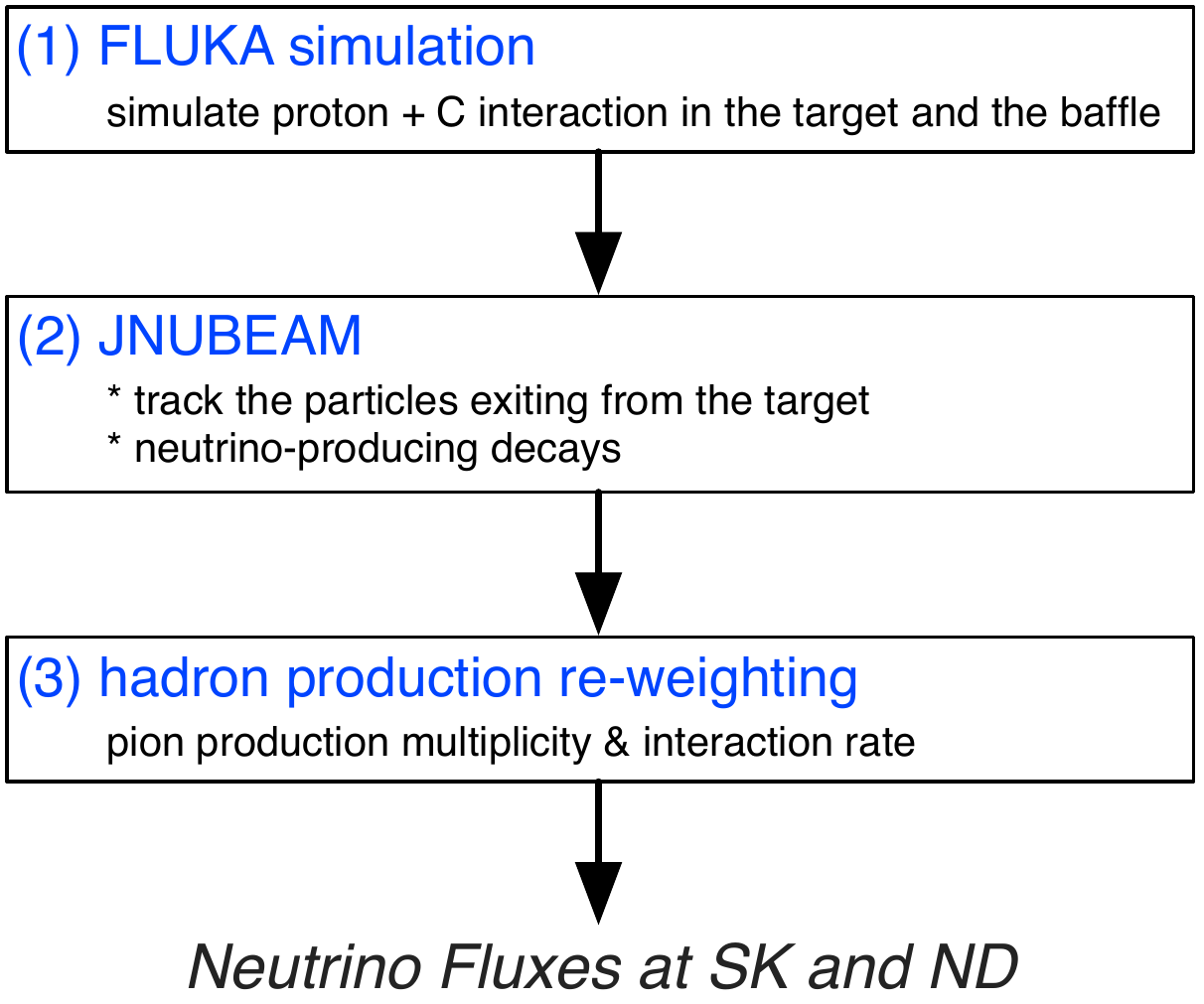}
  \end{center}
  \caption[A flow chart of how we predict the neutrino fluxes at SK and ND.]
  {\label{fig:fluxflow} Flow diagram of the flux prediction.
  }
\end{figure}

The general simulation procedure is outlined in Fig.~\ref{fig:fluxflow}.

\subsection{\label{sec:targetsim}Interaction of primary beam in the target}

\begin{figure}[ht]
\centering
\includegraphics[keepaspectratio=true,width=0.47\textwidth]{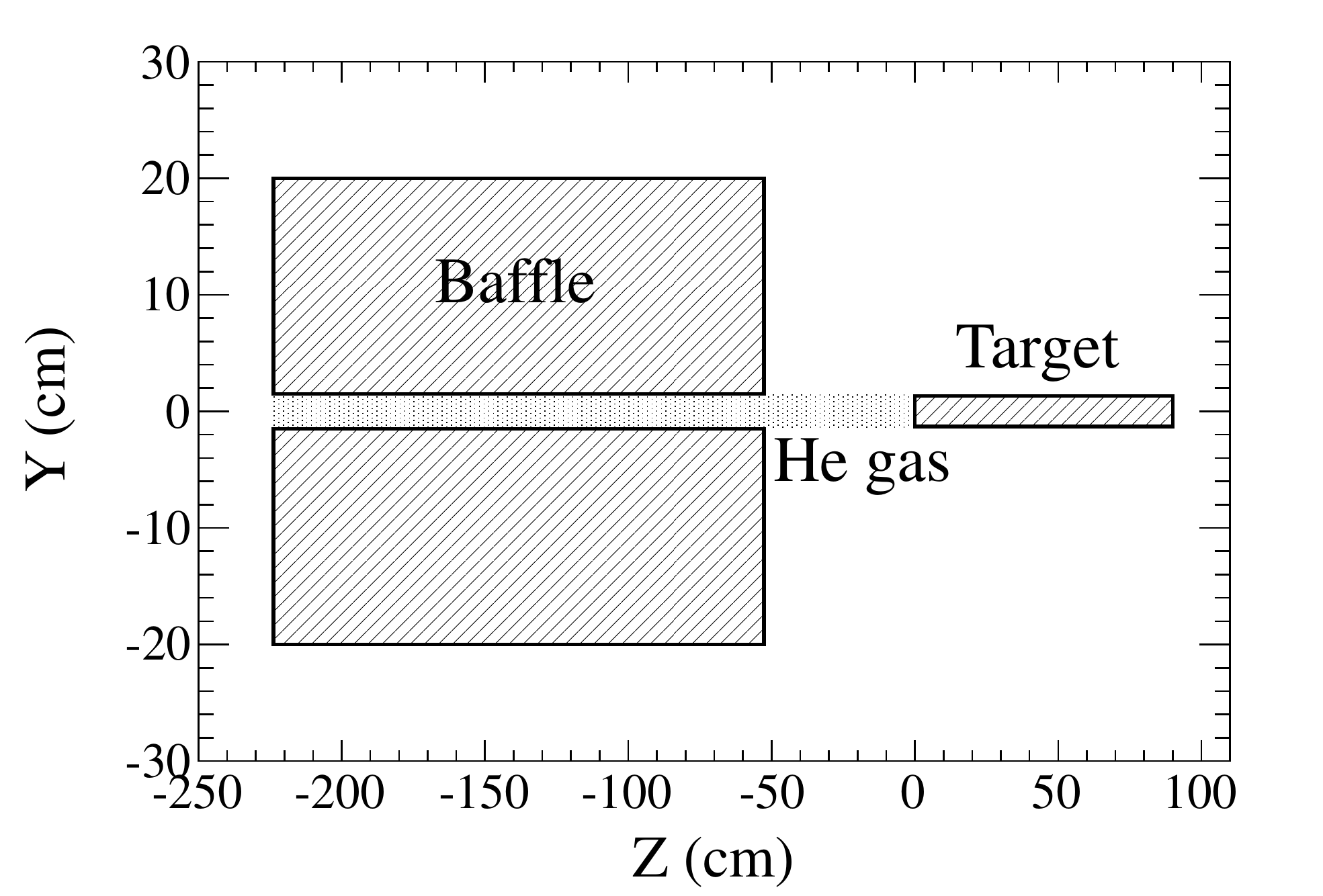}
\caption{A two dimensional view of the geometrical set-up in the FLUKA simulation of the baffle and the target.}
\label{fig:fluka_sim}
\end{figure}

The simulation of the interactions of the primary beam protons with
the graphite of the baffle and the target core is performed using
FLUKA2008. Incident protons are generated according to the measured proton beam spatial distribution and divergence. The kinetic energy is set to 30 GeV.
Figure~\ref{fig:fluka_sim} shows the two-dimensional projection of the simulated geometry.
The baffle is described as a graphite block with the dimensions $29\times 40\times
171.145 \mbox{ cm}^3$ and a 3.0 cm diameter cylindrical hole
through the center. The target is modeled as a graphite
cylinder 90 cm long and 2.6 cm in diameter. The volume
inside the baffle hole and between the baffle and the target is filled with He gas. 
The generated particles are traced until they emerge from the model geometry and then
information such as kinematic variables and hadron interaction history at that point is recorded.

\subsection{\label{sec:hornsim}Tracking inside horns and helium vessel.}

Particles are generated in JNUBEAM
according to the recorded information in the previous step, 
and are tracked through the horns and helium vessel.
The 2~mm thick graphite tube and 0.3~mm thick titanium case 
surrounding the target core are also modeled in JNUBEAM.
The interaction of generated particles with the materials in JNUBEAM
is modeled by GCALOR.

\subsubsection{Horn magnetic field}
As explained in Sec.~\ref{sec:tgthor}, a toroidal magnetic field is generated in the horns.
The field strength varies as $1/r$, where $r$ is the distance from the horn axis.
Since a low frequency pulsed current (3.6~ms full width) is loaded into the horn, 
the skin effect is small (the estimated skin depth is approximately 5~mm 
while the thickness of the inner conductor is 3~mm.). 
Therefore, we assume that the current flows in the conductor uniformly.
On this assumption, the magnetic field at radius $r$ in the inner conductor is calculated with Amp\`{e}re's Law as:
\begin{equation}
\label{eq_magfld}
  B(r) = \frac{\mu_{0} I}{2\pi r} \frac{r^{2} - a^{2}}{b^{2} - a^{2}}
\end{equation}
where $\mu_{0}$ is the magnetic permeability, $I$ is the current and $a$ and $b$ are, respectively, 
the inner and outer radii of the inner conductor.

\subsubsection{\label{subsubsec:nuproduct}Neutrino production}
The particles are tracked in the helium vessel, decay volume, and the surrounding 
concrete shield including the beam dump until they decay or their kinetic energy drops below 10 MeV (at which point unstable particles
are decayed). 
Decay products are also tracked except for neutrinos.
In JNUBEAM, $\pi^{\pm}$, $K^{\pm}$, $K_\mathrm{L}^{0}$ and $\mu^{\pm}$ decays 
listed in Table~\ref{tab:nu decay mode} are considered as neutrino sources.
The current best knowledge~\cite{PDG} on the branching ratios and $K_{\ell 3}^{\pm}$ ($K^+\rightarrow\pi^0 l^+\nu_l/K^-\rightarrow\pi^0 l^-\bar{\nu}_l \mbox{, }l = e, \mu$) decay form factors is used.
When a muon is generated from pion/kaon decay, its polarization information 
is stored. 
This polarization is then taken into account at the muon decays.

In order to save computing time, when a particle decays into neutrino(s), the 
neutrino(s) are forced to point in the direction of SK or a randomly chosen point
in the near detector planes. 
The neutrino energy in the center of mass frame is assigned based on the decay kinematics.
The neutrino is then boosted into the laboratory frame under the assumption that it points towards
the desired detector, and the probability of production
in the selected direction is stored as an event weight.
In addition to this probability, the neutrino flavor, energy and ancestors' truth information are stored.
The neutrino flux spectrum is obtained by weighting each event 
with the stored probability.
For neutrinos produced with energy less than 4 GeV, the storage of events is pre-scaled 
(and event weights are adjusted accordingly) to allow for sufficient statistics in the high energy
tail of the flux prediction without producing prohibitively large file sets.

\begin{table}[htb]
  \caption[Neutrino-producing decay modes considered in JNUBEAM.]
  {\label{tab:nu decay mode}
  Neutrino-producing decay modes considered in JNUBEAM and their branching ratio in percentage. 
  Decay modes for $\bar{\nu}_{\mu}$ and $\bar{\nu}_{e}$ are omitted in this table.
  The $\pi^{-}$, K$^{-}$ and $\mu^{-}$ modes are charge conjugates of the $\pi^{+}$, K$^{+}$ and $\mu^{+}$ modes, respectively.}
  \begin{center}
    \begin{tabular}{rlc}
      \hline
      \hline
       Particle & Decay Products & Branching Fraction ($\%$) \\ \hline
       $\pi^{+}$ & $\rightarrow \mu^{+}\nu_{\mu}$ & $99.9877$ \\
                 & $\rightarrow e^{+}\nu_{e}$ &  $1.23\times10^{-4}$\\
       K$^{+}$ &  $\rightarrow \mu^{+}\nu_{\mu}$ & $63.55$ \\
               &  $\rightarrow \pi^{0}\mu^{+}\nu_{\mu}$ & $3.353$ \\
               &  $\rightarrow \pi^{0}e^{+}\nu_{e}$ & $5.07$ \\
       K$_L^{0}$ & $\rightarrow \pi^{-}\mu^{+}\nu_{\mu}$ & $27.04$ \\
                 & $\rightarrow \pi^{-}e^{+}\nu_{e}$ & $40.55$ \\
       $\mu^{+}$ & $\rightarrow e^{+}\bar{\nu}_{\mu}\nu_{e}$ & $100$ \\
      \hline
      \hline
    \end{tabular}
  \end{center}
\end{table}


\subsection{\label{sec:hadronsim}The simulation of hadronic interactions}

As discussed in Sec.~\ref{sec:targetsim}, the hadronic interactions in the target
are modeled with FLUKA2008. Outside of the target, where GEANT3 controls the simulation, interactions are modeled with GCALOR. The chain of hadronic interactions for each simulated event
producing a neutrino is saved, and re-weighting based on hadron interaction measurements is applied
to the simulated events.

The hadron interaction data used are thin target data, described in Sec.~\ref{sec:hadrondata}, 
that include measurements of inelastic cross sections and differential hadron production.
Unlike the case of the thin target measurements, particles traversing the T2K target encounter a significant amount of material and can undergo multiple interactions. In addition particles can also interact with the material outside the target. A step-by-step re-weighting procedure is therefore applied to the hadronic interaction chain in each event.  The weights are applied to:
\begin{enumerate}
\item differential production of $\pi^{\pm}$, $K^{\pm}$ and $K^{0}_{L}$ in the
interactions of protons on the target materials (Sec.~\ref{sec:hadron_prod_tune}).
\item interaction rates for $p$, $\pi^{\pm}$ and $K^{\pm}$ that affect
the rate of interactions that produce hadrons, as well as the attenuation of 
hadrons that may decay to produce a neutrino (Sec.~\ref{sec:hadron_int_tune}).
\end{enumerate}

The hadrons are labeled as secondary hadrons if they
are produced in interactions of the original protons, and tertiary hadrons if they are
produced by interactions of hadrons other than the original proton. 
The breakdown of the predicted 
flux for a given flavor by the final hadron in the interaction chain is shown in 
Table~\ref{tab:finalhad}.  The $\nu_e$ and $\bar{\nu}_{e}$ originating from secondary
or tertiary pions are from subsequent muon decays.
A significant 
fraction of the fluxes come from tertiary pions and kaons, so it is important
to investigate hadron interaction data at both the T2K beam momentum and for lower momentum
hadrons.

\begin{table}[h]
\begin{center}
\caption{The fraction of the neutrino flux by the final hadron in the interaction 
chain after hadron interaction re-weighting is applied.}
\label{tab:finalhad}
\begin{tabular}{lrrrr}
\hline
\hline
              & \multicolumn{4}{c}{Flux percentage of each(all) flavor(s)} \\
Parent & \multicolumn{1}{c}{$\nu_{\mu}$} & \multicolumn{1}{c}{$\bar{\nu}_{\mu}$} &  \multicolumn{1}{c}{$\nu_{e}$} & \multicolumn{1}{c}{$\bar{\nu}_{e}$} \\ 
\hline
Secondary                     &                &               &               & \\
\hspace{0.1cm}  $\pi^{\pm}$   & $60.0(55.6)\%$ & $41.8(2.5)\%$ & $31.9(0.4)\%$ & $2.8(0.0)\%$  \\
\hspace{0.1cm}  $K^{\pm}$     & $4.0(3.7)\%$   & $4.3(0.3)\%$  & $26.9(0.3)\%$ & $11.3(0.0)\%$ \\ 
\hspace{0.1cm}  $K^{0}_{L}$   & $0.1(0.1)\%$   & $0.9(0.1)\%$  & $7.6(0.1)\%$ & $49.0(0.1)\%$ \\ 
\hline
Tertiary                     &                 &               &               & \\
\hspace{0.1cm} $\pi^{\pm}$    & $34.4(31.9)\%$ & $50.0(3.0)\%$ & $20.4(0.2)\%$  & $6.6(0.0)\%$  \\
\hspace{0.1cm}  $K^{\pm}$     & $1.4(1.3)\%$   & $2.6(0.2)\%$  & $10.0(0.1)\%$  & $8.8(0.0)\%$  \\
\hspace{0.1cm}  $K^{0}_{L}$   & $0.0(0.0)\%$   & $0.4(0.1)\%$  & $3.2(0.0)\%$  & $21.3(0.0)\%$  \\
\hline
\hline
\end{tabular}
\end{center}
\end{table}

\subsubsection{\label{sec:hadrondata}Data used for hadronic interaction re-weighting}
The pion and kaon differential production measurements used for
obtaining the T2K flux predictions are summarized in Table~\ref{tab:haddatat2k}.
\begin{table}[h]
\centering
\caption{Differential hadron production data relevant for the T2K neutrino flux predictions.}
\begin{tabular}{llll}
\hline \hline 
Experiment & Beam Mom. (GeV/c) & Target & Particles \\
\hline
NA61/SHINE \cite{Abgrall:2011ae}\cite{PhysRevC.85.035210} & 31          & C & $\pi^\pm$, $K^+$ \\
Eichten \textit{et  al.}~\cite{eichten} & 24        & Be, Al, ... & $p$, $\pi^\pm$, $K^\pm$\\
Allaby \textit{et al.}~\cite{allaby} & 19.2         & Be, Al, ... & $p$, $\pi^\pm$, $K^\pm$\\
BNL-E910 \cite{e910} & 6.4 -- 17.5                         & Be          & $\pi^\pm$ \\
\hline \hline
\end{tabular}
\label{tab:haddatat2k}
\end{table}

To predict the neutrino flux, T2K relies primarily on the measurements
of pion \cite{Abgrall:2011ae} and kaon \cite{PhysRevC.85.035210}
yields by the NA61/SHINE experiment at the CERN SPS. These data were taken with a thin (2 cm)
graphite target and the same proton beam energy as that of T2K. 
The results are based on the data collected in 2007 during
a first, limited statistics, run with about 6.7$\times 10^5$ registered events.
%
An additional data set, taken with the target removed, was used to account for  
the contamination by particles produced in interactions of the proton beam occurring outside the target.

Charged particles are identified by using the measurement of the specific energy loss ($dE/dx$)
and of the time-of-flight ($ToF$). 
The combined information of $ToF$ and $dE/dx$ is needed in the 1-4 GeV/c momentum range  
where different particle species have similar values for their specific energy loss.
A calibration of the mean $dE/dx$ as a function of the momentum with an accuracy  
of 0.1\% was required to limit the systematics on the particle identification
at the level of 1\%. 

\begin{figure*}
\centering
\subfloat[$\pi^{+}$]{
\includegraphics[width=0.32\textwidth]{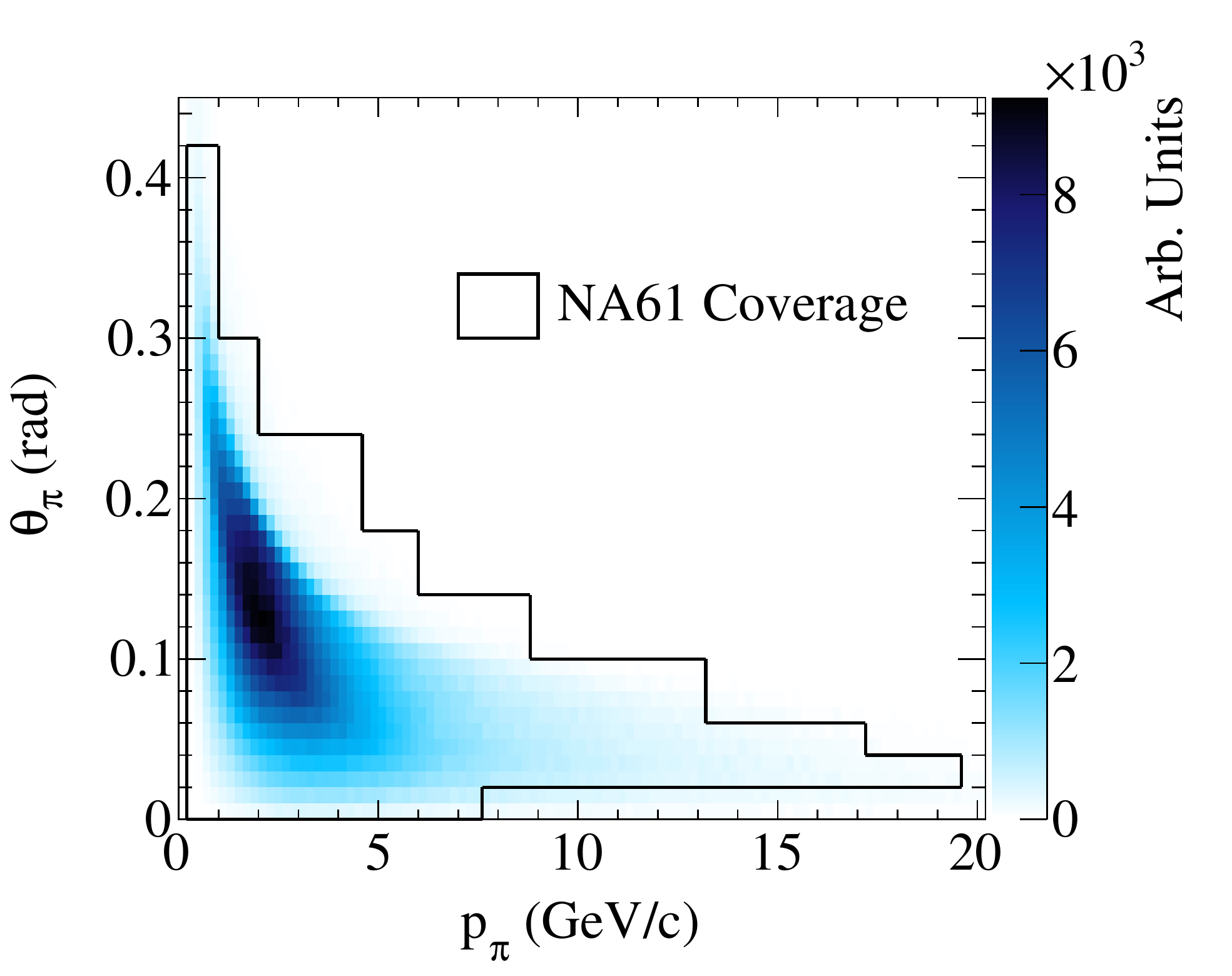}}
\subfloat[$\pi^{-}$]{
\includegraphics[width=0.32\textwidth]{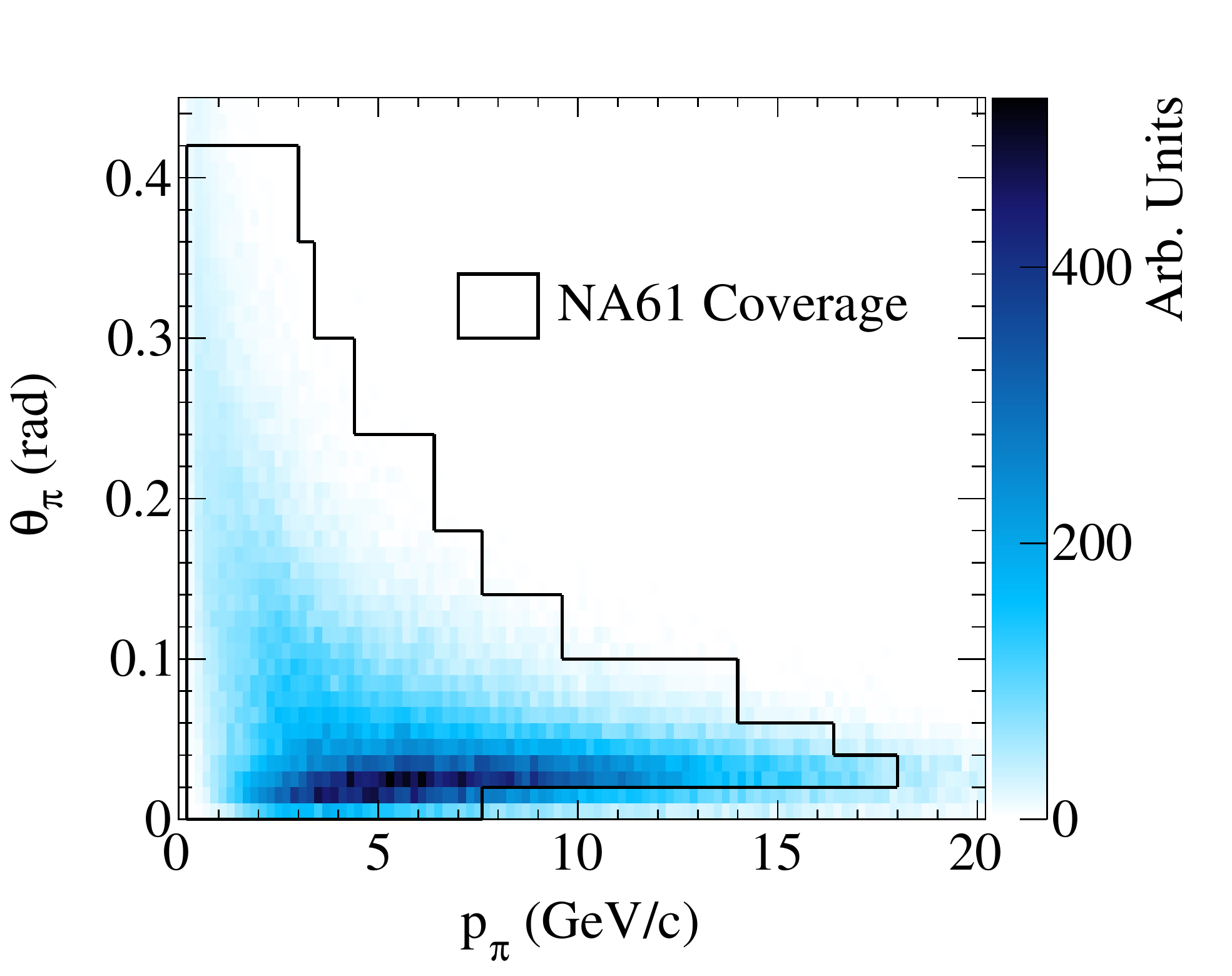}}
\subfloat[K$^{+}$]{
\includegraphics[width=0.32\textwidth]{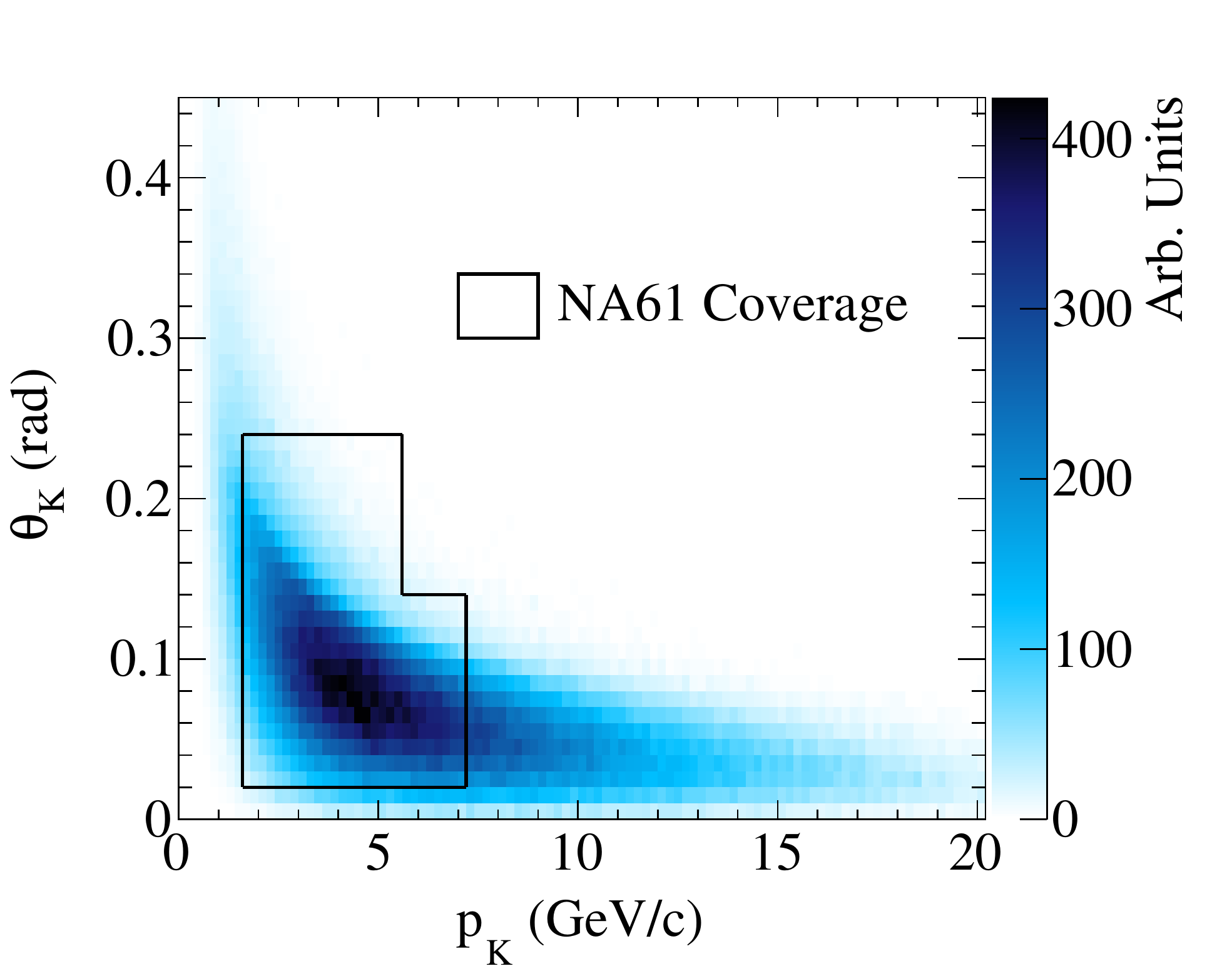}}
\caption{The phase space of pions and kaons contributing to the predicted neutrino
flux at SK, and the regions covered by NA61/SHINE measurements.}
\label{fig:na61_coverage}
\end{figure*}

Charged pion differential production cross sections were measured as a function of the pion laboratory
momentum in 10 intervals of the pion
laboratory polar angle relative to the proton beam direction, covering the range from 0 to 420 mrad. 
The considered momenta range from
0.2 GeV/c up to 19.6 GeV/c depending on the polar angle is illustrated in Fig.~\ref{fig:na61_coverage}.
For momenta above about 7.5 GeV/c a lower limit on the polar angle is set by the
limited detector acceptance in the forward region.
%
The experimental errors, dominated by the systematic uncertainties, are discussed in Sec.~\ref{subsec:pierrs}. 

The positive kaon production measurements 
were performed with a coarser data binning
and for a range of the kinematic variables which covers about 60$\%$ of the phase space relevant for T2K. 
Limitations were imposed by the available statistics and by the decreased 
sensitivity of the kaon identification at larger momenta as a consequence of 
the vanishing $K$/p and $K$/$\pi$ production ratios.
The maximum kinematic range considered is between
1.6 and 7.2 GeV/c in momentum and between 20 and 240 mrad for the polar angle (Fig.~\ref{fig:na61_coverage}).
The experimental errors on the $K^{+}$ production cross section, mainly dominated by the statistical uncertainties, 
are discussed in Sec.~\ref{subsec:kerrs}. 

The NA61/SHINE data cover most of the relevant hadron production phase space 
for the T2K flux, as illustrated in Fig.~\ref{fig:na61_coverage}, 
which shows the simulated momentum and production angle of pions and kaons 
that are produced in primary proton interactions and decay to contribute 
to the neutrino flux at SK.  
More than $90\%$ of the pion phase space is covered, 
and the K$^{+}$ data cover \textcolor{red}{60\%
}of the kaon phase space.

%
%
The importance of the NA61/SHINE future program of measurements 
is outlined in Sec.~\ref{subsec:sumhaderrs}.

The measurements of the differential kaon production by Eichten {\it et al.}~\cite{eichten}
and Allaby {\it et al.}~\cite{allaby} cover the forward production of high energy
kaons, which has not been measured yet by the NA61/SHINE experiment. These data are used to 
re-weight the model predictions in these regions. In addition, the
differential proton production measurements in these experiments are used to evaluate
systematic uncertainties in secondary nucleon production. 

The pion production data from the BNL-E910 experiment \cite{e910} is
used to evaluate systematic uncertainties associated with tertiary pion
production. 

\begin{table*}[ht]
\centering
\caption{Inelastic and production cross-section data used to re-weight 
hadron absorption probabilities.}
\begin{tabular}{lllll}
    \hline\hline
    Data &  Beam & Target & Beam Momentum (GeV/c) & Measurement \\
    \hline 
    Abrams \textit{et al.} \cite{Abrams} & $K^\pm$ & C, Cu & 1 -- 3.3 & $\sigma_\mtxtlbl{inel}$ \\
    Allaby \textit{et al.} \cite{Allaby:1970pv}\cite{Allaby1969500} & $\pi^-$, $K^-$ & C, Al, ... & 20 -- 65 & $\sigma_\mtxtlbl{inel}$ \\
    Allardyce \textit{et al.} \cite{Allardyce:1973ce} & $\pi^\pm$ & C, Al, ... & 0.71 -- 2 & $\sigma_\mtxtlbl{inel}$ \\
    Bellettini \textit{et al.} \cite{Bellettini1966609} & $p$ & C, Al, ... & 19.3, 21.5 & $\sigma_\mtxtlbl{inel}$ \\
    Bobchenko \textit{et al.} \cite{Bobchenko} & $\pi^-$, $p$ & C, Al, ... & 1.75 -- 9 & $\sigma_\mtxtlbl{inel}$ \\
    Carroll \textit{et al.} \cite{Carroll} & $\pi^\pm$, $K^\pm$, $p$ & C, Al, ... & 60 -- 280 & $\sigma_\mtxtlbl{prod}$ \\
    Cronin \textit{et al.} \cite{Cronin} & $\pi^-$ & C, Al & 0.73 -- 1.33 & $\sigma_\mtxtlbl{inel}$ \\
    Chen \textit{et al.} \cite{Chen} & $p$ & C, Al, ... & 1.53 & $\sigma_\mtxtlbl{inel}$ \\
    Denisov \textit{et al.} \cite{Denisov} & $\pi^\pm$, $K^\pm$, $p$ & C, Al, ... & 6 -- 60 & $\sigma_\mtxtlbl{inel}$ \\
    Longo  \textit{et al.} \cite{Longo} & $\pi^+$, $p$ & C, Al & 3 & $\sigma_\mtxtlbl{inel}$ \\
    NA61/SHINE \cite{Abgrall:2011ae}& $p$ & C & 31 & $\sigma_\mtxtlbl{prod}$ \\
    Vlasov \textit{et al.} \cite{Vlasov} & $\pi^-$ & C, Al & 2 -- 6.7  & $\sigma_\mtxtlbl{inel}$ \\
    \hline\hline
\end{tabular}
\label{tab:xprod}
\end{table*}

Measurements of the inelastic cross section for proton, pion, and
kaon beams with carbon and aluminum targets are used to re-weight 
particle interaction rates and absorption in the simulation. A summary of
these data is given in Table~\ref{tab:xprod}. The experiments typically
measure the inelastic cross section $\sigma_{inel}$ which is defined as 
the total cross section minus the elastic cross section.  Some 
experiments measure $\sigma_{prod}$, the production cross section, which is
defined here as:
\begin{equation}
\sigma_{prod} = \sigma_{inel} - \sigma_{qe}.
\label{eq:prod_xsec}
\end{equation}
Here, $\sigma_{qe}$ is the quasi-elastic scattering off of individual nuclei.
The production cross section represents the rate of interactions where hadrons are produced 
in the final state.


\subsubsection{\label{sec:hadron_prod_tune}Hadron differential production re-weighting}

The differential production re-weighting is evaluated using the differential multiplicity in the
momentum, $p$, of the produced particle and its angle, $\theta$, relative to the incident particle:
\begin{equation}
\frac{dn}{dp}(\theta,p_{in},A) = \frac{1}{\sigma_{prod}(p_{in},A)}\frac{d\sigma}{dp}(\theta,p_{in},A).
\label{eq:diffprod}
\end{equation}
The cross section $\sigma_{prod}(p_{in},A)$ depends on the incident particle momentum, $p_{in}$, 
and target nucleus, $A$.

\begin{figure}
\centering
\includegraphics[width=0.45\textwidth]{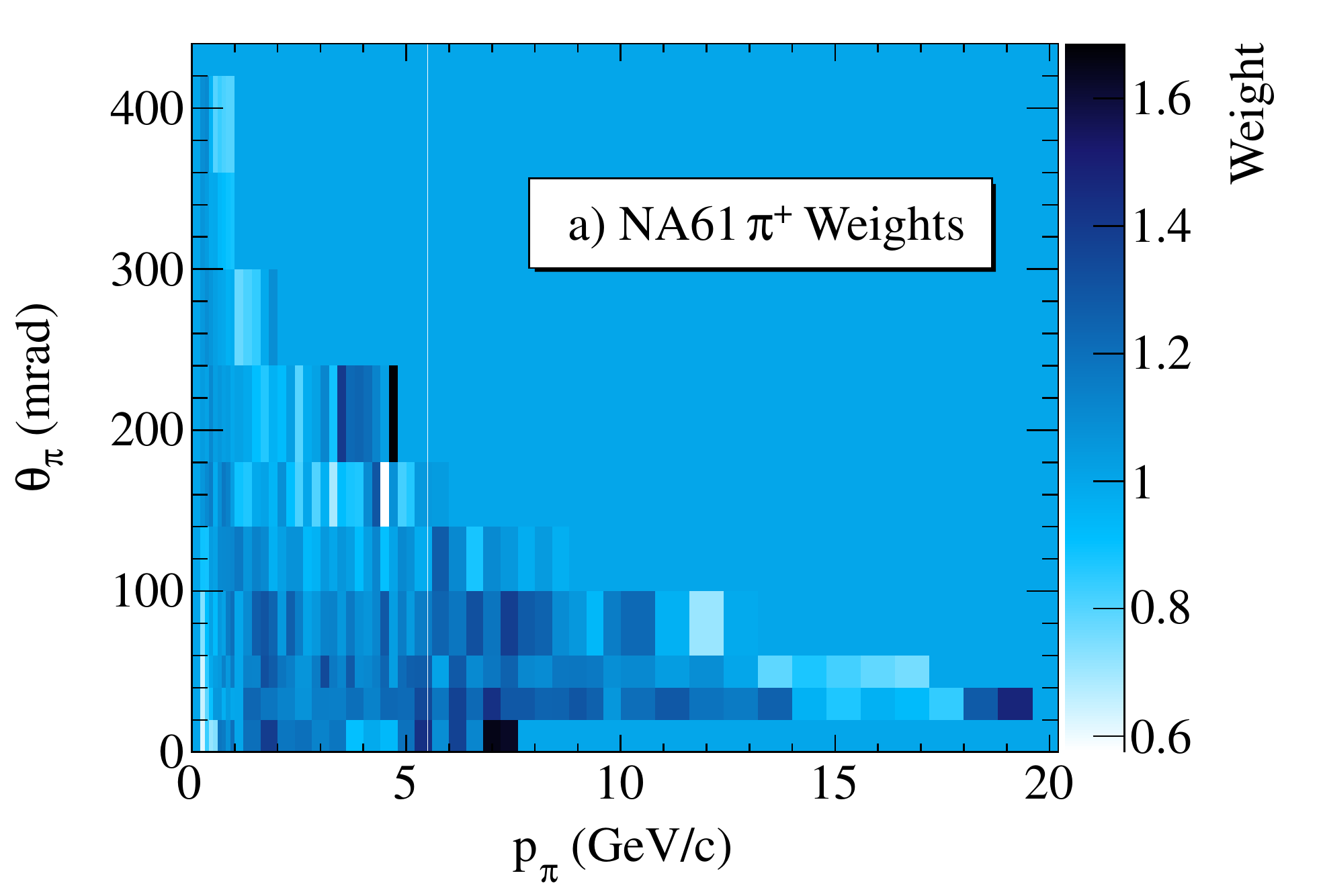}
\includegraphics[width=0.45\textwidth]{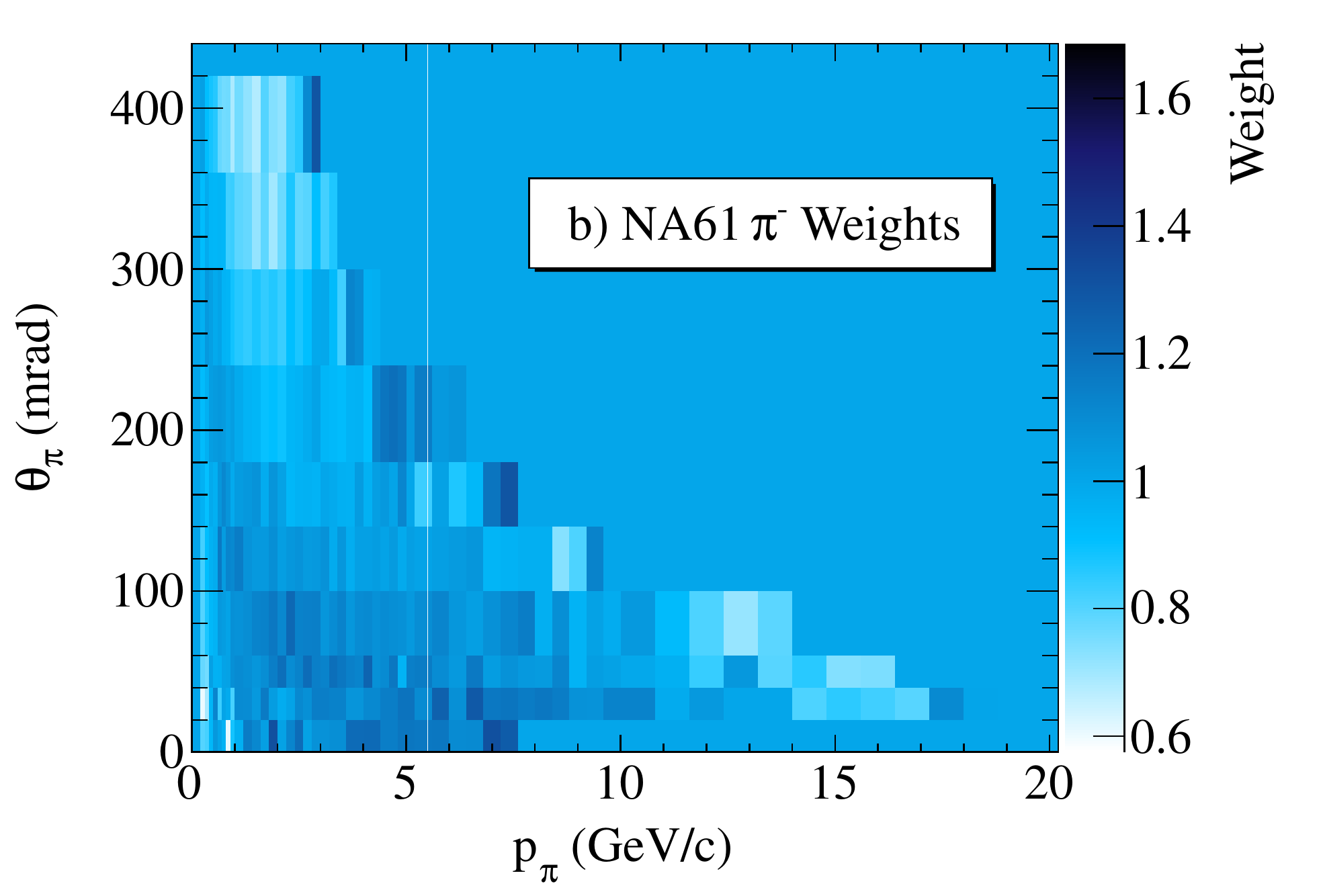}
\includegraphics[width=0.45\textwidth]{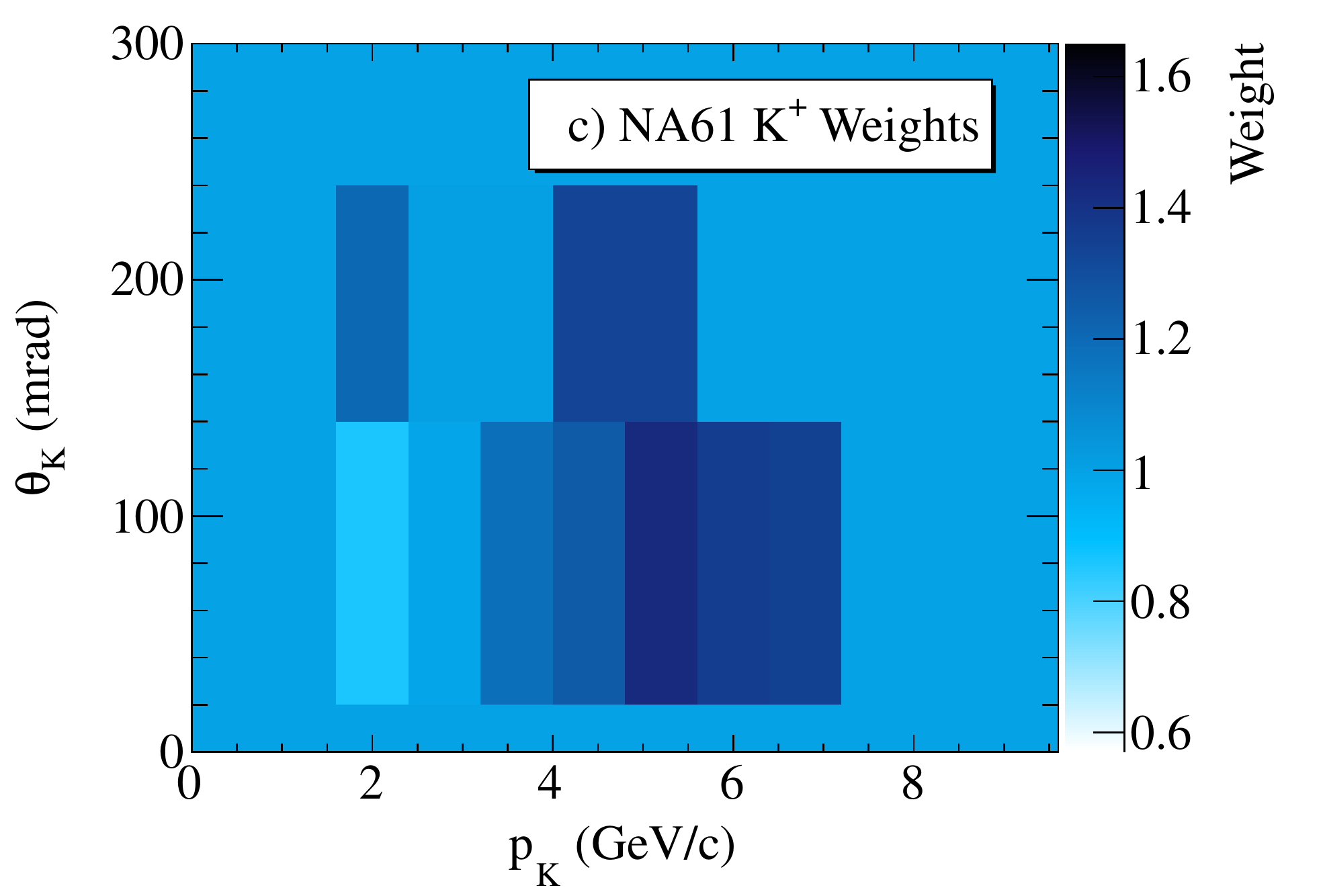}
\caption{The differential production weights from NA61/SHINE data for $\pi^{+}$ (top), 
$\pi^{-}$ (middle) and K$^{+}$ (bottom).}
\label{fig:na61tuning}
\end{figure}

The differential production weight that is applied to a given simulated interaction that produces hadrons
is the ratio of the production in data and simulation:
\begin{equation}
W(p_{in},A) = \frac{[\frac{dn}{dp}(\theta,p_{in},A)]_{data}}{{[\frac{dn}{dp}(\theta,p_{in},A)]_{MC}}}.
\label{eq:prodweight}
\end{equation}
For interactions of $31$ GeV/c protons on carbon that produce $\pi^{\pm}$ or $K^{+}$ in the phase space
covered by the NA61/SHINE data, the construction of the ratio in Eq.~\ref{eq:prodweight} is straightforward
since the differential production data provided is already in the form in Eq.~\ref{eq:diffprod}, at the correct
beam momentum, and on the correct target material.  The weights applied to differential production
in FLUKA simulated interactions are shown in Fig.~\ref{fig:na61tuning}.

The re-weighting of tertiary pion production from nucleon interactions requires extrapolations from the NA61/SHINE data to
lower incident nucleon momentum and other target materials, since tertiary production can happen in interactions
within the horns (aluminum).  Tertiary pions can also be produced in the interactions of secondary neutrons, in which
case data for the isospin symmetric reaction ($p+C\rightarrow \pi^{\pm}+X$ for
$n+C\rightarrow \pi^{\mp}+X$) are used to calculate weights.  
The same invariance is assumed for interactions on the Al nuclei, although
the isospin invariance of the nucleus is slightly broken.

The scaling of differential production rates to different incident nucleon momenta is carried out assuming 
Feynman scaling~\cite{Feynman69}.  The Feynman variable, $x_{F}$, is defined as:
\begin{equation}
x_{F} = \frac{p_{L}}{p_{L(max)}}, \label{eq:xf}
\end{equation}
where $p_{L}$ is the longitudinal momentum of the produced particle in the center of mass frame and 
$p_{L(max)}$ is the maximum allowed longitudinal momentum of the produced particle. The weights shown in 
Fig.~\ref{fig:na61tuning} are converted to an $x_{F},p_{T}$ dependence and applied to tertiary events
based on the $x_{F}$ and $p_{T}$ of those events.  This re-weighting method assumes perfect scaling, and the 
systematic effect is studied in Sec.~\ref{sec:hadronerrors} using data with lower incident particle momenta.

\begin{figure}
\centering
\includegraphics[width=0.47\textwidth]{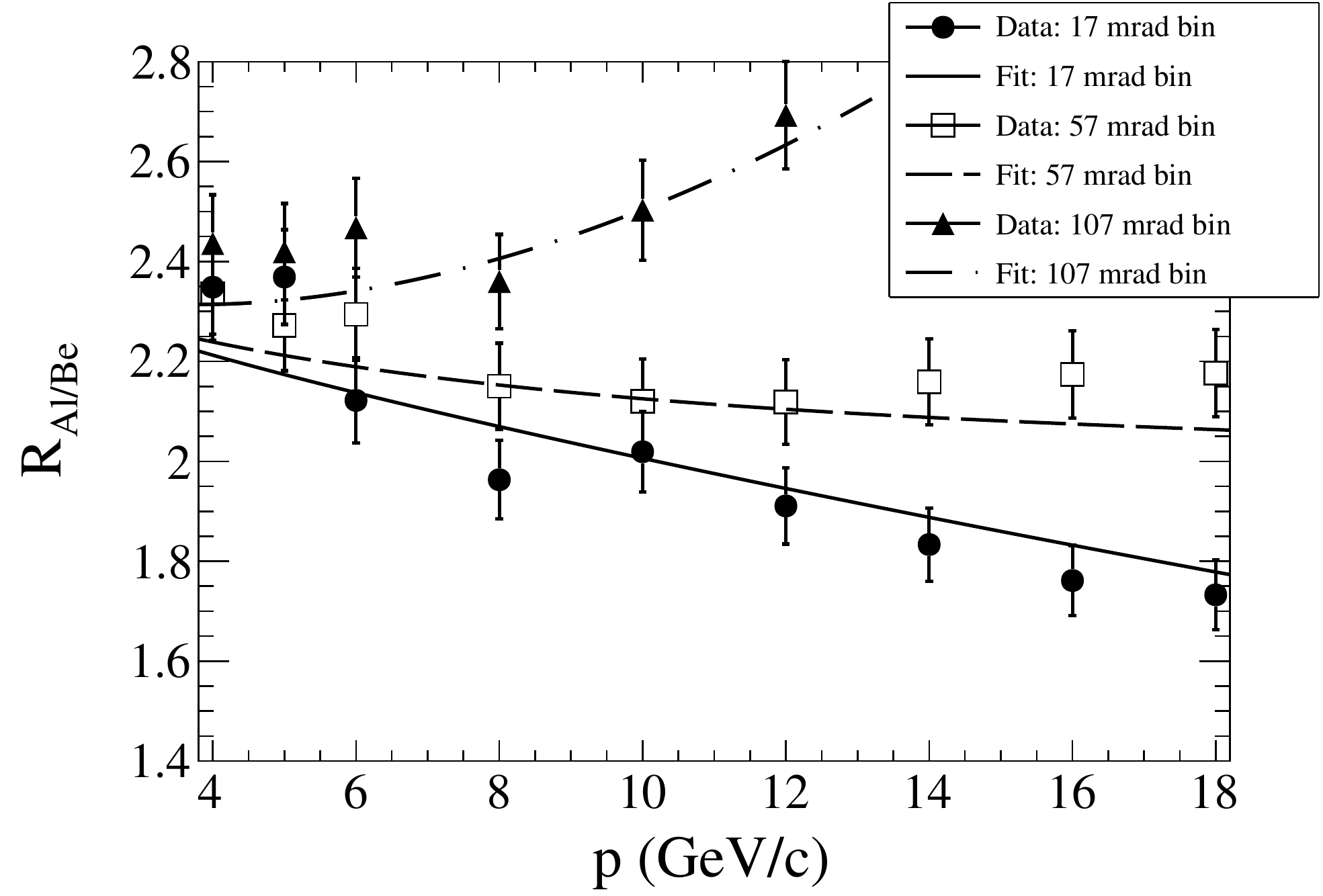}
\caption{Examples of the material scaling exponent $\alpha$ fit for a few angular bins in the
\cite{eichten} $K^+$ data.} 
\label{fig:a_scaling_fit}
\end{figure}

The NA61/SHINE data are also extrapolated from a carbon target to aluminum and used to re-weight interactions
in the horn material that are modeled in the GEANT3 (GCALOR) simulation.  The $A$-dependent scaling is carried
out using a parametrization proposed by Bonesini {\it et al.}~\cite{bmpt_paper} based on works by Barton {\it et al.}~\cite{barton} and Skubic {\it et al.}~\cite{skubic}:
\begin{equation}
E\frac{d^3\sigma(A_1)}{dp^3} = \left[\frac{A_1}{A_0}\right]^{\alpha(x_F,p_T)}E\frac{d^3\sigma(A_0)}{dp^3},
\end{equation}
where:
\begin{equation}
\alpha(x_F,p_T) = (a+bx_F+cx_F^2)(d+ep_T^2).
\end{equation}
The parameters $a$ through $e$ are determined by fitting the $A$-dependence in the data from 
Eichten {\it et al.}~\cite{eichten} and Allaby {\it et al.}~\cite{allaby}.  Examples of the 
fitted $A$ dependence for a few bins are shown in
Fig.~\ref{fig:a_scaling_fit}. In this figure, the ratio of the $K^+$ production from the Al target to 
that obtained from the Be target by \cite{eichten} is plotted at different momenta for three angular bins. 
The accuracy and precision of the scaling for the individual data points is discussed in Sec.~\ref{sec:hadronerrors}.
The fitted parameter values along with the values reported in \cite{bmpt_paper} are listed in 
Table~\ref{tab:a_scaling_params}.

\begin{table}
\begin{center}
\caption{Parameters for material scaling.}
\begin{tabular}{lccccc}
\hline
\hline
             &   a   &   b   &  c   &  d   &  e   \\
\hline
 Bonesini {\it et al.}~\cite{bmpt_paper} & 0.74  & -0.55 & 0.26 & 0.98 & 0.21 \\
 Fit to $\pi$ data & 0.75  & -0.52 & 0.23 & 1.0 (fixed)  & 0.21 \\
 Fit to  $K$ data  & 0.77  & -0.32 & 0.0  & 1.0 (fixed)  & 0.25 \\  
\hline
\hline
\end{tabular}
\label{tab:a_scaling_params}
\end{center}
\end{table}

\begin{figure}
\centering
\includegraphics[width=0.45\textwidth]{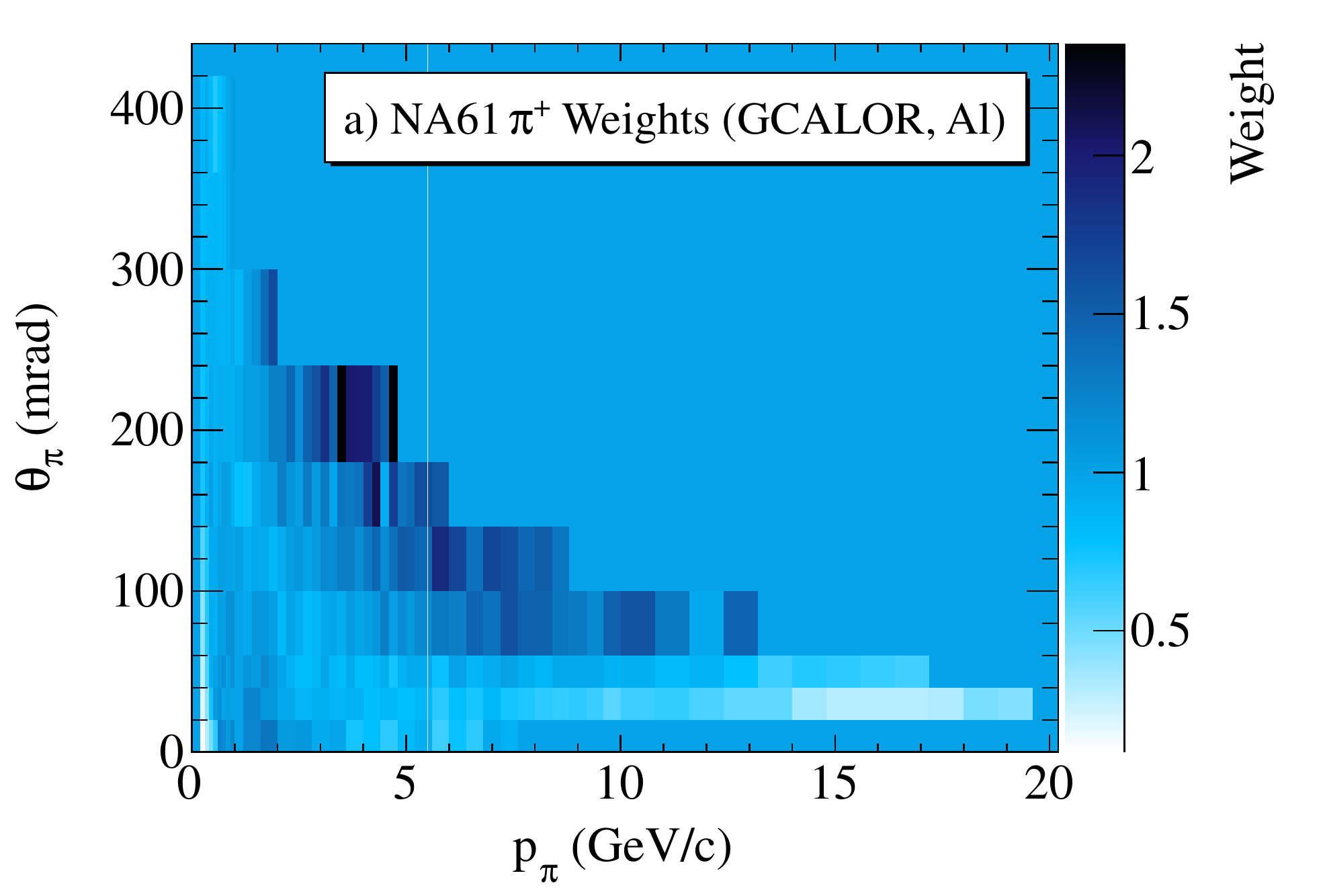}
\includegraphics[width=0.45\textwidth]{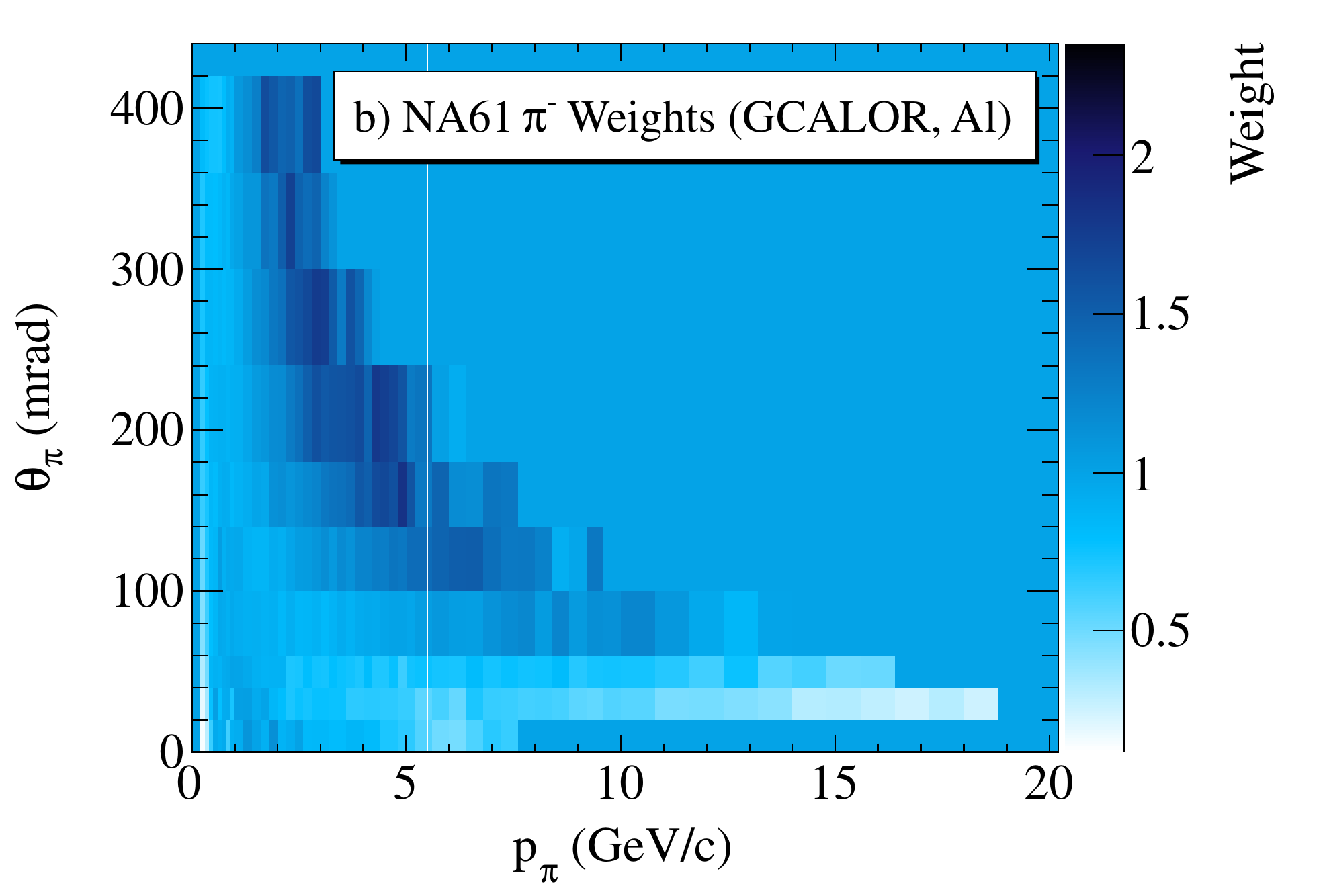}
\caption{The differential production weights for GCALOR from $A$-scaled NA61/SHINE data for $\pi^{+}$ (top), 
$\pi^{-}$ (bottom).}
\label{fig:gcalor_pi_tuning}
\end{figure}

The NA61/SHINE pion production data are scaled to aluminum using the parameters in
 Table~\ref{tab:a_scaling_params}, and the resulting weights applied to the production
in GCALOR are shown in Fig.~\ref{fig:gcalor_pi_tuning}.  The weights are calculated for
GCALOR, since the simulation of interactions in the horn material is done with GEANT3.

\begin{figure*}
\centering
\includegraphics[width=0.45\textwidth]{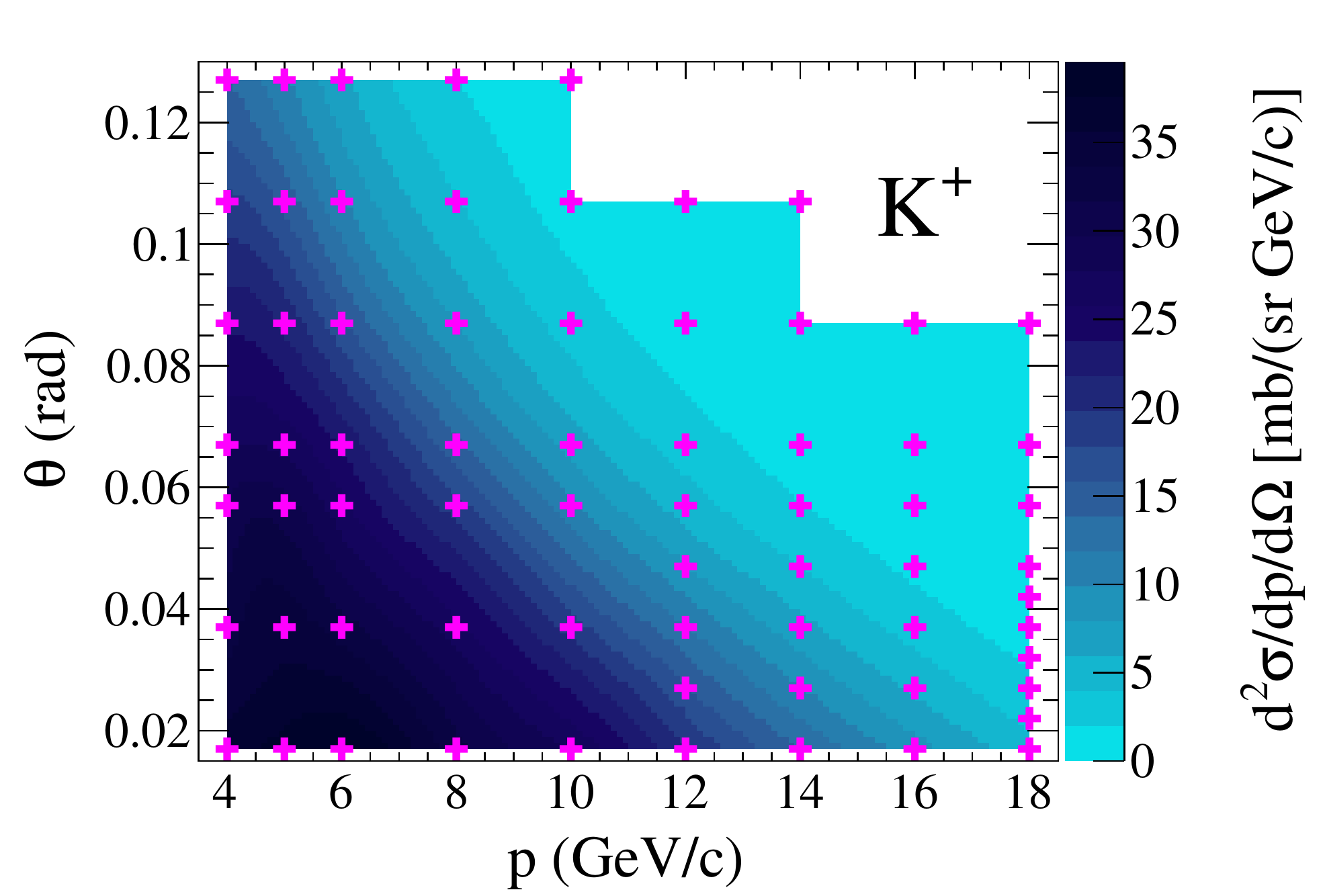}
\includegraphics[width=0.45\textwidth]{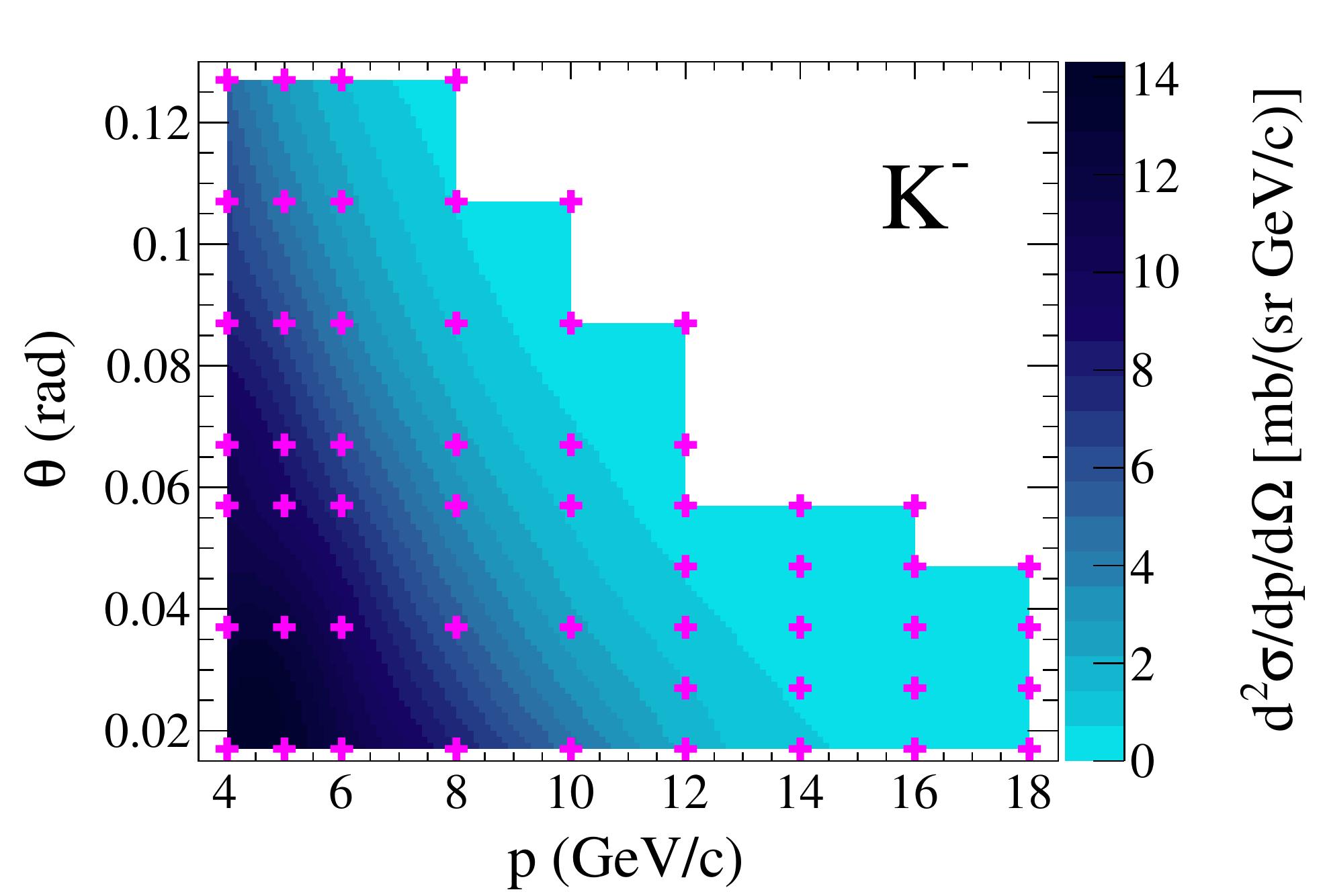}
\includegraphics[width=0.45\textwidth]{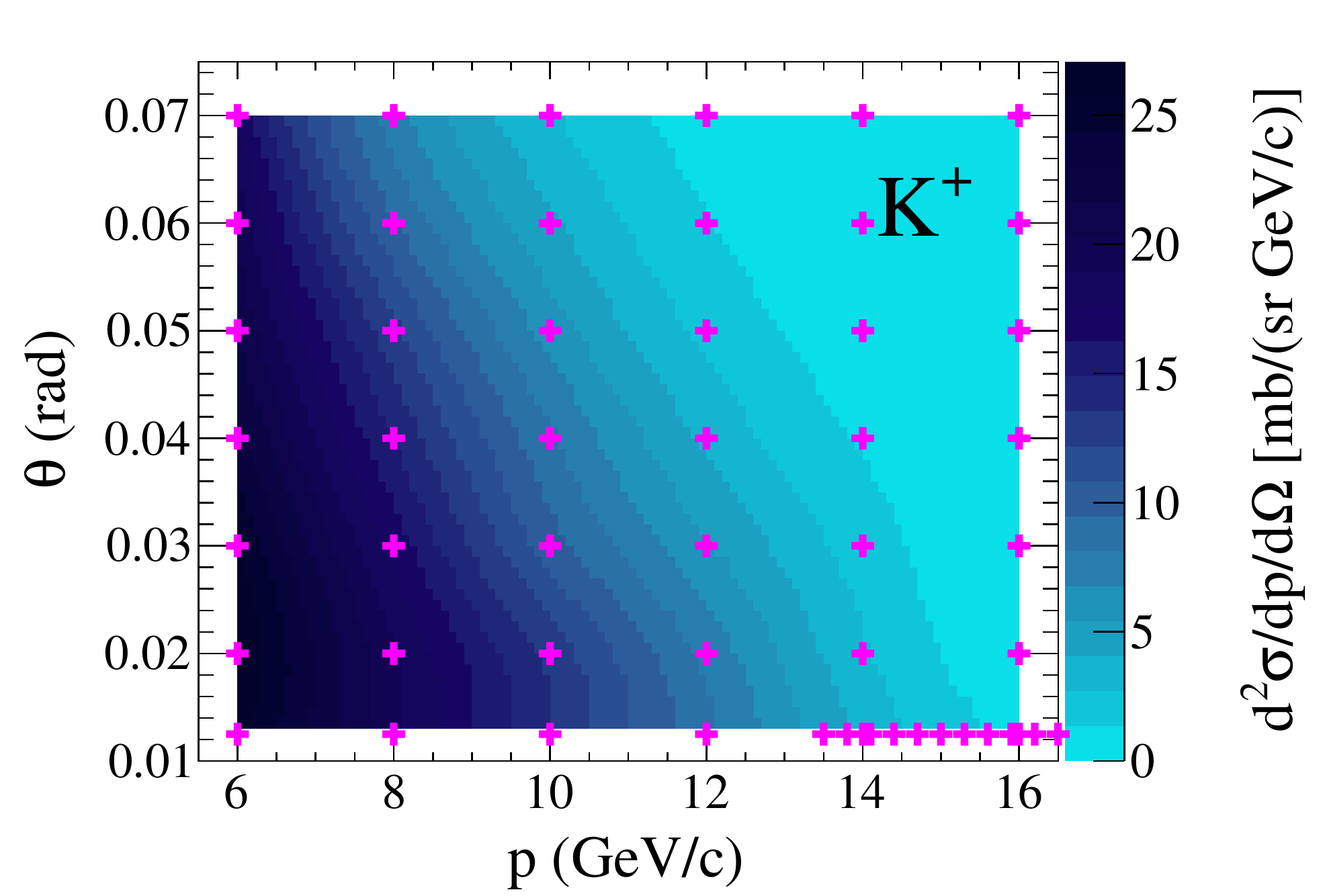}
\includegraphics[width=0.45\textwidth]{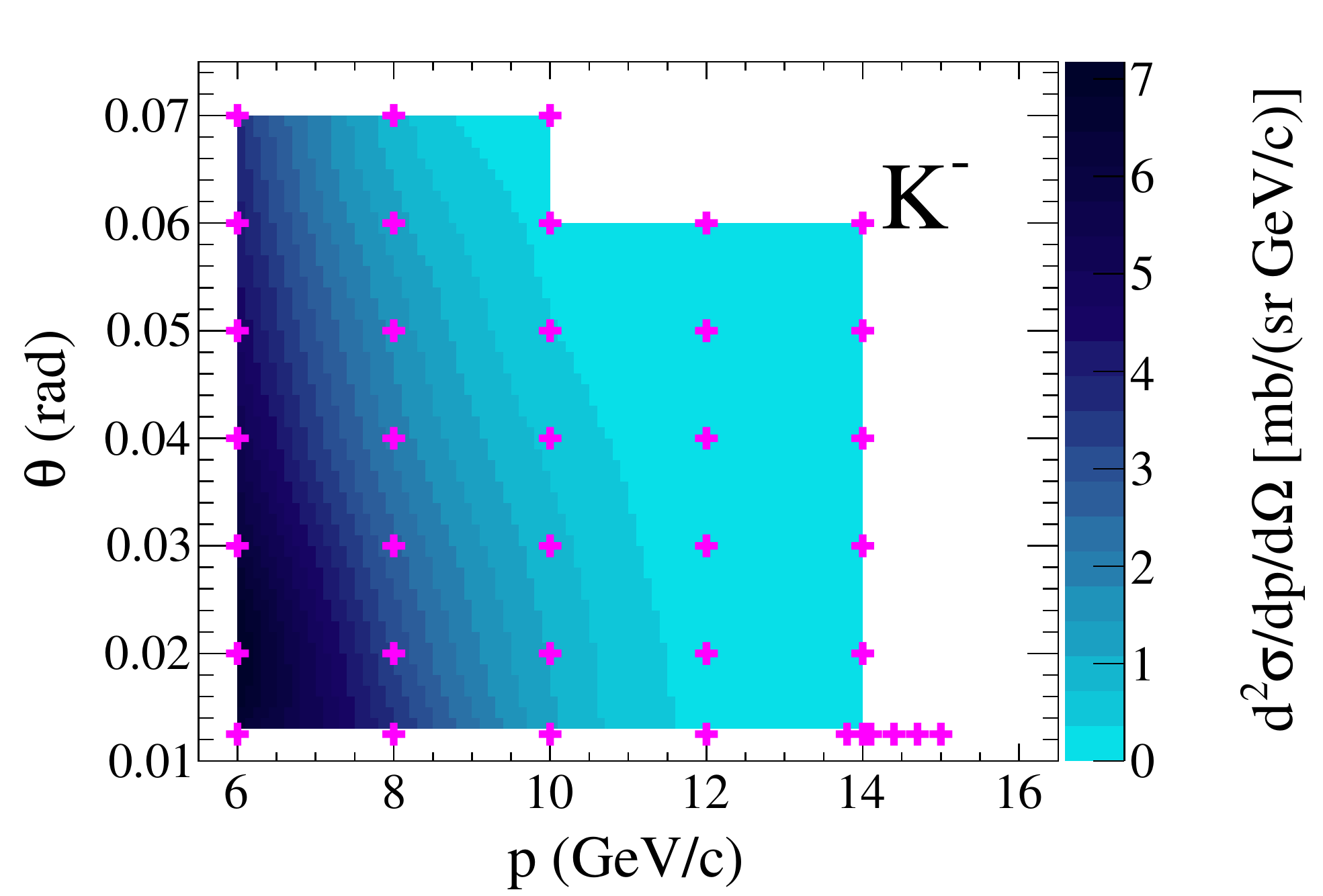}
\caption{The interpolated kaon production double differential cross section measurements of Eichten {\it et al.}~\cite{eichten} (top) and Allaby {\it et al.}~\cite{allaby} (bottom). The markers indicate the locations of the data points.}
\label{fig:kaon_spline_xsec}
\end{figure*}

The re-weighting of $K^{+}$ and $K^{-}$ production in the phase space not covered by NA61/SHINE is carried out using the Eichten {\it et al.}~\cite{eichten} 
and Allaby {\it et al.}~\cite{allaby} kaon production data.  Since these data sets only 
measure the differential production at points that cover a small momentum and angular range,
it is necessary to interpolate between the data points to find the weights for intermediate
simulated production.  A bi-cubic spline interpolation is performed to each data set separately, 
and the resulting differential production cross sections are shown in Fig.~\ref{fig:kaon_spline_xsec}.

\begin{figure}
\centering
\includegraphics[width=0.45\textwidth]{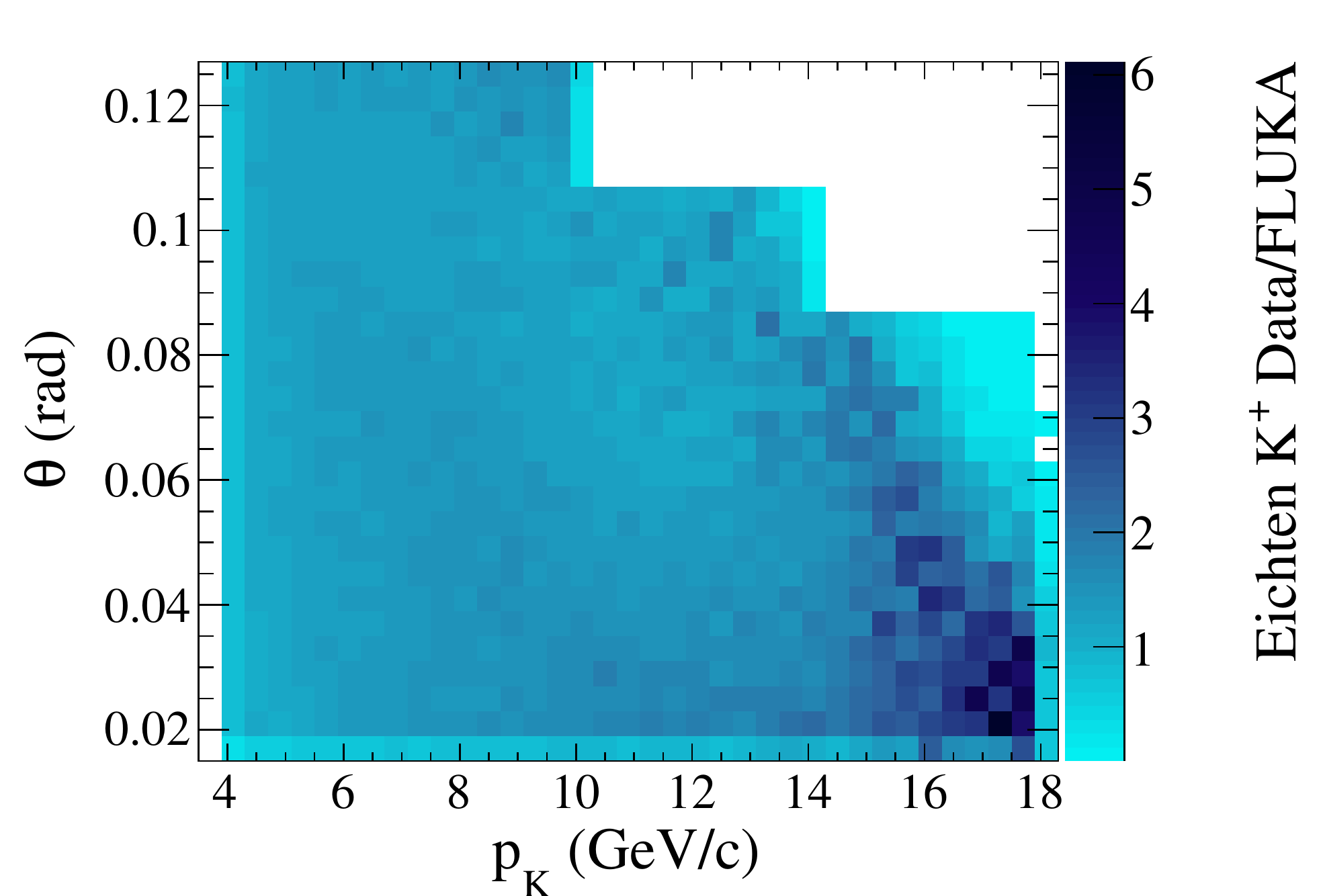}
\includegraphics[width=0.45\textwidth]{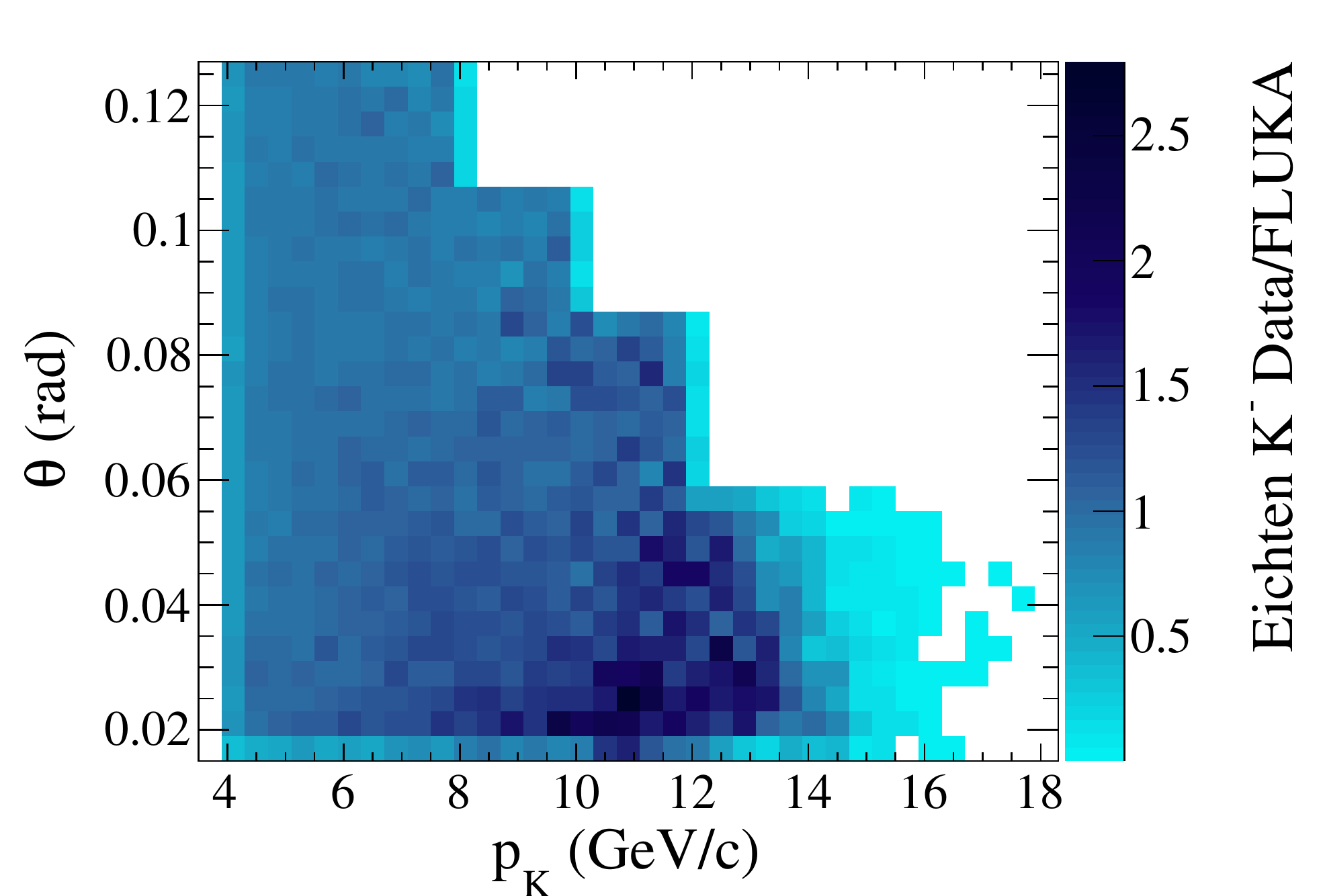}
\includegraphics[width=0.45\textwidth]{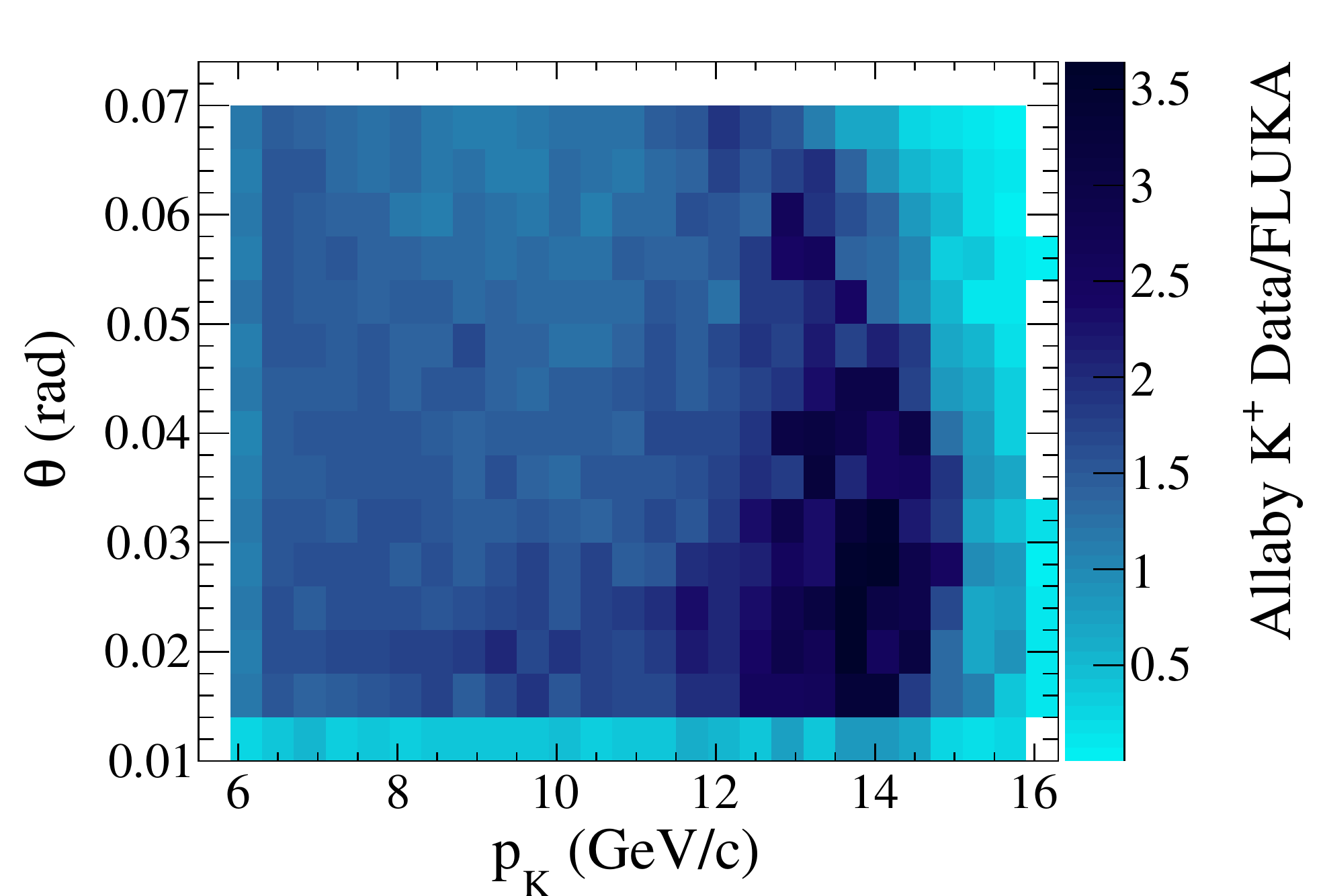}
\includegraphics[width=0.45\textwidth]{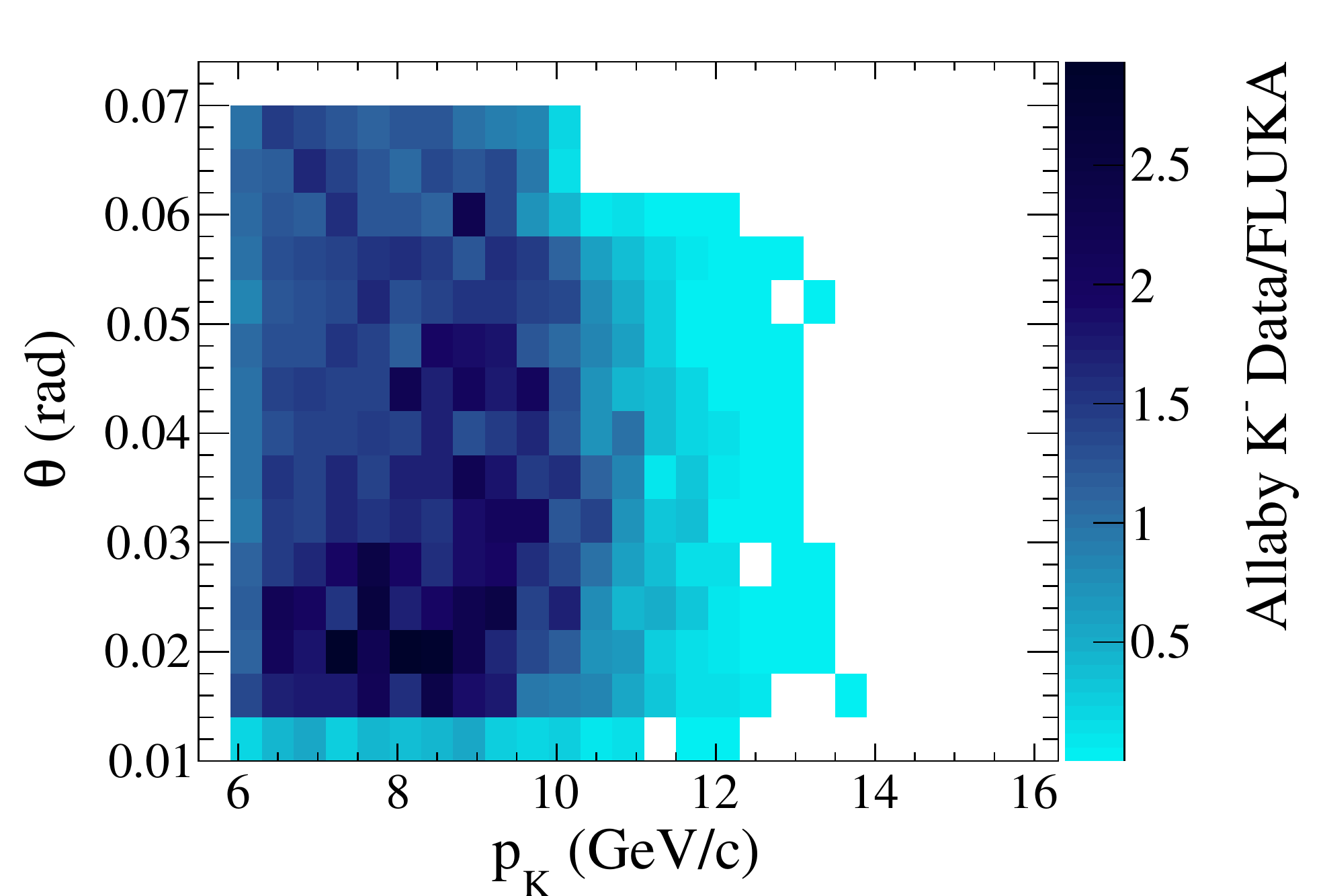}
\caption{Ratios of the interpolated charged kaon double differential production multiplicity measurements from Eichten and Allaby over the FLUKA predicted yields from Be for 24 GeV/c and 19.2 GeV/c proton beam momenta, respectively.}
\label{fig:kaon_spline_tune}
\end{figure}

Since these data sets do not include points on carbon, the data on Be are compared to 
the FLUKA prediction for Be at the same incident particle momentum as the data set.  The ratios
of the data and FLUKA predictions are evaluated and the corresponding distributions of
weights from each data set are shown in Fig.~\ref{fig:kaon_spline_tune}.

The weights in Fig.~\ref{fig:kaon_spline_tune} are converted to the $x_{F}-p_{T}$ basis
and applied 
assuming $x_{F}$ scaling.  The Eichten data are used in regions covered by
that data set, but not covered by the NA61/SHINE $K^{+}$ data.  The Allaby data are used in regions covered by that data set, but  
not covered by either NA61/SHINE $K^{+}$ data or the Eichten data. For regions not covered by
any data, no re-weighting is applied and the effect is studied as part of the uncertainty, as discussed
in Sec.~\ref{subsec:kerrs}. 

The $K^0_L$ multiplicity is calculated from the Eichten and Allaby data using a simple quark parton 
model (QPM)~\cite{QPM_Feynman, QPM_Bjorken}. Assuming the following conditions on the number densities of sea and valence quarks
\begin{subequations}
\label{eq:quarkcond}
\begin{equation}
u_{s} = \bar{u}_{s} = d_{s} = \bar{d}_{s}\mbox{, } s_{s} = \bar{s}_{s}
\end{equation}
\begin{equation}
n \equiv u_{v}/d_{v} = 2,
\end{equation}
\end{subequations}
a relation between the number of produced $K_{L}^{0}$ ($K_{S}^{0}$), $K^{+}$, and $K^{-}$ can be established: 
\begin{equation} \label{eq:QPM}
  N(K_{L}^{0})
  = N(K_S^{0})
  = \frac{N(K^{+}) + 3N(K^{-})}{4}.
\end{equation}

\begin{figure}
\centering
\includegraphics[width=0.45\textwidth]{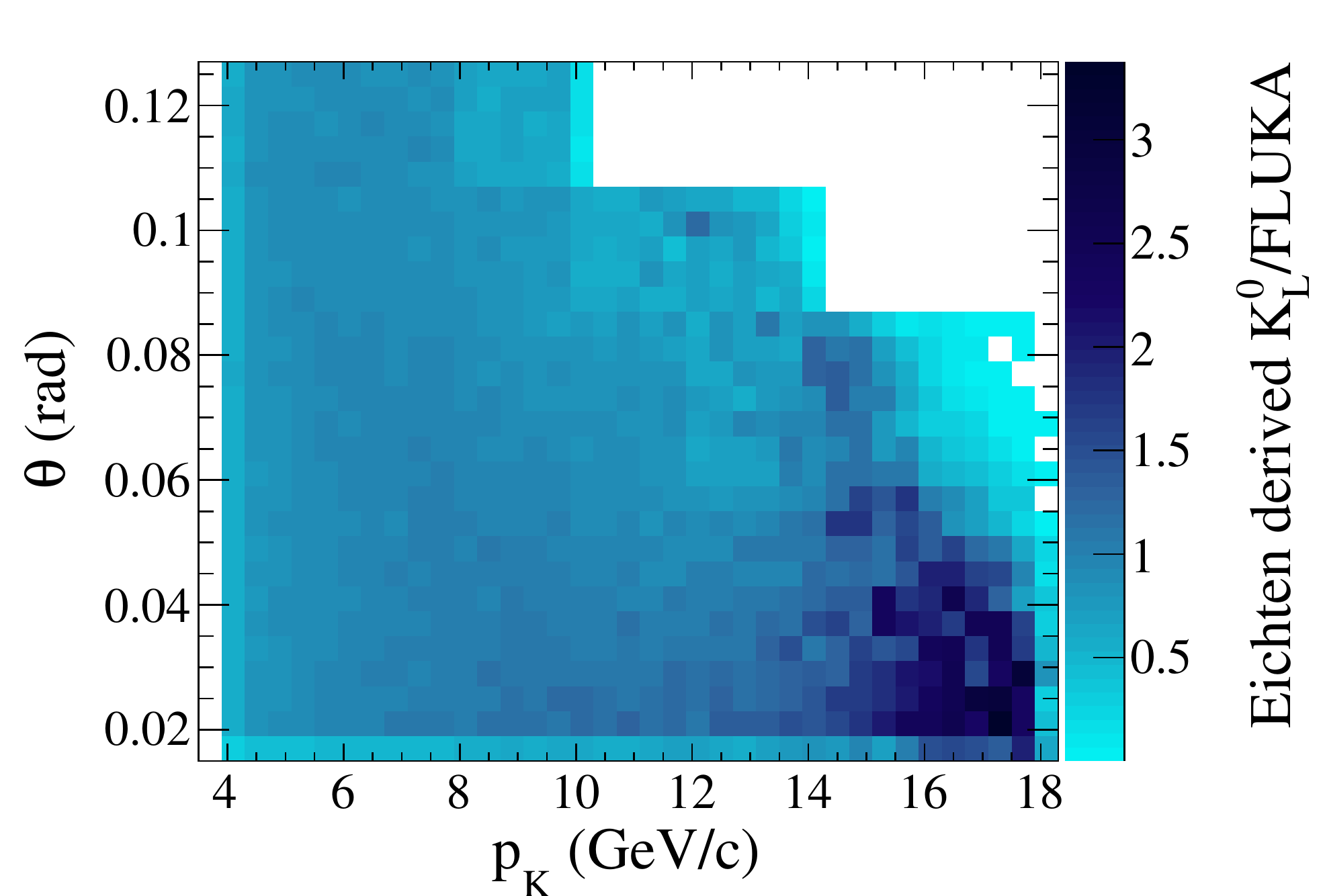}
\includegraphics[width=0.45\textwidth]{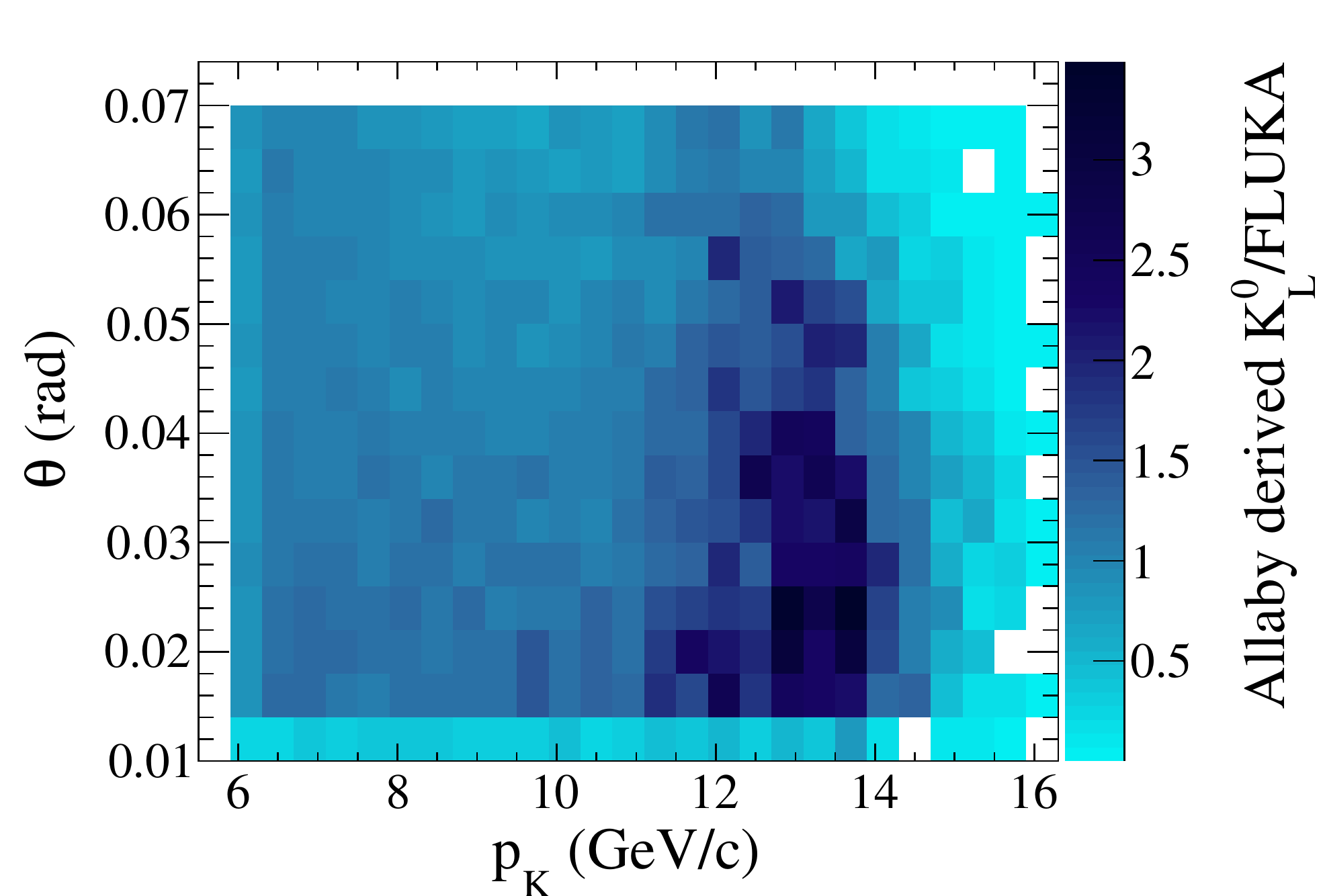}
\caption{Ratios of the interpolated $K^0_L$ double differential production multiplicity derived from Eichten and Allaby over the FLUKA predicted yields from Be for 24 GeV/c and 19.2 GeV/c proton beam momenta, respectively.}
\label{fig:klong_tune}
\end{figure}

After calculating the $K^0_L$ production according to Eq.~\ref{eq:QPM}, the $K^0_L$ multiplicity is re-weighted in the same manner as in the case of $K^{\pm}$.  The weights are shown in Fig.~\ref{fig:klong_tune}.

Although Eq.~\ref{eq:QPM} is only strictly valid 
for proton-proton collisions ($n=2$), 
the effect of proton-neutron ($n=1$) interactions leads to 
only small changes in the flux predictions that are $<1\%$. 
It is, therefore, not considered at this time. 

\subsubsection{\label{sec:hadron_int_tune}Hadron interaction rate re-weighting}

\begin{figure}
\centering
\includegraphics[width=0.45\textwidth]{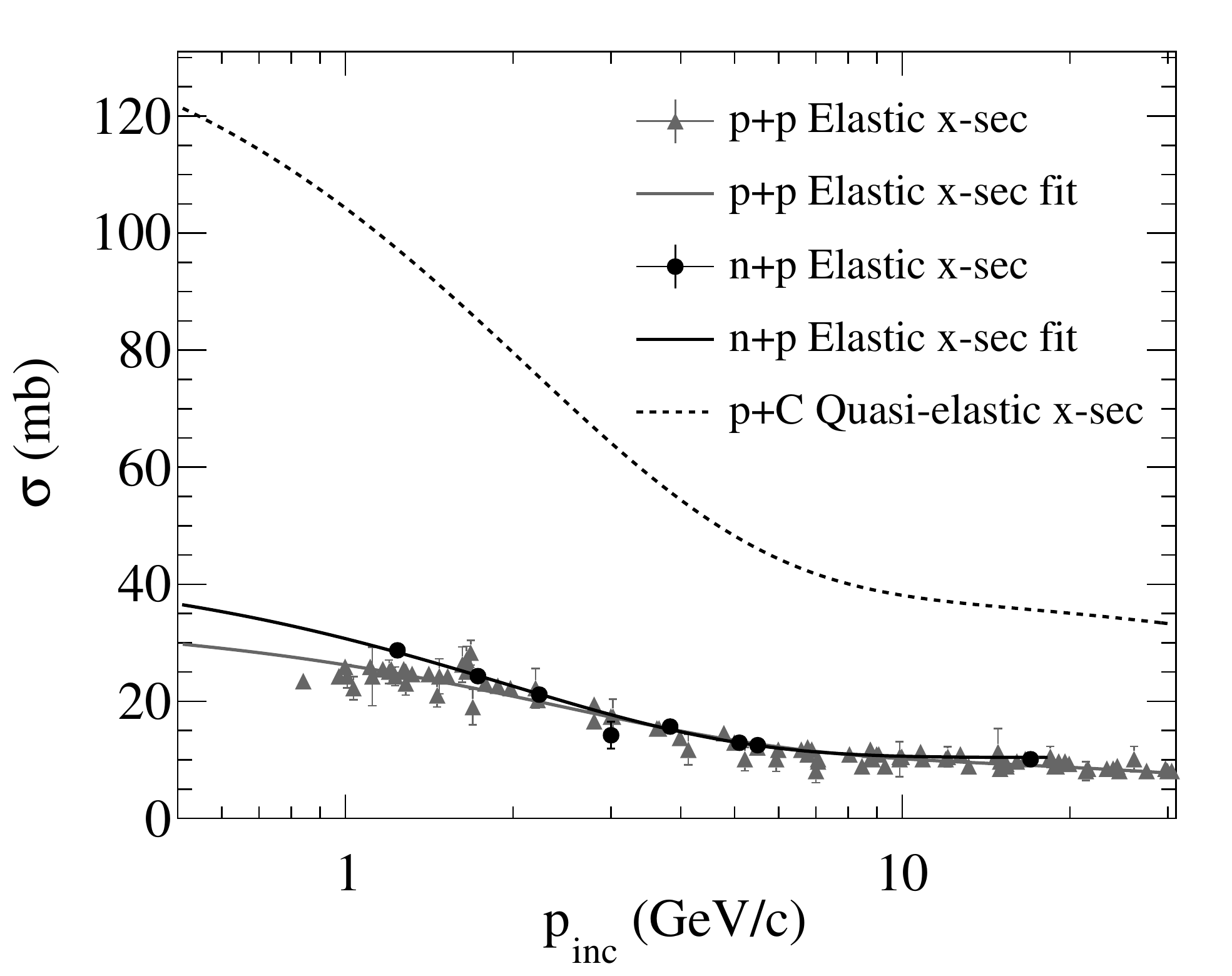}
\caption{The elastic cross sections for protons scattering on protons and neutrons and the derived quasi-elastic cross section for a carbon target.}
\label{fig:pC_qe_xsec}
\end{figure}

In addition to re-weighting the multiplicity of hadrons produced in interactions of nucleons,
it is necessary to re-weight the rate at which hadrons interact.  The quantity that is re-weighted is 
the production cross section defined in Eq.~\ref{eq:prod_xsec}.  Many experiments, however, measure
the inelastic cross section which includes the quasi-elastic component.  To carry out re-weighting
based on the $\sigma_{prod}$, the quasi-elastic cross section must be subtracted from the measurements.
The quasi-elastic cross section is extrapolated from hadron+nucleon scattering
data using a modification of the empirical dependence derived by Bellettini {\it et al.}~\cite{Bellettini1966609}:
\begin{equation}
\sigma_{qe} = 0.8(\sigma^{el}_{hp}+\sigma^{el}_{hn})A^{1/3}.
\end{equation}
Here $\sigma^{el}_{hp}$ and $\sigma^{el}_{hn}$ are the elastic cross sections of the hadron $h$ on the
proton and neutron respectively.  The formula is modified from Bellettini {\it et al.} to include the
average of the elastic cross section on the proton and neutron instead of the proton only.  The 
quasi-elastic cross section evaluated for proton interactions on carbon in this manner is shown in
Fig.~\ref{fig:pC_qe_xsec}.  The value of $\sigma_{qe}=33.1$~mb is slightly higher than the value that NA61/SHINE 
derived, $\sigma_{qe}=27.9\pm1.5$ $(sys)$~mb~\cite{Abgrall:2011ae}, using Glauber
model calculations.  As discussed in Sec.~\ref{subsec:xsecerrs}, the uncertainties on the weights 
are conservatively set to the magnitude of the quasi-elastic correction used to derive the production
cross section.

The re-weighting of the interaction rate models the change in the survival probability of particles as the
cross section changes, as well as the change in rate at a given interaction point.  The probability that
a particle with hadron production cross section of $\sigma_{prod}$ travels a distance $x$ and 
interacts in the infinitesimal length $\Delta x$ to produce hadrons is:
\begin{eqnarray}
P(x;\sigma_{prod}) &=& \int_{x}^{x+\Delta x} \sigma_{prod}\rho e^{-x'\sigma_{prod}\rho} dx' \\
&=& \Delta x \sigma_{prod}\rho e^{-x\sigma_{prod}\rho}.
\end{eqnarray}
Here, $\rho$ is the density of nuclear targets in the material.  When the production cross section changes,
$\sigma_{prod} \rightarrow \sigma_{prod}'$, the weight applied to model the change is the ratio of the probabilities:
\begin{eqnarray}
W &=& \frac{P(x;\sigma_{prod}')}{P(x;\sigma_{prod})} \\
  &=& \frac{\sigma_{prod}'}{\sigma_{prod}} e^{-x(\sigma_{prod}'-\sigma_{prod})\rho}.
\label{eq:intlwght}
\end{eqnarray}
The first factor in Eq.~\ref{eq:intlwght} is the change in interaction probability at that point, while the second factor is the attenuation of the particle flux over the distance traveled.  For a particle that decays before interacting, the weight applied is:
\begin{equation}
W = e^{-x(\sigma_{prod}'-\sigma_{prod})\rho}.
\end{equation}

\begin{figure*}
\centering
\includegraphics[width=0.45\textwidth]{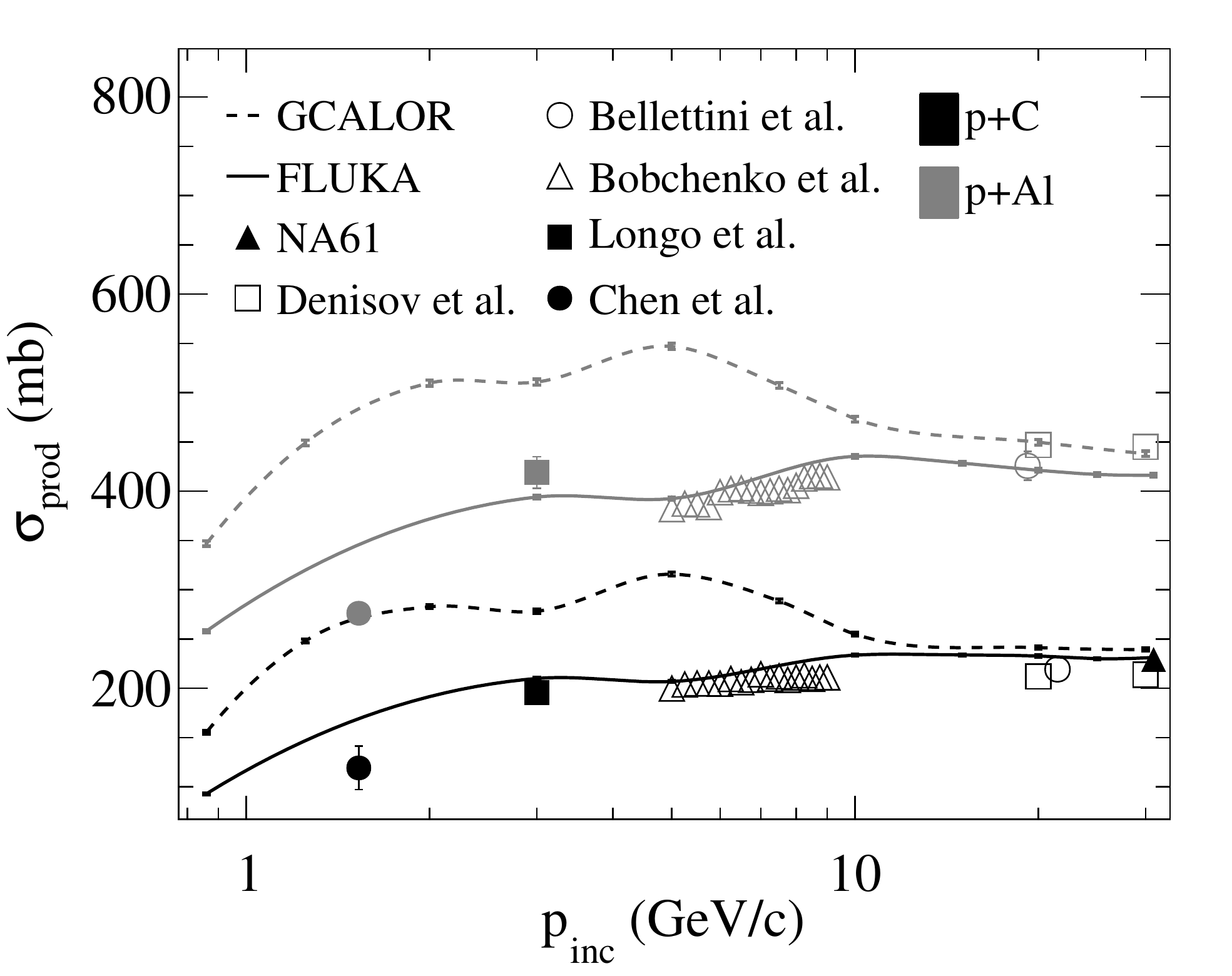}
\includegraphics[width=0.45\textwidth]{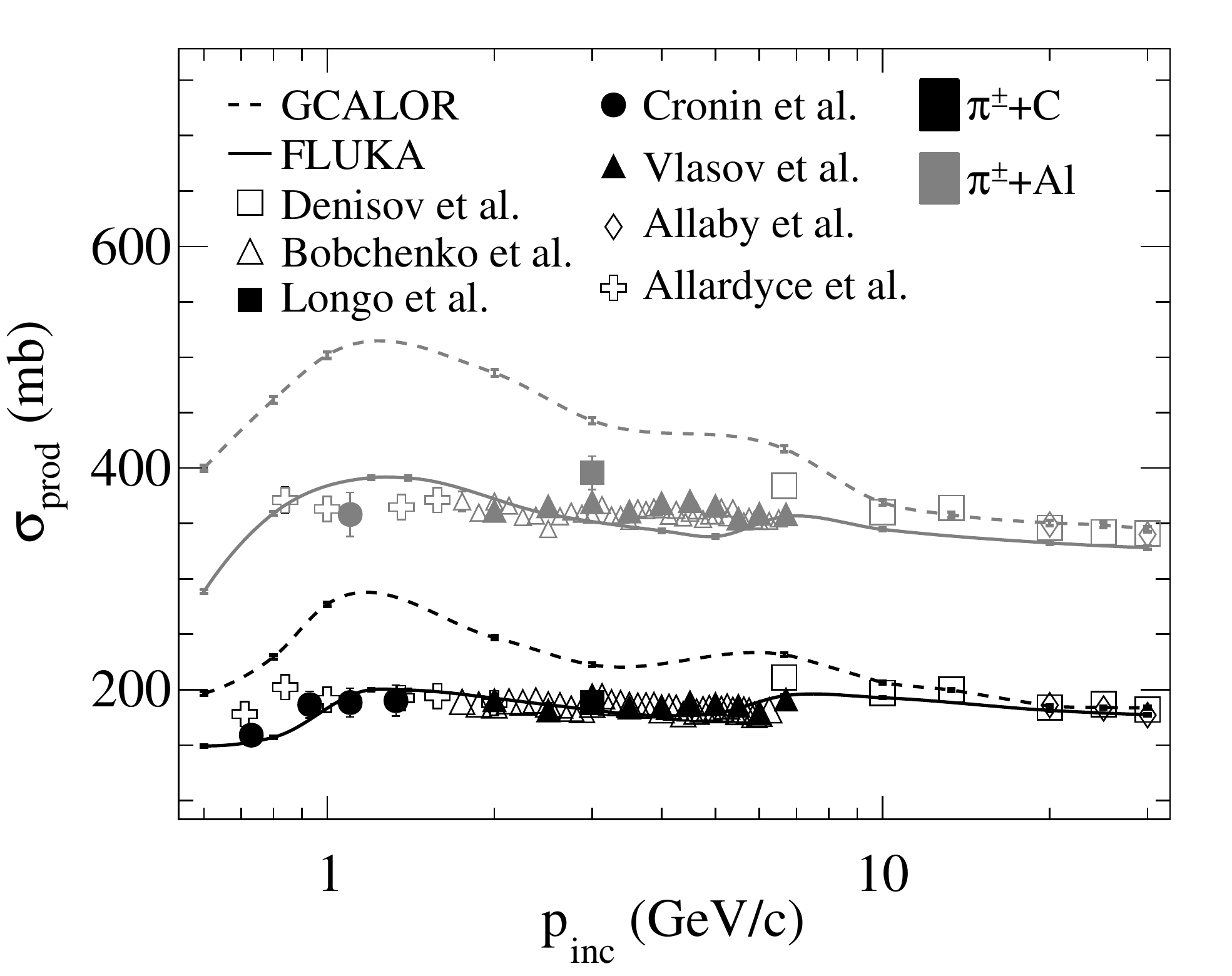}
\includegraphics[width=0.45\textwidth]{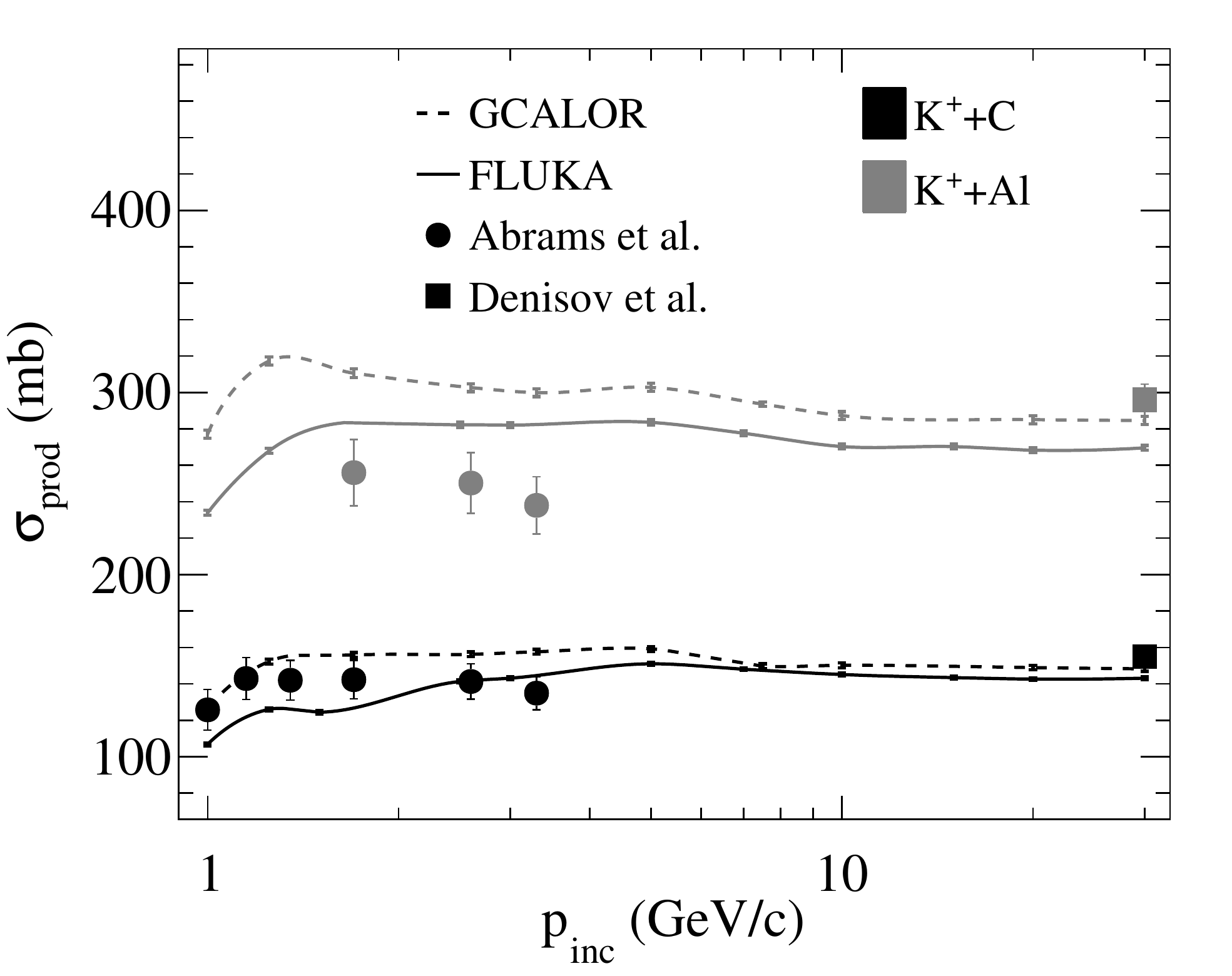}
\includegraphics[width=0.45\textwidth]{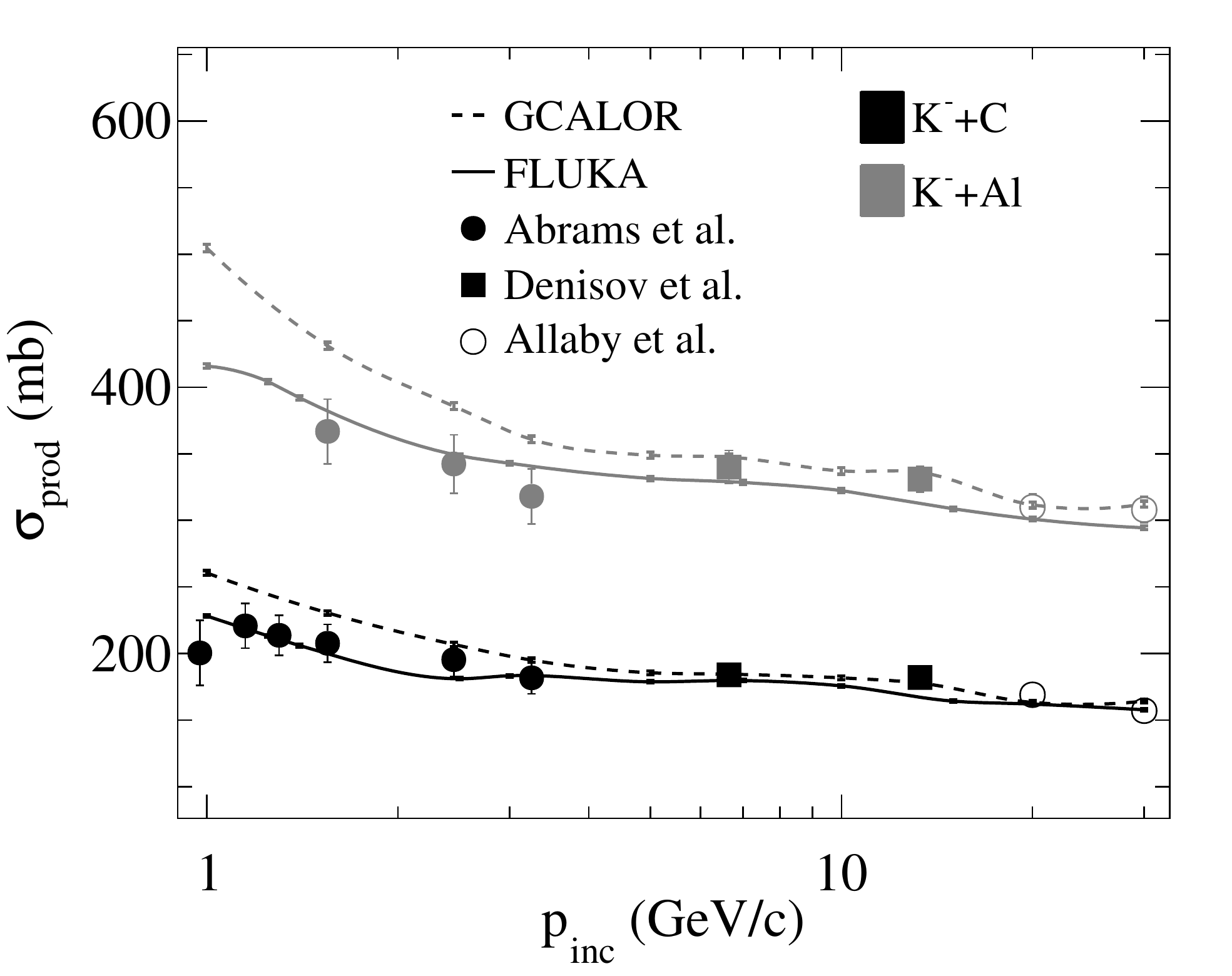}
\caption{Comparisons of $\sigma_{prod}$ measurements and the values used in the simulation (solid line for FLUKA
and dashed line for GCALOR), for incident
protons (top left) and charged pions (top right), $K^{+}$ (bottom left) and $K^{-}$ (bottom right).}
\label{fig:xsec_comp}
\end{figure*}

The comparison of data and simulated cross sections in Fig.~\ref{fig:xsec_comp} shows that FLUKA is in 
good agreement with the data, while GEANT3 (GCALOR) has significant disagreement at low incident particle
momenta. 
For both the simulated cross section and the data, where applicable, the quasi-elastic cross sections are subtracted. 
Therefore, no weights are applied to the FLUKA simulation of interactions in the target, but the 
GEANT3 (GCALOR) production cross sections are re-weighted to the FLUKA value.

\subsubsection{Hadron interaction re-weighting summary}
\begin{figure*}[ht]
\centering
\includegraphics[width=0.45\textwidth]{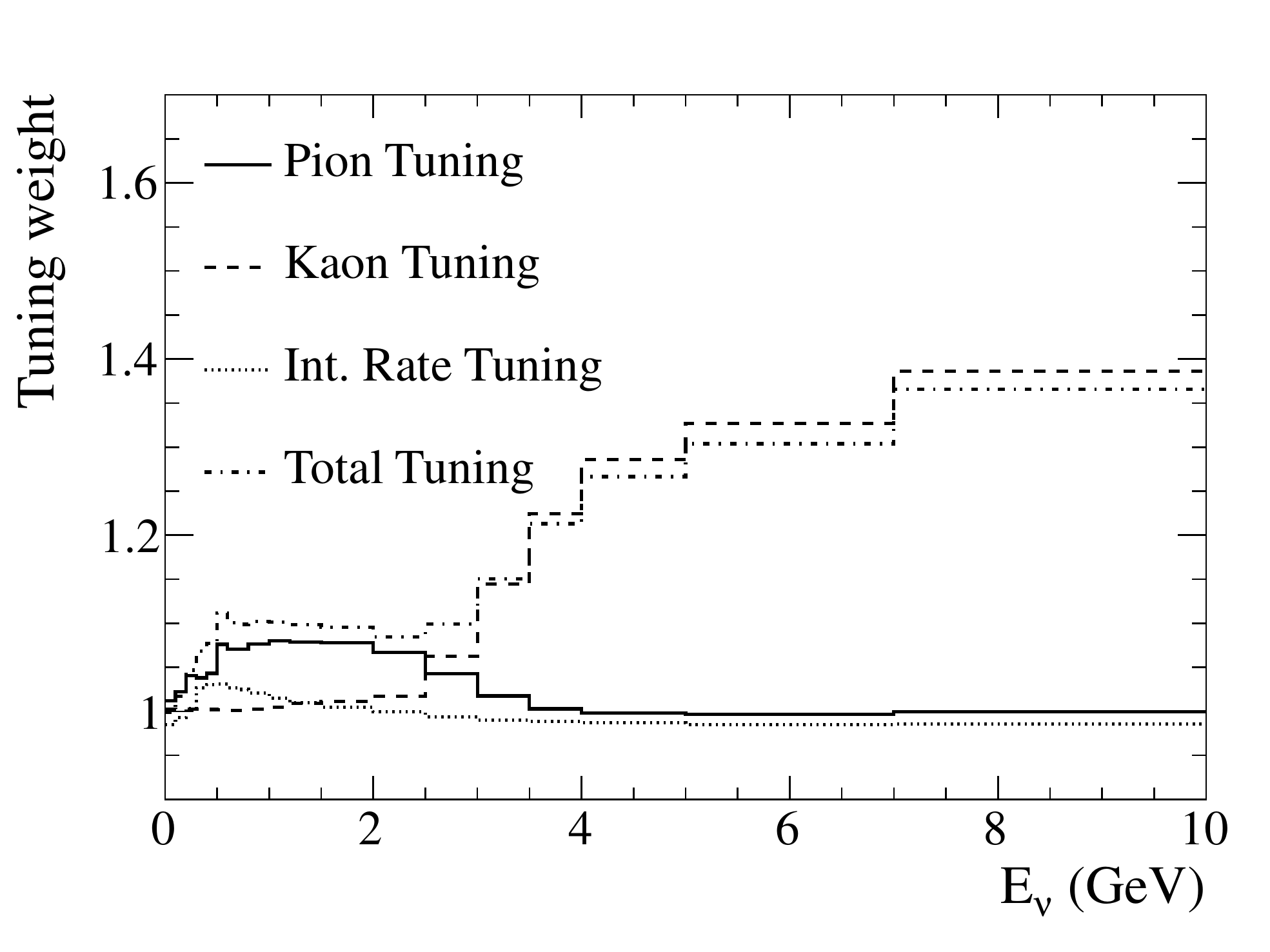}
\includegraphics[width=0.45\textwidth]{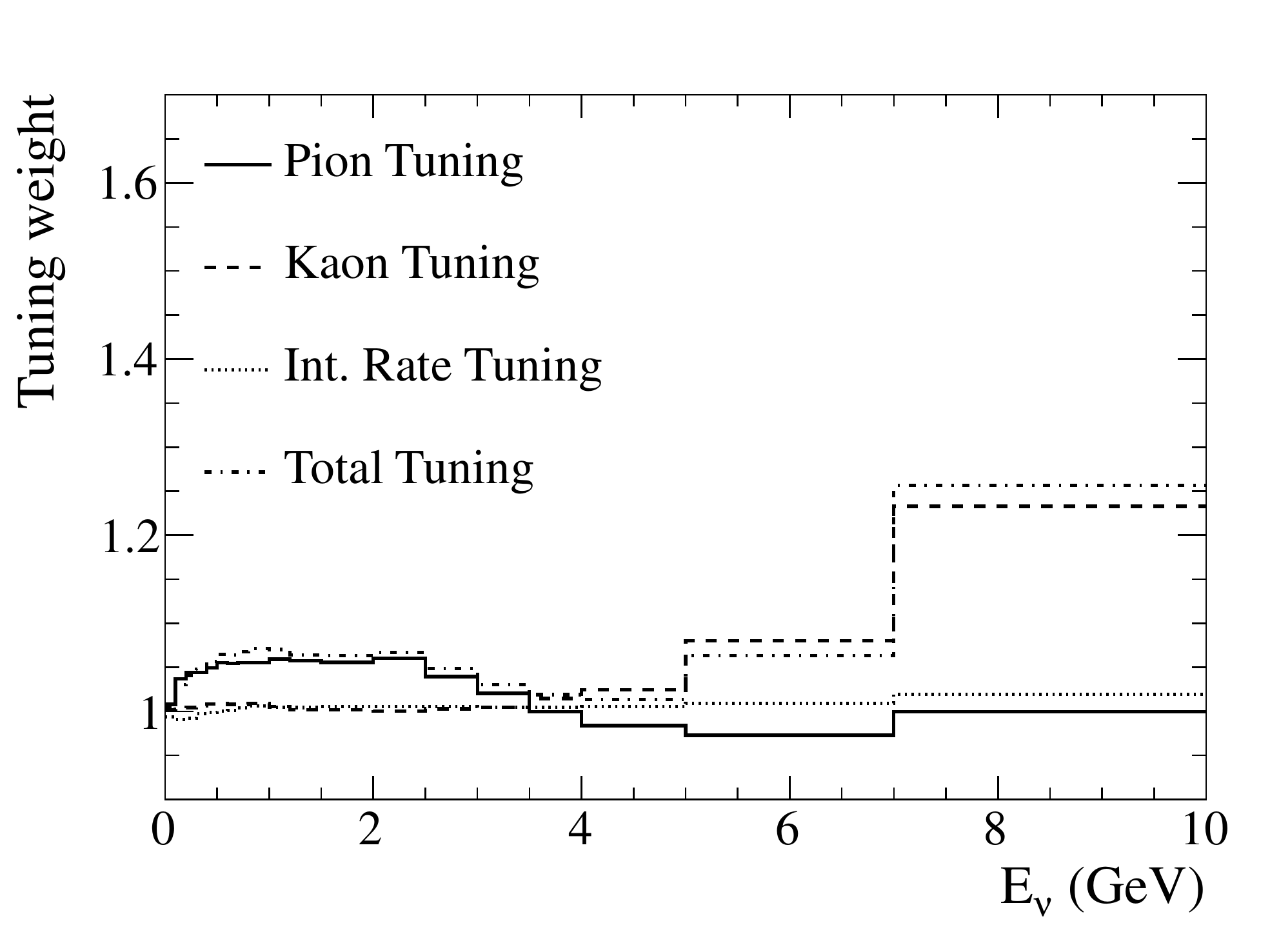} \\
\includegraphics[width=0.45\textwidth]{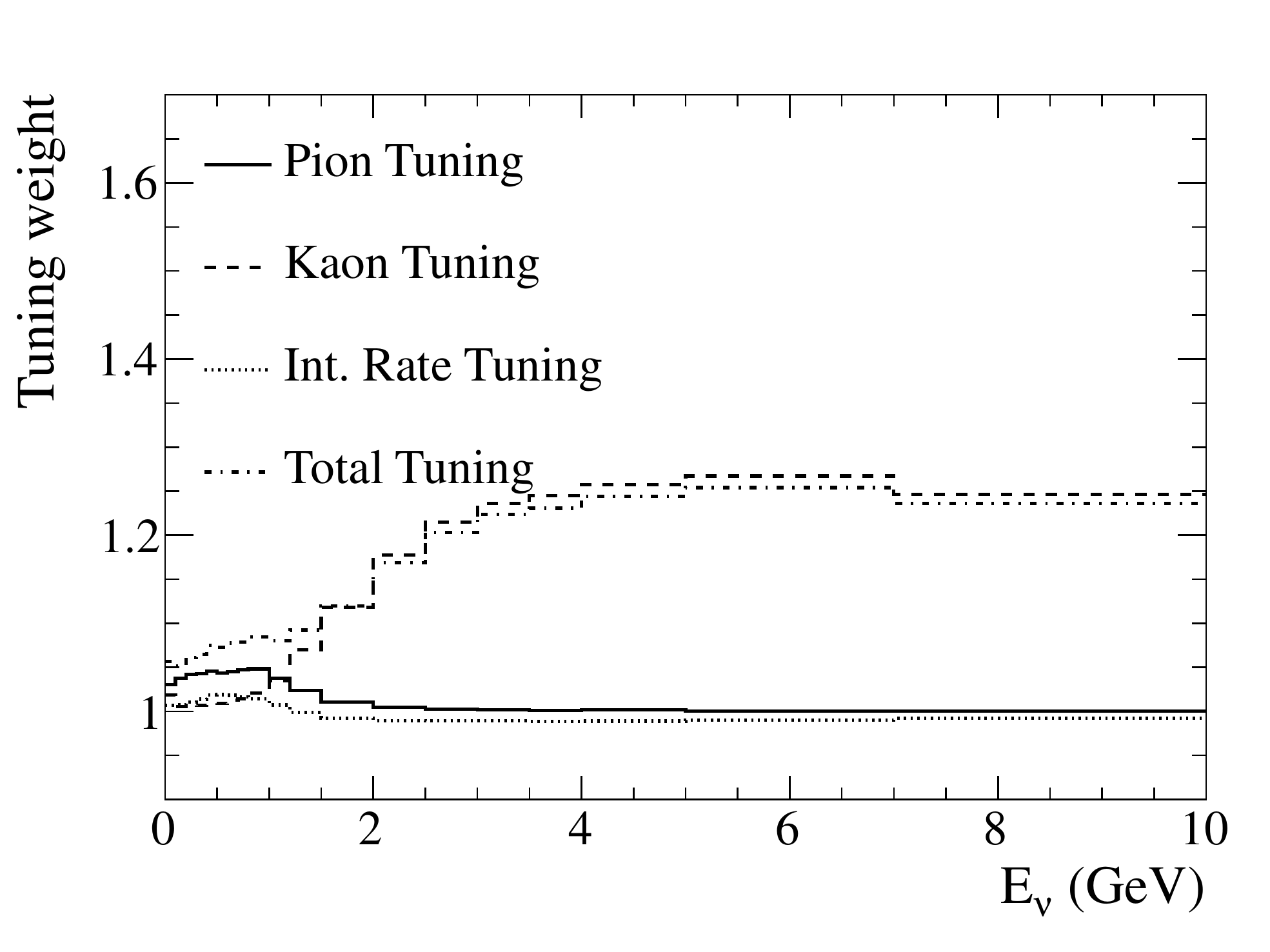}
\includegraphics[width=0.45\textwidth]{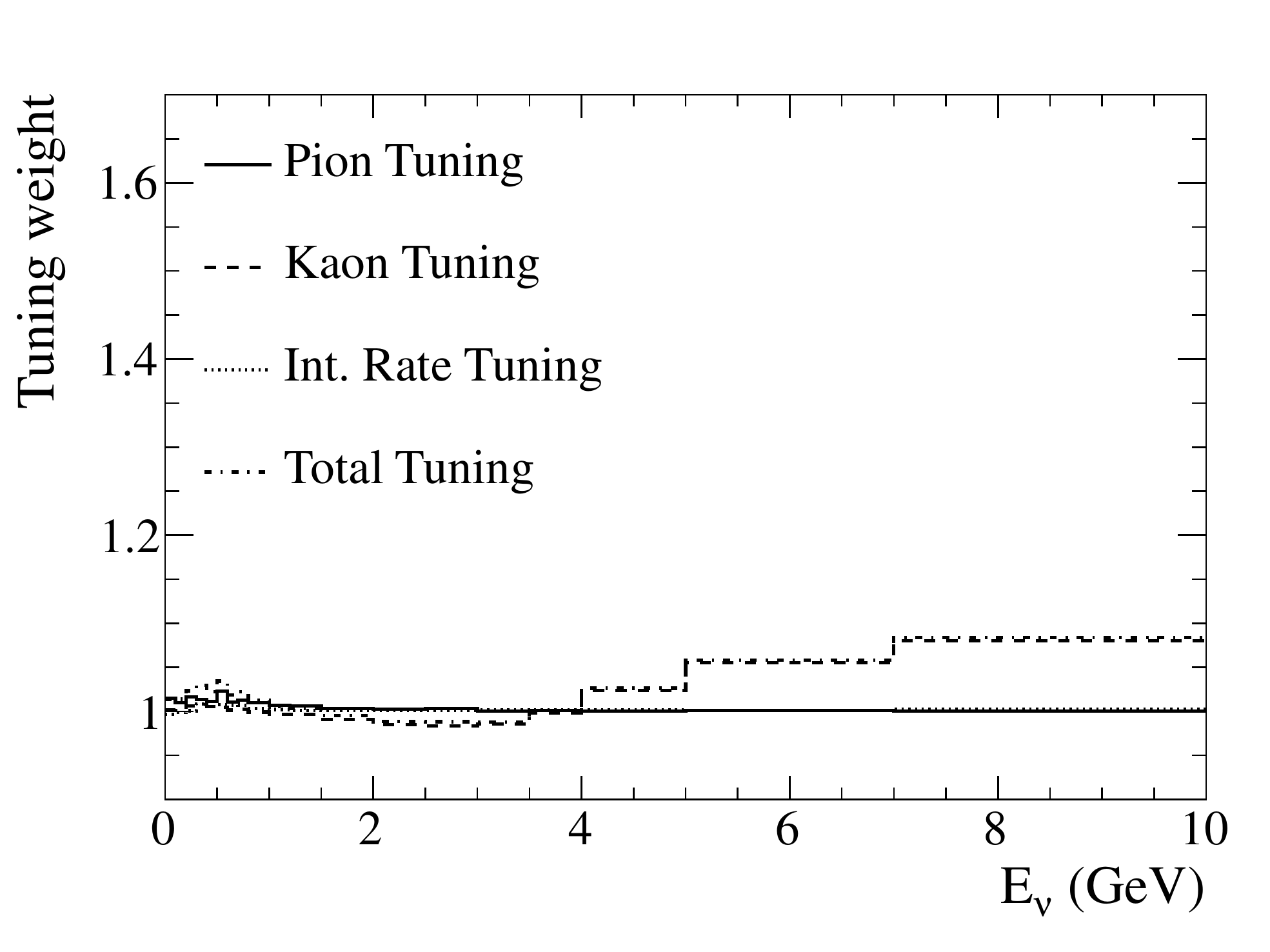} 
\caption{Ratio of the hadron interaction re-weighted flux over the nominal flux for $\nu_{\mu}$ (upper left),
$\bar{\nu}_{\mu}$ (upper right), $\nu_{e}$ (lower left), $\bar{\nu}_{e}$ (lower right)}
\label{fig:flux_tuning}
\end{figure*}

The hadron multiplicity re-weighting described in Sec.~\ref{sec:hadron_prod_tune} and the hadron interaction
rate re-weighting described in Sec.~\ref{sec:hadron_int_tune} are applied to the simulation to achieve the
hadron interaction re-weighted flux prediction for T2K.  The effect of the weights are seen by taking the 
energy dependent ratio of the flux with and without the weights applied, as shown in Fig.~\ref{fig:flux_tuning}.
The pion multiplicity re-weighting has the largest effect at low energies, while the
kaon multiplicity re-weighting is dominant at high energies.

\subsection{Summary of the T2K flux prediction}
\begin{figure*}
\centering
\includegraphics[width=0.45\textwidth]{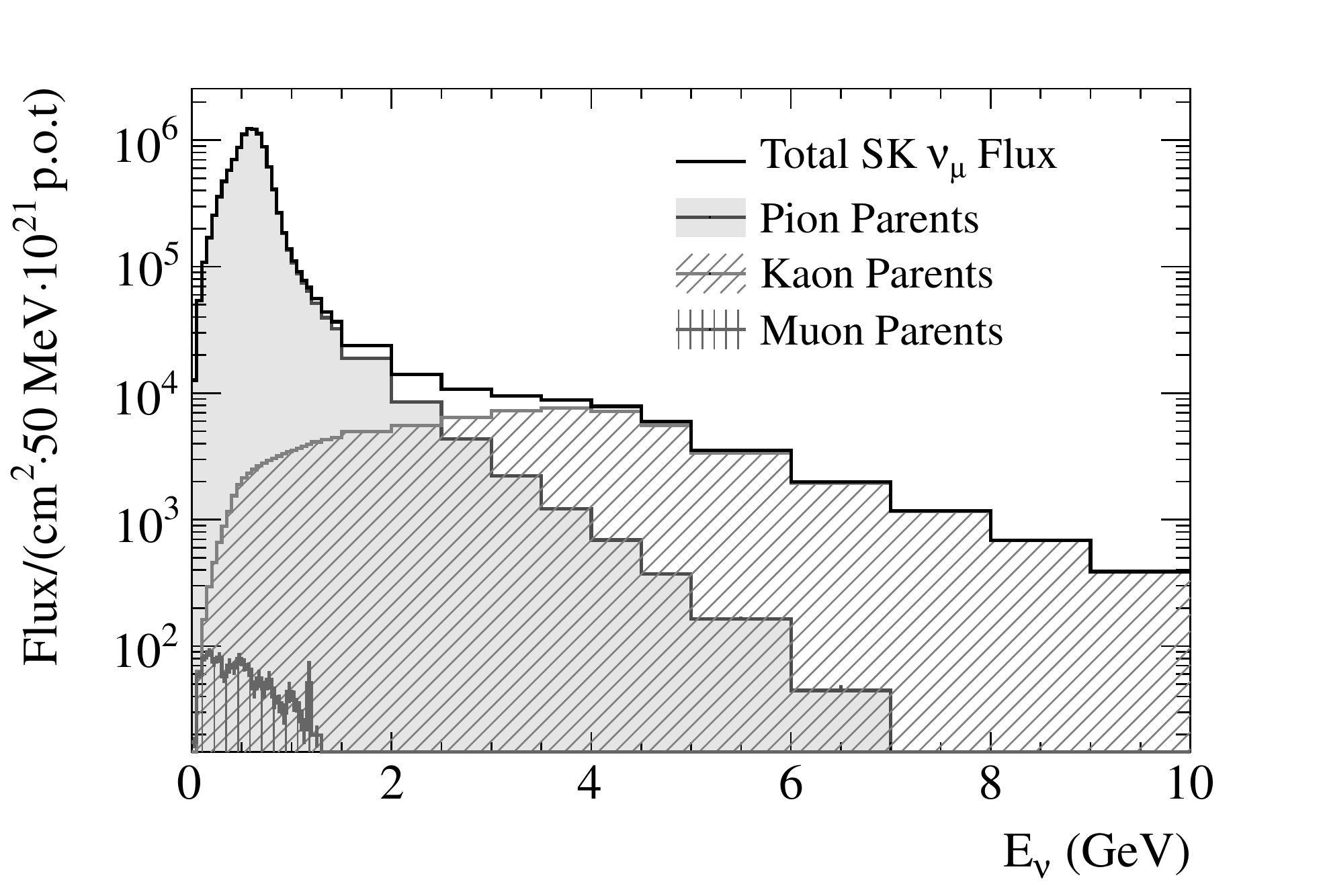}
\includegraphics[width=0.45\textwidth]{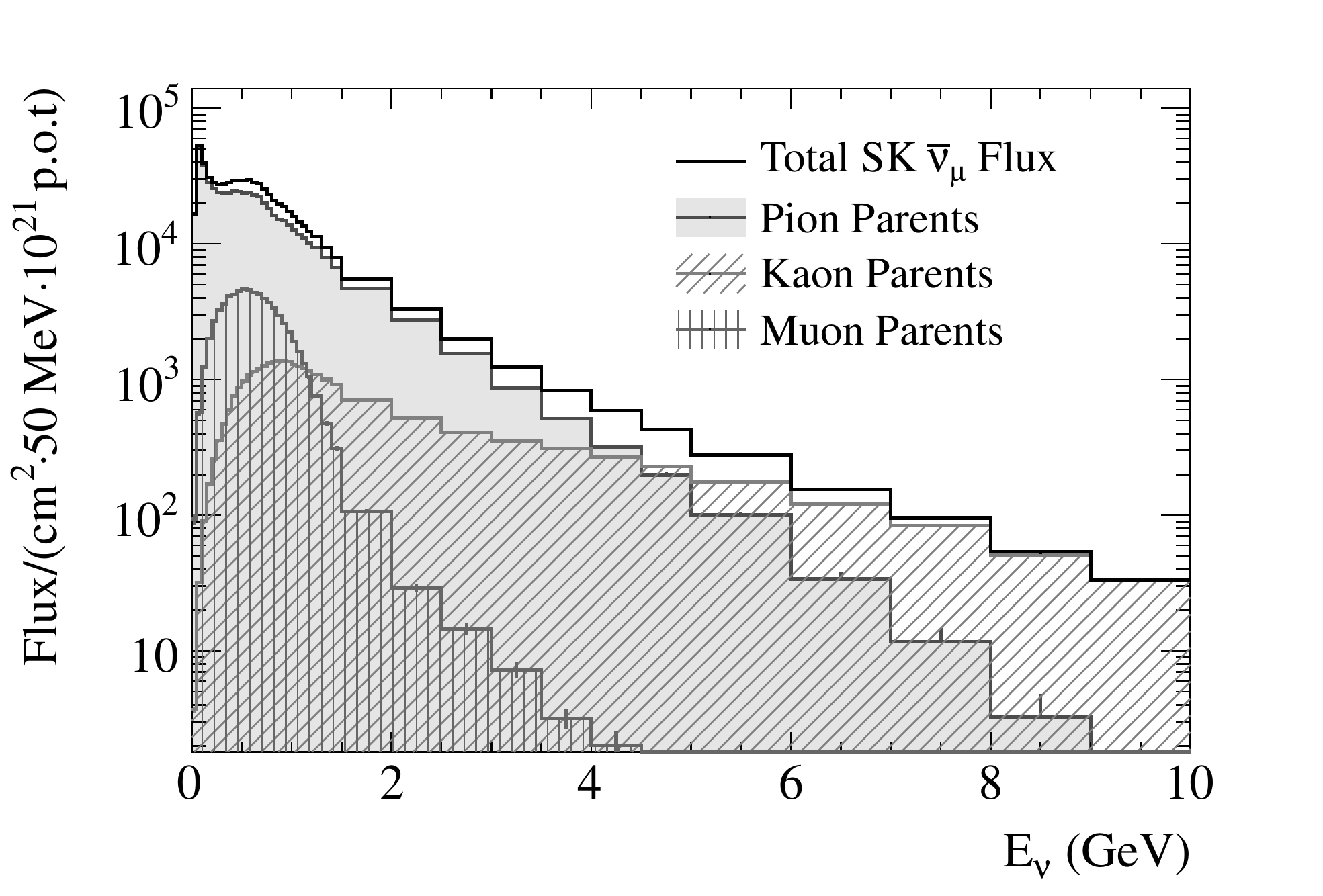}
\includegraphics[width=0.45\textwidth]{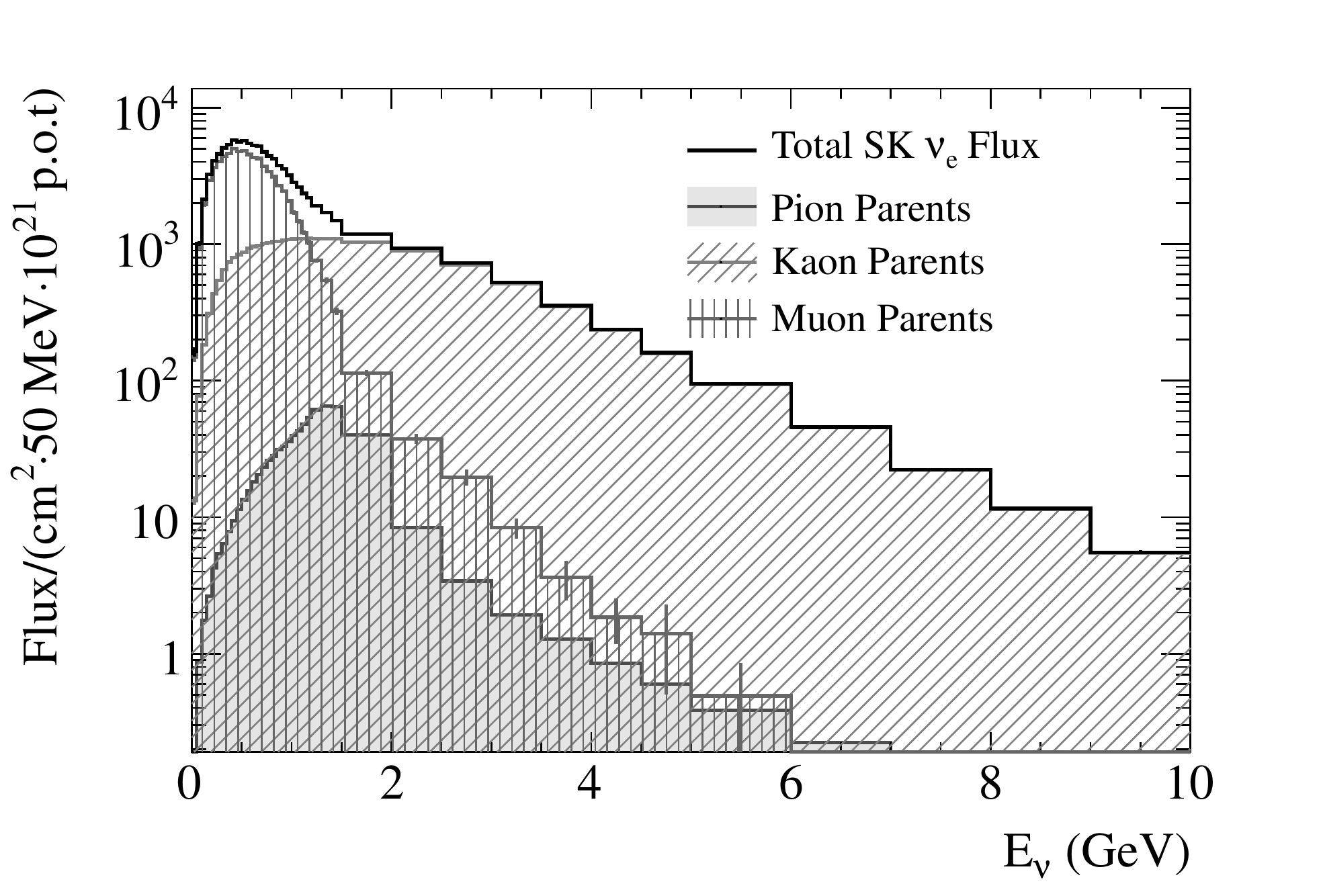}
\includegraphics[width=0.45\textwidth]{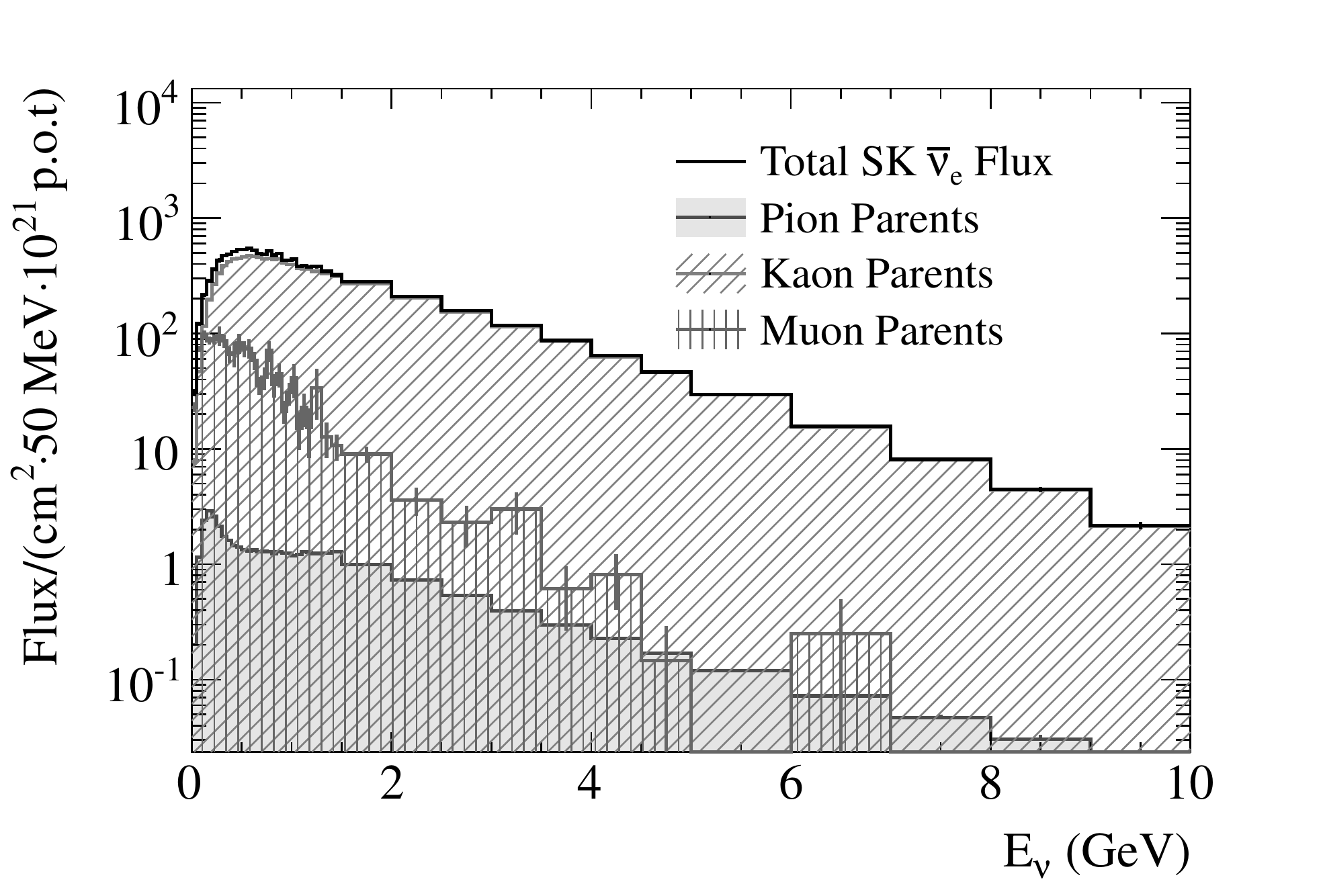}
\includegraphics[width=0.45\textwidth]{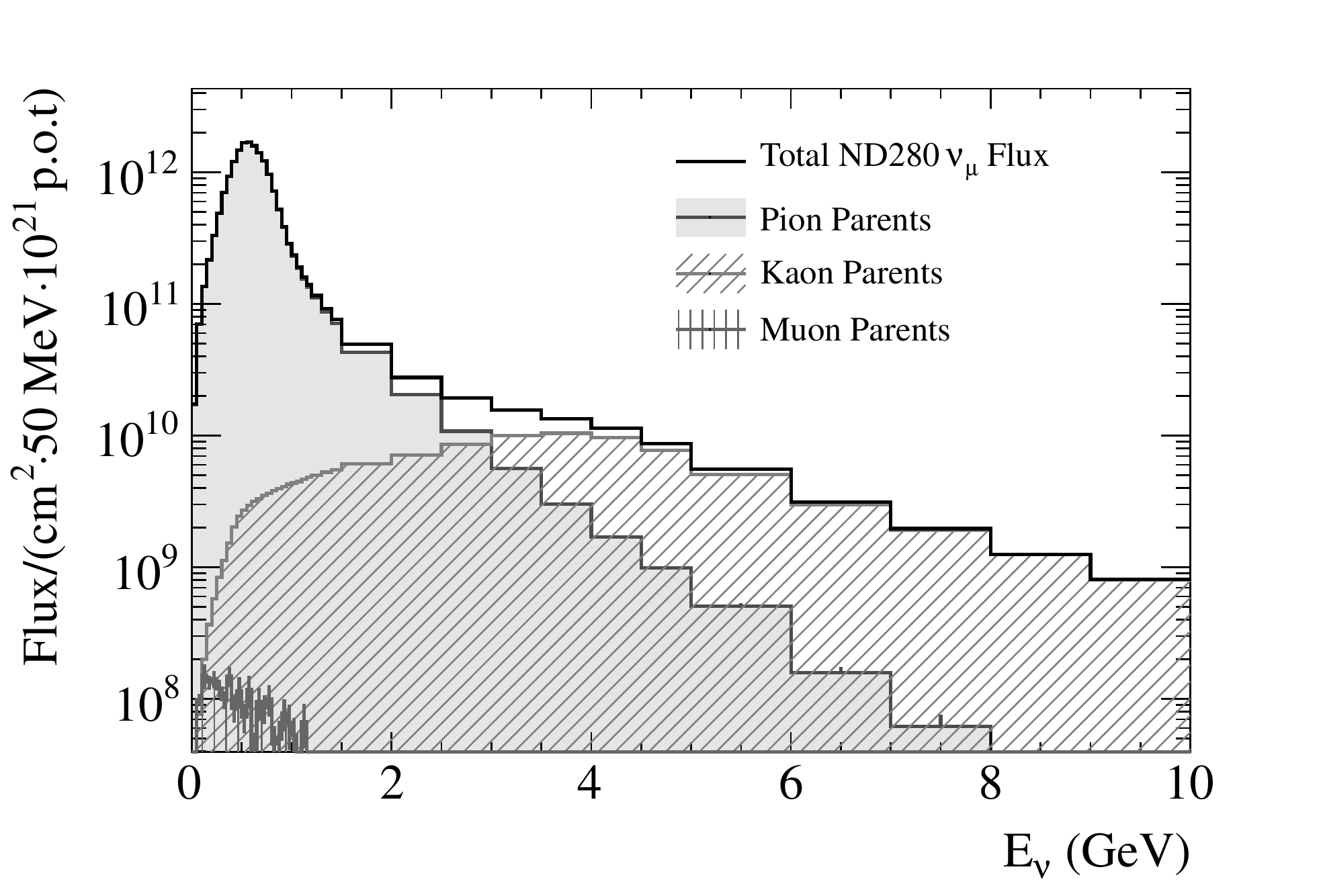}
\includegraphics[width=0.45\textwidth]{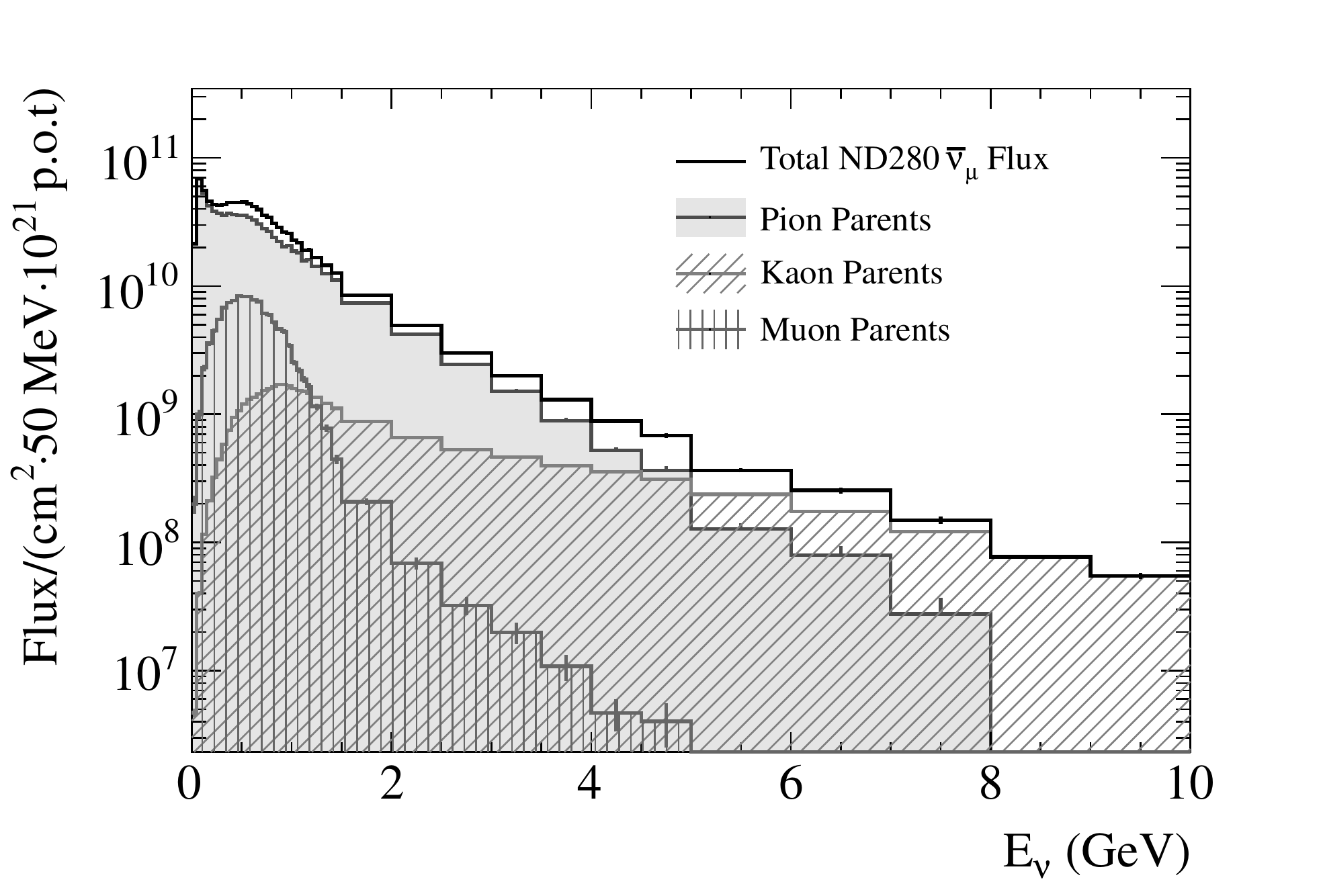}
\includegraphics[width=0.45\textwidth]{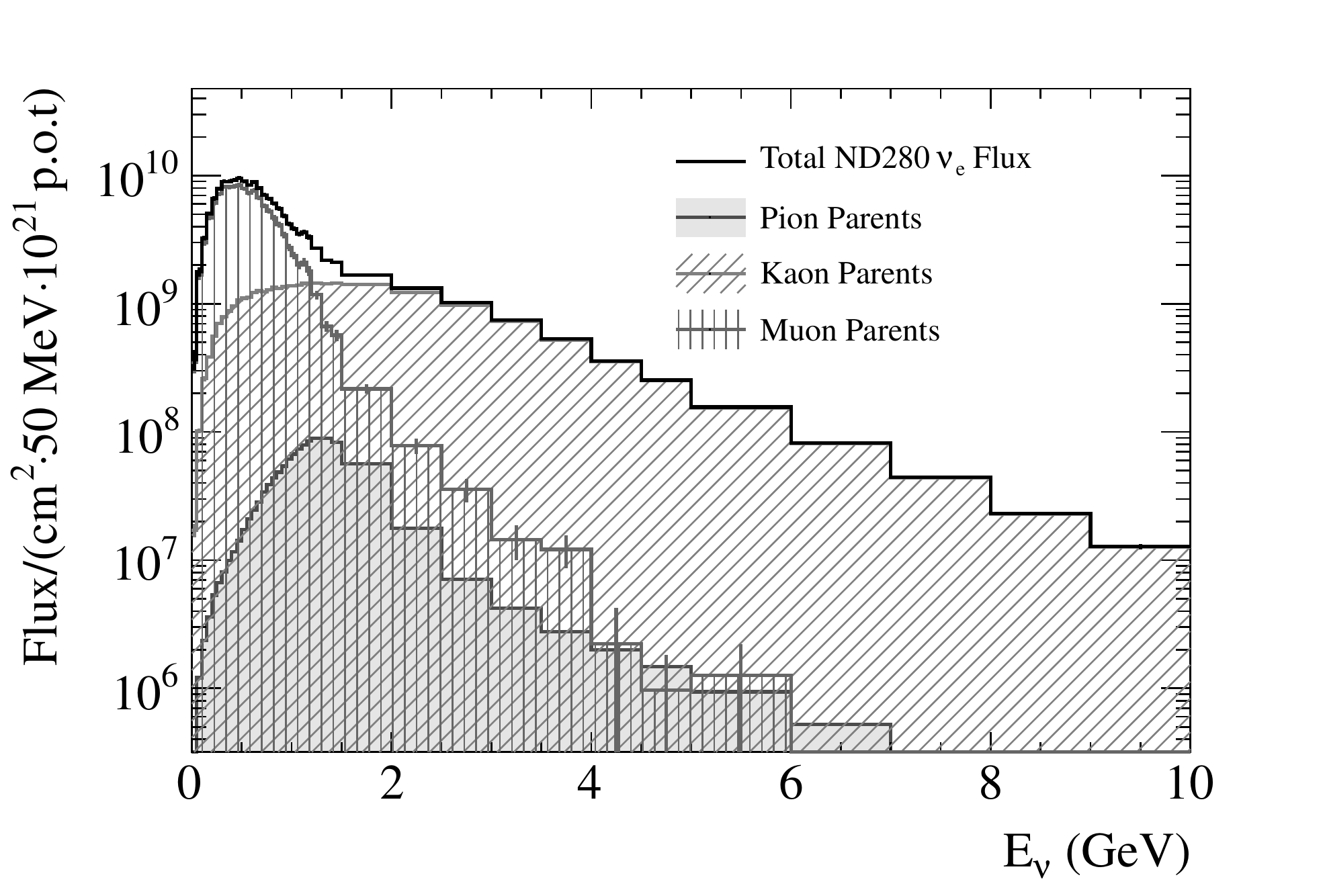}
\includegraphics[width=0.45\textwidth]{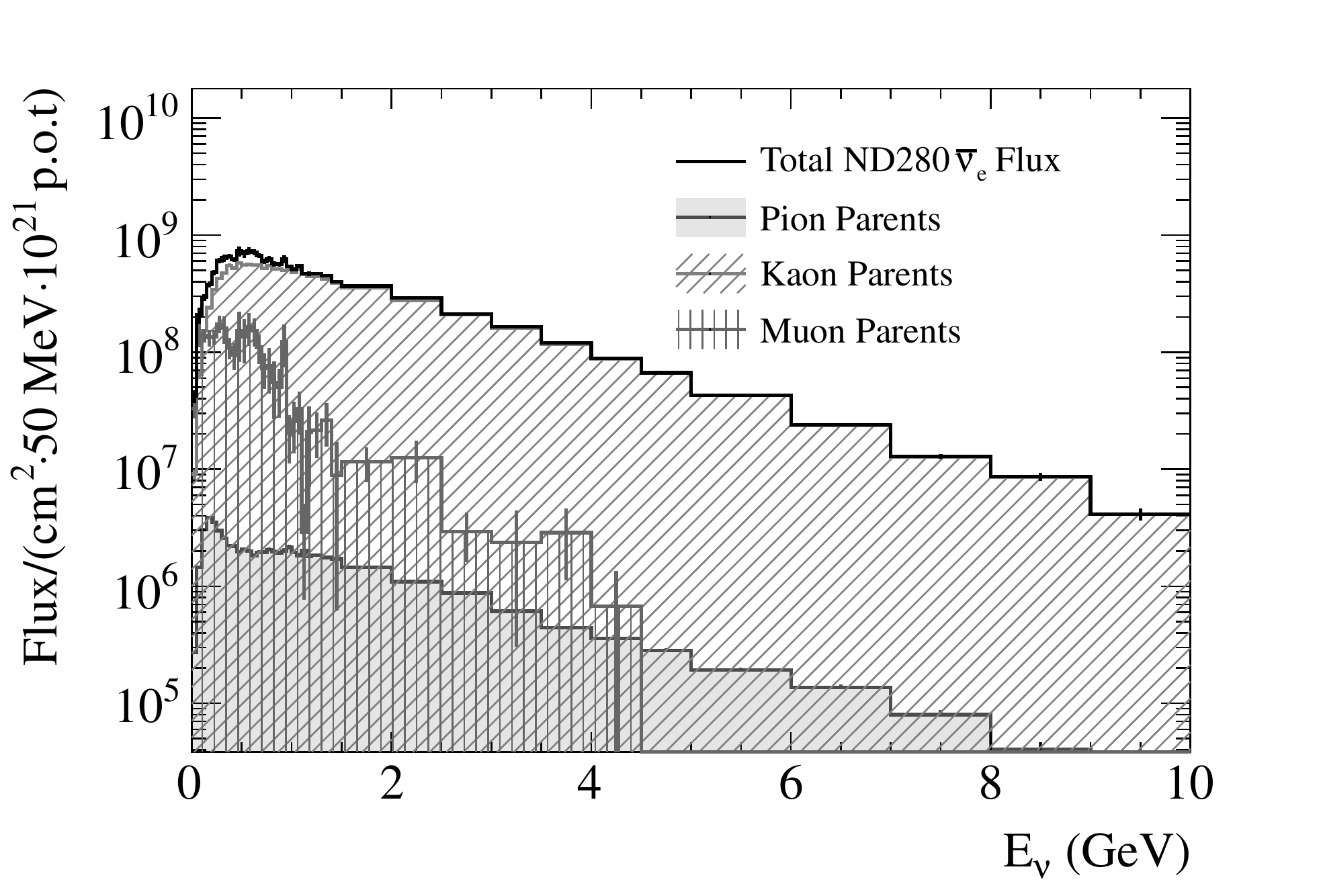}
\caption{The flux predictions for the SK far detector and ND280 near detector 
broken down by the neutrino parent particle type. 
\textcolor{red}{The error bars, which are too small to be seen 
in most energy bins, are due to the MC statistical error.}}
\label{fig:flux_pred_par}
\end{figure*}

The T2K flux is predicted using the simulation including the re-weighting of hadron interaction 
probabilities outlined
here.  The flux is predicted for each neutrino flavor at the far and near detectors.
Figure~\ref{fig:flux_pred_par} shows the flux predictions for both SK and the ND280 off-axis detector broken down by the
parent particle that decays to the neutrino.  The relative fractions of each flavor 
in the SK flux prediction for $0-1.5$, $1.5-3.0$ and $>3.0$ GeV energy ranges after re-weighting is applied are 
listed in Table~\ref{tab:flavor_frac}.  The 
$\nu_e$ flux, which constitutes an irreducible background for the study of $\nu_{\mu}\rightarrow\nu_{e}$ 
oscillations, accounts for less than $1\%$ of the flux below $1.5$ GeV, and the $\bar{\nu}_{\mu}$ contamination
is $\sim5\%$. 
In the intermediate ($1.5-3.0$) GeV energy bin, the relative fraction of $\bar{\nu}_{\mu}$ increases as
the flux becomes more dominated by forward going pions that are not focused, which include $\pi^{-}$ that decay to 
$\bar{\nu}_{\mu}$.
The $\nu_{e}$ fraction
also increases as the contribution from kaon decays becomes dominant.  For the high energy bin ($>3.0$ GeV),
the fraction of $\bar{\nu}_{\mu}$ flux decreases, since the contribution from the decay of focused kaons becomes
significant.

\begin{table}[h]
\begin{center}
\caption{The fraction of the total flux by flavor in bins of the neutrino energy. The
fractions in parentheses are relative to the total flux over all neutrino energies.}
\label{tab:flavor_frac}
\begin{tabular}{lccc}
\hline
\hline
              & \multicolumn{3}{c}{Energy Range (GeV)} \\
Flavor & $0-1.5$ & $1.5-3.0$ &  $>3.0$ \\ 
\hline
$\nu_{\mu}$         & $0.9363(0.8570)$ & $0.7719(0.0391)$ & $0.8821(0.0372)$    \\
$\bar{\nu}_{\mu}$   & $0.0542(0.0496)$ & $0.1729(0.0087)$ & $0.0795(0.0034)$    \\
$\nu_{e}$           & $0.0085(0.0078)$ & $0.0451(0.0023)$ & $0.0304(0.0013)$    \\
$\bar{\nu}_{e}$     & $0.0010(0.0009)$ & $0.0100(0.0005)$ & $0.0080(0.0003)$    \\
\hline
\hline
\end{tabular}
\end{center}
\end{table}

\section{\label{sec:fluxerrors}Uncertainties on the flux prediction }
In this section, we discuss uncertainties on the flux prediction. 
The neutrino flux uncertainties arising from hadron production uncertainties (Sec.\ref{sec:hadronerrors}), 
proton beam and off-axis angle uncertainties (Sec.\ref{sec:pbeamoffaxiserrors}), 
target and horn alignment uncertainties (Sec.\ref{sec:alignerrors}), 
horn current and magnetic field uncertainty (Sec.\ref{sec:hcurrerrors}) are considered. 

The uncertainties on the flux prediction are studied by varying underlying inputs to the flux 
simulation (the hadron production model, the proton beam profile, the horn currents, etc.) and
evaluating the effect on the predicted flux.  Two approaches are used.  

Where an error source
includes a number of correlated underlying parameters (degrees of freedom), re-weighting methods are used when possible.  
The underlying parameters are varied according to their covariance, and the flux prediction
is re-weighted with each of $N$ sets (typically 500 or more) of the parameter values.  The effect on the flux is evaluated by
constructing a covariance matrix from the $N$ re-weighted versions of the flux prediction:
\begin{equation}
V_{ij} = \frac{1}{N}\sum_{k=1}^{k=N}(\phi^{i}_{nom}-\phi^{i}_{k})(\phi_{nom}^{j}-\phi^{j}_{k}).
\end{equation}
Here the $\phi^{i}_{nom}$ are the nominal flux and $i$ specifies the neutrino energy bin, flavor
and detector at which the flux is evaluated.  The $\phi^{i}_{k}$ are the corresponding bins of the
\textit{k\textsuperscript{th}} re-weighted version of the flux.  Flux uncertainties evaluated with this method are the hadron
interaction uncertainties and the proton beam profile uncertainties.  

The second method for evaluating uncertainties is applied for uncertainties represented by variations
of the flux due to changes in a single underlying parameter.  For these uncertainties the flux is
typically re-simulated for variations of the parameter at $\pm1\sigma$.  As with the previous method
a covariance matrix is calculated:
\begin{equation}
V_{ij} = \frac{1}{2}[(\phi^{i}_{nom}-\phi^{i}_{+})(\phi_{nom}^{j}-\phi^{j}_{+})+(\phi^{i}_{nom}-\phi^{i}_{-})(\phi_{nom}^{j}-\phi^{j}_{-})].
\end{equation}
The $\phi^{i}_{+}$ and $\phi^{i}_{-}$ are the re-simulated flux for $+1\sigma$ and $-1\sigma$ variations of
the underlying parameter.

The combined uncertainty on the flux prediction is simply represented by the sum of the covariances from
each independent source of uncertainty described in the following text.  Since the flux is evaluated as a
covariance between bins in neutrino energy, neutrino flavor, and neutrino detector, the covariance 
between the flux prediction at the near and far detectors is included.  The covariance can be used directly
in an extrapolation method, or to calculate the uncertainty on a far-to-near ratio.

\subsection{\label{sec:hadronerrors}Hadron interaction uncertainties}
The systematic uncertainties associated with the hadronic interactions come
from a variety of sources.  One of them is the experimental uncertainties in the data. Another is the scaling of the differential production yields to different incident particle momenta (see Section~\ref{sec:hadron_prod_tune}). In addition, the systematic
effects associated with the extrapolation of the differential particle
yields to different target materials must be considered. It is also
necessary to estimate the contribution from the regions of particle
production phase space not covered by the data. Finally, the
systematic uncertainties associated with the total interaction rate
(production cross section) of particles in a given material must be 
evaluated.


\subsubsection{Pion production uncertainties} 
\label{subsec:pierrs}

The uncertainty on the pion production multiplicity modeling arises from a number of sources:
\begin{enumerate}
\item The uncertainty on the NA61/SHINE data used to re-weight the pion production multiplicity
\item The uncertainty on the incident particle momentum scaling used to apply the NA61/SHINE data to
interactions with lower momentum incident nucleons
\item The uncertainty from phase space that is not covered by the NA61/SHINE data points
\end{enumerate}

\begin{figure}
\centering
\includegraphics[width=0.45\textwidth]{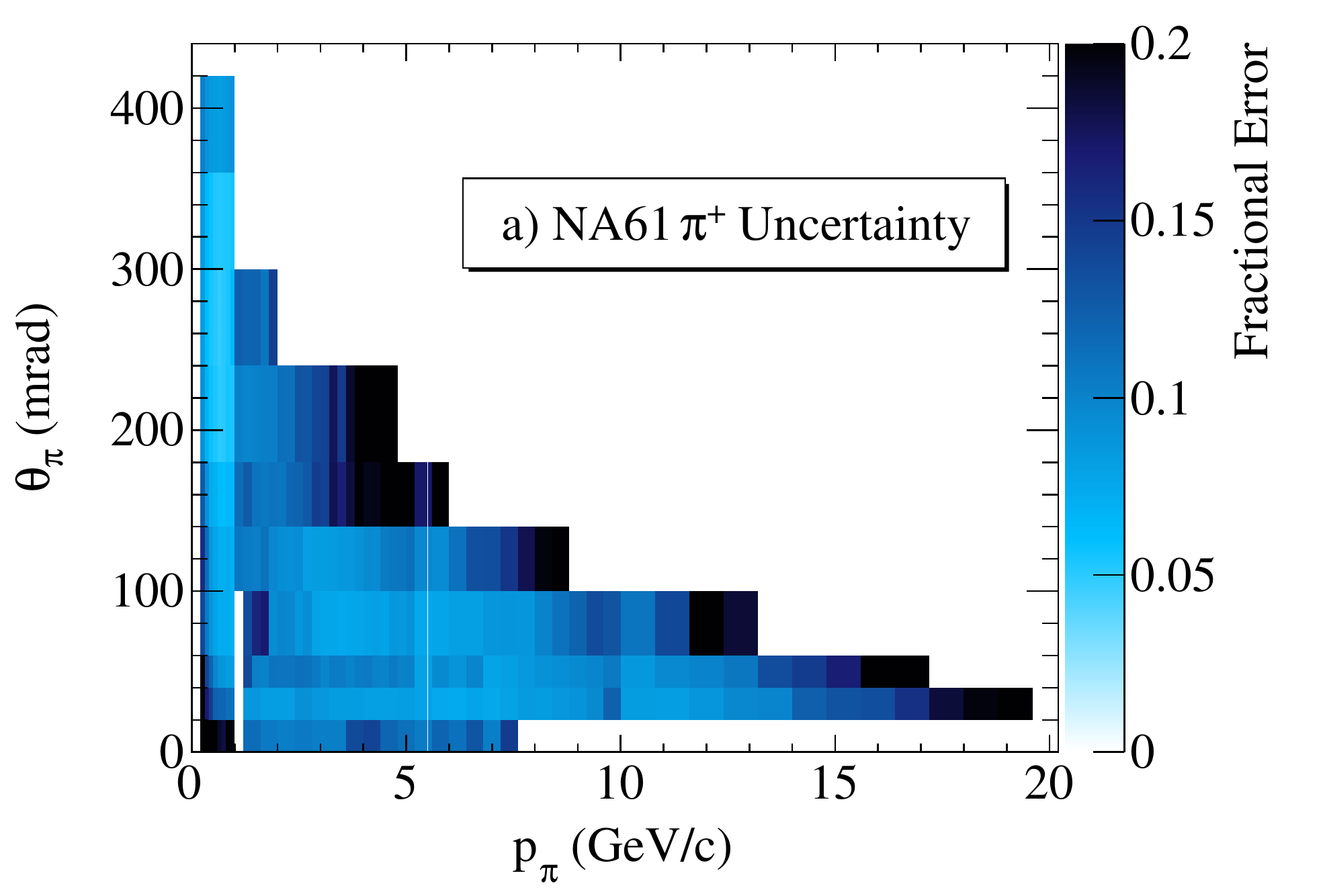}
\includegraphics[width=0.45\textwidth]{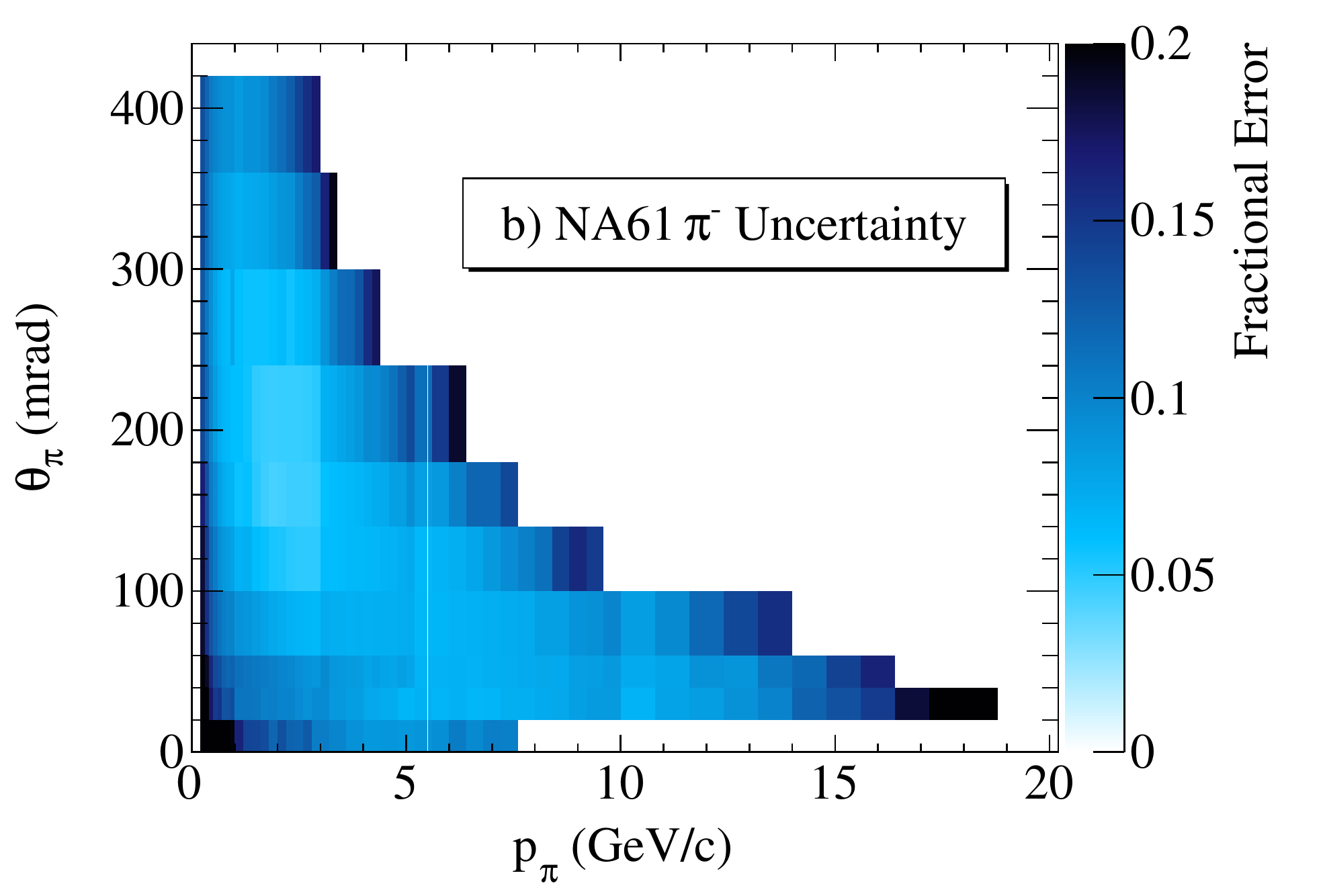}
\caption{The fractional error on the NA61/SHINE measurements in each of the $p-\theta$ bins.
The gap at $\pi^{+}$ momentum of 1.0-1.2 GeV/$c$ is a region with no NA61/SHINE data points. }
\label{fig:na61_errors}
\end{figure}

The uncertainty from the NA61/SHINE pion multiplicity data points is dominated by the systematic uncertainties, which are described in detail elsewhere~\cite{Abgrall:2011ae}. Figure~\ref{fig:na61_errors} shows the total errors including statistical errors for each of the NA61/SHINE $p-\theta$ bins. The total errors are typically 5 to 10\% in the most important regions of the phase space. The dominant sources of uncertainty are the correction for the feed-down from strange particle decays and particle identification. For most sources of uncertainty, the systematic effect is assumed to be correlated across all NA61/SHINE bins. This is a reasonable assumption for the feed-down error given the correlated model dependence of the strange particle production. For the particle identification error, it is assumed that bins of similar momenta are more correlated, and the systematic errors are modeled with a ranged correlation:
\begin{align}
C_{i,j} = & 1-\left(\frac{p_{i}-p_{j}}{6\; \textrm{GeV/c}}\right)^2 &  \:for \:|p_{i}-p_{j}|\leq6\;\textrm{GeV/c} \\
        = & 0  & \:for \:|p_{i}-p_{j}|>6\;\textrm{GeV/c}
\end{align}
Here $p_{i}$ and $p_{j}$ are the central value for the momentum in each bin.  The functional form
with a range of 6 GeV/c was chosen since it gives a reasonable model for the correlations and propagates
the errors conservatively.  

The NA61/SHINE data are also used to re-weight pion production from the interactions of nucleons in the horn conductor aluminum after 
A-dependent scaling has been applied. 
For the scaled data points, additional errors of 5\% (correlated between $p-\theta$ bins) and 5\% (uncorrelated
between $p-\theta$ bins) are applied to account for the scaling uncertainty described in Section~\ref{subsec:kerrs}.

The error associated with scaling the NA61/SHINE pion multiplicity to lower incident nucleon momenta is 
studied by carrying out an alternative method of re-weighting tertiary events.  Proton on Be data from the
BNL-E910 experiment at beam momenta of 12.3~GeV/$c$ and 17.5~GeV/$c$ provide an alternative source for re-weighting
interactions at lower incident momenta.  The data are scaled from Be to C using the method outlined in 
Section~\ref{sec:hadron_prod_tune}.  Since the BNL-E910 data are more coarsely binned in $p-\theta$ than the
NA61/SHINE, data at each momentum are separately fit with the empirical parametrization developed by 
Bonesini \textit{et al.}~\cite{bmpt_paper} (BMPT) for 400 and 450 GeV/$c$ proton on Be differential production data. 
The BMPT parametrization uses the radial scaling variable $x_{R}$ and the transverse momentum of the produced particle $p_{T}$.
The $x_{R}$ variable, is defined as:
\begin{equation}
x_R = \frac{E^{cm}}{E^{cm}_{max}},
\end{equation}
where $E^{cm}$ is the energy of the produced particle in the center of mass frame, and $E^{cm}_{max}$ is the maximum energy that
the particle can have.
Taylor {\it et al.}~\cite{taylor_radscale} found that the invariant cross section when parametrized in $x_{R}$ and $p_{T}$ 
does not depend on the total center of mass energy $\sqrt{s}$ (so called radial scaling) for $\sqrt{s}\gtrsim10$ GeV, while Feynman
scaling with $x_{F}$ (Eq.~\ref{eq:xf}) only holds at higher $\sqrt{s}$.
The parametrization for the production of 
positively charged pions and kaons in proton collisions on nuclei is:
\begin{align}
E\frac{d^3\sigma}{dp^3} =& A(1-x_{R})^{\alpha}(1+Bx_{R})x_{R}^{-\beta}\times \nonumber \\ 
                         & [1+\frac{a}{x_{R}^{\gamma}}p_{T}+\frac{a^{2}}{2x_{R}^{\delta}}p_{T}^{2}]e^{-a/x_{R}^{\gamma}p_{T}}.
\end{align}
The ratio of positive to negative hadron production was also found to be well described by simple parametrizations:
\begin{equation}
r(\pi)=r_{0}(1+x_{R})^{r_{1}},
\end{equation}
\begin{equation}
r(K)=r_{0}(1-x_{R})^{r_{1}}.
\end{equation}

\begin{figure}
\centering
\includegraphics[width=0.45\textwidth]{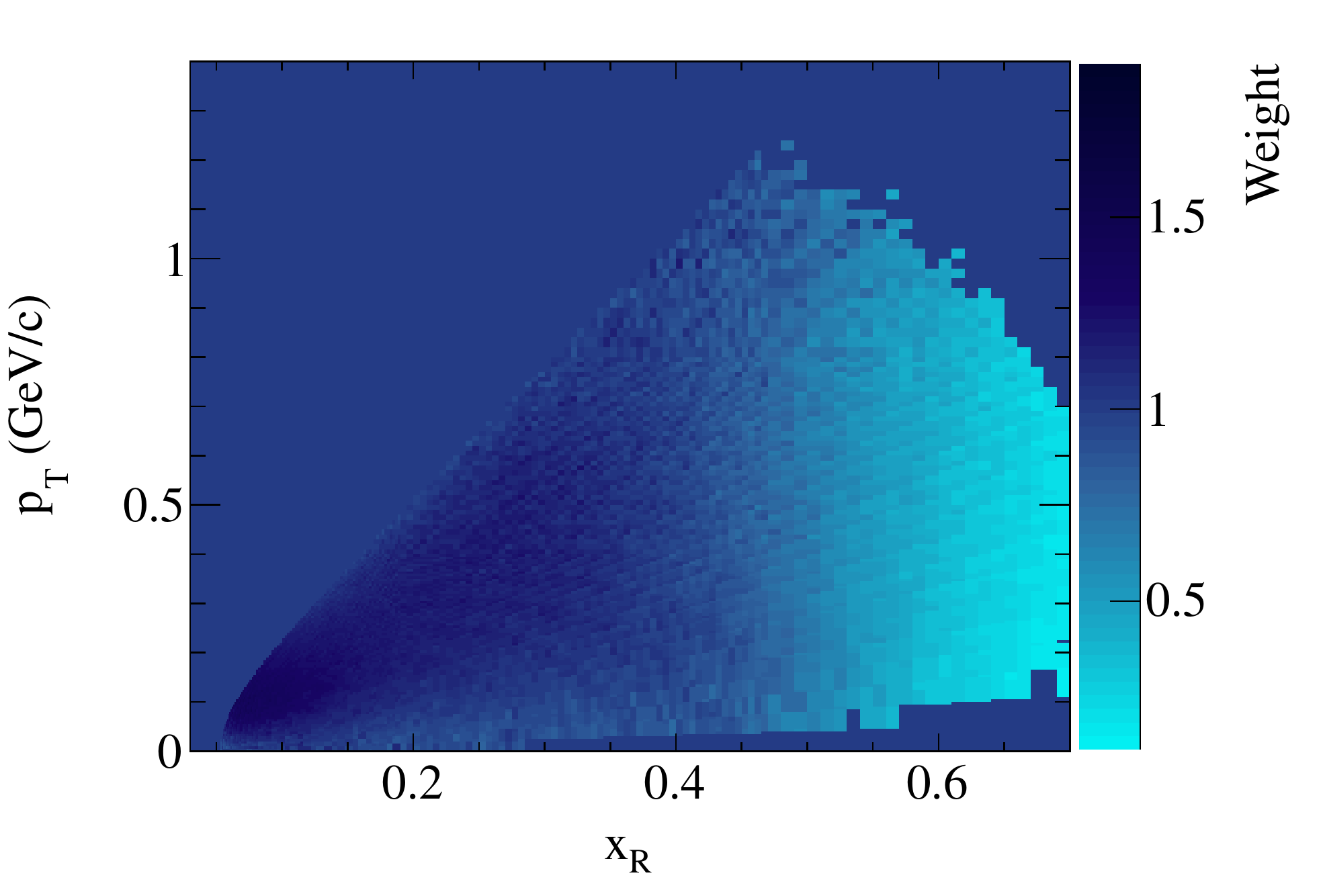}
\includegraphics[width=0.45\textwidth]{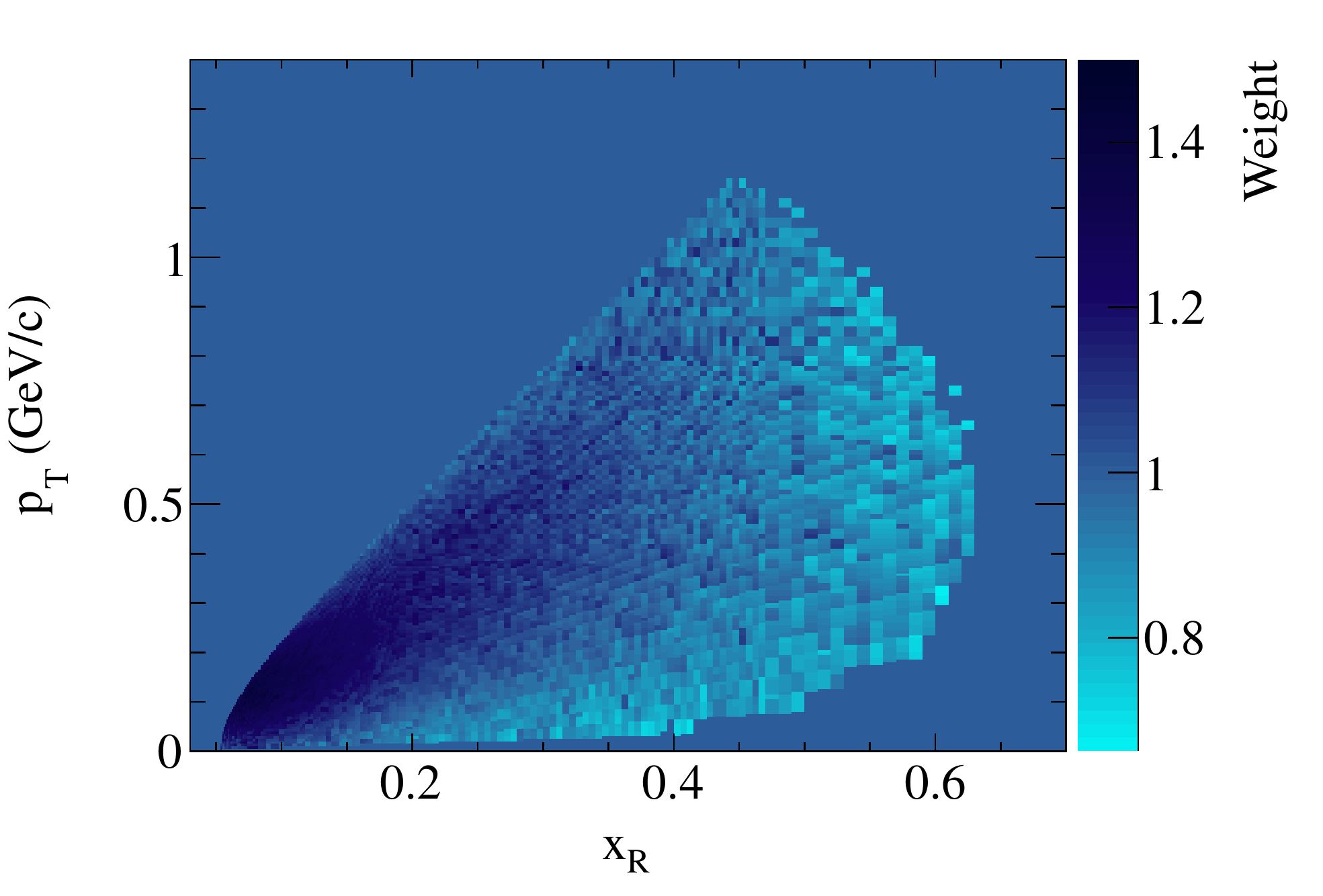}
\caption{The weights for FLUKA p+C interactions derived from the BMPT fits of BNL-E910 $\pi^{+}$ (top)
and $\pi^{-}$ (bottom) multiplicity data with 17.5 GeV/$c$ protons.}
\label{fig:e910_tuning}
\end{figure}

The BMPT parametrization is found to work well for the BNL-E910 data and provides a smooth interpolation of the data points.
Separate fits are done for the 12.3 GeV/$c$ and 17.5 GeV/$c$ data to allow breaking of the $x_{R}$ 
scaling in the BMPT parameters.  Tuning weights are calculated by taking the ratio of the BMPT fit to 
data over the FLUKA prediction, and plotted in the $x_{R}-p_{T}$ space, as shown in Fig.~\ref{fig:e910_tuning}.
If a simulated interaction has an incident particle momentum between the BNL-E910 data sets, a linear interpolation
of the weights of the two data sets in the incident particle momentum is applied.  Similarly a linear interpolation
is applied for interactions with incident particle momenta between the 17.5 GeV/$c$ BNL-E910 and NA61/SHINE data. 
The alternative method described here varies from the default method in that it allows for a breaking of the
$x$ scaling and it uses data at lower incident particle momenta to guide the breaking of the $x$ scaling.  The 
uncertainty on the flux is estimated by applying the two methods
of re-weighting tertiary events and taking the difference in the predicted flux.

\begin{figure}[h]
\centering
\includegraphics[width=0.45\textwidth]{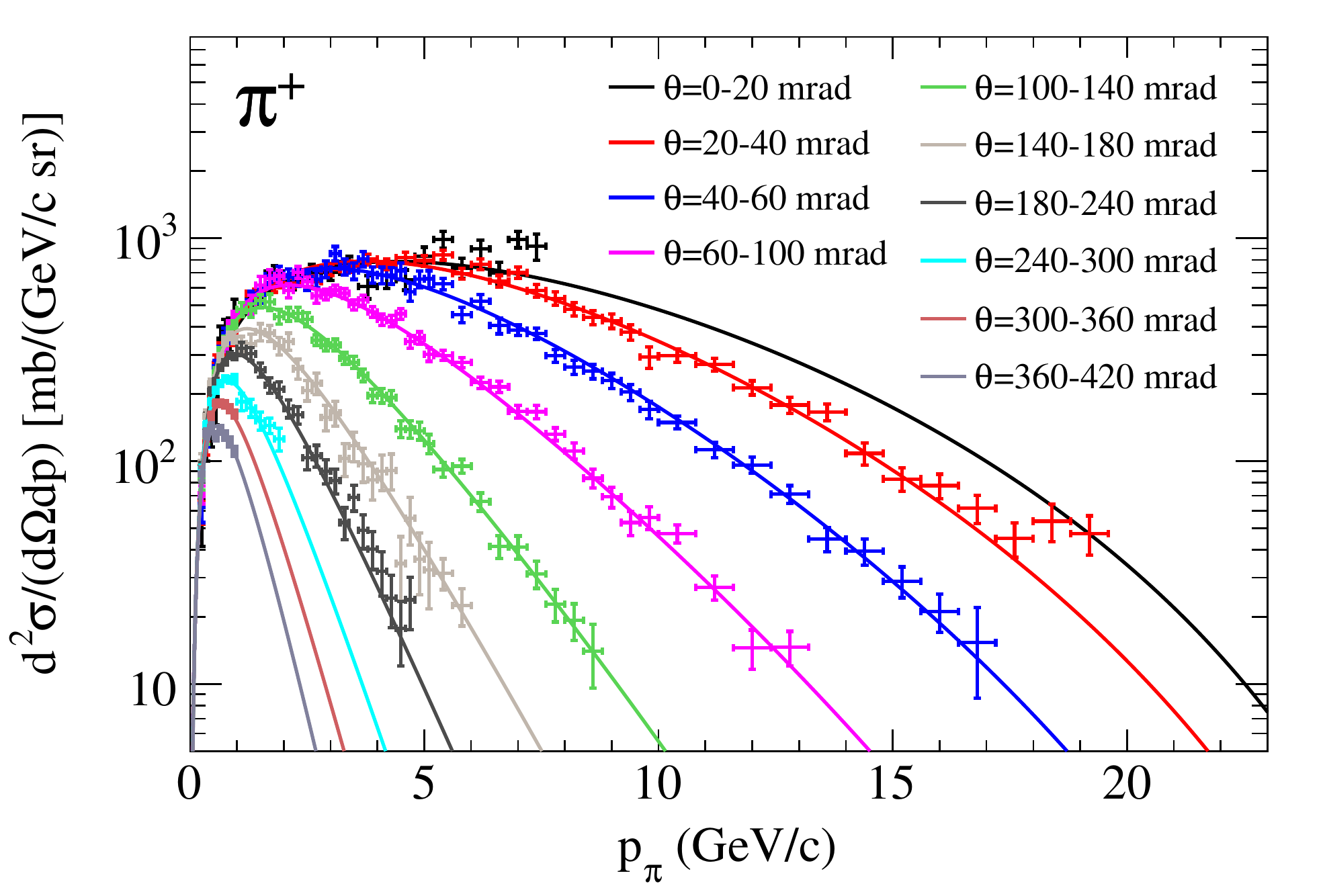}
\includegraphics[width=0.45\textwidth]{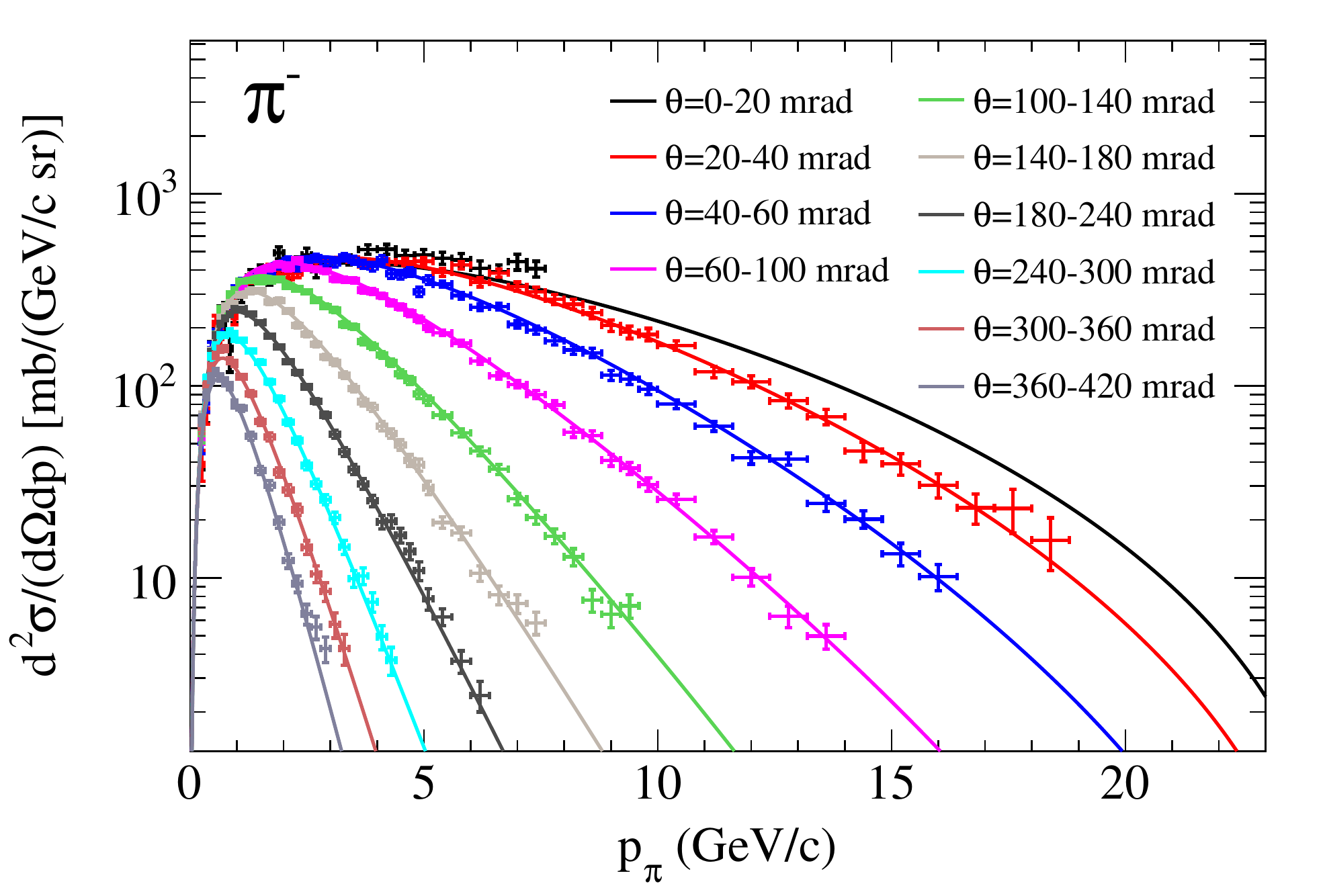}
\caption{The BMPT fits to the NA61/SHINE pion production data.}
\label{fig:na61_bmpt_fit}
\end{figure}

The NA61/SHINE data cover most of the phase space for secondary pions that contribute to the T2K neutrino flux.
To study the effect of pion multiplicities in the uncovered region, the NA61/SHINE data are fitted with the 
BMPT parametrization, which is used to extrapolate the data into the uncovered region.  To improve the 
agreement between the fits, the $\pi^{+}$ and $\pi^{-}$ are fitted separately. Figure~\ref{fig:na61_bmpt_fit}
shows that the BMPT parametrization is able to reasonably fit the NA61/SHINE data and the parameter
values are listed in Table~\ref{table:na61_bmpt_parms}.  The uncertainty in the FLUKA
model in the uncovered region is estimated as the change in the flux when the production is re-weighted by 
the BMPT fits in the uncovered region.  In addition, the uncertainty on the flux due to the uncertainty
on the fitted BMPT parameters is included.

\begin{table}
\begin{center}
\caption{Fitted BMPT parameters from NA61/SHINE pion data.}
\label{table:na61_bmpt_parms}
\begin{tabular}{lccccc}
\hline
\hline
BMPT &  &   \\ 
Parameter & $\pi^{+}$  & $\pi^{-}$  \\ 
\hline
$A$ (mb/GeV$^{2}$)  & $188\pm15$  & $90.8\pm2.7$  \\
$B$           & $-0.661\pm0.379$  & $-1.15\pm0.07$ \\
$\alpha$      & $3.40\pm0.35$     & $1.89\pm0.13$  \\
$\beta$       & $0.303\pm0.029$   & $0.461\pm0.012$ \\
$a$ (GeV$^{-1}$) & $5.37\pm0.14$  & $5.19\pm0.045$  \\
$\gamma$      & $0.245\pm0.018$   & $0.194\pm0.005$ \\
$\delta$      & $0.799\pm0.053$   & $0.783\pm0.017$ \\
$r_0$         & -                 & $1.10\pm0.031$ \\
$r_1$         & -                 & $1.95\pm0.17$ \\
\hline
\hline
\end{tabular}
\end{center}
\end{table}

\begin{figure*}[ht]
\centering
\includegraphics[width=0.3\textwidth]{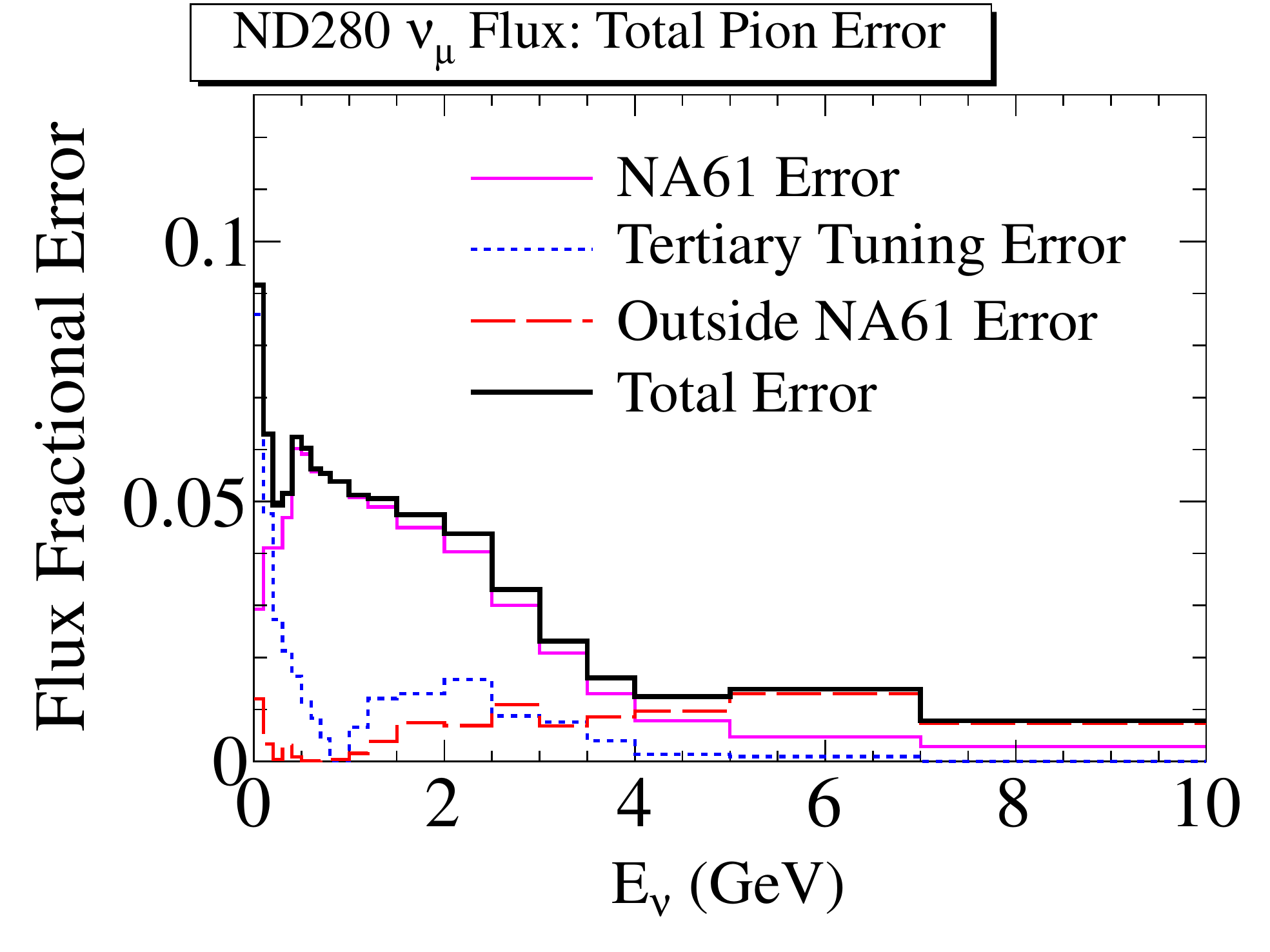}
\includegraphics[width=0.3\textwidth]{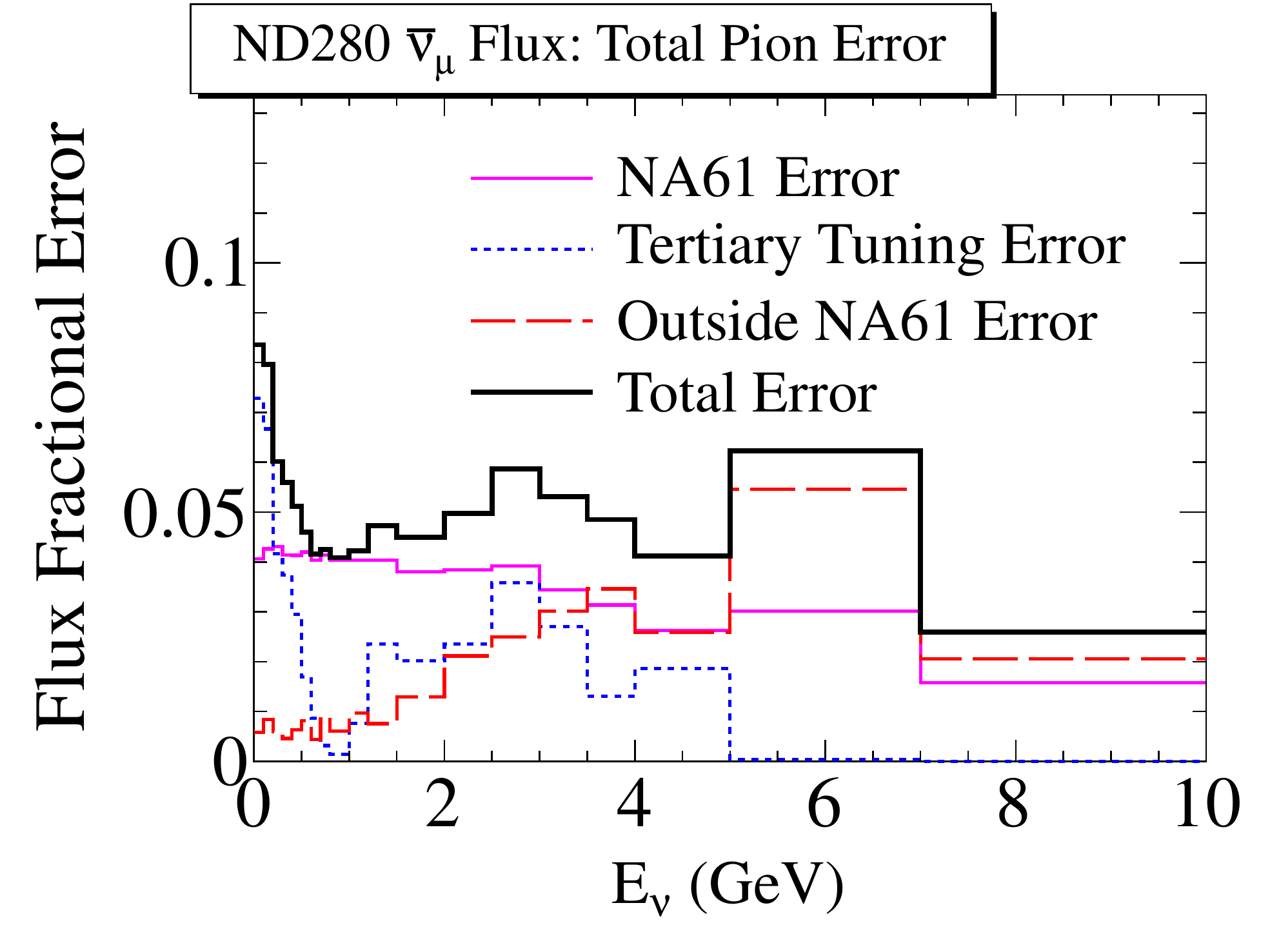}
\includegraphics[width=0.3\textwidth]{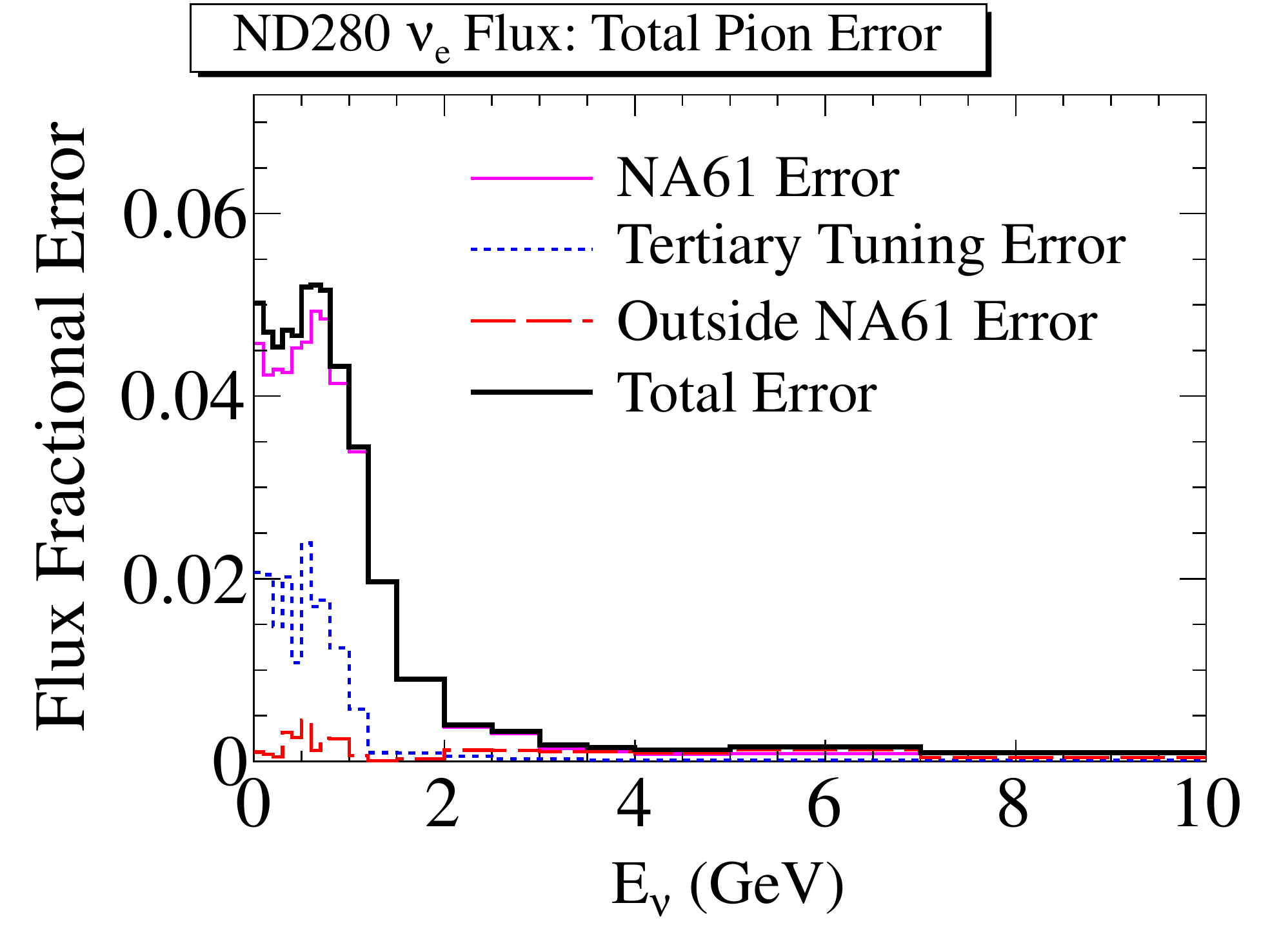}\\
\includegraphics[width=0.3\textwidth]{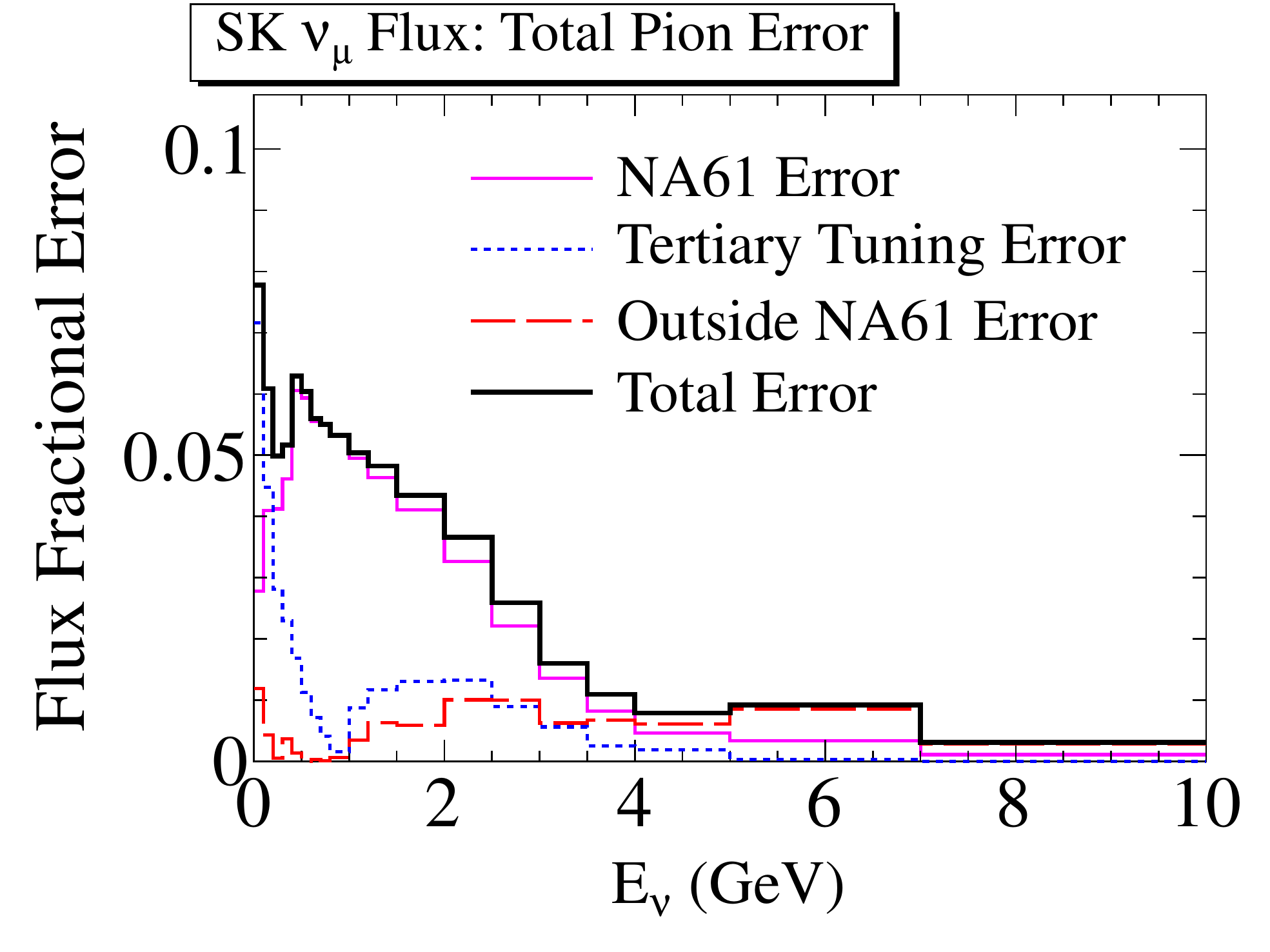}
\includegraphics[width=0.3\textwidth]{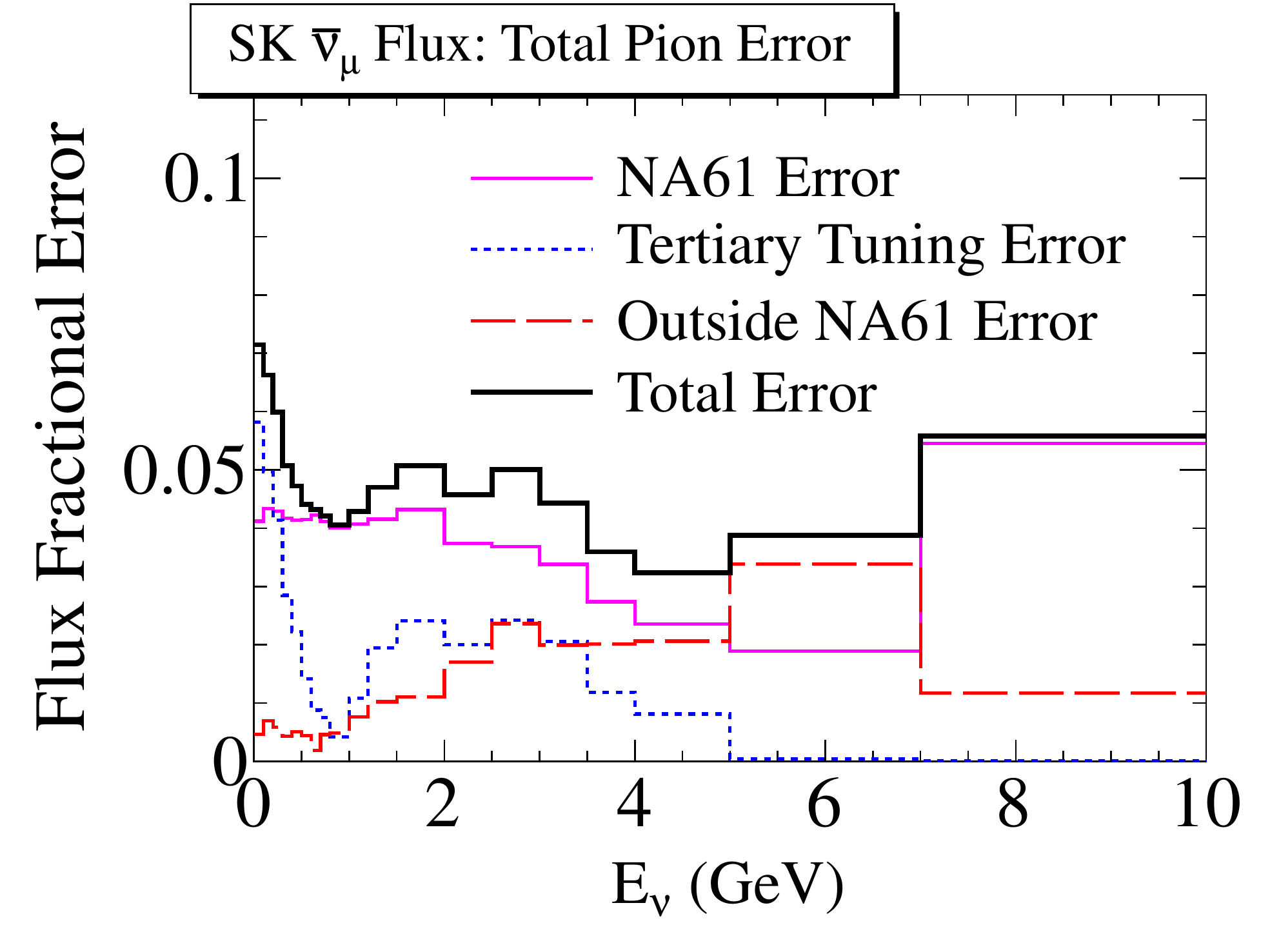}
\includegraphics[width=0.3\textwidth]{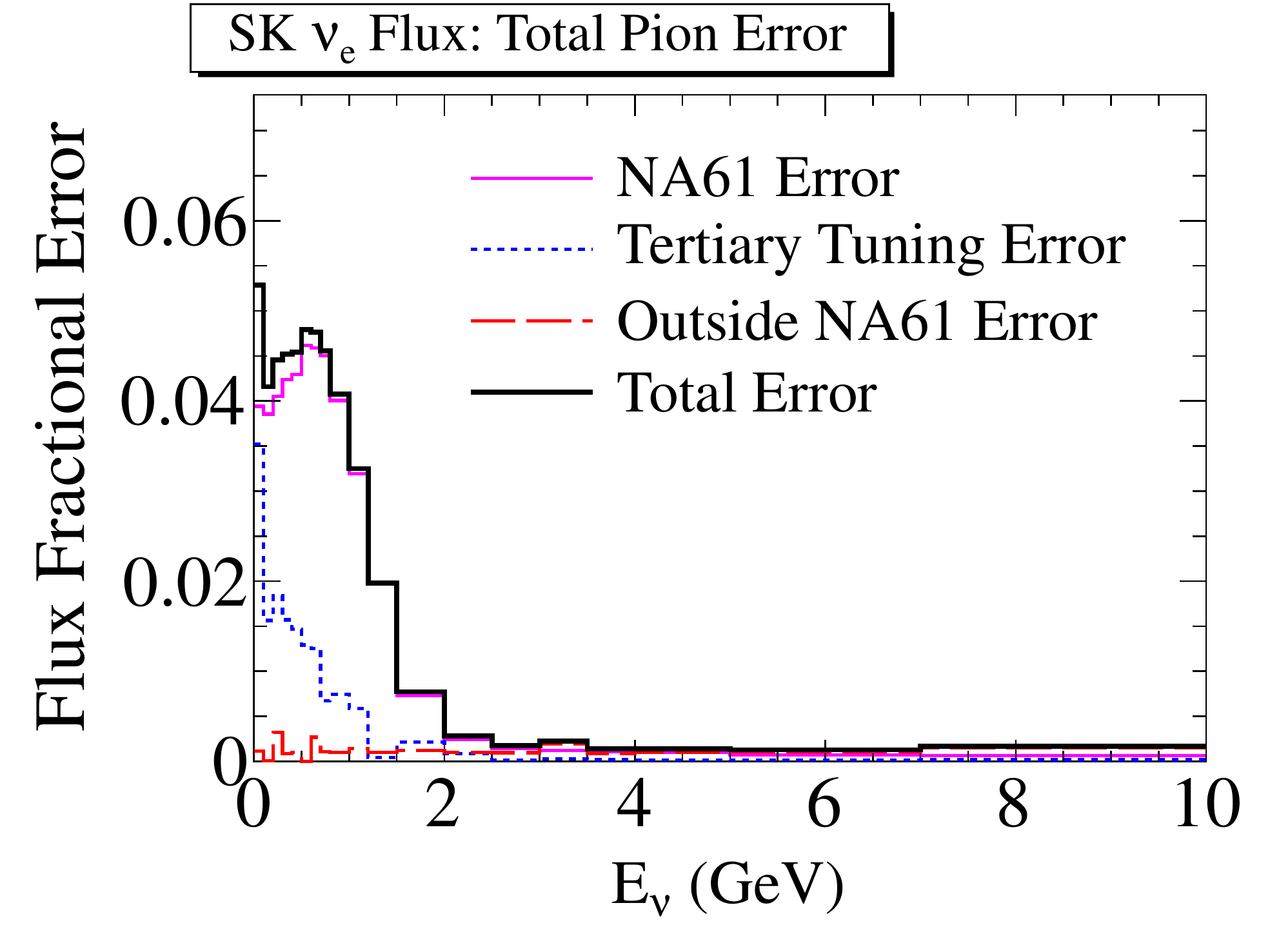}
\caption{Fractional flux error due to pion production as a function of neutrino energy, for each
flavor and at the near and far detectors.}
\label{fig:pion_errs}
\end{figure*}

The total uncertainty on the T2K flux prediction due to the modeling of pion production arises from the
sources outlined here and the magnitude of the uncertainty on the flux is summarized in Fig.~\ref{fig:pion_errs}.
The dominant source of uncertainty for the $\nu_{\mu}$ and $\nu_{e}$ flux predictions near the flux peak is
from the uncertainty on the NA61/SHINE pion multiplicity data points.

\subsubsection{Kaon production uncertainties} 
\label{subsec:kerrs}

Similarly to the pion case, the uncertainty on the kaon production multiplicity modeling comes from a number of sources:
\begin{enumerate}
\item The uncertainty on the data used to re-weight the kaon production multiplicity
\item The uncertainty on the incident particle momentum scaling used to apply the data to
interactions with lower momentum incident nucleons
\item The uncertainty from phase space that is not covered by the data points
\item The uncertainty on the kaon production outside of the target
\end{enumerate}

The uncertainties associated with the experimental kaon production data are divided into three categories.  The first is the uncertainty in the overall normalization, $\sigma_N$, for each data set. This uncertainty is fully correlated between different momentum and angular bins and, in the case of Eichten {\it et al.} and Allaby {\it et al.} measurements, for $K^+$ and $K^-$ data sets. In the second category are the uncertainties, $\sigma_{\Delta p \Delta \theta}$, which are uncorrelated between different data bins. These are typically statistical uncertainties. In the final category are the uncertainties in normalization for a given angular bin, $\sigma_{\Delta\theta}$. These are treated as fully correlated for all momentum bins in each $\Delta \theta$ for both $K^+$ and $K^-$ data, but are taken to be uncorrelated between different angular bins. The magnitudes of the uncertainties in these three categories are summarized in Table~\ref{tab:kdataerr}.

\begin{table}
\centering
\caption{Summary of the fractional uncertainties in the kaon production data. The uncertainty in the overall normalization is $\sigma_N$. The uncertainty for a given data bin is $\sigma_{\Delta p \Delta \theta}$. The uncertainty in the normalization for a given angular bin is $\sigma_{\Delta\theta}$.}
\label{tab:kdataerr}
\begin{tabular}{lrrr}
\hline \hline
 & \multicolumn{1}{c}{$\sigma_N$} & \multicolumn{1}{c}{$\sigma_{\Delta p \Delta \theta}$} & \multicolumn{1}{c}{$\sigma_{\Delta\theta}$} \\
\hline
NA61/SHINE              &  $2.3\%$  & $11-24\%$   &  $-$    \\
Eichten \textit{et al.} &  $15\%$   & $4\%$       &  $5\%$  \\
Allaby \textit{et al.}  &  $10\%$   & $2-5\%$     &  $10\%$ \\
\hline \hline
\end{tabular}
\end{table}

In the case of the NA61/SHINE $K^+$ data, the systematic uncertainties (apart from the overall normalization) are treated as uncorrelated between different data bins. This is due to the fact that the dominant uncertainties for each bin are statistical. These uncertainties vary in the region $10-22\%$ depending on the momentum bin, while the systematic uncertainties are around $4\%$ for most of the bins. 

\begin{figure}
\centering
\includegraphics[width=0.45\textwidth]{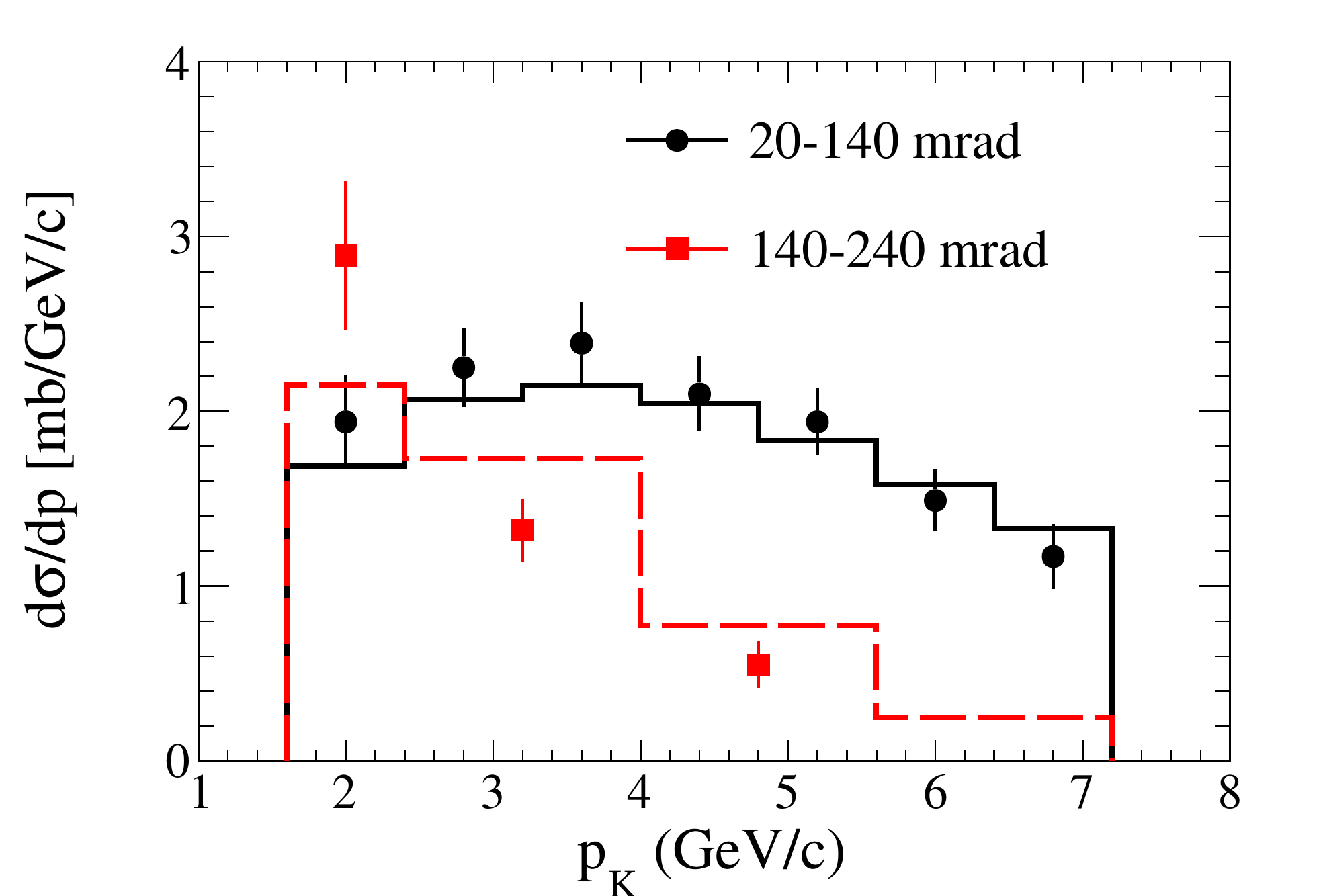}
\caption{The BMPT fits to the NA61/SHINE $K^+$ data.}
\label{fig:bmptfit_na61_kp}
\end{figure}

\begin{figure}
\centering
\includegraphics[width=0.45\textwidth]{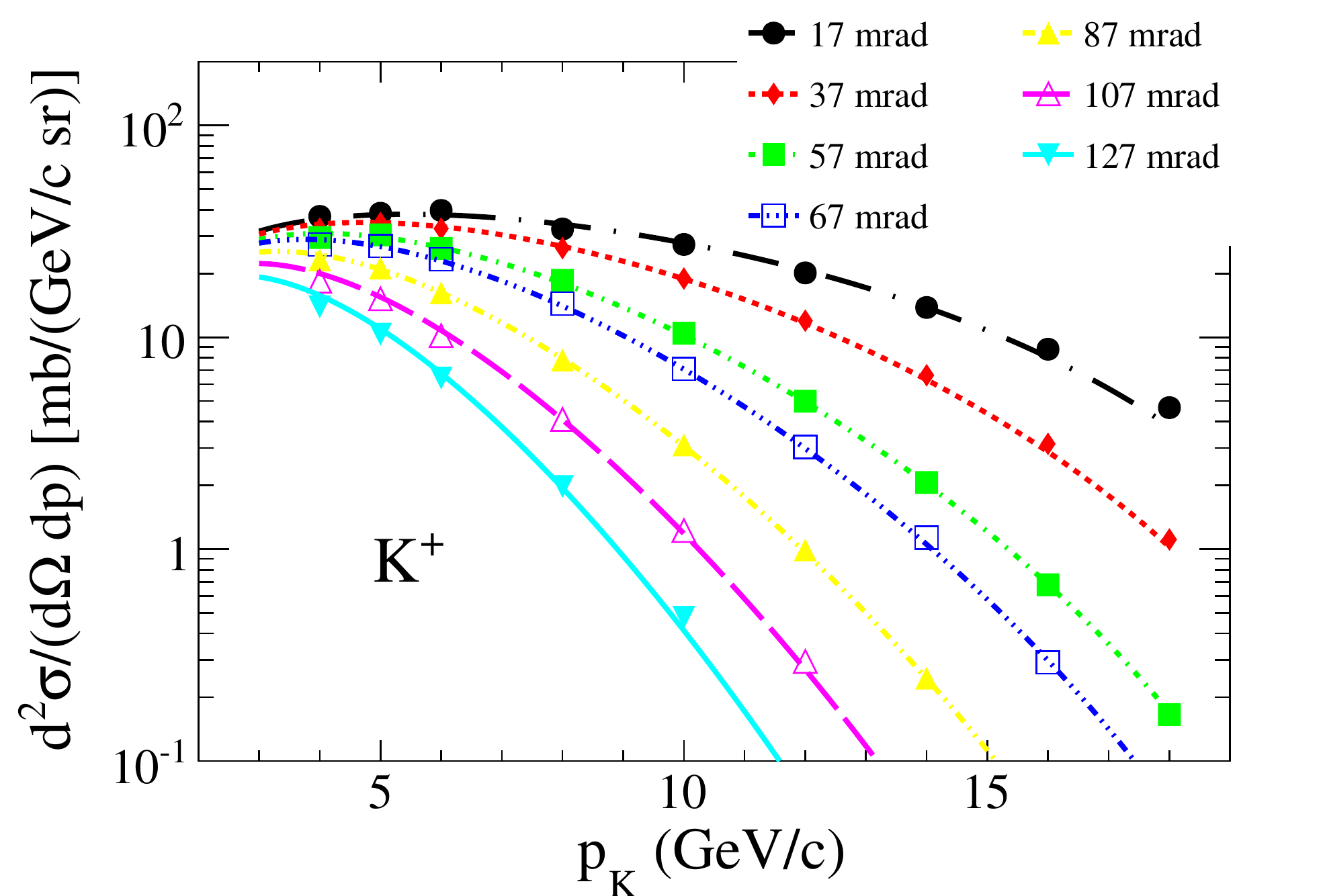}
\includegraphics[width=0.45\textwidth]{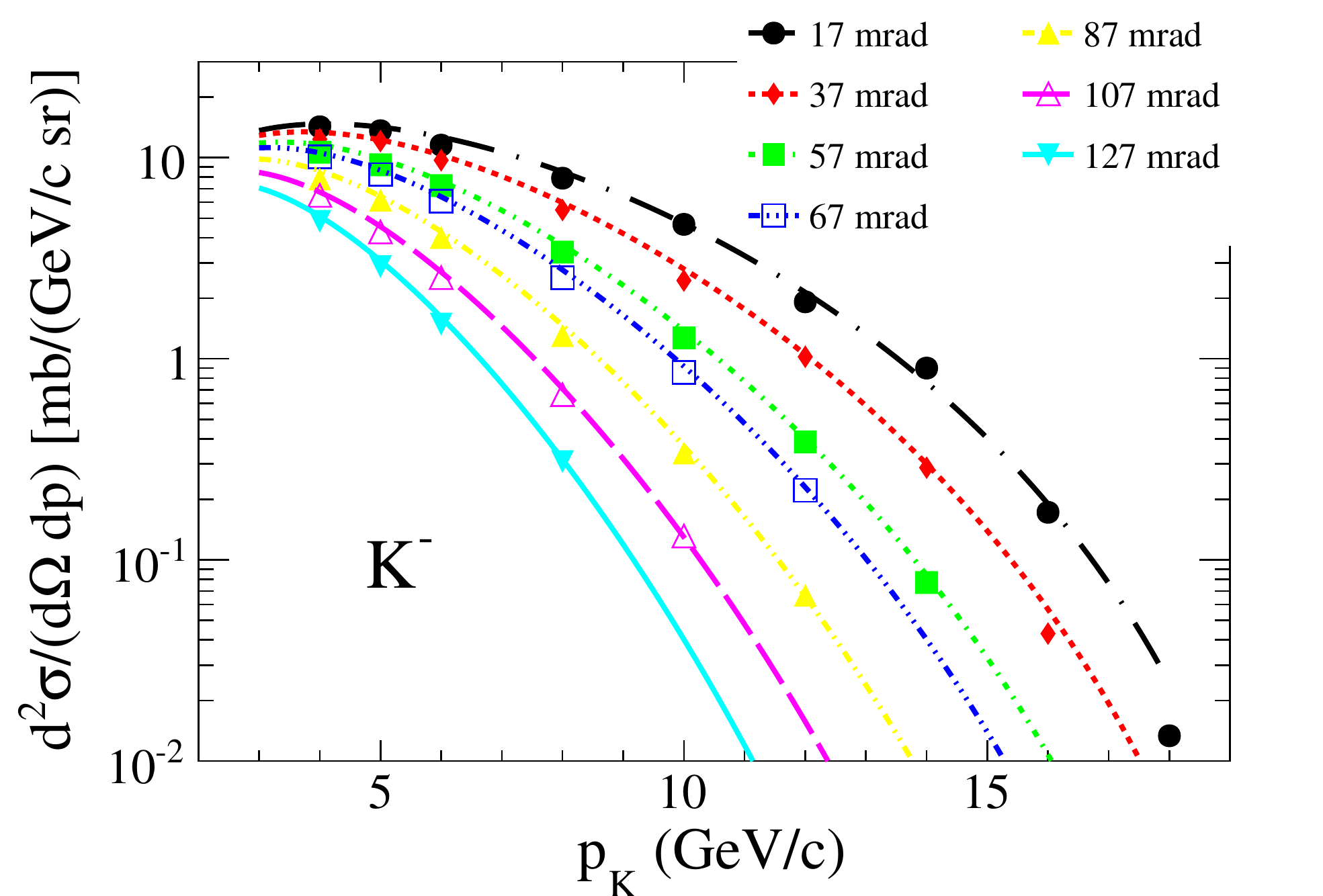}
\caption{The BMPT fits to the kaon data of Eichten {\it et al.} \cite{eichten}.}
\label{fig:bmptfit_eich_k}
\end{figure}

\begin{figure}
\centering
\includegraphics[width=0.45\textwidth]{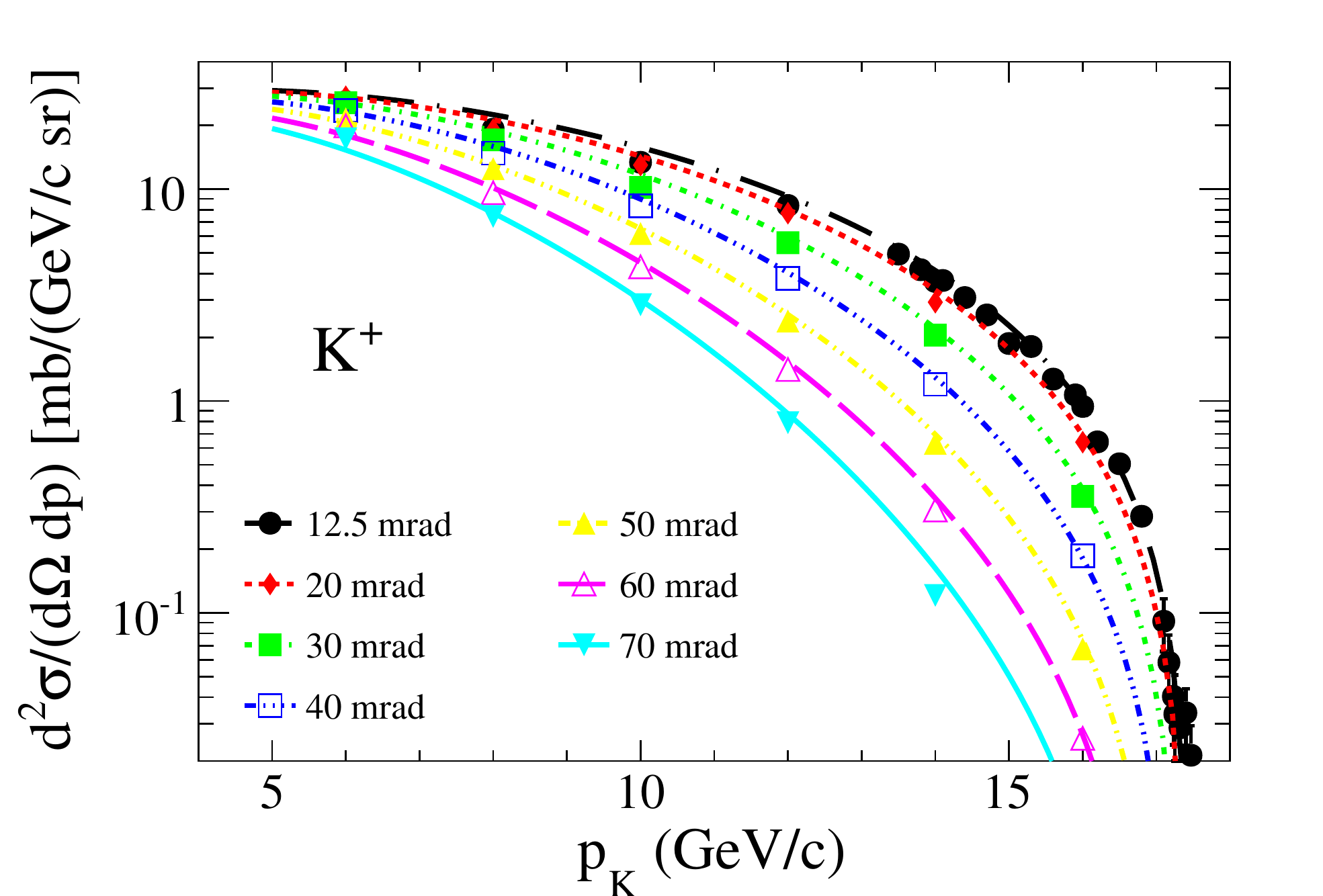}
\includegraphics[width=0.45\textwidth]{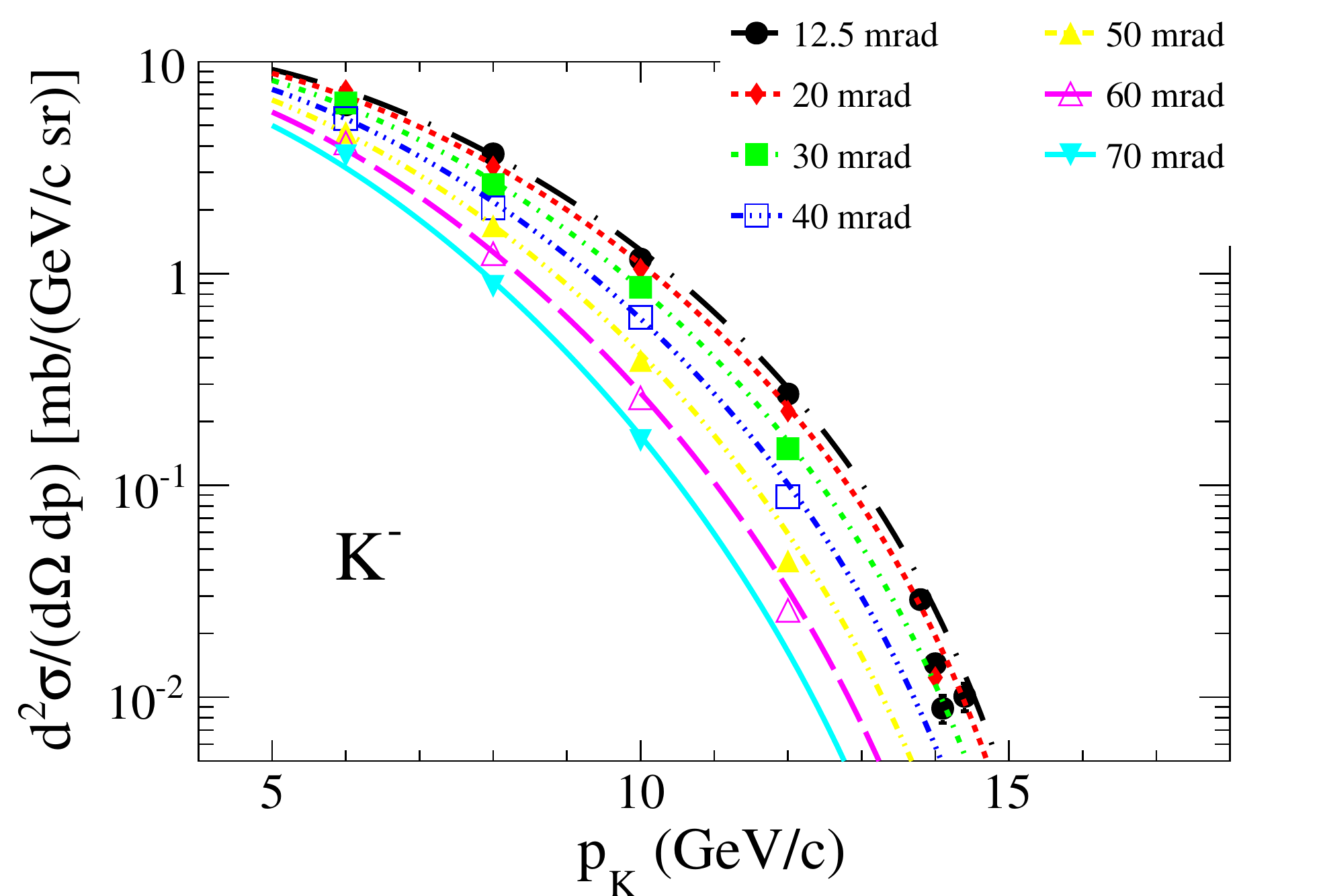}
\caption{The BMPT fits to the kaon data of Allaby {\it et al.} \cite{allaby}.}
\label{fig:bmptfit_alla_k}
\end{figure}

A coarse momentum and angular binning of the data had to be adopted for the NA61/SHINE $K^+$ data due to limited statistics. The sensitivity of the predicted neutrino flux to this choice of binning has been studied by modeling the shape of the $K^+$ production multiplicity within a given bin with the BMPT parameterization. The parameters in this parameterization have been determined from a combined fit to the kaon production data of the NA61/SHINE, Eichten \textit{et al.}, and Allaby \textit{et al} (see Figs.~\ref{fig:bmptfit_na61_kp} -- \ref{fig:bmptfit_alla_k}).
The change in the predicted flux when such shape information is included is treated as an additional systematic uncertainty.

\begin{figure}
\centering
\includegraphics[keepaspectratio=true,width=0.47\textwidth]{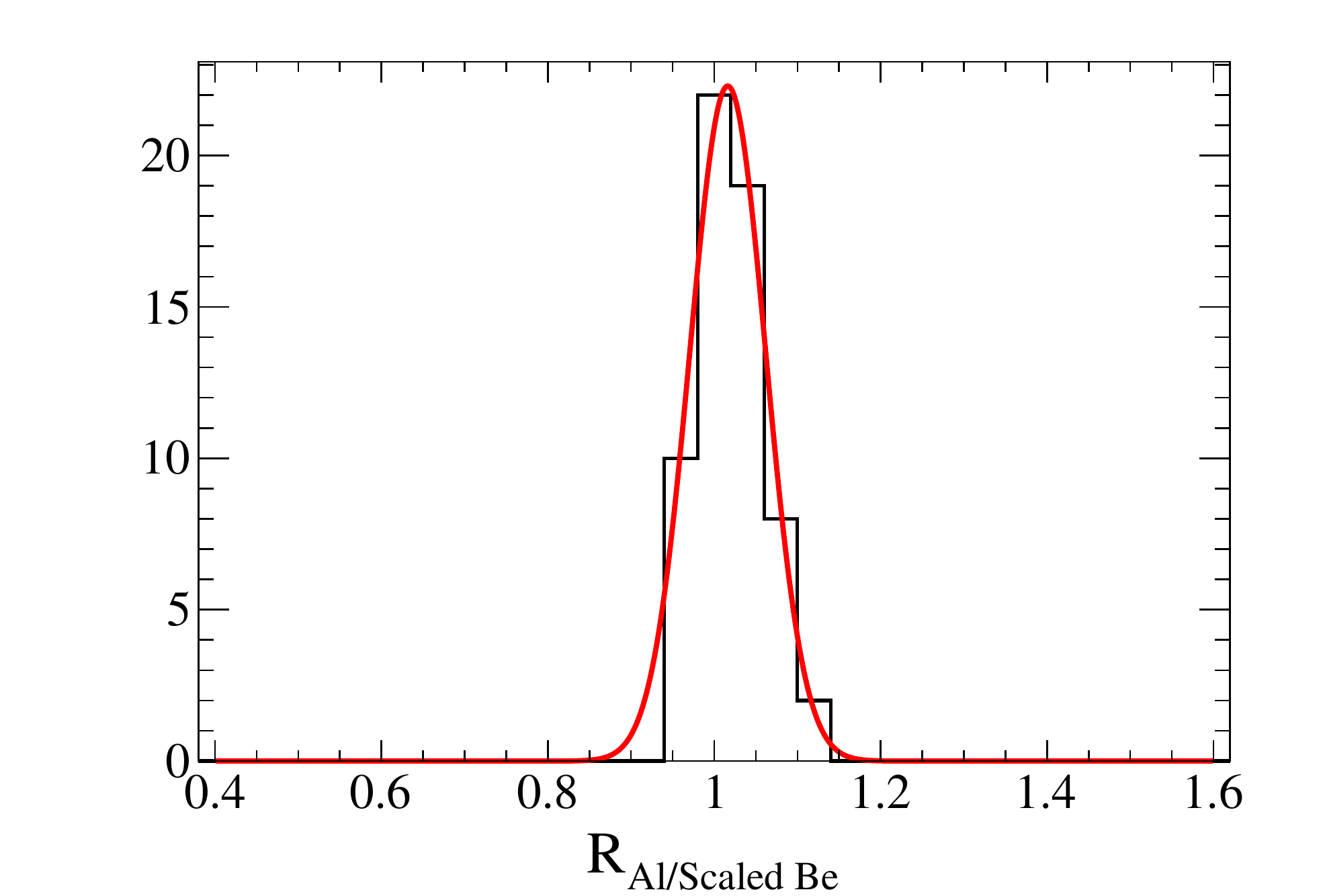}
\caption{Distribution of the ratios of Al to scaled Be yields, $R_\mtxtlbl{Al/Scaled Be}$, for the \cite{eichten} $K^+$ data. The fit (gaussian function), which extracts the mean and RMS of the distribution, is overlaid.}
\label{fig:eich_kp_A_sclerr}
\end{figure}

Since the data of Eichten \textit{et al.} and Allaby \textit{et al.} used in the re-weighting of the kaon multiplicity are for Be targets, uncertainties due to scaling of the differential yields from Be to C are applied in their case. 
These uncertainties are estimated based on the discrepancy between the measurements obtained with the Al targets by these two experiments and the expectations derived by scaling their yields from Be to Al following the procedure  outlined in Section~\ref{sec:hadron_prod_tune}. Two types of uncertainties are assigned: one, $\sigma^A_\mtxtlbl{bias}$, based on the average discrepancy observed for all the data bins and the other, $\sigma^A_\mtxtlbl{RMS}$ based on the RMS deviation from the mean value. To estimate these, the distributions of the ratios of Al to scaled Be yields, $R_\mtxtlbl{Al/Scaled Be}$, are checked for each meson type and each data set and the mean and RMS are extracted. An example of one such distribution for the $K^+$ data of Eichten \textit{et al.} is shown in Fig.~\ref{fig:eich_kp_A_sclerr}. Based on these distributions, $5\%$ is assigned to both $\sigma^A_\mtxtlbl{bias}$ and $\sigma^A_\mtxtlbl{RMS}$. The former is treated as a normalization type of an uncertainty correlated between the Eichten \textit{et al.} and Allaby \textit{et al.} data sets, while the latter is applied as an uncorrelated uncertainty for each data point.

The uncertainty in data scaling for different incident beam energies is assigned based on the change in the predicted neutrino fluxes when an alternative scaling variable $x_R$ is adopted in place of $x_F$. In addition, the scaling is checked with the data by re-scaling the measurements of Allaby \textit{et al.} from the 19.2 GeV/c to 24 GeV/c incident beam momentum and then comparing them directly with those of Eichten \textit{et al}. The discrepancy between the two is then included as an additional source of systematic uncertainty.

To assign the uncertainty on the parts of the production phase space outside of the ones covered by the data, the BMPT parameterization is used. Its predictions are compared to those of FLUKA and the difference is assigned as a systematic uncertainty. The uncertainties in the data used to determine the parameters in the function are also propagated. 

For the kaon production from the interactions in the Al around the target, the uncertainties are evaluated based on the comparison of the Eichten \textit{et al.} measurements to the GCALOR predictions. GCALOR predictions are also compared to the NA61/SHINE $K^+$ data after those are re-scaled to account for the difference in the target materials. The discrepancy between the data and the model predictions is treated as the systematic uncertainty. 

\begin{figure*}[ht]
\centering
\includegraphics[width=0.3\textwidth]{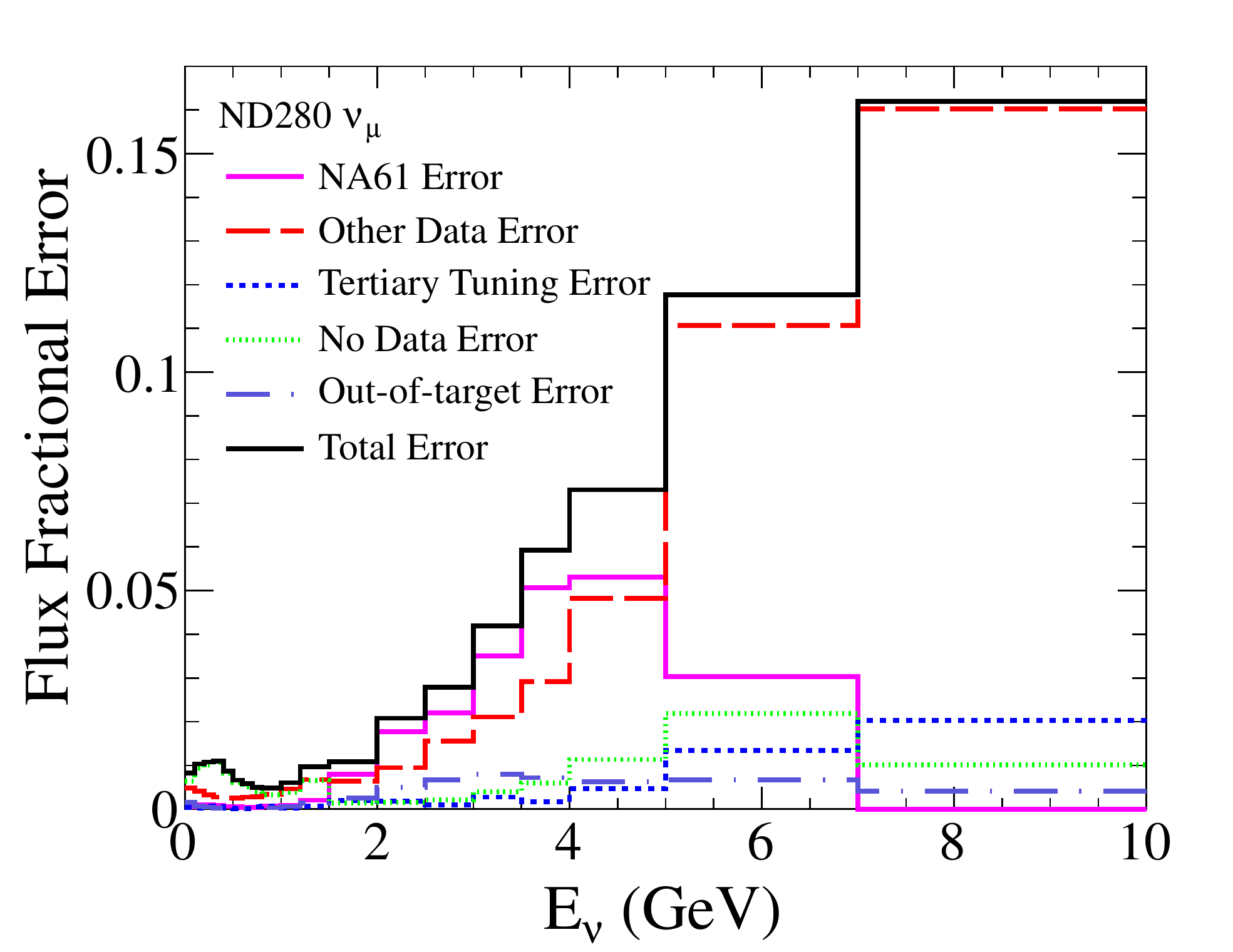}
\includegraphics[width=0.3\textwidth]{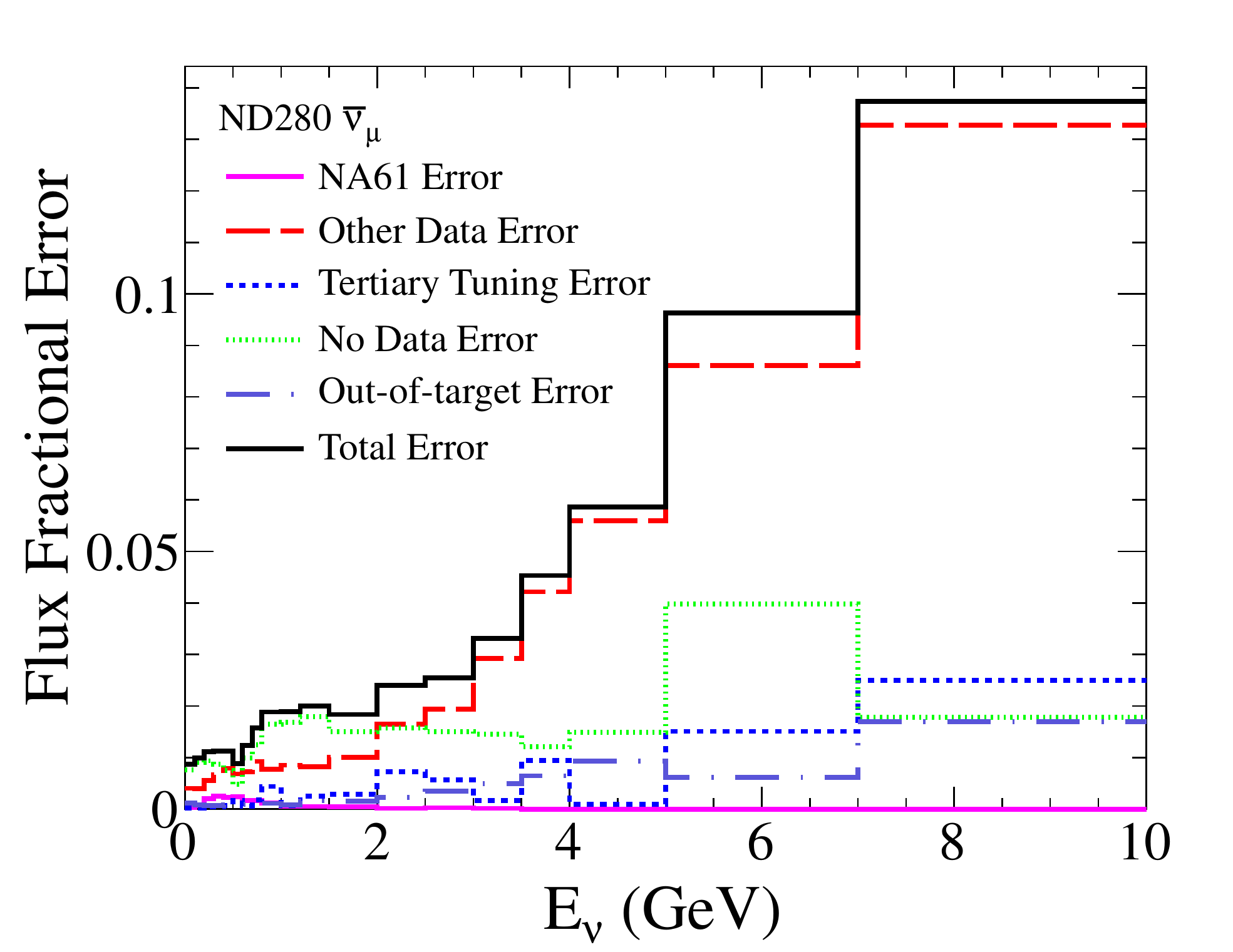}
\includegraphics[width=0.3\textwidth]{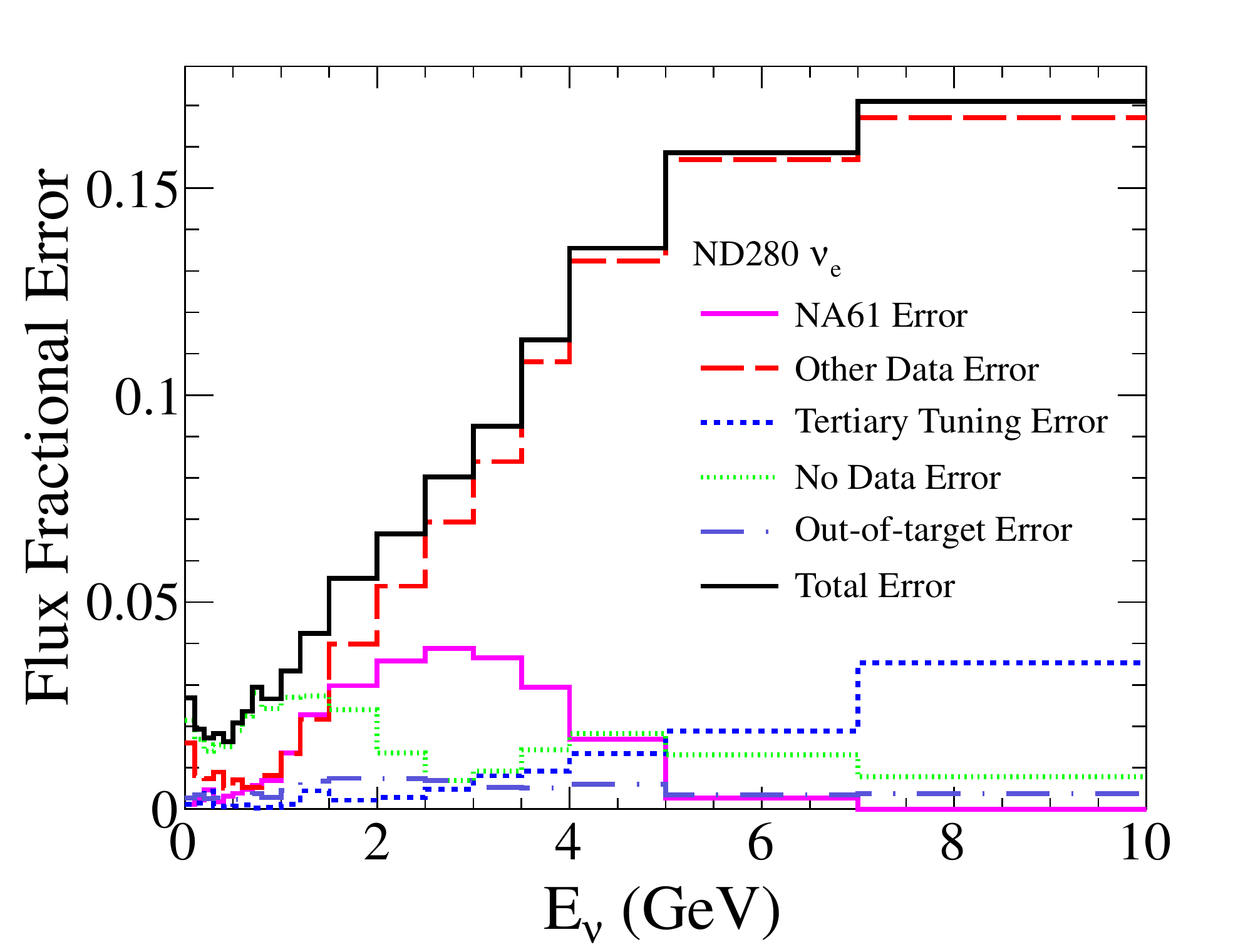}\\
\includegraphics[width=0.3\textwidth]{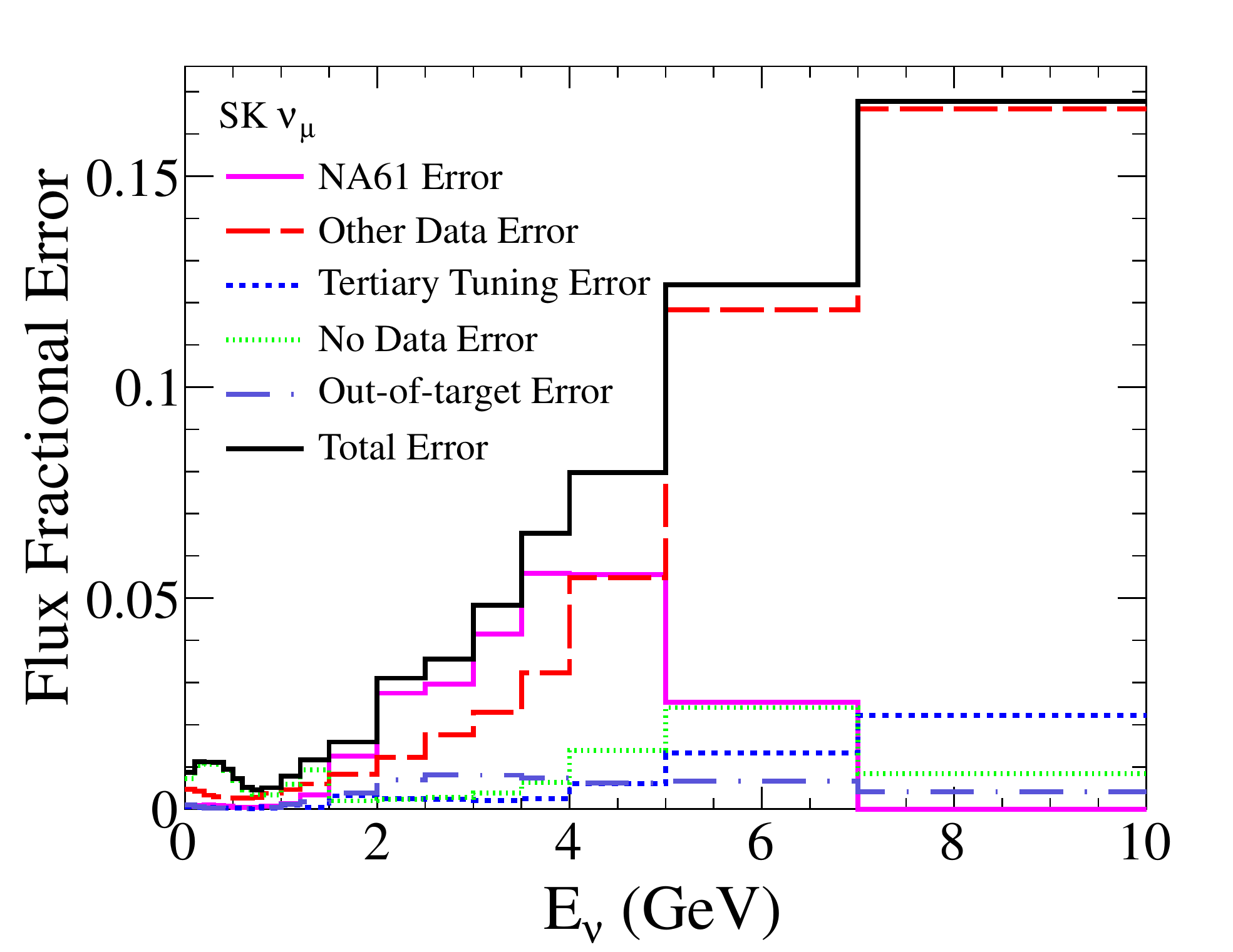}
\includegraphics[width=0.3\textwidth]{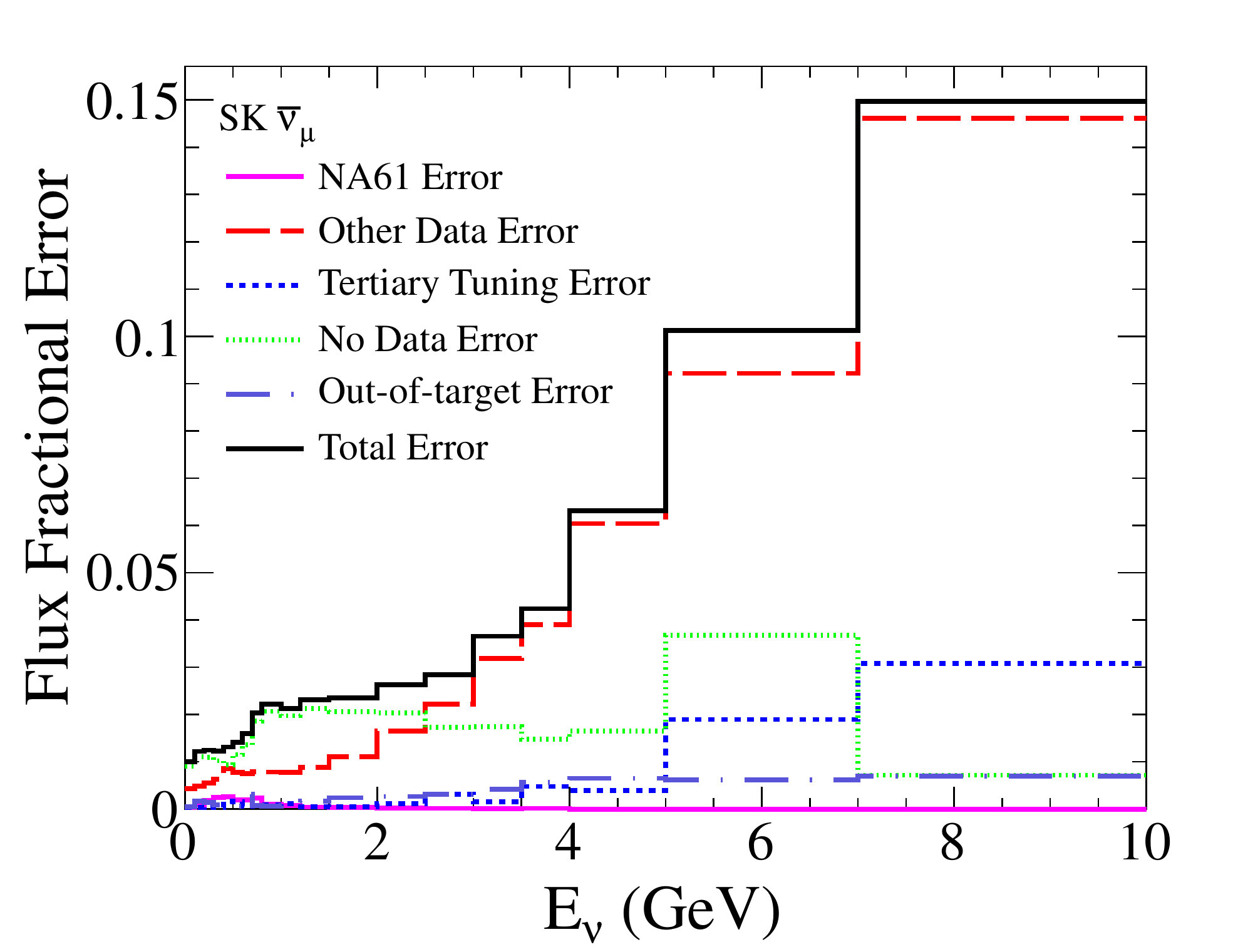}
\includegraphics[width=0.3\textwidth]{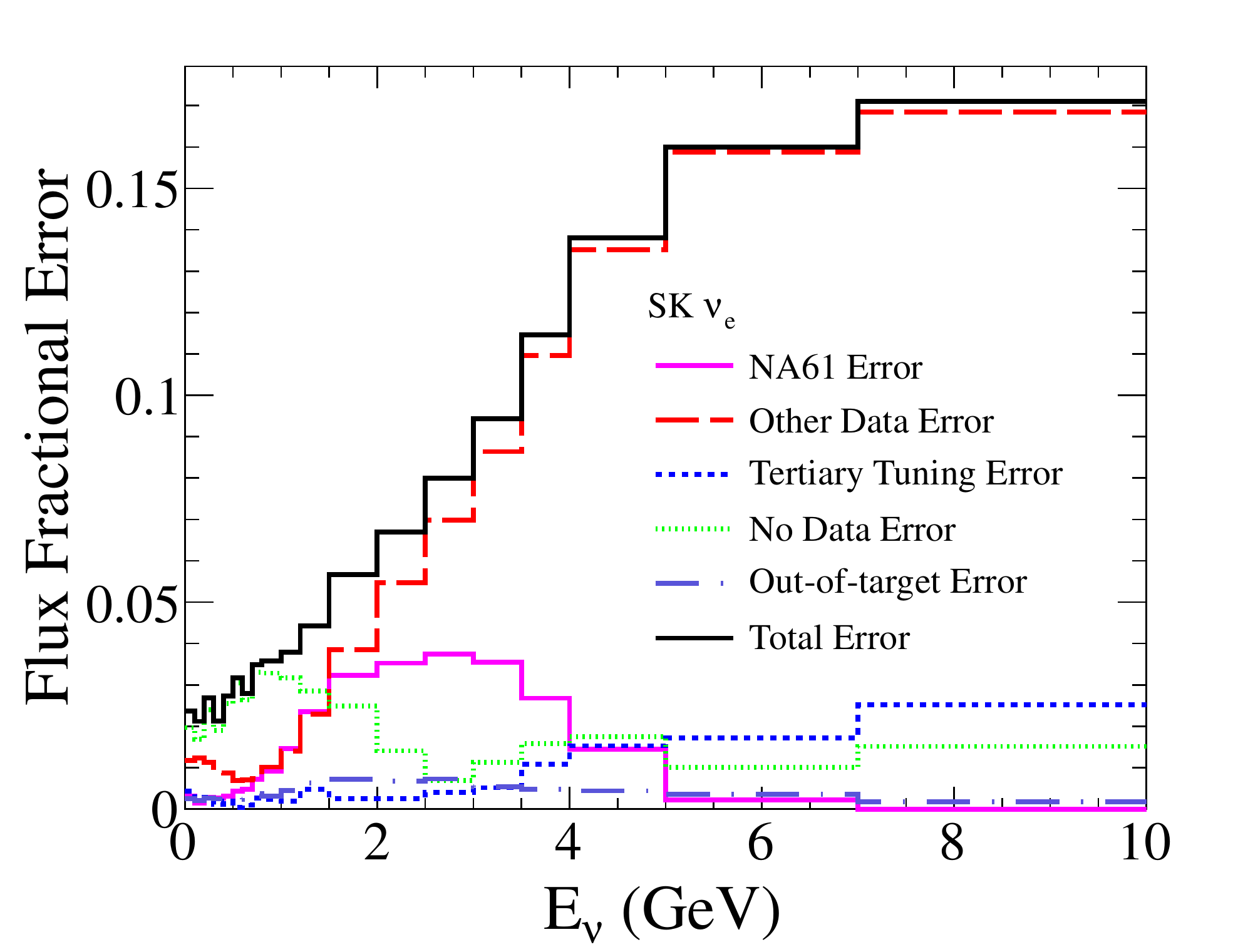}
\caption{Fractional flux error due to kaon production as a function of neutrino energy, for each
flavor and at the near and far detectors.}
\label{fig:kaonerrors}
\end{figure*}

The different contributions to the uncertainty associated 
with the kaon production are shown in Fig.~\ref{fig:kaonerrors}.

\subsubsection{Secondary nucleon production uncertainties} 
\label{subsec:secpnerrs}
\begin{figure*}[ht]
\centering
\subfloat[Secondary protons]{
\includegraphics[width=0.47\textwidth]{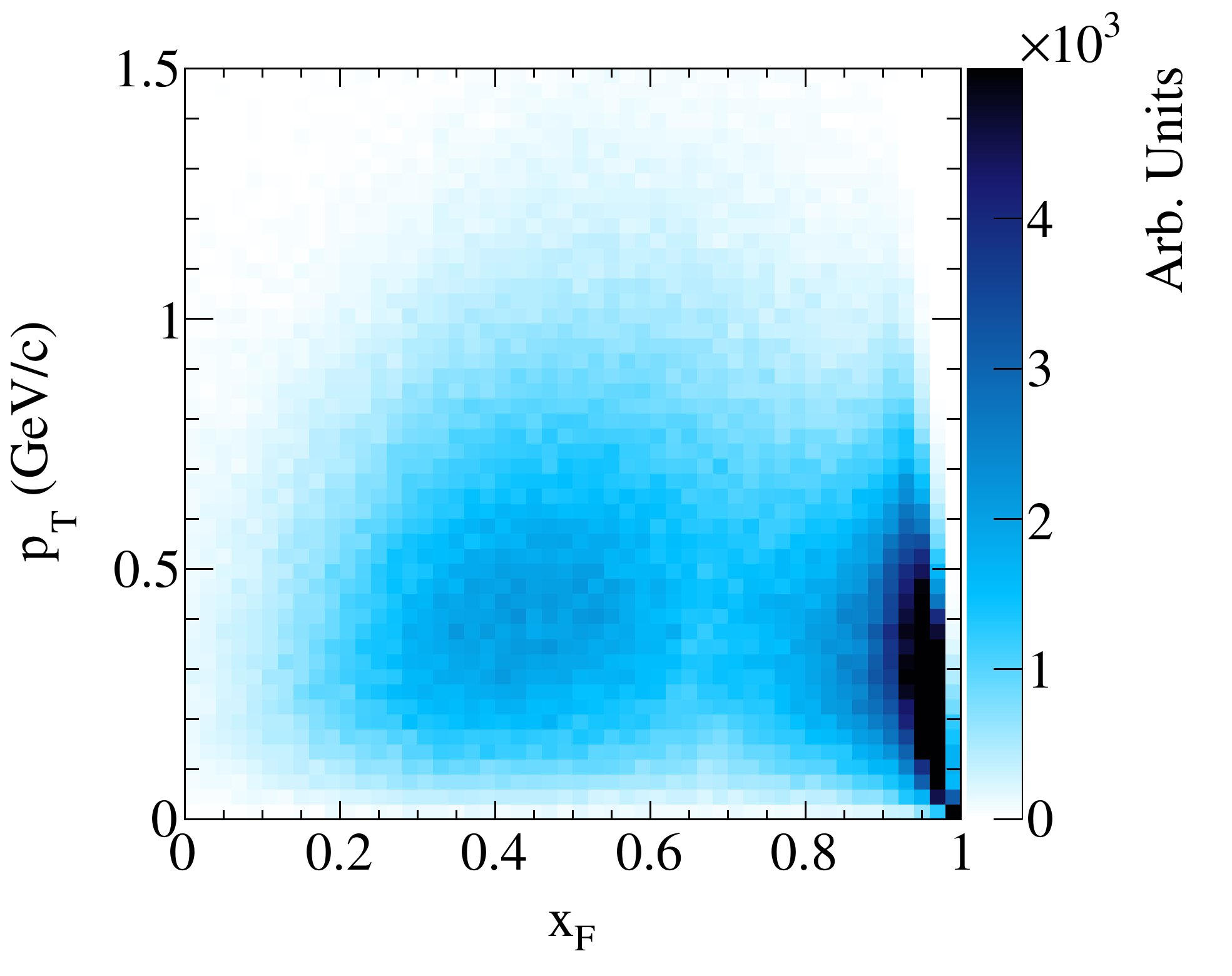}
\label{fig:secnucl_prot}}
\subfloat[Secondary neutrons]{
\includegraphics[width=0.47\textwidth]{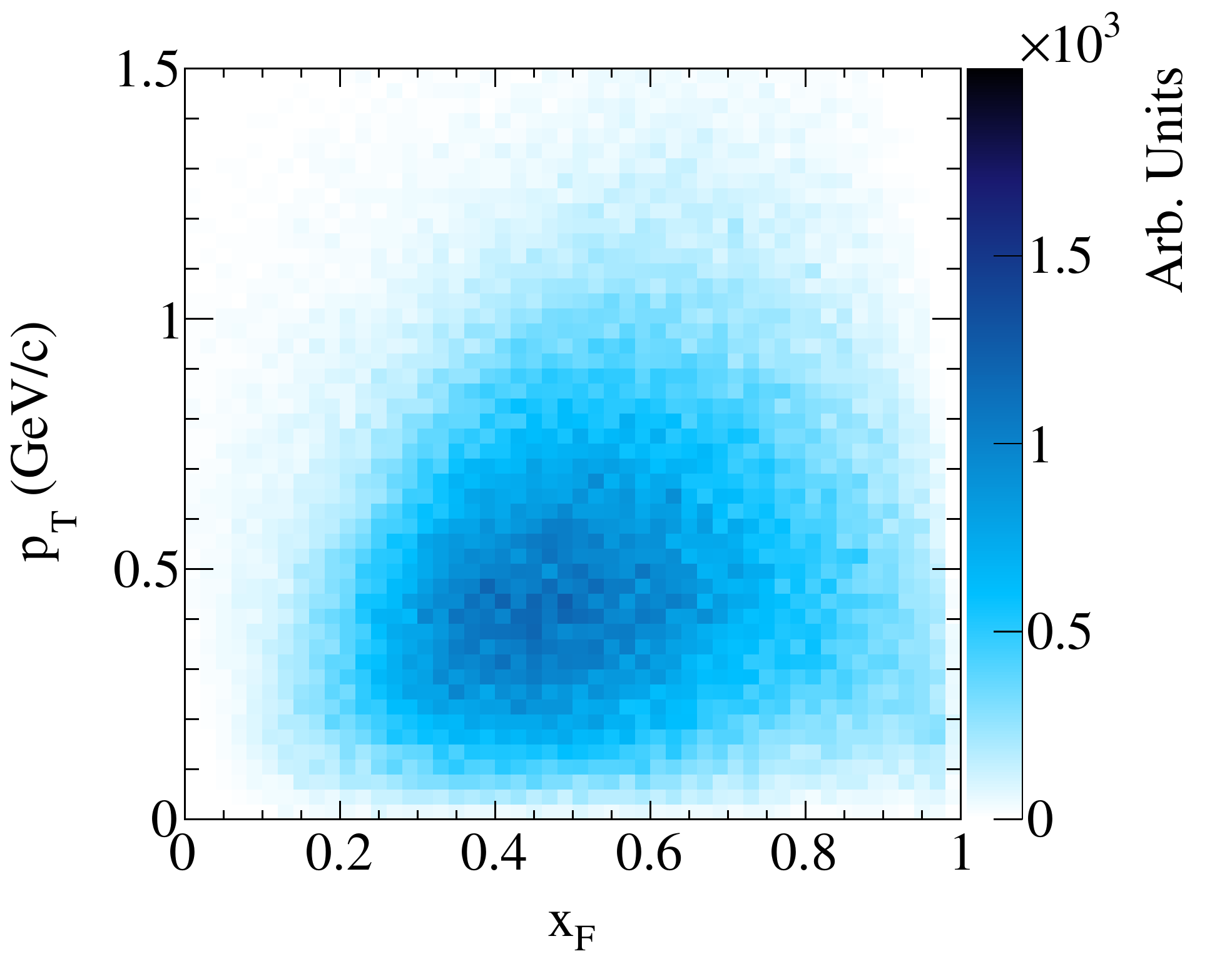}
\label{fig:secnucl_neut}}
\caption[Distribution of secondary protons and neutrons contributing to the neutrino flux at SK]{Distribution of secondary protons
 and neutrons contributing to the neutrino flux at SK, evaluated with the FLUKA hadron interaction model.}
\label{fig:secnucl}
\end{figure*}

Interactions of the secondary protons (neutrons) inside the target contribute about $16\%$ ($5\%$) to the neutrino flux. The $x_F - p_T$ phase space of the contributing protons and neutrons are shown in Fig.~\ref{fig:secnucl_prot} and Fig.~\ref{fig:secnucl_neut}, respectively. 

\begin{figure}
\centering
\includegraphics[width=0.47\textwidth]{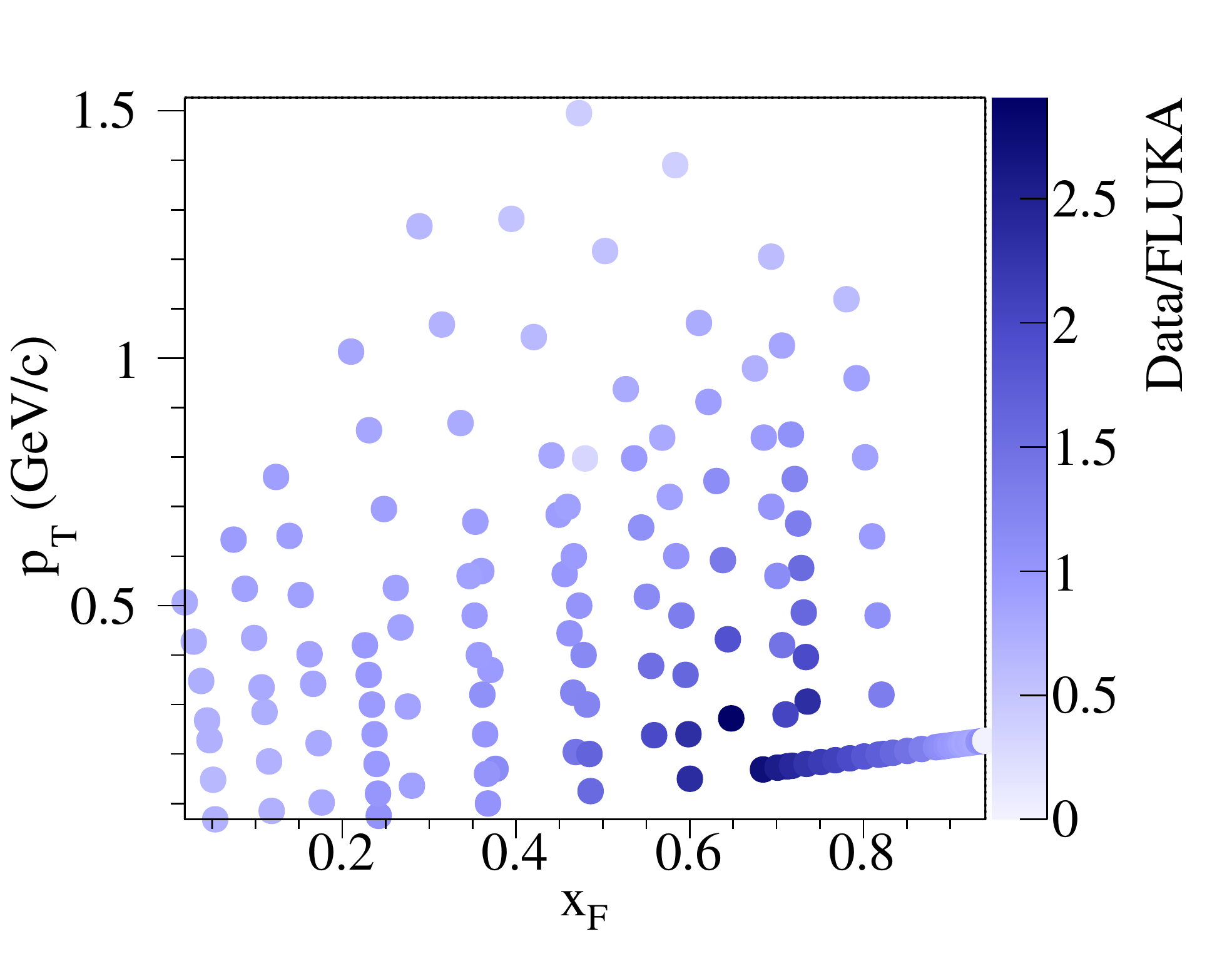}
\caption{Ratio of the secondary proton measurements from Eichten {\it et al.} and Allaby {\it et al.} and the
FLUKA modeling of secondary protons. Each circle is a point from the data sets. }
\label{fig:sec_prot_dataOMC}
\end{figure}

There are two components in the proton contribution: one for $x_F \leq 0.9$ and the other for $x_F>0.9$. This is not the case for neutrons where only those with $x_F<0.9$ contribute significantly. The high $x_F$ protons are produced in quasi-elastic scattering or scattering where soft pions are produced, while the contribution from $x_F \leq 0.9$ is due to hadronic production. The evaluation of the uncertainty for the secondary nucleon production is, consequently, separated into two regions.

In the region with $x_F \leq 0.9$ the uncertainty for the secondary proton production is evaluated based on the discrepancy
 between the FLUKA model and the proton production measurements of Eichten {\it et al.}~\cite{eichten} and 
Allaby {\it et al.}~\cite{allaby}. 
As shown in Fig.~\ref{fig:sec_prot_dataOMC}, the FLUKA model underestimates the production in the low $p_T$ region with $x_{F}>0.5$. 
The uncertainty in the neutrino flux prediction is calculated by weighting the FLUKA secondary proton and neutron production with the 
ratio of 
data to the FLUKA model, and is $<10\%$, as illustrated in Fig. ~\ref{fig:had_prod_errs}.
This is a conservative estimate of the uncertainty
since no constraint on the average multiplicity of nucleons is applied in the re-weighting procedure.  A re-weighting method that requires
an average nucleon multiplicity of 1 in the region of the phase space where $N/\bar{N}$ production is not kinematically allowed would
lead to a smaller estimate of the uncertainty, and will be considered in the future.

In the region of proton production with $x_F > 0.9$, the incident protons undergo collisions with small momentum transfer.  When studying variations in the flux due to changes in the secondary nucleon scattering, care is taken to ensure that the hard nucleon multiplicity remains unity, since no additional nucleons are produced. Due to the lack of relevant data, a 100\% uncertainty is assigned on the proton production multiplicity in this region, but the effect on the flux is still relatively small since these nucleons are forward-going and carry most of the original proton momentum.

\subsubsection{Production cross-section uncertainties} 
\label{subsec:xsecerrs}
\begin{figure}
\centering
\includegraphics[width=0.47\textwidth]{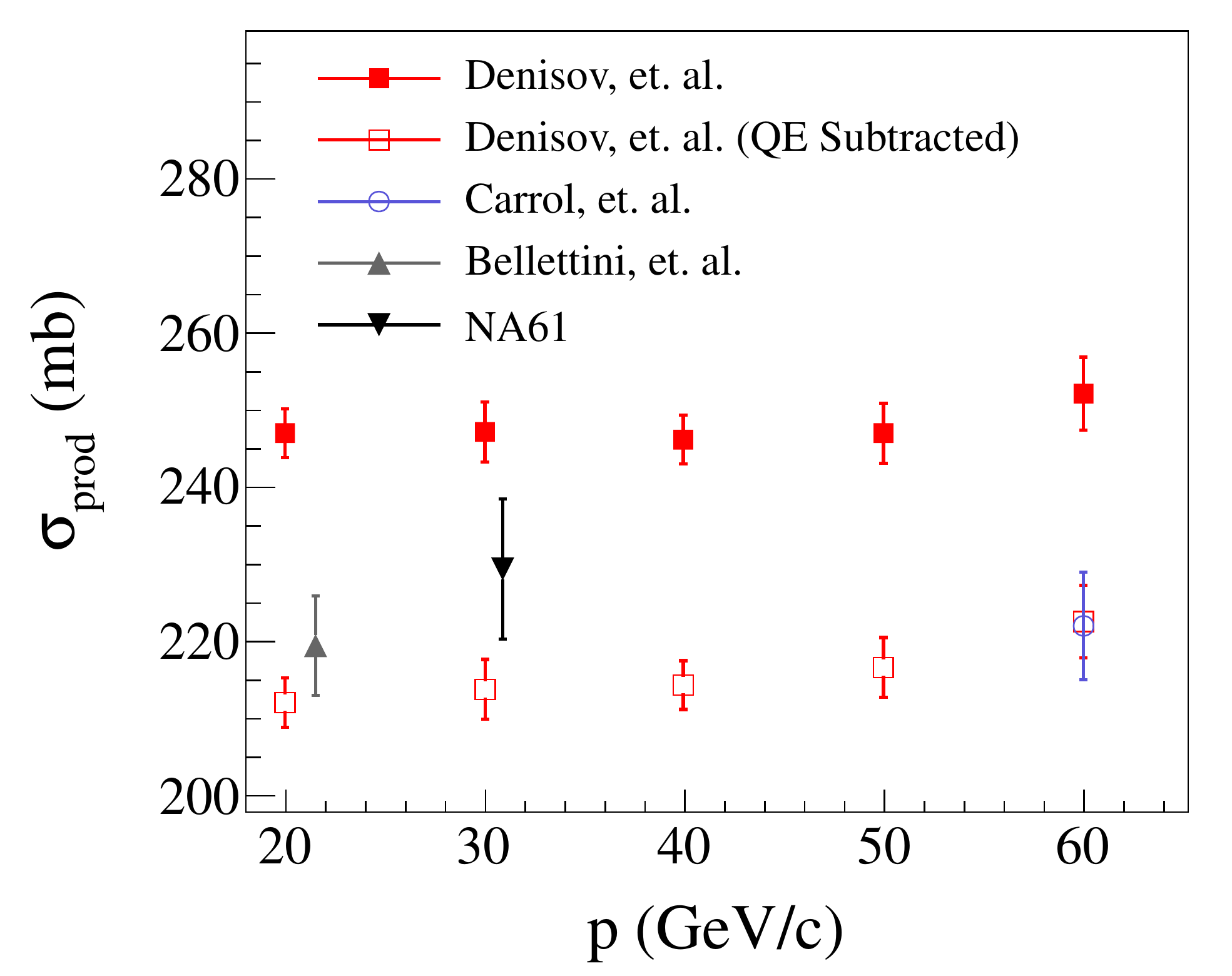}
\caption{Production cross-section measurements for protons on graphite targets for momenta 20--60 GeV/c.
The data from Denisov {\it et al.} are shown with and without the quasi-elastic estimate subtracted since
the quantity that is measured is ambiguous.}
\label{fig:prod_xsecdata}
\end{figure}

The systematic uncertainty in the production cross section is conservatively taken to be represented by the magnitude of the quasi-elastic correction, $\sigma_\mtxtlbl{qe}$, applied to the total inelastic cross section for a given particle and at given beam energy. This is based on an apparent discrepancy between the cross-section measurements for protons of Denisov \textit{et al.} \cite{Denisov} and those of Bellettini \textit{et al.} \cite{Bellettini1966609}, Carroll \textit{et al.} \cite{Carroll}, and NA61/SHINE \cite{Abgrall:2011ae}, which may be indicative of the difficulty in understanding whether experiments measure the inelastic or production cross sections. These data are plotted in Fig.~\ref{fig:prod_xsecdata}. 

For the measurement of Bellettini \textit{et al.}, the quasi-elastic contribution of 30.4 mb \cite{Abgrall:2011ae} has been subtracted from the reported value of 254 mb. In addition, the measurements by Denisov \textit{et al.} are also shown after an estimate of the quasi-elastic contribution has been subtracted from the reported values. The fact that after the subtraction the agreement between all of the four experiments is better can be interpreted as that the magnitude of the discrepancy is roughly similar to the size of the quasi-elastic cross section. A conservative approach is therefore taken by using $\sigma_\mtxtlbl{qe}$ as the systematic uncertainty.


\subsubsection{Summary of the hadron production uncertainties and prospect from future measurements} 
\label{subsec:sumhaderrs}

\begin{figure*}[ht]
\centering
\includegraphics[width=0.45\textwidth]{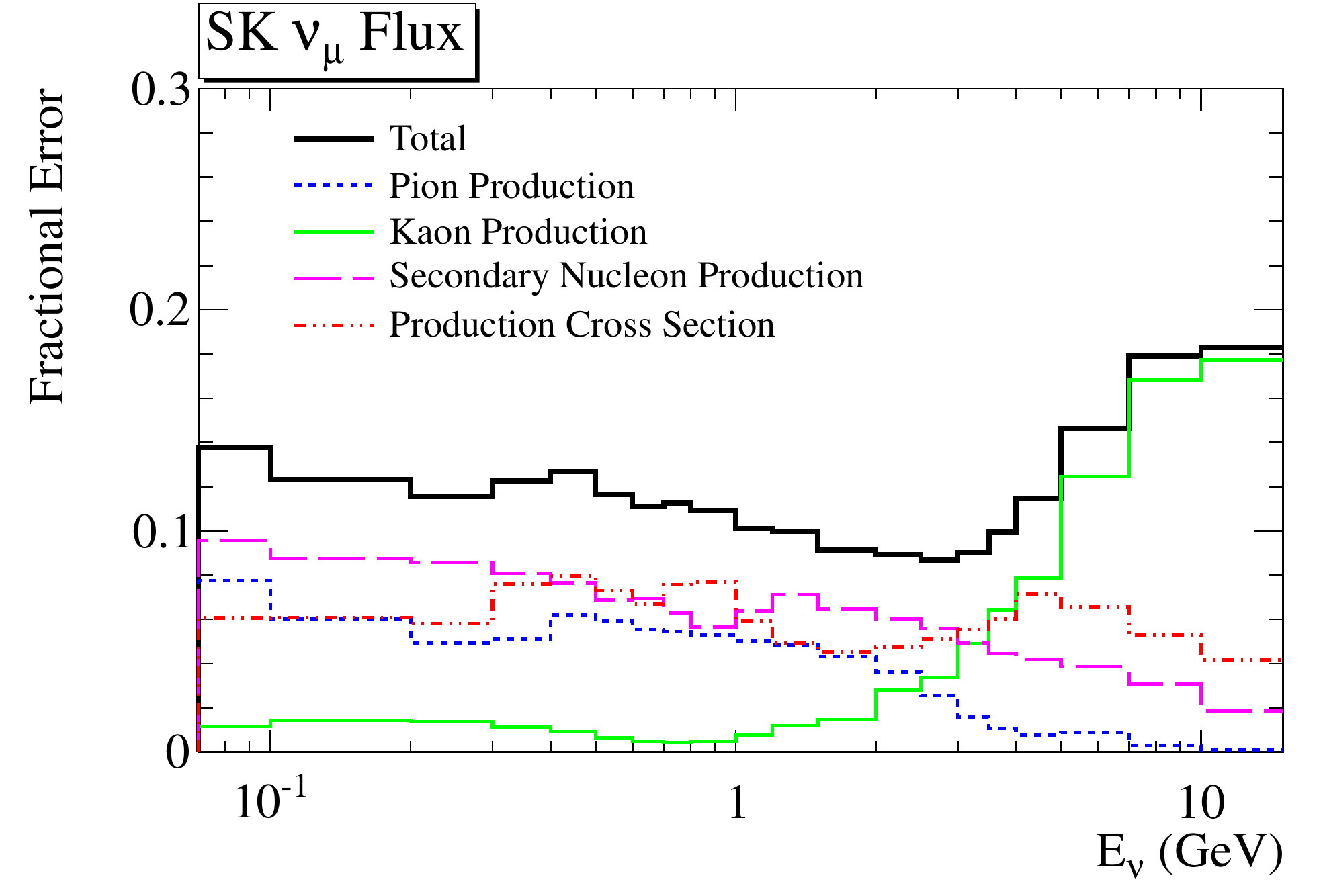}
\includegraphics[width=0.45\textwidth]{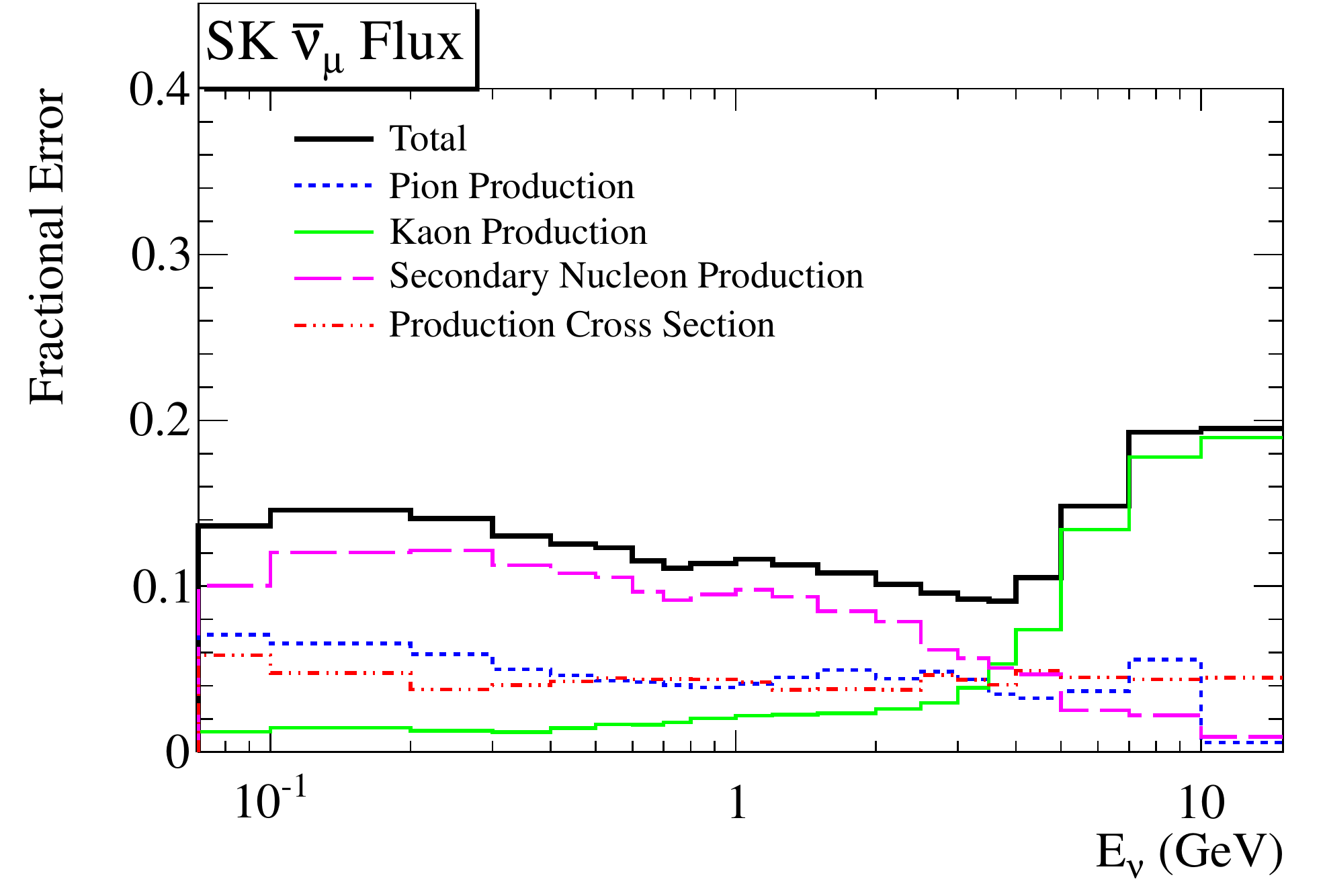}
\includegraphics[width=0.45\textwidth]{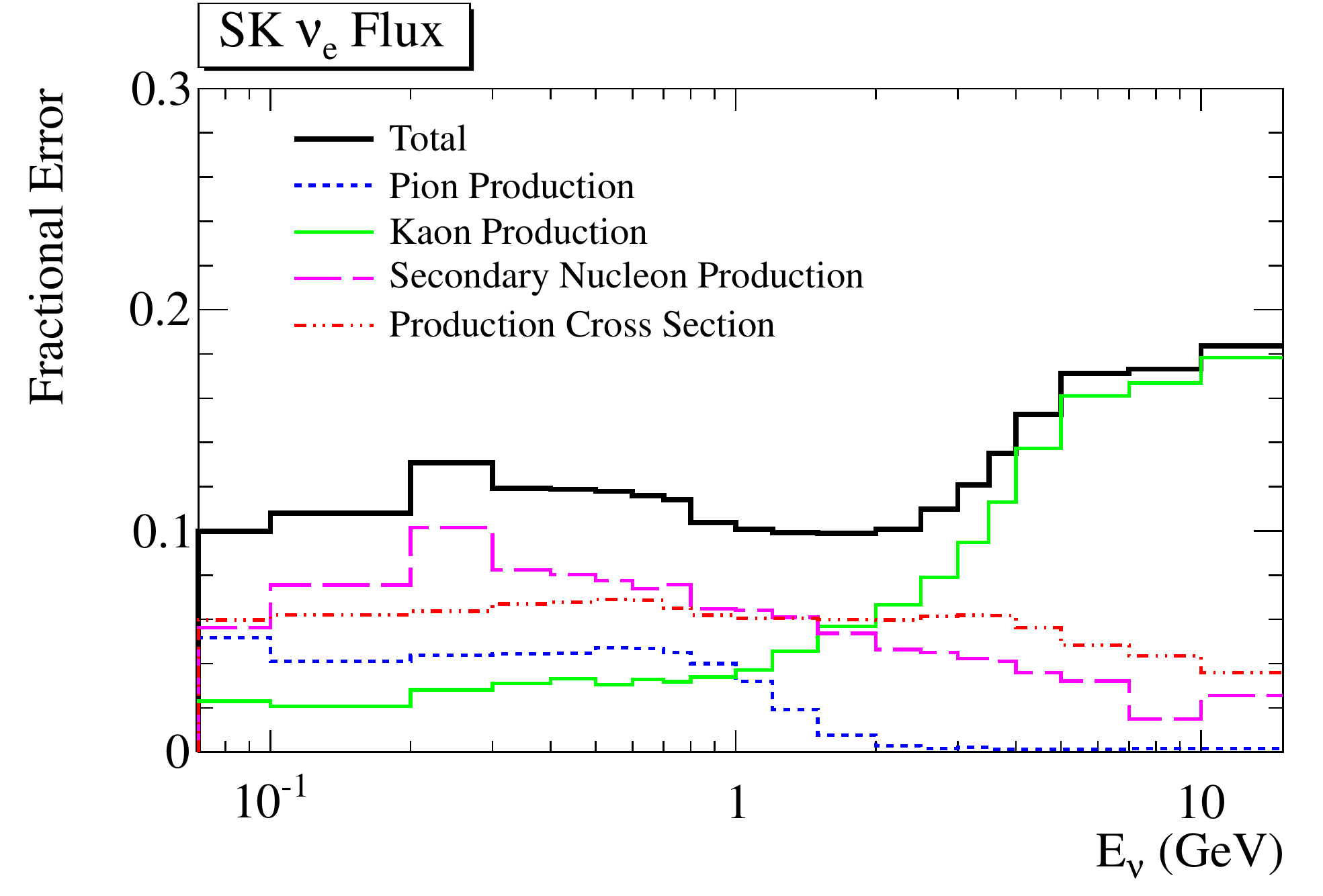}
\includegraphics[width=0.45\textwidth]{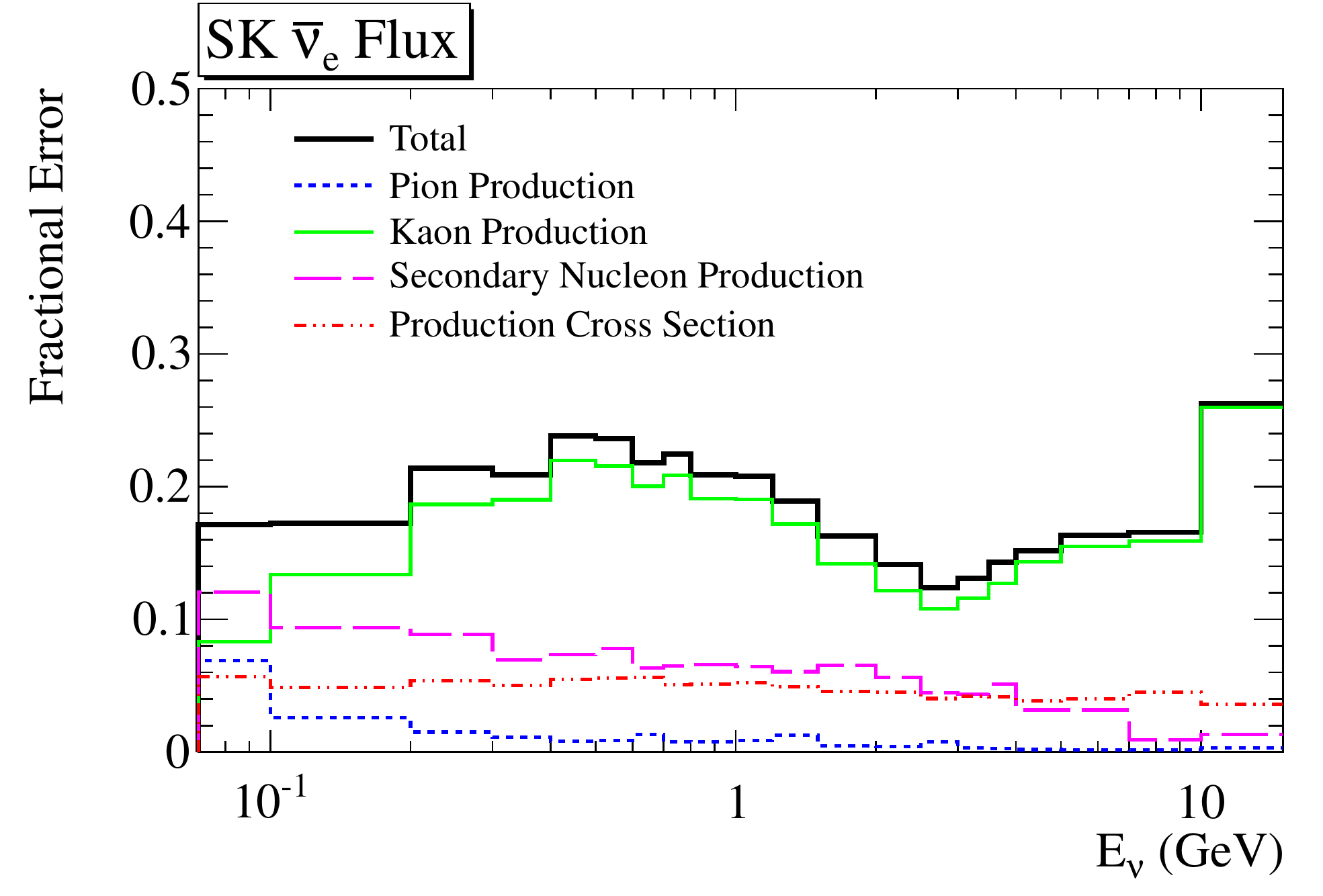}
\caption{Fractional flux error due to hadron production uncertainties.}
\label{fig:had_prod_errs}
\end{figure*}

The uncertainty on the SK flux as a function of neutrino energy due to hadronic interaction uncertainties
 is shown in Fig.~\ref{fig:had_prod_errs}.  
The uncertainties at the off-axis near detector are similar.  
At low energy,
the largest sources of uncertainty in the $\nu_{\mu}$ flux are from the secondary nucleon production
and production cross sections. 
At high energy, the flux uncertainty is instead dominated by the experimental errors on the kaon production.

The results of the next set of measurements from NA61/SHINE will reduce the
overall uncertainty on the neutrino flux prediction.
Higher statistics thin target data have been collected with an upgraded detector
configuration that increases the small angle acceptance.  These data will have reduced uncertainties
and cover the full phase space of interest for T2K.  In particular, the kaon production measurement
will be significantly improved.
The pion production uncertainty is already well controlled by the NA61/SHINE measurement, and the additional
data will have reduced uncertainties and slightly larger phase space coverage.
One of the major source of systematics, the contamination of pions from the decays of strange particles, 
will be further reduced by the NA61/SHINE measurement of $\Lambda$ and $K^{0}_{S}$ production rates.

\begin{figure}
\centering
\includegraphics[width=0.42\textwidth]{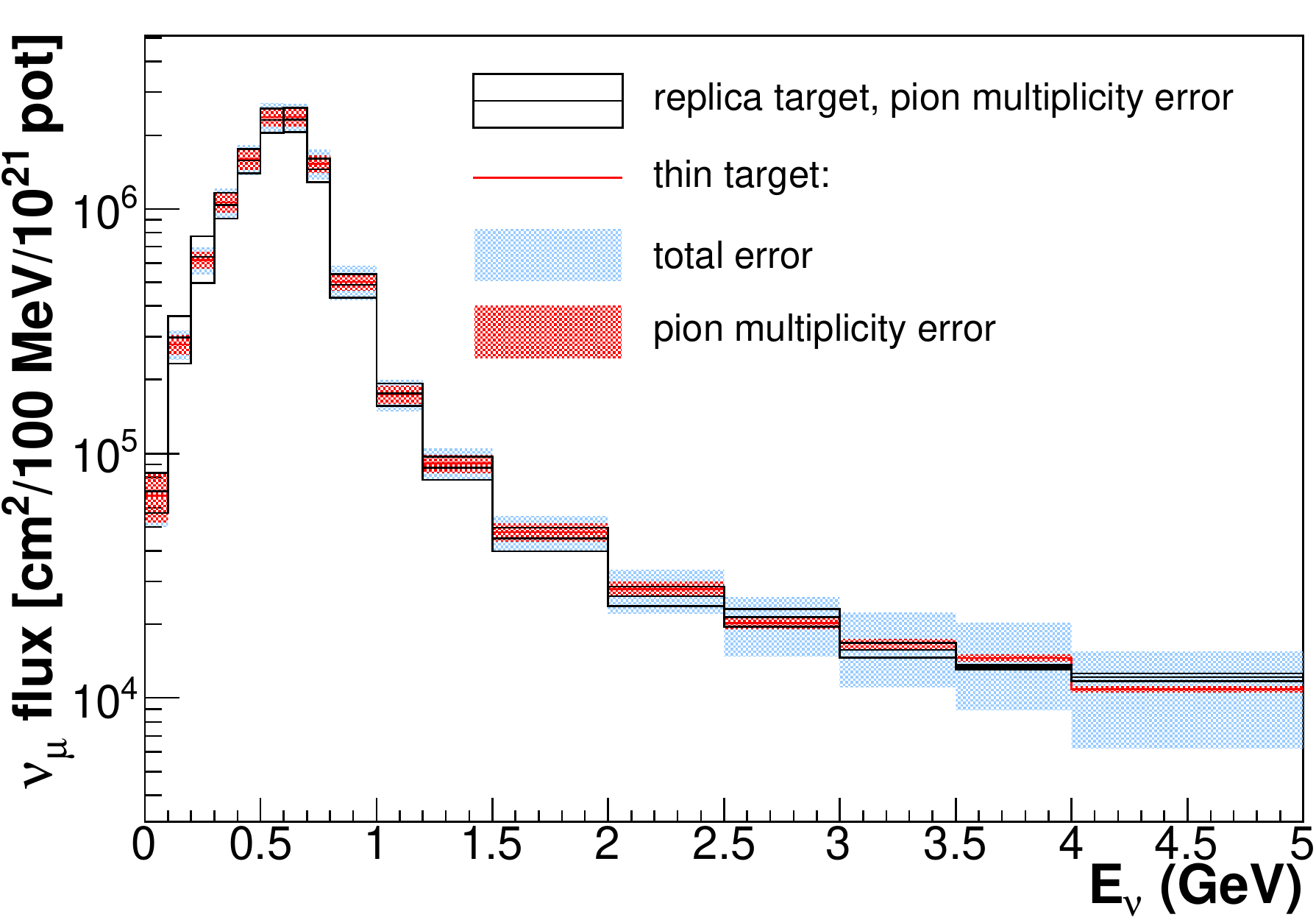}
\caption{Re-weighted $\nu_{\mu}$ flux at the far detector based on the NA61/SHINE thin target and
replica target measurements.}
\label{fig:na61_replica_rw}
\end{figure}

The ultimate precision on the flux prediction 
will finally be achieved through the measurements of hadron emission from the same (replica)  
target as the one used by T2K. 
With precise replica target measurements it will be possible 
to reduce the uncertainties related to the hadron production 
via reinteractions inside the target.
NA61/SHINE has already performed a pilot analysis using low statistics replica target data~\cite{na61_replica}
to establish the method for re-weighting the production of pions emitted from the T2K target.
%
Fig.~\ref{fig:na61_replica_rw} shows
the neutrino flux calculated using the re-weighting of the positively charged pion 
production based on the replica target data compared to the flux obtained
with the re-weighting based on the NA61/SHINE thin target measurements.
   
\subsection{\label{sec:pbeamoffaxiserrors}Proton beam and off-axis angle uncertainties}
\subsubsection{\label{sec:pbeamerrors}Proton beam systematic uncertainties}

The proton beam is generated in the simulation according to the measured primary proton orbit and optics parameters as described in Sec.~\ref{sec:pbeamprop}.

\begin{figure}[ht]
  \begin{center}
    \includegraphics[width=.45\textwidth]{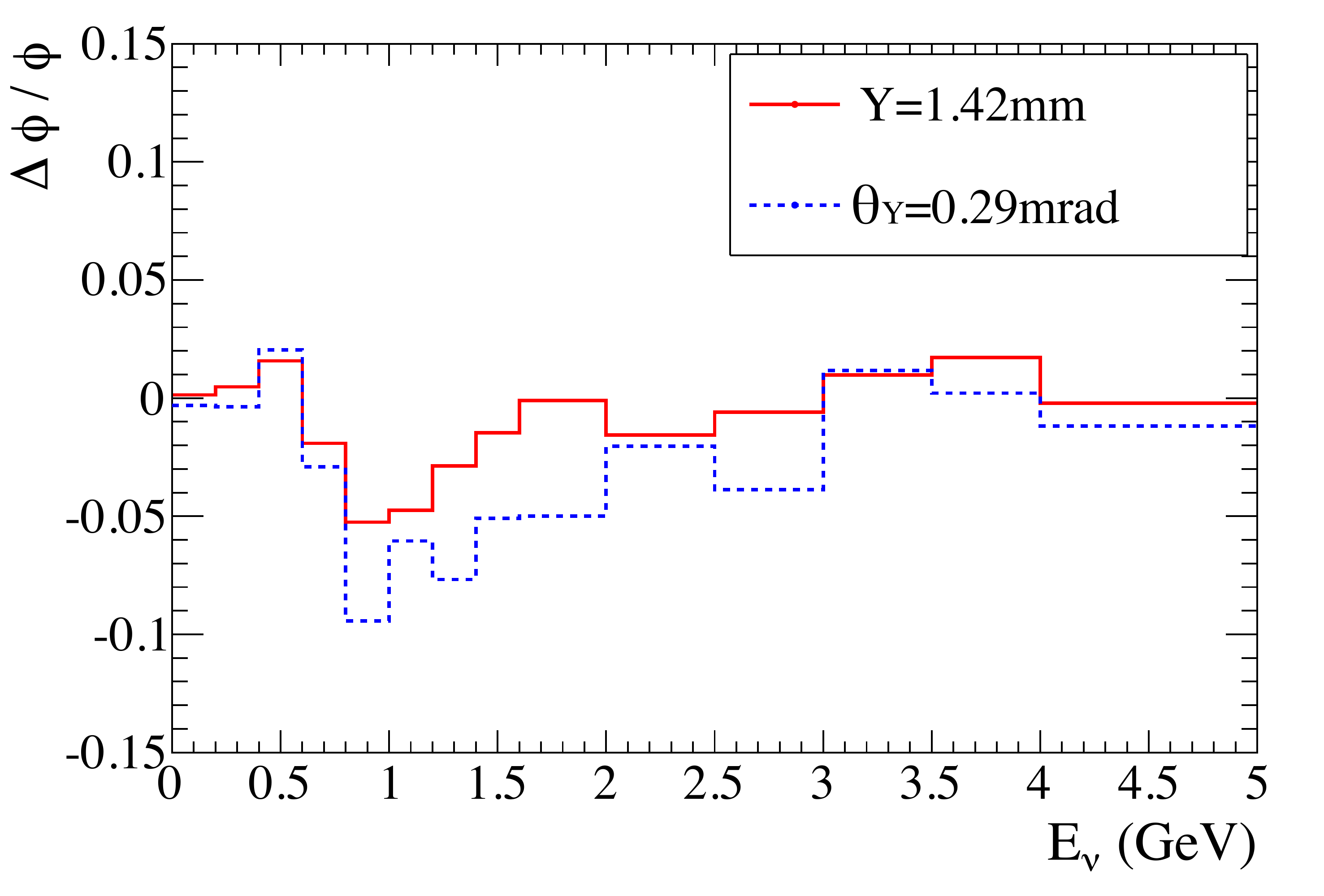}
  \end{center}
  \caption{
  An example of the fractional change of SK $\nu_{\mu}$ flux when the beam center position ($Y$) and center angle ($\theta_{Y}$) 
  measured in Run 1 are changed by 1$\sigma$, i.e. set to 1.42 mm and 0.29 mrad, respectively.}
  \label{pbeam_flxcmpr}
\end{figure}

To study the effects of the systematics errors in the proton beam measurements on the neutrino flux,
 those parameters were changed within the errors listed in Table~\ref{tab:pcenter_err}. 
The correlation among different parameters was taken into account. 
It was found that only the systematic errors for the vertical center position ($Y$) and center angle ($\theta_{Y}$) of the 
beam have a sizable effect on the neutrino flux prediction. This is because these parameters effectively change 
the off-axis angle at the far detector, which is displaced from the beam axis predominantly in the vertical direction. 
As an example, Fig.~\ref{pbeam_flxcmpr} shows the flux change when ($Y$, $\theta_{Y}$) are changed by their error sizes.
Therefore, only these errors are considered in the evaluation of the flux uncertainty.

%
%

A large number of flux samples were prepared with ($Y$, $\theta_{Y}$) thrown according to correlated uncertainties listed in Table~\ref{tab:pcenter_err}.
In order to avoid re-running JNUBEAM for these different sets of $Y$ and $\theta_{Y}$, a special sample was 
generated with a large emittance in the $Y-\theta_{Y}$ phase space and then weighted to reproduce each thrown pair of ($Y$, $\theta_{Y}$). 


%
%

The absolute flux normalization uncertainty arises from the errors 
on the proton beam intensity measured by CT5, i.e. 2\%
as described in Sec.~\ref{sec:pmon}.



\subsubsection{\label{sec:offaxiserrors}Neutrino beam direction (off-axis angle) systematic uncertainties}

The neutrino beam direction is measured by INGRID and the results are summarized in Table~\ref{tab:bmdir_ingrid}.

The neutrino flux uncertainty due to the uncertainty in the off-axis angle is evaluated by looking 
at a variation of the neutrino flux when the SK and ND280 detectors are moved 
by 0.44 mrad (Sec.~\ref{sec:secmon}) in JNUBEAM.
To save computational time, the neutrino flux predictions for the moved detectors are calculated by using the
nominal flux predictions and rescaling the energy and weight of each neutrino for the moved detector position
using the stored parent particle information.

\begin{figure}[]
  \begin{center}
    \includegraphics[keepaspectratio=true,width=80mm]{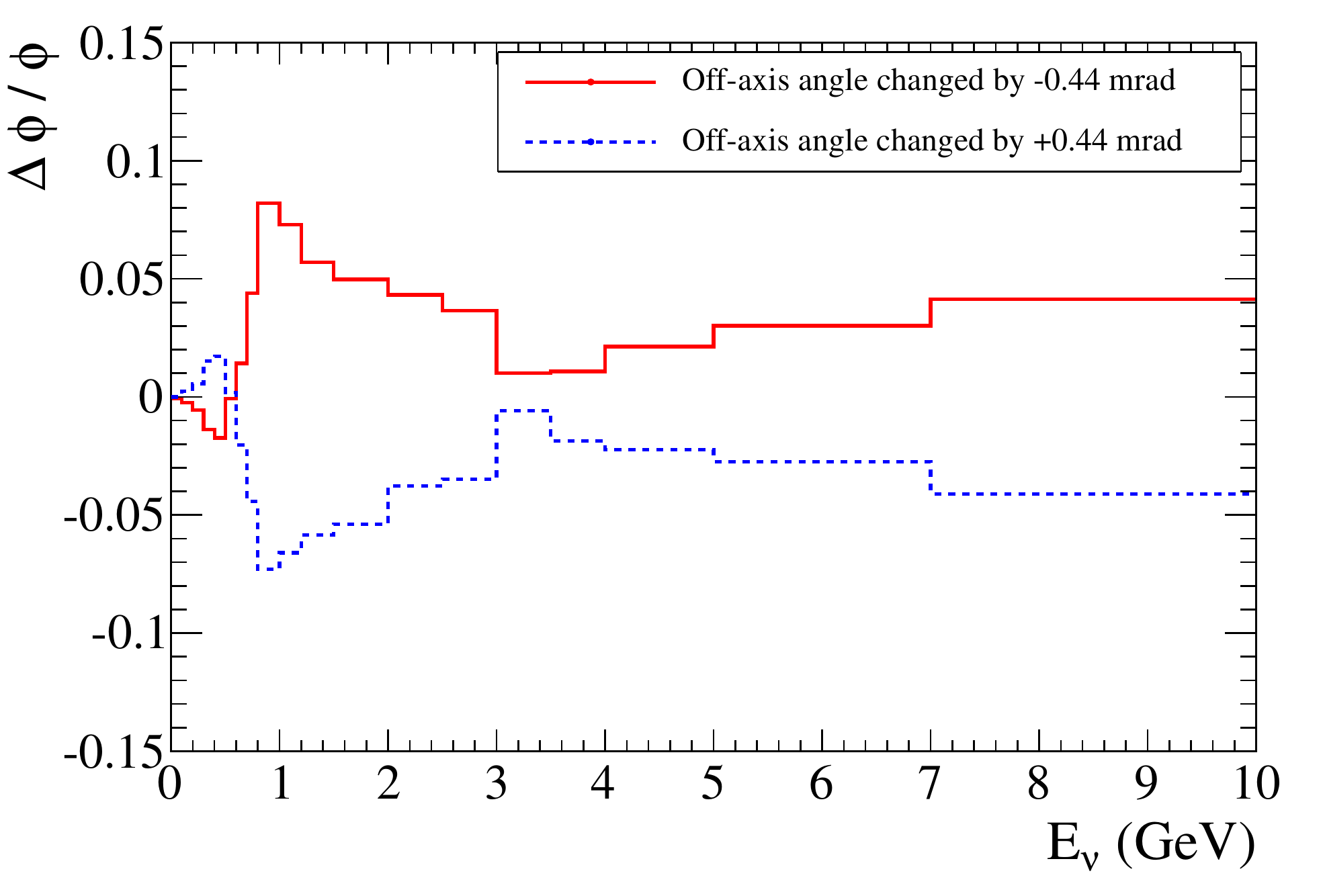}
    \includegraphics[keepaspectratio=true,width=80mm]{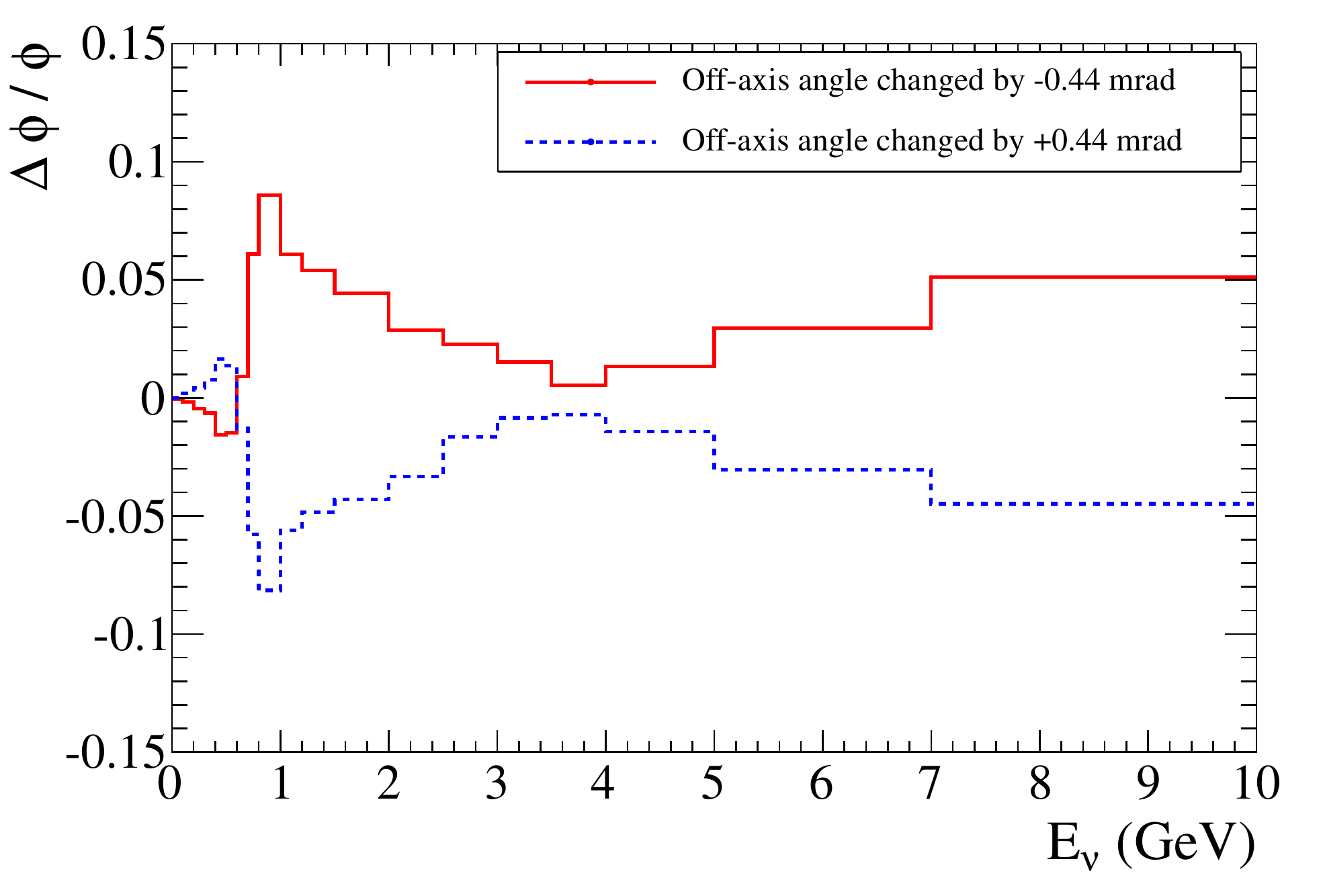}
    \end{center}


  \caption {\label{fig:err enu beamdir}
  Fractional change of the $\nu_{\mu}$ flux at ND280 (top) and SK (bottom) 
  corresponding to the systematics error of the off-axis angle. }
\end{figure}

Figure~\ref{fig:err enu beamdir} shows the variation of the neutrino flux due to the off-axis angle uncertainty.
The flux variations at the SK and the ND280 off-axis detector are similar to each other.

\subsection{\label{sec:alignerrors}Target and horn alignment uncertainties}
The systematic uncertainties associated with the target and horn alignments, discussed in Section~\ref{sec:alignment}, are summarized in Table~\ref{tab:targhorn_align}. The effects of the target alignment were studied by rotating the target in JNUBEAM by 1.3 (0.1) mrad in the horizontal (vertical) plane. This configuration results in a few percent change in the predicted neutrino flux, which is included as the systematic uncertainty. 

\begin{table}[ht]
\centering
\caption{Summary of the horn and target alignment uncertainties.}
\label{tab:targhorn_align}
\begin{tabular}{lcccc}
\hline \hline
                & \multicolumn{1}{c}{Target} & \multicolumn{1}{c}{Horn 1} & \multicolumn{1}{c}{Horn 2} & \multicolumn{1}{c}{Horn 3} \\
\hline
$\delta x$ (mm) & -- & 0.3 & 0.3 & 0.3 \\
$\delta y$ (mm) & -- & 1.0 & 1.0 & 1.0 \\
$\delta s$ (mm) & -- & 1.0 & 1.0 & 1.0 \\
$\delta\theta_\mtxtlbl{H}$ (mrad) & 1.3 & 0.2 & 0.2 & 0.2 \\
$\delta\theta_\mtxtlbl{V}$ (mrad) & 0.1 & 0.2 & 0.2 & 0.2 \\
\hline \hline
\end{tabular}
\end{table}

In the case of the horn position alignment uncertainties, the effects of horn movements along each coordinate axis were studied. Out of the three directions only the uncertainty in $y$ results in a sizable change (at a few percent level) in the predicted flux. Since the dominant contribution to this systematic uncertainty is an overall uncertainty in the relative alignment between the primary beamline and the secondary beamline, it is treated as fully correlated between the horns. For the case of the horn angular alignment uncertainties, the effects of horn rotations in both the horizontal and vertical plane by 0.2 mrad were studied. Only rotations of the first horn, however, showed any significant effect on the predicted neutrino flux.

\begin{figure}[ht]
\centering
\includegraphics[width=0.45\textwidth]{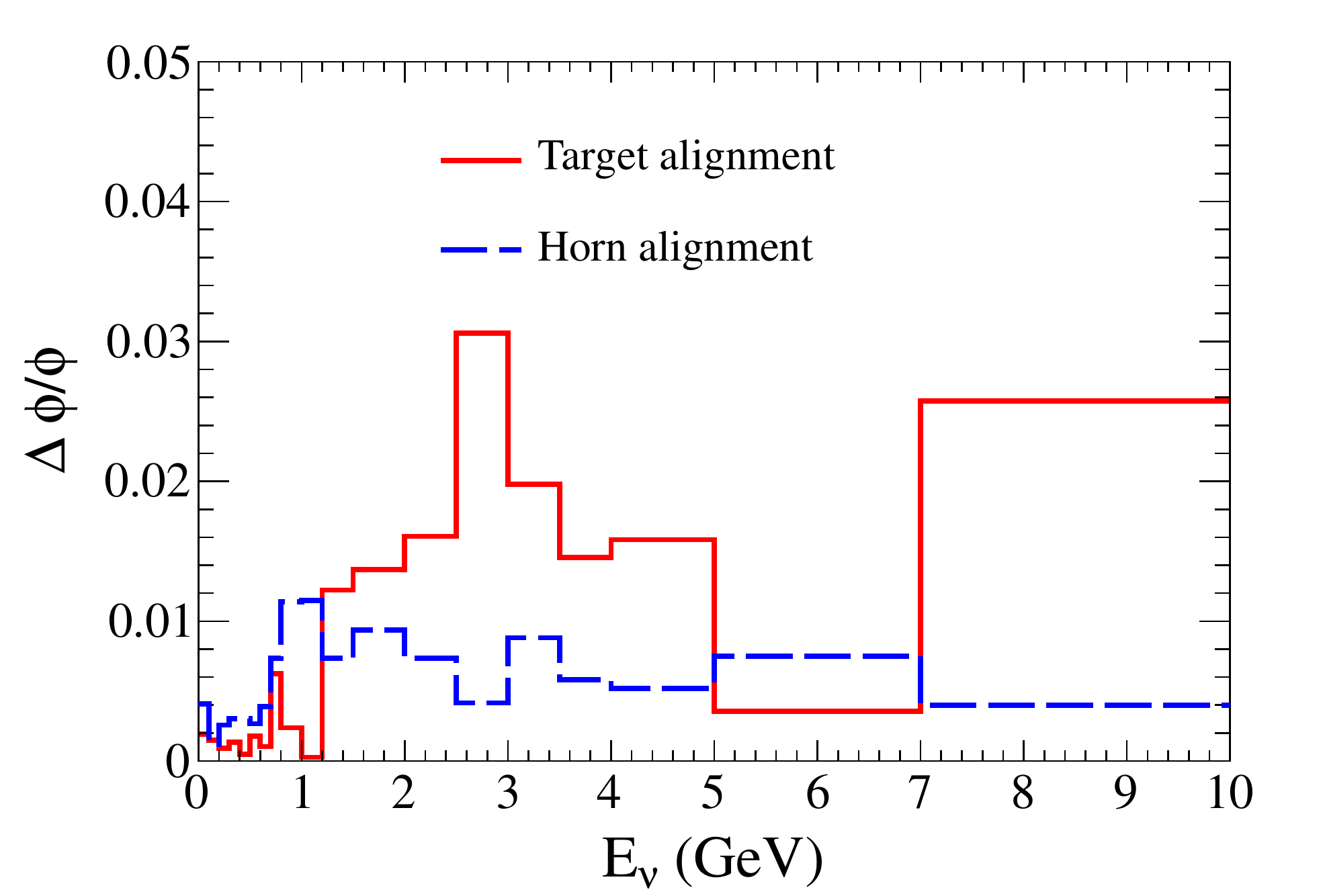}
\includegraphics[width=0.45\textwidth]{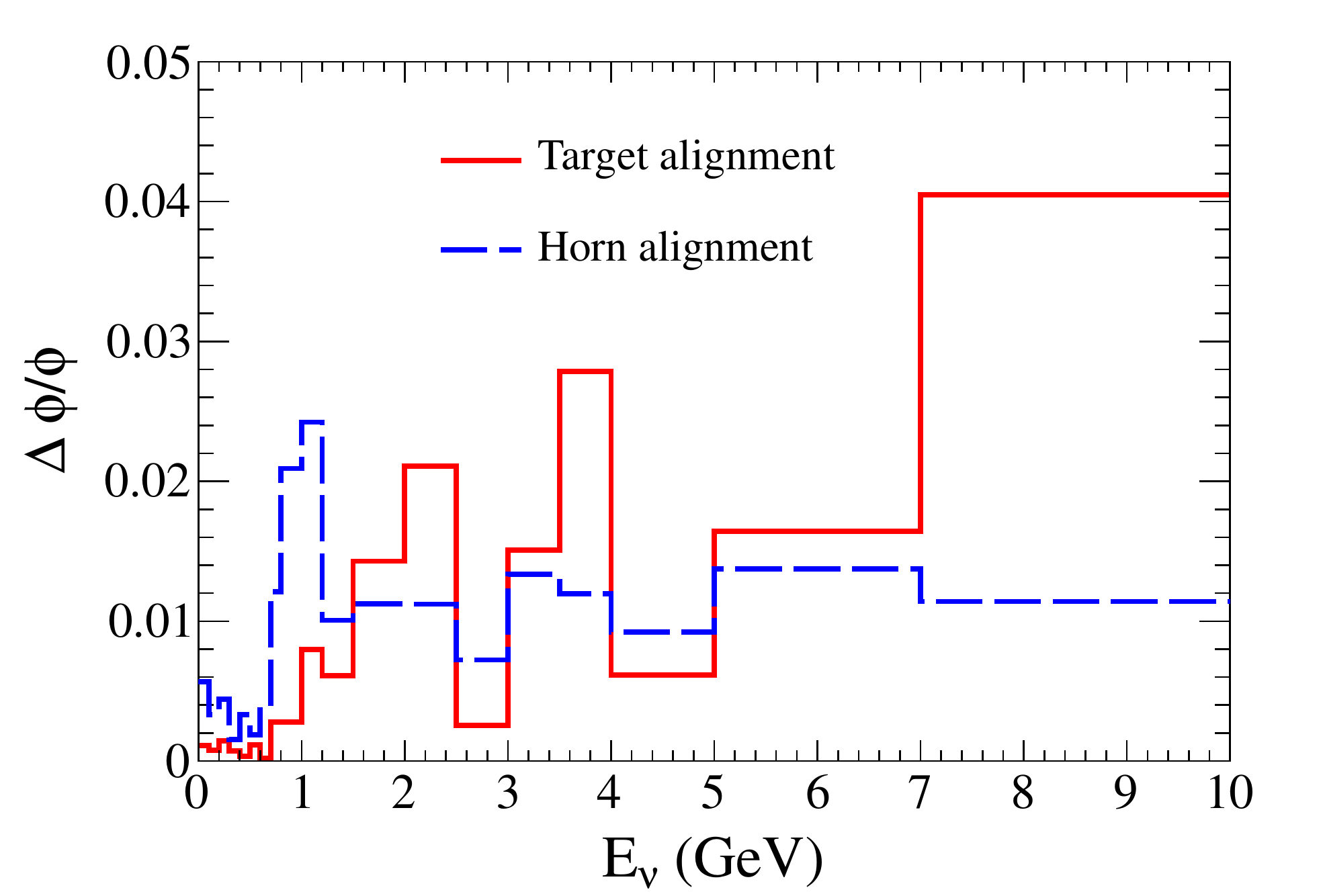}
\caption{Fractional uncertainties due to the target and horn alignment in the $\nu_\mu$ flux for ND280 (top) and SK (bottom).}
\label{fig:targ_horn_align}
\end{figure}

The effects of the systematic uncertainties in the target and horn alignments on the predicted $\nu_\mu$ fluxes at ND280 and SK are summarized in Fig.~\ref{fig:targ_horn_align}. For neutrinos with energies below 7 GeV the fractional uncertainties due to these sources are under $3\%$.

\subsection{\label{sec:hcurrerrors}Horn current and magnetic field uncertainties}

As described in Sec.~\ref{sec:tgthor}, the total uncertainty
of the horn current measurement is 1.3\% and the measured
magnetic field strength is consistent with the expected one 
within 2\%. Therefore, we adopted 2$\%$ (5 kA) as the total uncertainty 
on the absolute field strength. This results in 2\% uncertainty at most
in the neutrino flux. 

The anomalous field shown at Table~\ref{devtable} is also simulated by JNUBEAM. The effect on neutrino flux is less than 1\% for energies up to 1 GeV, 
and less than 4\% for energies greater than 1 GeV. 


\subsection{Summary of flux uncertainties}
\begin{figure*}[ht]
\centering
\includegraphics[width=0.45\textwidth]{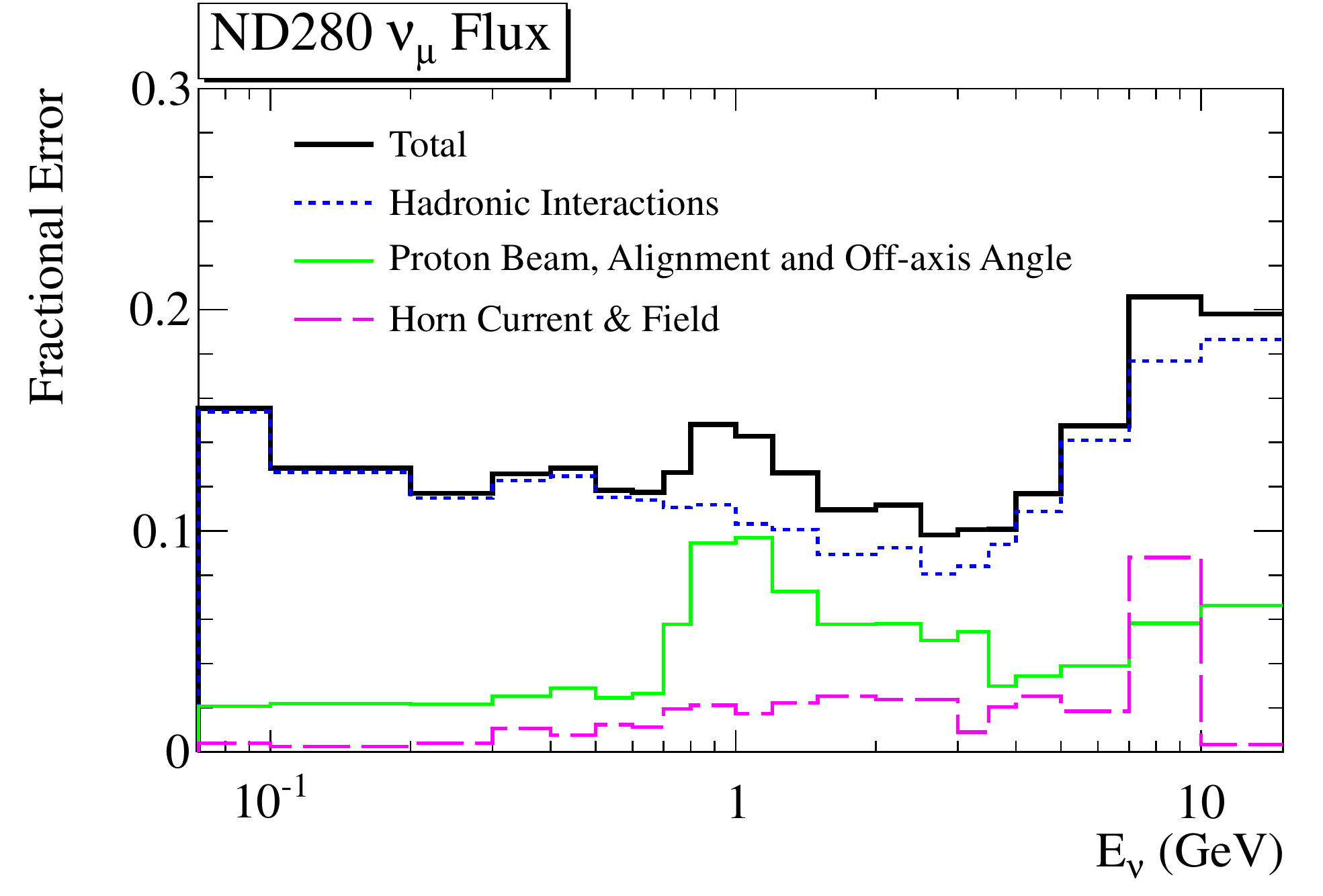}
\includegraphics[width=0.45\textwidth]{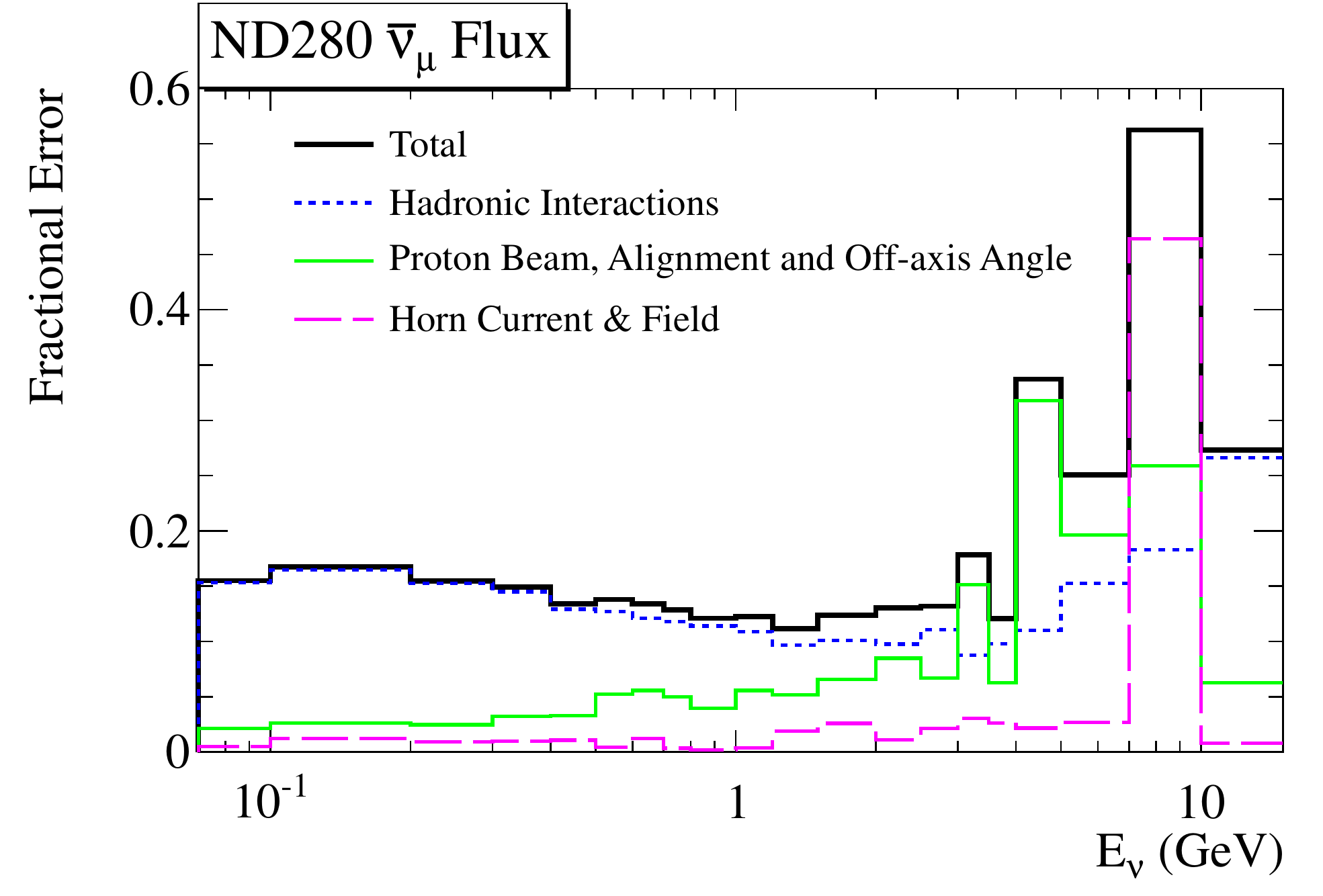}
\includegraphics[width=0.45\textwidth]{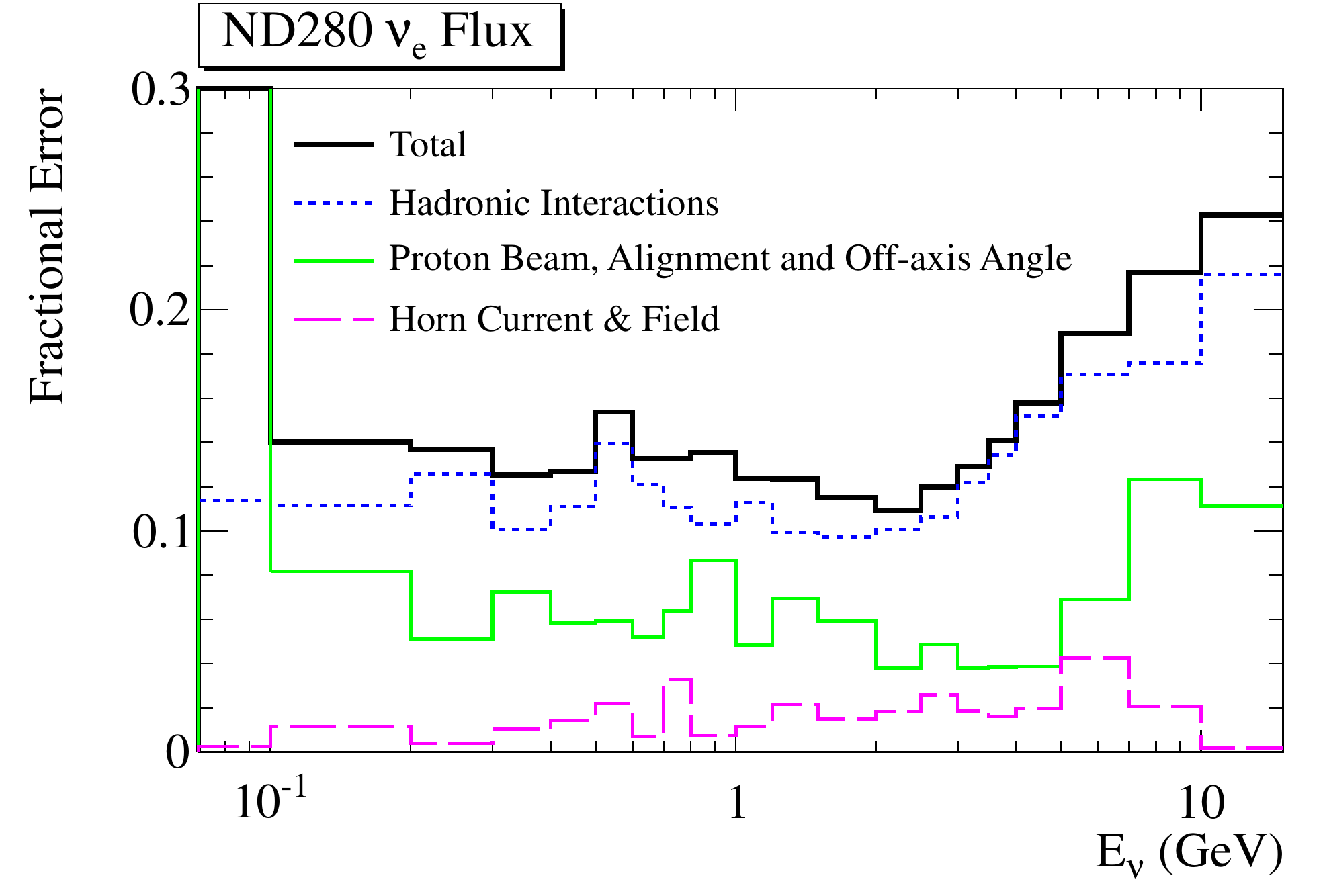}
\includegraphics[width=0.45\textwidth]{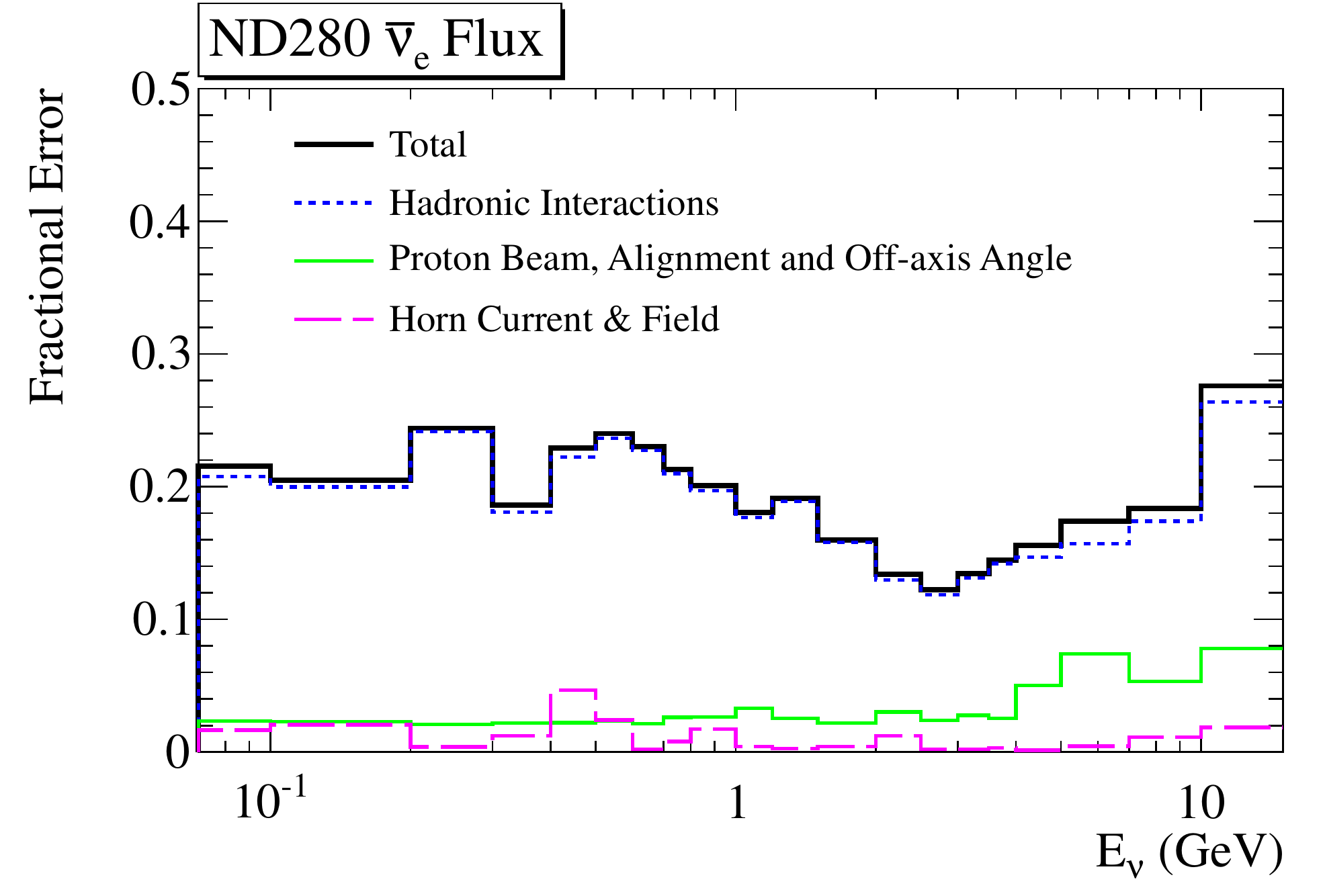}
\includegraphics[width=0.45\textwidth]{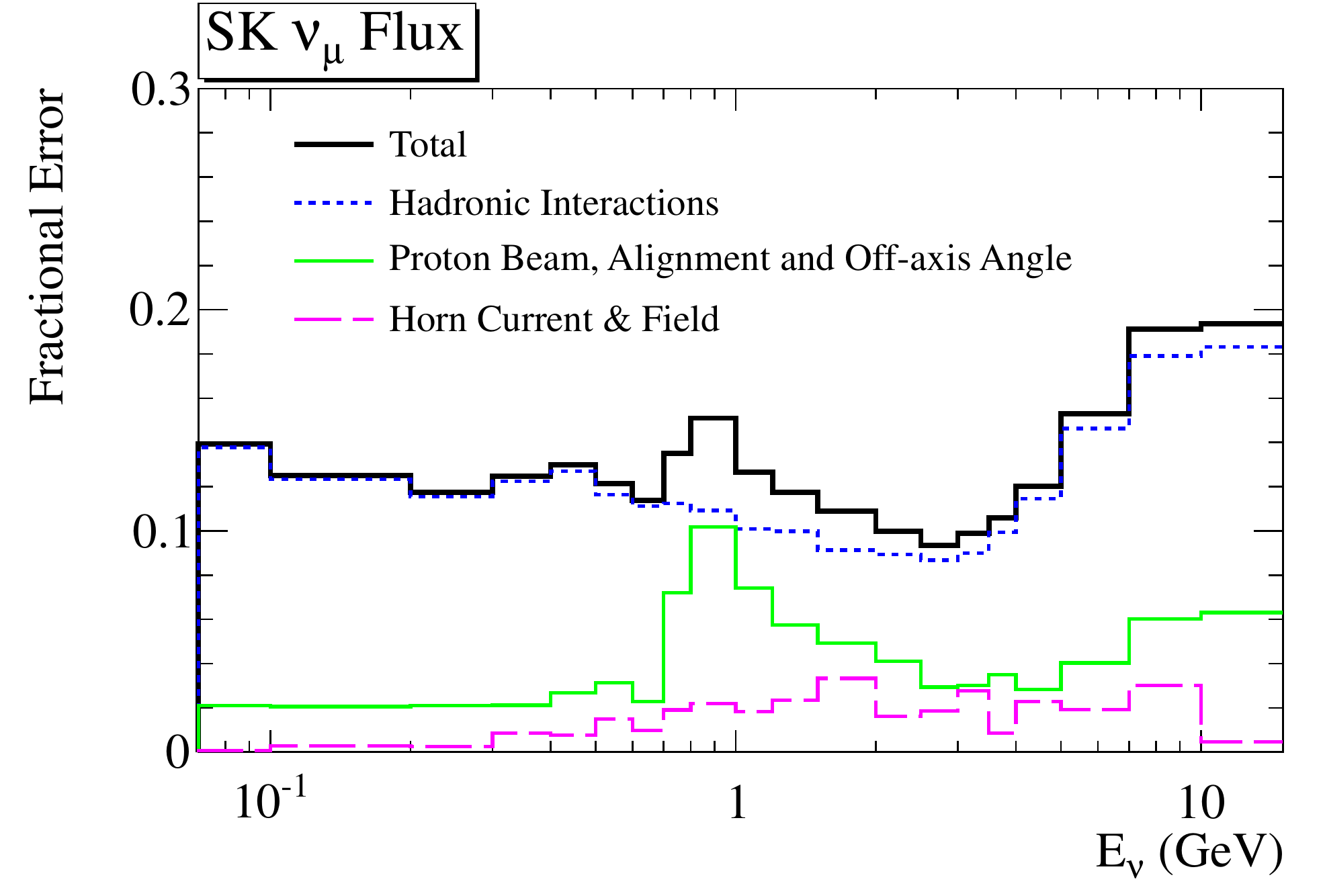}
\includegraphics[width=0.45\textwidth]{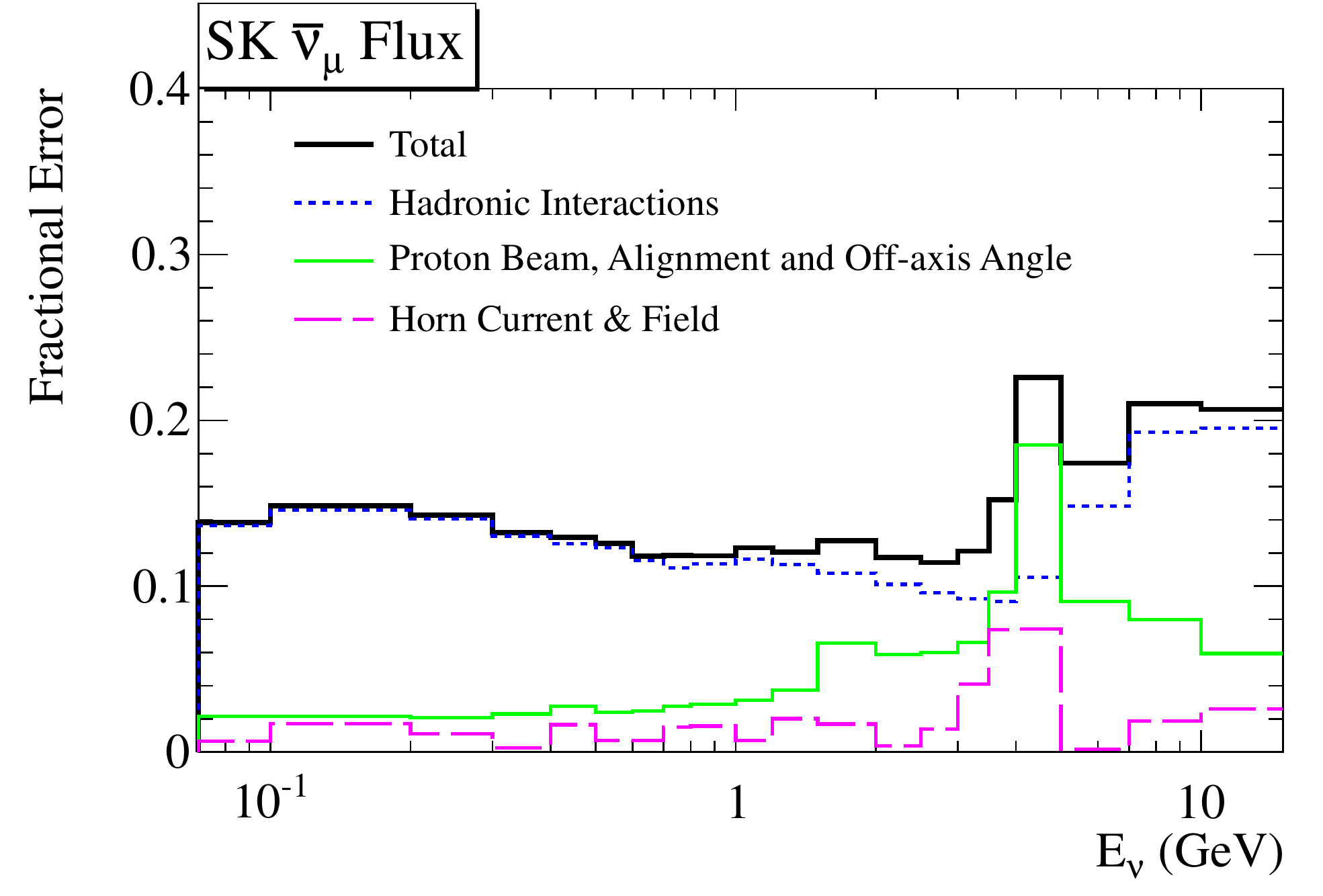}
\includegraphics[width=0.45\textwidth]{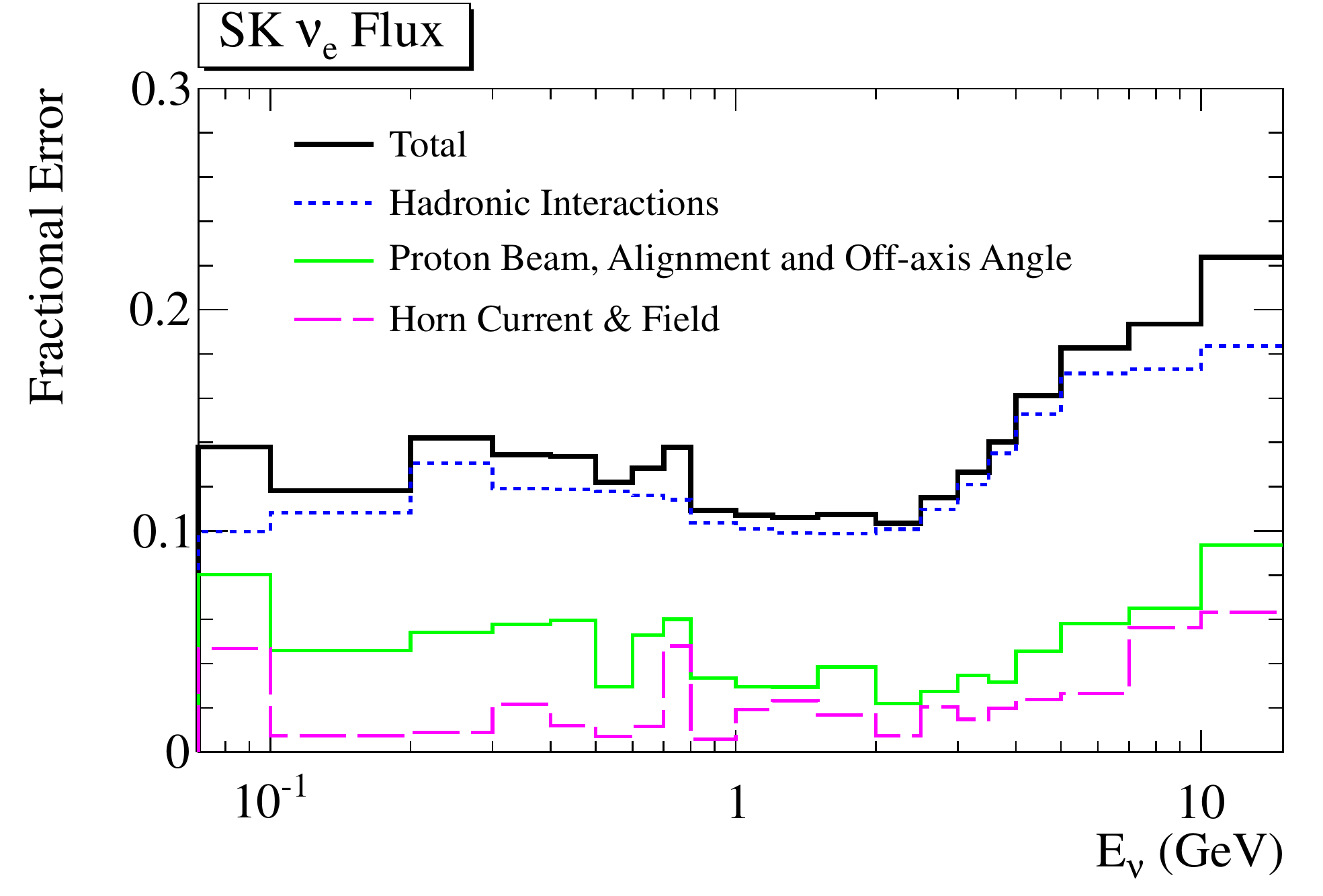}
\includegraphics[width=0.45\textwidth]{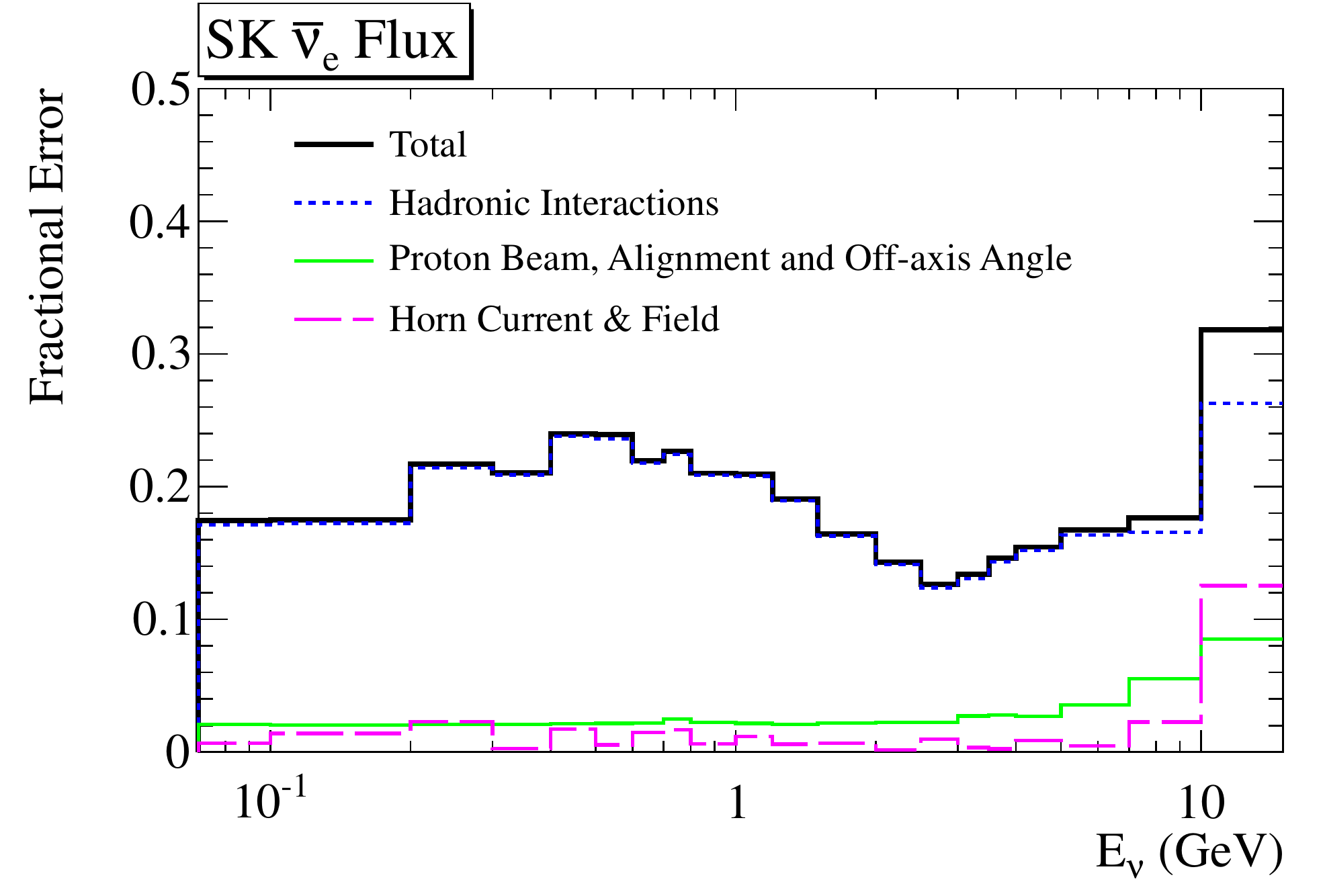}
\caption{Fractional flux error including all sources of uncertainties.}
\label{fig:flux_total_errs}
\end{figure*}

The total flux uncertainty as a function of neutrino energy, as shown in Fig.~\ref{fig:flux_total_errs},
is dominated by the hadron interaction uncertainties, with a significant contribution to the
uncertainty around the flux peak arising from the off-axis angle and proton beam uncertainties.
Shifts in the off-axis angle and proton beam tend to shift the peak position of the flux in energy.

\begin{figure*}
\centering
\includegraphics[width=0.9\textwidth]{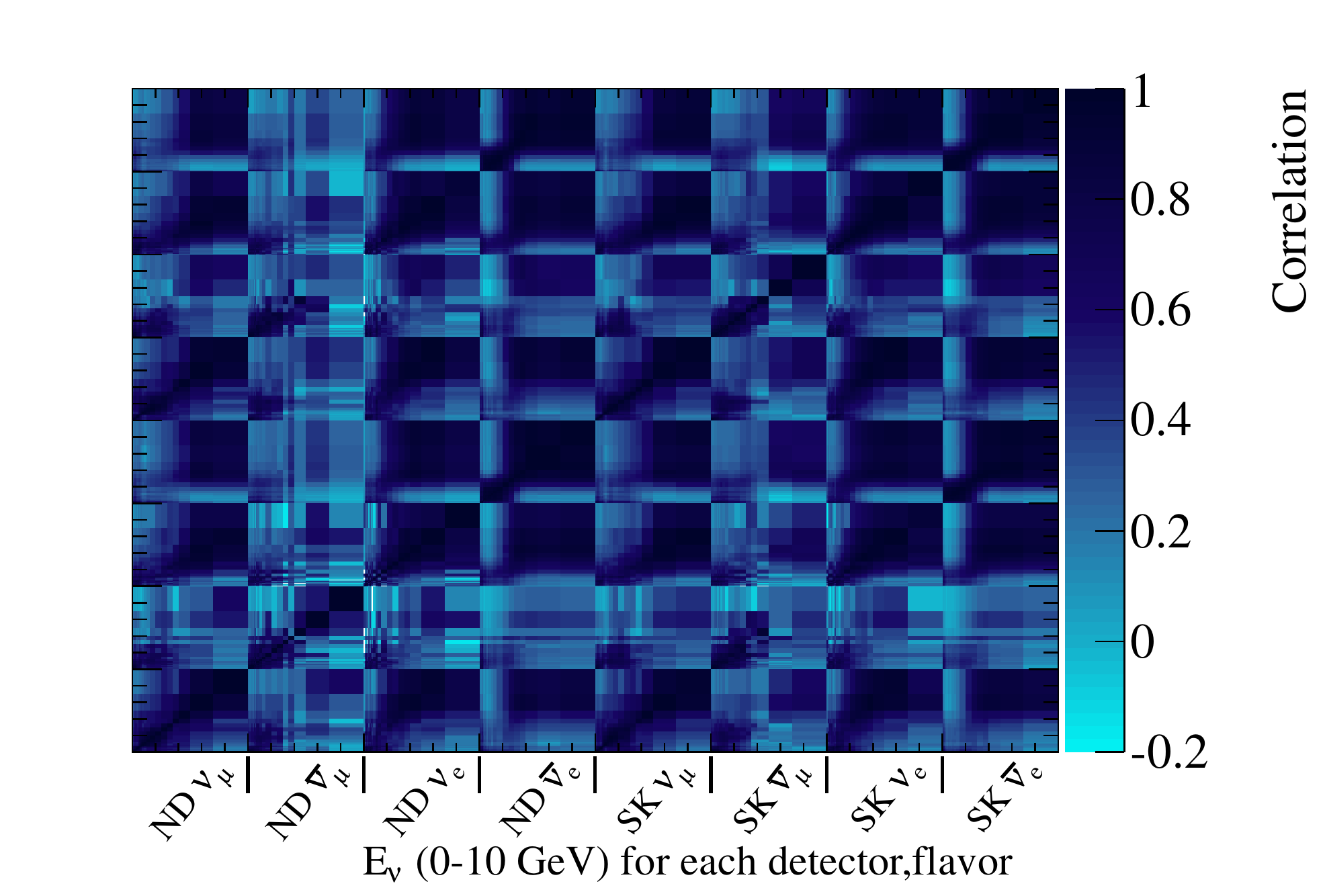}
\caption{Correlations of the flux for a given flavor, energy and detector.  The binning on the
y-axis is identical to the binning on the x-axis.}
\label{fig:flux_correlations}
\end{figure*}

The flux correlations for each neutrino flavor and energies from $1-10$ GeV at the near and far detector are
shown in Fig.~\ref{fig:flux_correlations}.  The correlations between the near and far detector are significant 
for the $\nu_{\mu}$ flux.  It is also true that the $\nu_{\mu}$ and $\nu_{e}$ fluxes have significant correlations
through the hadron interaction uncertainties, indicating that measurements of the $\nu_{\mu}$ flux can constrain
the $\nu_{e}$ contamination in the beam.

\begin{figure}
\centering
\includegraphics[width=0.45\textwidth]{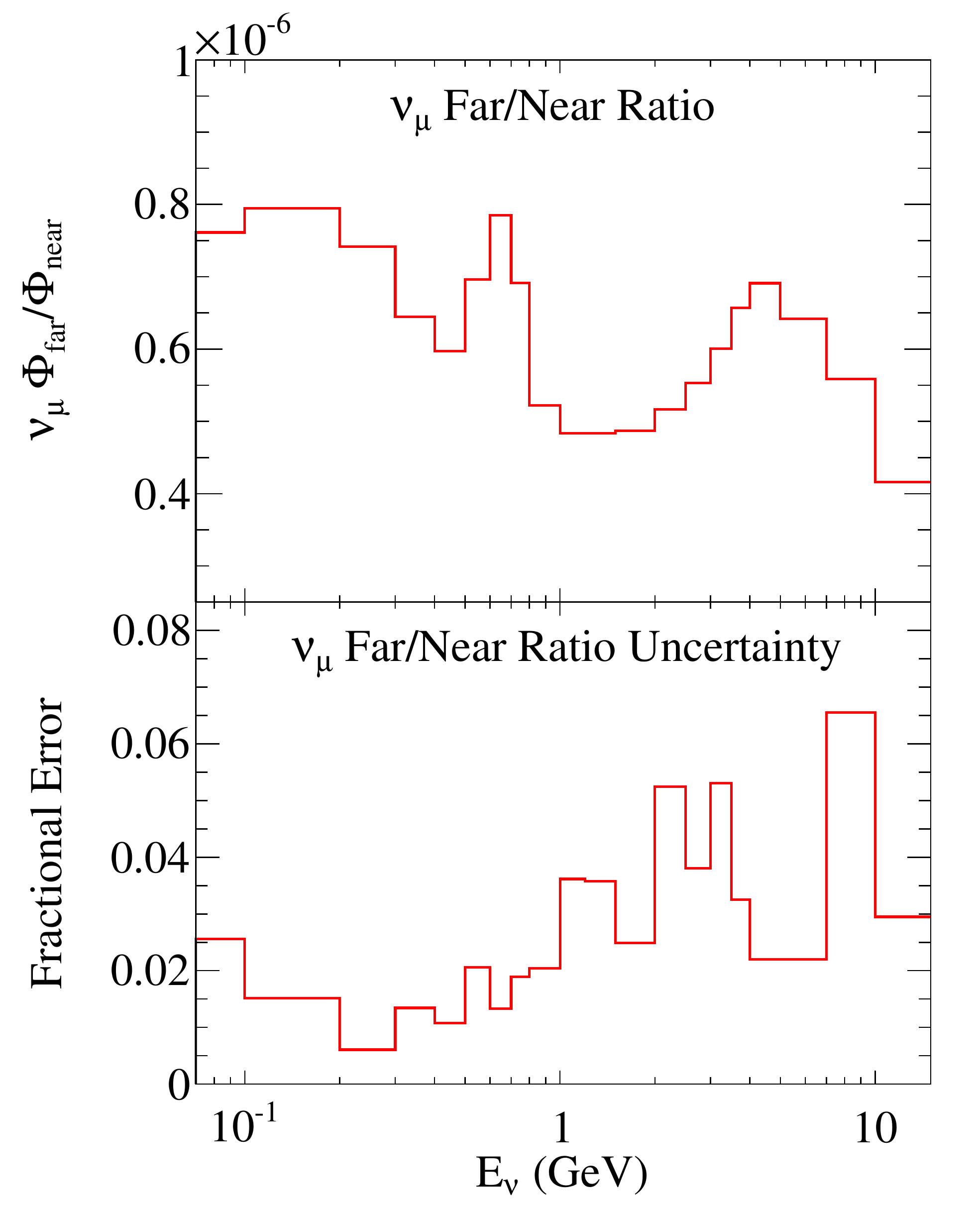}
\caption{The far/near ratio for the $\nu_{\mu}$ flux prediction (top) and the uncertainty on the ratio (bottom).}
\label{fig:fn_ratio_unc}
\end{figure}

Typically, the flux prediction is used in an analysis where it is combined with near detector data to 
predict the flux at the far detector.  The uncertainty on the ratio of the flux predictions at the far and 
near detectors is an estimate of how the uncertainty is propagated in an analysis where the flux
is measured at the near detector.  As shown in Fig.~\ref{fig:fn_ratio_unc}, the uncertainty on the far/near
ratio for the $\nu_{\mu}$ flux prediction is less than 2\% near the flux peak and less than 6\% for all
energies.  The structure in the far/near ratio itself arises from the fact that the near detector sees a line
source of neutrinos and hence a range of off-axis angles, while the far detector sees a point source and only
a single off-axis angle. 

\section{\label{sec:fluxvsdata}Flux prediction and T2K neutrino data }
The flux prediction described here, in combination with the NEUT~\cite{neut} neutrino
interaction generator, is used to predict the event rates at the near and far neutrino
detectors.  Comparisons of the predictions with the near detector data probe the accuracy of
the flux model.

\subsection{\label{sec:ingriddata}The INGRID direction and rate measurements }

As described in Sec.~\ref{sec:secmon}, INGRID measures the event rate at each neutrino detector module and reconstructs the
neutrino beam profile~\cite{Abe:2011xv}. 
The peak of the neutrino beam profile is a direct measurement of the neutrino beam direction.

\begin{figure}[htb]
\centering
\includegraphics[width=0.45\textwidth]{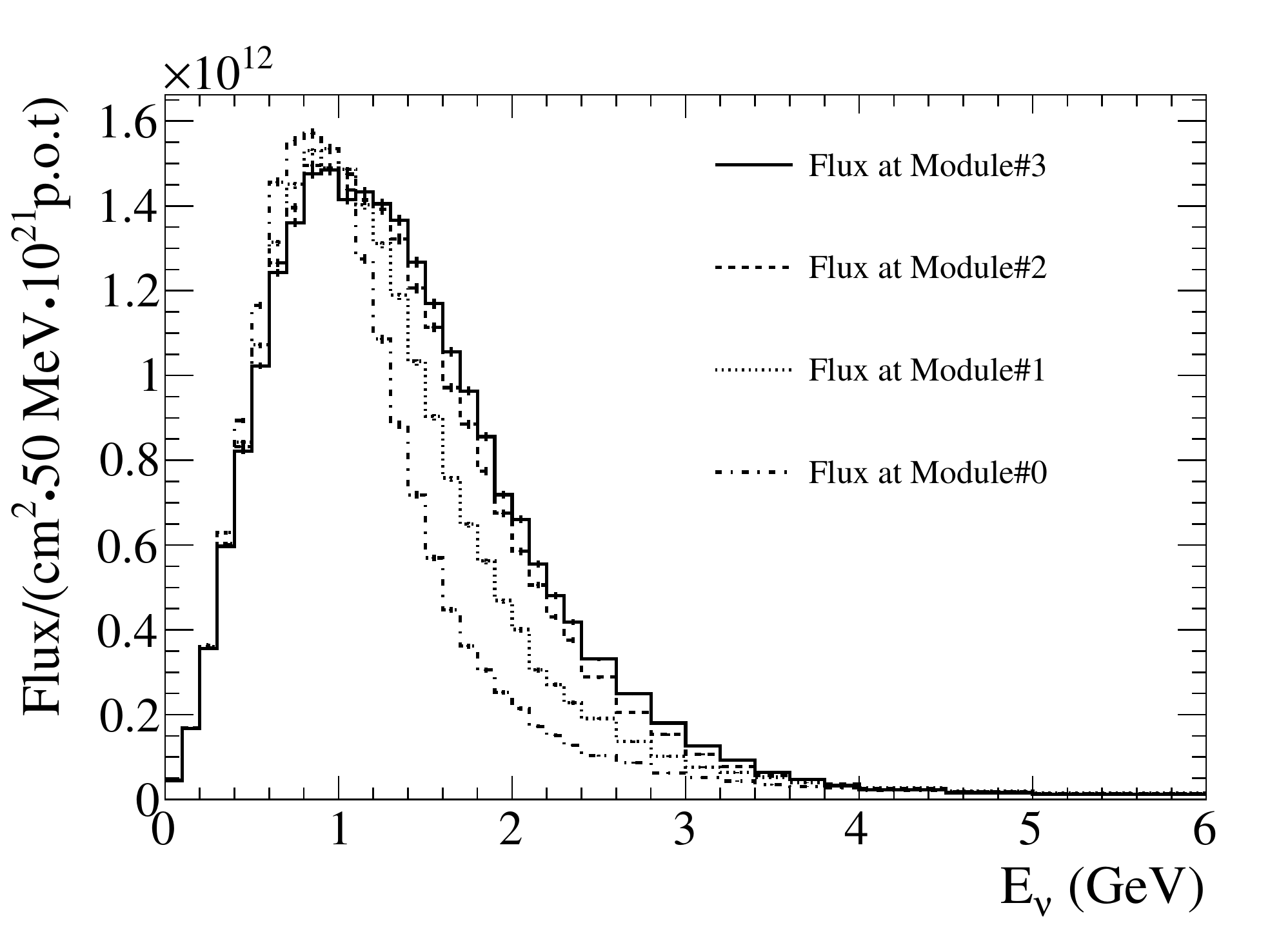}
\caption{The predicted flux at each of the horizontal INGRID detector modules from the center module (3) to the edge module (0).}
\label{fig:ingrid_flux}
\end{figure}

\begin{figure}[htb]
\centering
\includegraphics[width=0.45\textwidth]{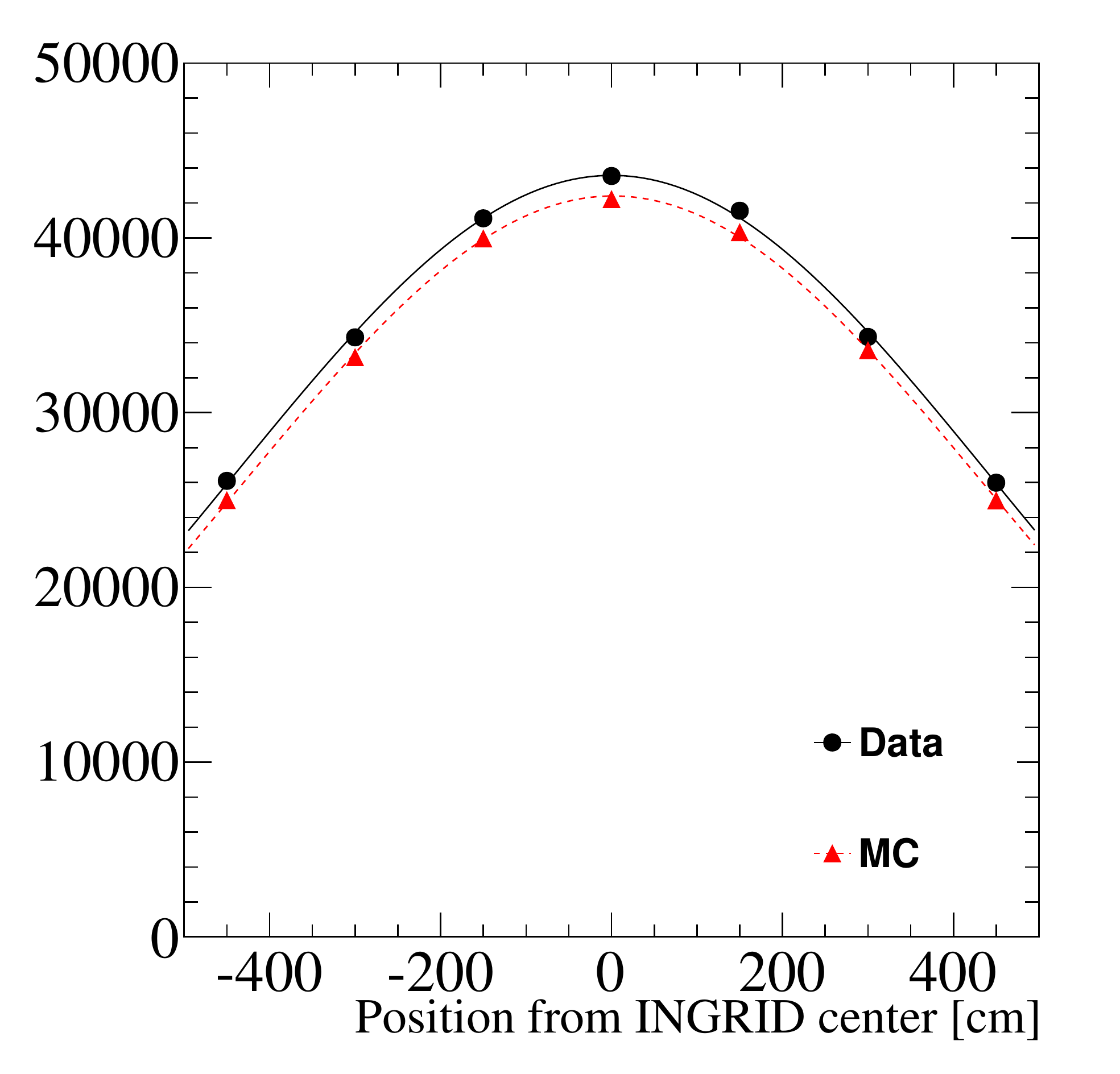}
\caption{The accumulated horizontal neutrino beam profile 
reconstructed by INGRID for the Run 1 period.  The profile of the number of events
at each detector module is fitted with a Gaussian function. Systematic errors
are not shown in this plot.}
\label{fig:ingrid_profile_run1}
\end{figure}

The flux at each module is calculated, as illustrated in Fig.~\ref{fig:ingrid_flux}, which shows how the $\nu_{\mu}$ flux prediction
varies across the horizontal modules.
The neutrino interaction rates at each detector module are predicted using the flux
prediction, the NEUT interaction generator, and a Geant4-based detector simulation.
Figure~\ref{fig:ingrid_profile_run1} shows the predicted and measured accumulated neutrino 
beam profile and
Table~\ref{tab:ingrid_data_mc} summarizes the comparison of the predicted and measured beam 
center and rate for the Run 1 data taking period.  For this period, the proton beam was
aimed slightly off center of the target in the $y$-direction. 
Therefore an offset is expected in the INGRID profile center.
The predictions agree well with the measurements of the neutrino interaction
rate and profile center.

\begin{table}[htb]
  \caption{\label{tab:ingrid_data_mc}
Summary of the predicted and measured INGRID beam center and rate for the Run 1 period.
The systematic uncertainty only includes the detector efficiency uncertainty and does not include 
flux or neutrino interaction uncertainties.} 
\begin{center}
\scalebox{0.83}[0.83]{
\begin{tabular} {lll}
\hline
\hline
 & Data & Prediction  \\
\hline
Rate [events/POT] & 1.59 $\times 10^{-14}$ & 1.53 $\times 10^{-14}$ \\
Horizontal center [mrad] &  0.009$\pm$0.052(stat.)$\pm$0.336(syst.) & 0.064  \\
Vertical center [mrad]   & -0.314$\pm$0.055(stat.)$\pm$0.373(syst.) & -0.477 \\
\hline
\hline
\end{tabular}
}
\end{center}
\end{table}

\subsection{\label{sec:nd280data}The ND280 inclusive $\nu_{\mu}$ measurement }

\begin{figure}
\centering
\includegraphics[width=0.45\textwidth]{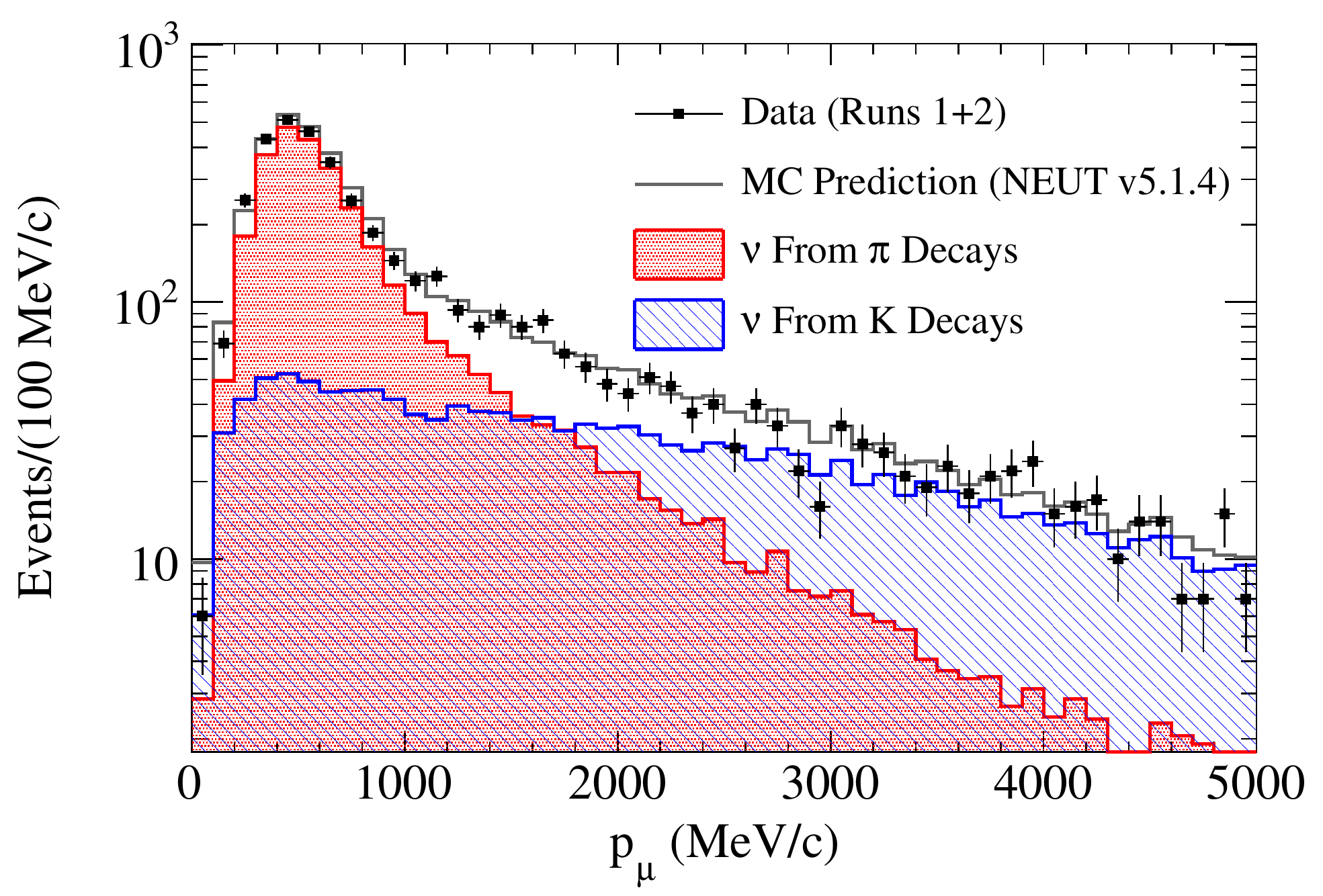}
\includegraphics[width=0.45\textwidth]{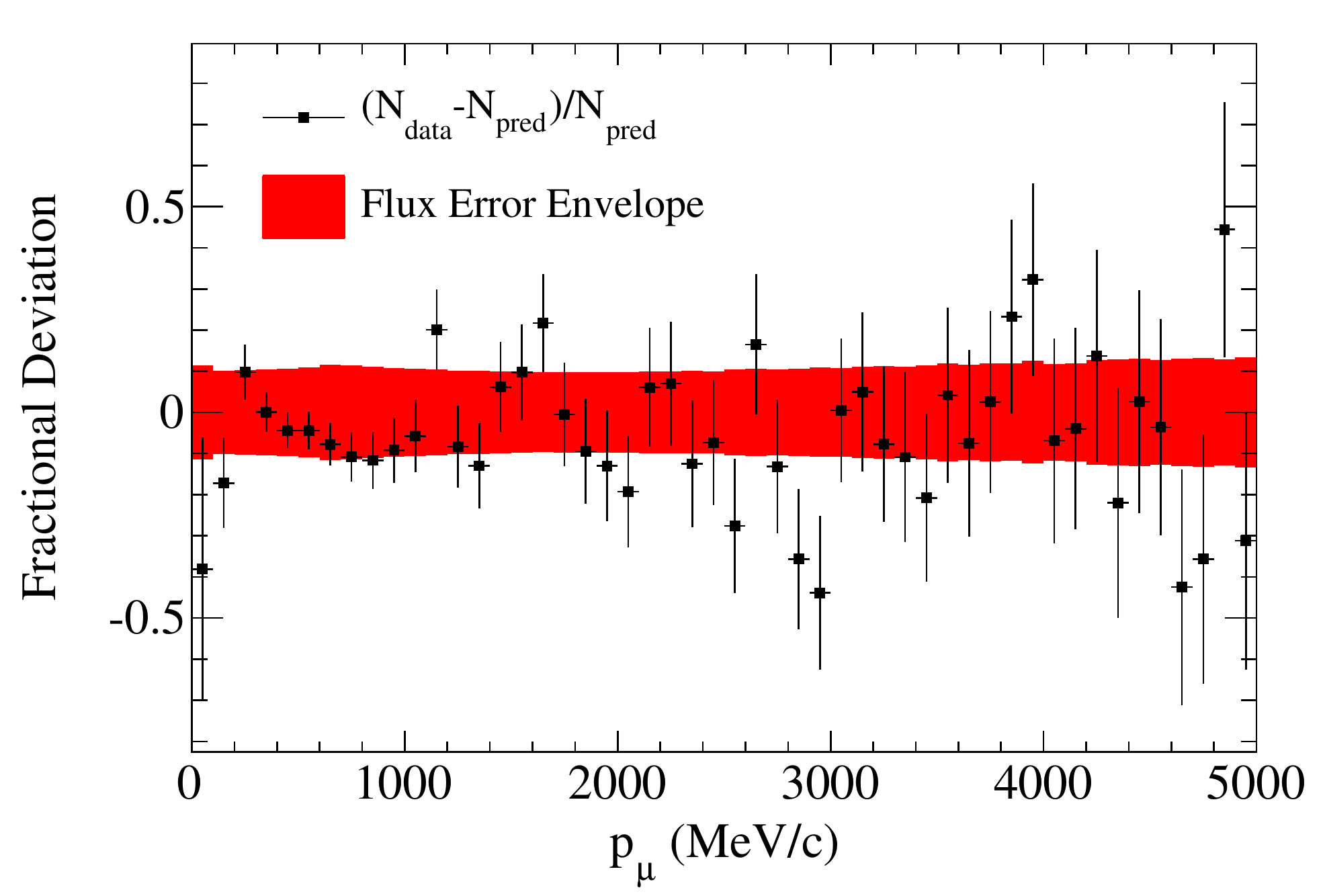}
\caption{The predicted and measured muon momentum spectrum at ND280 
for the inclusive selection (top) and the fractional flux 
uncertainty (not including neutrino interaction uncertainties nor the detector systematic error) and deviations of the data from the prediction 
on that sample (bottom).}
\label{fig:nd280_incl_numu}
\end{figure}

The rate of neutrino interactions in the off-axis ND280 near detector is predicted using the flux prediction
described here, the NEUT neutrino interaction generator (version 5.1.4), and a GEANT4 
Monte Carlo simulation of the ND280 detector.  An inclusive $\nu_{\mu}$ selection is applied to the interactions at ND280 by
searching for events with
a negatively charged track originating in the fiducial volume of the first fine grained detector that is tracked
by the immediately downstream time projection chamber and identified as muon-like by dE/dx.
The predicted muon momentum distribution for this selection is compared to the measured distribution from data collected
in Runs 1 and 2, as shown
in Fig.~\ref{fig:nd280_incl_numu}.  The interactions from neutrinos produced in pion decays tend to produce events
with lower muon momentum (since the neutrino energy is typically smaller), while neutrinos from kaon decays 
are the dominant contribution for interactions with higher muon momenta.  The predicted and measured spectra
show good agreement within the uncertainty of the flux prediction, which is $\sim10\%$ for all muon momenta.
The ratio of the total number of measured events relative to the prediction is:
\begin{equation}
R_{data/MC} = 0.956\pm0.014(stat.)\pm0.098(flux)
\end{equation}
\textcolor{red}{Even though there are additional 
neutrino interaction model and detector systematic
error uncertainties, which are not quoted here, the data and our prediction 
show good agreement.}

\section{\label{sec:conclusion}Conclusion}

In this paper, we have described the neutrino flux prediction 
in the T2K experiment. 
The predicted neutrino flux and energy spectrum are based on 
hadron production data, including NA61/SHINE measurements, 
the proton beam profile measurements in T2K, and measurements of the horn 
magnetic fields. 
The systematic uncertainties on the neutrino flux are based on 
uncertainties from these experimental measurements that are inputs to 
the flux prediction.
Taking into account possible correlations between the systematic uncertainties for 
different angular and momentum bins in the hadron production data, 
we estimate the uncertainties on the neutrino flux including 
correlations between neutrinos of different energy and at different detectors.  
The total systematic uncertainty at the peak energy is approximately 15~\% 
for both the near and far detector where the dominant source is the hadron interaction uncertainties. 
The uncertainty on the ratio of the flux predictions at the far and near detectors for $\nu_{\mu}$ 
flux is less than 2~\% near the flux peak and less than 6~\% for all energies. 

The predicted flux with simulated neutrino interactions 
is compared with the measurements at the near detectors. 
The measurements of the beam direction and event rate are consistent with 
the prediction.

\begin{acknowledgments}
We thank the J-PARC accelerator team for the superb accelerator performance 
and the J-PARC center for the continuous support of the T2K experiment.
We are grateful to the NA61/SHINE collaboration 
and FLUKA team for their assistance to make our data-driven flux prediction.
We acknowledge the support of MEXT, Japan; 
NSERC, NRC and CFI, Canada;
CEA and CNRS/IN2P3, France;
DFG, Germany; 
INFN, Italy;
National Science Centre, Poland;
RAS, RFBR and the Ministry of Education and Science
of the Russian Federation; 
MEST and NRF, South Korea;
MICINN and CPAN, Spain;
SNSF and SER, Switzerland;
STFC, U.K.; NSF and 
DOE, U.S.A.
We also thank CERN for their donation of the UA1/NOMAD magnet 
and DESY for the HERA-B magnet mover system.
In addition, participation of individual researchers
and institutions in T2K has been further supported by funds from: ERC (FP7), EU; JSPS, Japan; Royal Society, UK; 
DOE Early Career program, and the A. P. Sloan Foundation, U.S.A.

\end{acknowledgments}

\bibliographystyle{apsrev4-1}
\bibliography{FluxPaper}

\end{document}